\documentclass[a4paper,fleqn,usenatbib]{mnras}

\usepackage{mathptmx}
\usepackage{txfonts}
\usepackage{xcolor}

\usepackage[T1]{fontenc}
\usepackage{ae,aecompl}

\usepackage{graphicx}	% Including figure files
\usepackage{lscape}
\usepackage{savesym}
\savesymbol{iint}
\usepackage{wasysym}
\restoresymbol{WAS}{iint}

\title[A Survey for High-Mass Eclipsing Binaries]{A Survey for High-Mass Eclipsing Binaries}

\author[F. Pozo Nu\~nez et al.]{
F. Pozo Nu\~nez$^{1,2}$\thanks{E-mail: francisco.pozon@gmail.com},
R. Chini$^{1,3}$,
A. Barr Dom\'{i}nguez$^{4}$,
Ch. Fein$^{1}$, M. Hackstein$^{1}$,
\newauthor~Pietrzy\'{n}ski, Grzegorz$^{2}$, and M. Murphy$^{5}$
\\
\\
$^{1}$Astronomisches Institut, Ruhr--Universit\"at Bochum,
Universit\"atsstra{\ss}e 150, 44801 Bochum, Germany.\\
$^{2}$Centrum Astronomiczne im. Mikolaja Kopernika, PAN, Bartycka 18, 00-716 Warsaw, Poland.\\
$^{3}$Instituto de Astronom\'{i}a, Universidad Cat\'{o}lica del Norte, Avenida Angamos 0610, Casilla 1280,Antofagasta, Chile.\\
$^{4}$Centro de Investigaci\'{o}n Multidisciplinario de la Araucan\'{i}a, Facultad de Ingenier\'{i}a, Universidad Autonoma de Chile, Avenida \\ Alemania 01090, Temuco, Chile.\\
$^{5}$Departamento de F\'{i}sica, Universidad Cat\'{o}lica del Norte, Avenida Angamos 0610, Casilla 1280 Antofagasta, Chile
}

\date{Accepted 2019 October 16. Received 2019 October 13; in original form 2019 March 01}

\pubyear{2019}

\begin{document}
\label{firstpage}
\pagerange{\pageref{firstpage}--\pageref{lastpage}}
\maketitle

\begin{abstract}
We report results from a search for Galactic high-mass eclipsing binaries. The photometric monitoring campaign was performed in Sloan $r$ and $i$ with the robotic twin refractor RoBoTT at the Universit\"atssternwarte Bochum in Chile and complemented by Johnson $UBV$ data. Comparison with the SIMBAD database reveals 260 variable high-mass stars. Based on well-sampled light curves we discovered 35 new eclipsing high-mass systems and confirm the properties of six previously known systems. For all objects, we provide the first light curves and determine orbital periods through the Lafler-Kinman algorithm. Apart from GSC\,08173-0018 and Pismis\,24-13 ($P = 19.47\,d$ and $20.14\,d$) and the exceptional short-period system TYC\,6561-1765-1 ($P = 0.71\,d$), all systems have orbital periods between 1 and 9 days. We model the light curves of 26 systems within the framework of the Roche geometry and calculate fundamental parameters for each system component. The Roche lobe analysis indicates that 14 systems have a detached geometry while 12 systems have a semi-detached geometry; seven of them are near-contact systems. The deduced mass ratios $q = M_2/M_1$ reach from 0.4 to 1.0 with an average value of 0.8. The similarity of masses suggests that these high-mass binaries were created during the star formation process rather than by tidal capture.
\end{abstract}

\begin{keywords}
stars: fundamental parameters -- stars: formation
-- binaries: eclipsing -- binaries: spectroscopic -- Galaxy: open clusters and associations
\end{keywords}

%%%%%%%%%%%%%%%%%%%%%%%%%%%%%%%%%%%%%%%%%%%%%%%%%%

%%%%%%%%%%%%%%%%% BODY OF PAPER %%%%%%%%%%%%%%%%%%

\section{Introduction}

There is growing evidence that a large fraction of the total stellar population consists of binary and multiple stars (\citealt{2013ARA&A..51..269D}). The role of such systems for astrophysics is tremendous because they provide the most accurate data about the basic stellar properties like mass, radius, luminosity, projected rotational velocity or abundances. At the same time, binary and multiple systems are crucial for studying gas dynamics, accretion processes, and interaction of stellar winds. Most notably, the formation of binaries and multiple systems remain major unsolved problems (e.g. \citealt{2003IAUS..212...80Z}; \citealt{2008MNRAS.389..925T}; \citealt{2014AAS...22321401B}). Interestingly, high-mass stars seem to have the highest multiplicity rate among all stars which increases the chance to be observed as eclipsing binary (EB) systems. As a consequence early-type EBs are not only a unique laboratory to study the formation and evolution of stars but -- due to their brightness -- they have even the potential for measuring accurate distances to nearby galaxies (e.g. \citealt{2013IAUS..289..173B}).

During recent years many stars which were treated as singles for many decades turned out to be double or multiple systems when taking a closer look. This is true for the immediate solar neighbourhood ($d < 25$\,pc) where well known F-type stars are found to form binary or hierarchical systems \citep[e.g.][]{2003A&A...397..159H,2012ApJS..203...30F,2013CEAB...37..295C,2017ApJ...836..139F}, rising the multiplicity fraction for solar type stars to about 58\%. For A-type stars, \citet{2014MNRAS.437.1216D} derive a companion star fraction of $68.9 \pm 7$\,\%.

Recent surveys for O-type stars have revealed multiplicity fractions of 50\% to more than 80\%.
\citep[e.g.][]{2012MNRAS.424.1925C,2012Sci...337..444S,2013A&A...550A.107S,2014ApJS..211...10S,2016ApJS..224....4M} and led to the assumption that only a small fraction of stars above $20\,M_\odot$ are born as single stars. Multiplicity surveys for B-type stars are less frequent \citep[e.g.][]{2012ASPC..462...87S,2012MNRAS.424.1925C,2015A&A...580A..93D} but it seems that their properties such as period and mass-ratio follow those of O-type stars. Obviously, although the fraction of high-mass stars in a galaxy is a minor portion of the total stellar content their multiplicity fraction appears to be the highest among all stellar masses. This peculiarity is not yet understood and requires further investigation.

Eventually, orbital period and mass-ratio provide significant constraints on models of star formation and evolution. Thus, apart from knowing the masses of the components in a multiple system, orbital parameters are important to understand the formation of high-mass stars. The multiplicity surveys for high-mass stars have revealed mass ratios $q = M_2/M_1$ close to unity and typical periods of only a few days suggesting that the evolution of members of multiple systems must differ significantly from that of a single high-mass star. This holds for the formation scenarios (fragmentation, disk-instabilities), the evolution (mass-transfer, merging) and the corresponding final states (blue stragglers, X-ray binaries, gamma-ray bursts). \citet{2013A&A...550A.107S} estimated that more than 50\% of the current O-star population will exchange mass with its companion within a binary system. Assuming a constant star formation rate, \citet{2014ApJ...782....7D} found that 8\% of a sample of early-type stars are the product of a merger event in a close binary system. In total they estimate that 30\% of massive main-sequence stars are the product of binary interaction.

Concerning the formation of multiple high-mass stars the sequence of events, from the fragmentation of cloud cores to the dynamical interaction and orbit migration of stellar components, seems well established. Still, theoretical predictions for the formation of multiple systems and the evolution of their orbits into their present configuration of a tight inner binary and possibly a much more distant third component vary widely, depending on the initial conditions, physical processes involved or modelled and computational details. The most important aspects are briefly reviewed in the following:

\begin{itemize}

\item \emph{Cascade fragmentation}. Fragmentation is believed to be the dominant mechanism forming binary and multiple stars. \cite{2007ApJ...656..959K} present radiation-hydrodynamic simulations of collapse and fragmentation to investigate whether massive protostellar cores form a small or large number of protostars. Interestingly, it is found that radiation feedback from the first few accreting protostars inhibits the formation of further fragments within the cloud core. In other words, the majority of the collapsing mass accretes onto one or a few objects. In a second step, further fragmentation may occur in massive, self-shielding disks which are driven to gravitational instability by rapid accretion. Whether the accreting gas is transported to the central star or increases the mass of the fragment inside the disk depends on the model calculations. In any case these results demonstrate that massive cores with observed properties are not likely to fragment into many stars.

\item \emph{Accretion}. Binary protostars produced by fragmentation during the isothermal cloud collapse have separations between $10^2$ and $10^4$ AU (e.g. \citealt{2003A&A...411...91S}). The final properties of a multiple system will be determined by the ongoing accretion of gas. Numerous hydrodynamical simulations show that accretion leads to similar-mass components (\citealt{2003A&A...397..159H}; \citealt{2007A&A...463..683S}).

\item \emph{N-body dynamics}. Calculations (e.g. \citealt{2005A&A...439..565G} and references therein) show that it is not possible to reproduce the observed multiplicity fraction through the dynamical evolution of star clusters that are born with a single-star population. While dynamical interactions may disrupt many wide binaries they are neither able to pair stars efficiently nor to change the properties of close binaries significantly. Taking into account also the observed separations of multiple systems the authors argue that of every 100 star-forming cores in a molecular cloud about 60 will produce binaries while 40 will produce triple systems. Among the latter group 25 are long-lived hierarchical systems while 15 will decay into 15 binaries and 15 single stars - predominantly of low mass which will be ejected. Thus both observations and theory lead to the conclusion that star-forming cores must typically produce only 2 or 3 physically bound stars. This result is in contrast to many numerical simulations that predict the formation of $5 - 10$ stars per core.

    \cite{2004MNRAS.351..617D} performed particle hydrodynamics and $N-$body simulations of fragmenting cores. They find that 80\% of the forming stars are members of multiples systems, with component separations between 1 and 1000\,AU; this fraction is an increasing function of primary mass. The multiple systems consist of binaries and triples, where the mass ratio within binaries attain typical values of $0.5 - 1$. With increasing time dynamics disrupt mostly wide and/or low-mass companions with low binding energies, but has little effect on the inner subsystems with high-mass components.

\item \emph{Dynamical interactions and orbital decay}. The formation of close binary systems by dynamical interactions and orbital decay was studied by \cite{2002MNRAS.336..705B}. Their hydrodynamical star formation calculation shows that close binaries with separations $\le 10$\,AU need not be formed directly by fragmentation but could be the result of a combination of dynamical interactions in unstable multiple systems and the orbital decay of initially wider binaries. Orbital decay may be due to gas accretion and/or the interaction of a binary with its circum-binary disk. The interesting aspect of this scenario is the fact that close binaries with roughly equal-mass components are formed. This is because dynamical exchange interactions and the accretion of gas with high specific angular momentum drive mass ratios towards unity. Another consequence of dynamical interactions is the tendency that stars of higher mass should have a higher frequency of close companions.

    \cite{2008MNRAS.389..925T} compared the statistics of triple and quadruple stars. According to the observations the properties of multiple stars are not compatible with the dynamical decay of small clusters which would imply that $N-$body dynamics is not the dominant process of their formation. In contrast, (cascade) fragmentation possibly followed by migration of inner and/or outer orbits could explain the observations of triple and quadruple stars.

\item \emph{Kozai cycles with tidal friction}.
    Many binary stars have separations of only a few stellar radii. This implies that their orbits must have shrunk by $1 - 2$ orders of magnitude after their formation because the radius of a star also shrank considerably from its birth to the main sequence. Orbital shrinkage was investigated by \cite{2007ApJ...669.1298F} who considered the effects of secular perturbations from a distant companion star (so-called Kozai cycles) and tidal friction. \cite{1962AJ.....67..591K} discovered that the amplitude of both the eccentricity and the inclination of the inner system oscillates -- independent of the strength of the perturbation from the outer body. The oscillation amplitude only depends on the initial mutual inclination between inner and outer binaries; for the extreme case of an initial inclination of $90^\circ$ the maximum eccentricity becomes 1, i.e. the two inner bodies will collide. Separations less than a few stellar radii -- as they may occur during some phase of a Kozai cycle -- imply a tidal friction that drains energy from the orbit, reducing the semi-major axis even more. \cite
    {2007ApJ...669.1298F} find that binaries with orbital periods of $0.1 - 10$ days are produced from binaries with much longer initial periods (10 to $\sim 10^5$ days). This result is consistent with the observation that short-period binaries are often accompanied by a third star.

\end{itemize}

The huge progress provided by spectroscopic and interferometric surveys can be complemented by photometric searches for high-mass EBs because they can provide system properties that are not accessible through spectroscopy alone due to the unknown inclination of the systems. Therefore, finding high-mass EBs is a major step ahead in our understanding of the properties of these exotic systems. Interestingly, to our knowledge, the literature describes only 18 high-mass EBs (\citealt{2001MNRAS.326.1149R}, \citealt{2001A&A...369..561F}, \citealt{2003A&A...405.1063S}, \citealt{2006MNRAS.371...67S}, \citealt{2006ApJ...639.1069H}, \citealt{2006MNRAS.367.1450N}, \citealt{2013A&A...557A..13B} (hereafter called Paper\,I), \citealt{2014ApJS..213...34K}, \citealt{2015ApJ...811...85K}, \citealt{2017A&A...600A..33M}). The three high-mass EBs systems from Paper\,I -- HD\,319702 (O+B), CPD-51\,8946 (O+B), and Pismis\,24-1 (O+O) -- were discovered in the framework of our long-term photometric Bochum Galactic Disk Survey (GDS; \citealt{2015AN....336..590H}, \citealt{2015AN....336..677K}).

The present paper reports the complete high-mass star results from the GDS which comprise 473 stars crudely classified as OB in the literature. Among those are 263 variable sources that we analyze for the presence of EBs on the basis of their light curves. For the detected EBs we complement the GDS standard photometry in Sloan $r$ and $i$ by new $UBV$ data and present photometric spectral types, light curves and periods for all systems.

\begin{table*}
\begin{center}
\caption{Characteristics of the observed high-mass eclipsing binaries.}
\label{tab:objectlist}
\begin{tabular}{clllcl}
\hline\hline
\noalign{\smallskip}
             \multicolumn{1}{c}{No.}
           & \multicolumn{1}{c}{Name}
           & \multicolumn{1}{c}{RA}
           & \multicolumn{1}{c}{Dec}
           & \multicolumn{1}{c}{spectral designations}
           & \multicolumn{1}{c}{references}\\
             \multicolumn{1}{c}{}
           & \multicolumn{1}{c}{}
           & \multicolumn{2}{c}{J2000}
           & \multicolumn{1}{c}{}
           & \multicolumn{1}{c}{}\\
             \hline
             \noalign{\smallskip}
1  & CPD\,$-$\,24\degr\,2836       & 07 45 33.18  & -24 42 03.29  & O9-B0                 & $a$       \\
2  & CPD\,$-$\,26\degr\,2656       & 07 50 18.04  & -27 15 03.40  & B1:V:, OB:, B0V, B5V  & $a,b,c,d$ \\
3  & TYC\,6561-1765-1              & 07 52 20.09  & -27 04 56.23  & O9.5                  & $c$       \\
4  & CD\,$-$\,31\degr\,5524        & 08 05 55.87  & -31 48 30.62  & OB, O9/9.5        	  & $o,p$     \\
5  & CPD\,$-$\,42\degr\,2880       & 08 44 26.45  & -43 10 02.48  & O9.5-B2               & $a$       \\
6 & CPD\,$-$\,45\degr\,3253       & 08 55 38.09  & -46 21 38.47  & B0:, OB$^0$?, O9:     & $a,e,f$   \\
7 & GSC\,08156-03066              & 08 56 25.90  & -48 24 35.47  & OB$^0$                & $e$       \\
8 & GSC\,08173-00182              & 09 09 53.72  & -48 51 56.20  & OB$^+$                & $e$       \\
9 & TYC\,8175-685-1               & 09 22 39.17  & -49 02 10.04  & OB$^+$                & $e$       \\
10 & ALS\,18551                    & 10 58 17.00  & -61 11 54.00  & O4.5                  & $h$       \\
11 & CPD\,$-$\,39\degr\,7292       & 17 13 24.46  & -39 47 42.68  & OB$^+$, B1V, B1IIIn?  & $i,j,k$   \\
12 & Pismis\,24-4                  & 17 24 40.49  & -34 12 06.51  & O9-B0?                & $g$       \\
13 & CD\,$-$\,29\degr\,14032       & 17 49 09.40  & -29 14 33.20  & B1III                 & $k$       \\
14  & SS\,117				       & 07 27 31.64  & -13 38 41.30  & OB$^-$				  & $l$       \\
15  & CD\,$-$\,33\degr\,4174        & 07 48 38.55  & -33 51 00.60  & OB, OB$^+$            & $m,l$     \\
16  & CD\,$-$\,28\degr\,5257	       & 08 00 03.26  & -28 50 25.91  & O8, O9.5V		      & $d,n$     \\
17 & LS\,1221				       & 08 59 50.56  & -47 24 31.39  & OB$^-$, OB$^0$        & $o,e$     \\
18 & CD\,$-$\,51\degr\,10200       & 16 28 09.45  & -51 31 08.07  & OB                    & $o$		  \\
19 & GSC\,07380-00198		       & 17 36 15.59  & -33 31 28.70  & OB$^-$                & $q$       \\
20 & ALS\,17569				       & 08 59 19.42  & -48 19 51.50  & OB?                   & $e$       \\
21 & HD\,300214              	   & 09 45 25.92  & -54 22 09.47  & O, B7                 & $o,r$     \\
22 & TYC\,8958-4232-1              & 10 58 45.47  & -61 10 43.00  & O5Ifp                 & $s$       \\
23 & HD\,308974				       & 11 41 31.39  & -61 50 03.59  & OB                    & $t$       \\
24 & CPD\,$-$\,54\degr\,7198       & 16 12 27.00  & -54 29 43.80  & OB                    & $u$       \\
25 & HD\,152200                    & 16 53 51.63  & -41 50 32.55  & O9.7IV(n)             & $s$       \\
26 & TYC\,6265-2079-1		       & 18 22 03.13  & -15 46 21.98  & OB, B3/B4	          & $v,w$     \\
27 & BD\,$-$\,024786               & 18 56 03.64  & -02 37 35.36  & OBe                   & $x$       \\
28 & ALS\,15204                    & 10 43 41.24  & -59 35 48.18  & O7.5V                 & $h$       \\
29  & V\,467\,Vel $^*$   	       & 08 43 49.81  & -46 07 08.78  & O6.5V       	      & $h$       \\
30 & HD\,92607 $^*$ 		       & 10 40 12.43  & -59 48 10.10  & O9IV        	      & $h$       \\
31 & CPD\,$-$\,59\degr\,2603 $^*$  & 10 44 47.31  & -59 43 53.23  & O7.5V, B0V(n)         & $y$       \\
32 & CPD\,$-$\,59\degr\,2628 $^*$  & 10 45 08.23  & -59 40 49.48  & O9.5V, B0.5V(n)       & $h$       \\
33 & CPD\,$-$\,59\degr\,2635 $^*$  & 10 45 12.72  & -59 44 46.17  & O8V, O9.5V  	      & $h$       \\
34 & V662\,Car $^*$            	   & 10 45 36.32  & -59 48 23.37  & O5V, B0:V             & $h$       \\
35 & EM\,Car $^*$	 			   & 11 12 04.50  & -61 05 42.94  & O7.5V, O7.5V          & $h,z$     \\
36 & HD\,115071 $^*$	 		   & 13 16 04.80  & -62 35 01.47  & O9.5III, B0Ib         & $h$       \\
37 & HD\,152219 $^*$			   & 16 53 55.61  & -41 52 51.47  & O9.5III               & $h$       \\
38 & CD\,$-$\,41\degr\,11042 $^*$  & 16 54 19.85  & -41 50 09.36  & O9.2IV, B1:V          & $h$       \\
39 & TYC\,7370-460-1 $^*$		   & 17 18 15.40  & -34 00 05.94  & O6V, O8V              & $h$       \\
40 & HDE\,323110 $^*$			   & 17 21 15.79  & -37 59 09.58  & ON9Ia                 & $h,z$     \\
41 & Pismis\,24-13 $^*$           & 17 24 45.79  & -34 09 39.94  & O6V                    & $h$       \\
\hline
\end{tabular}
\end{center}
Notes:
$a$:~\cite{1975A&AS...21..193N}, $b$:~\cite{1966ArA.....4...65L}, $c$:~\cite{1981A&AS...45..193P}, $d$:~\cite{1983MNRAS.205..241R}, $e$:~\cite{1977AJ.....82..474M}, $f$:~\cite{1988JRASC..82..276S}, $g$:~\cite{2007ApJS..168..100W}, $h$:~\cite{2016ApJS..224....4M}, $i$:~\cite{1991ApJS...76.1033D}, $j$:~\cite{1982A&AS...50..261D}, $k$:~\cite{1993ApJS...89..293V}, $l$:~\cite{1977ApJS...33..459S}, $m$:~\cite{1992AJ....104..590O}, $n$:~\cite{2006MNRAS.372.1407C}, $o$:~\cite{1971PW&SO...1a...1S}, $p$:~\cite{1974A&AS...16..445S}, $q$:~\cite{1977ApJ...215..106M}, $r$:~\cite{2003AJ....125.2531R}, $s$:~\cite{2014ApJS..211...10S}, $t$:~\cite{1969ArA.....5..181L}, $u$:~\cite{1968AJ.....73..590D}, $v$:~\cite{1961AJ.....66..186S}, $w$:~\cite{2009ApJ...696.1278P}, $x$:~\cite{1963LS....C04....0N},
$y$:~\cite{2001MNRAS.326.1149R}, $z$:~\cite{2003AcA....53..341P}. $^*$ stars from the Galactic O-Star Catalogue (\citealt{2016ApJS..224....4M}) identified as EBs in the present work.
\end{table*}

\section{Observations}

\subsection{Sample selection}

\begin{figure}
\includegraphics[width=1.0\columnwidth]{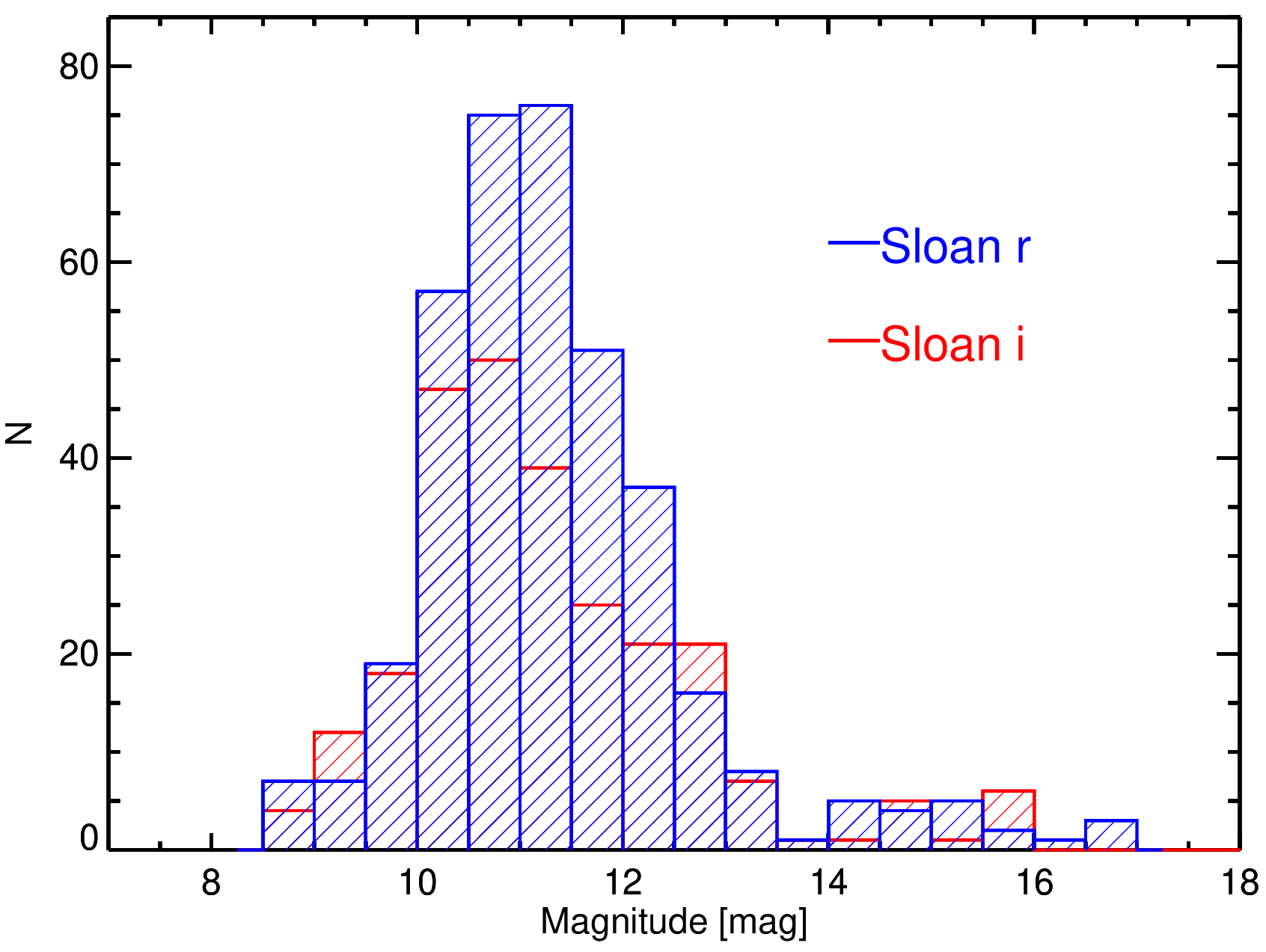}
\includegraphics[width=1.0\columnwidth]{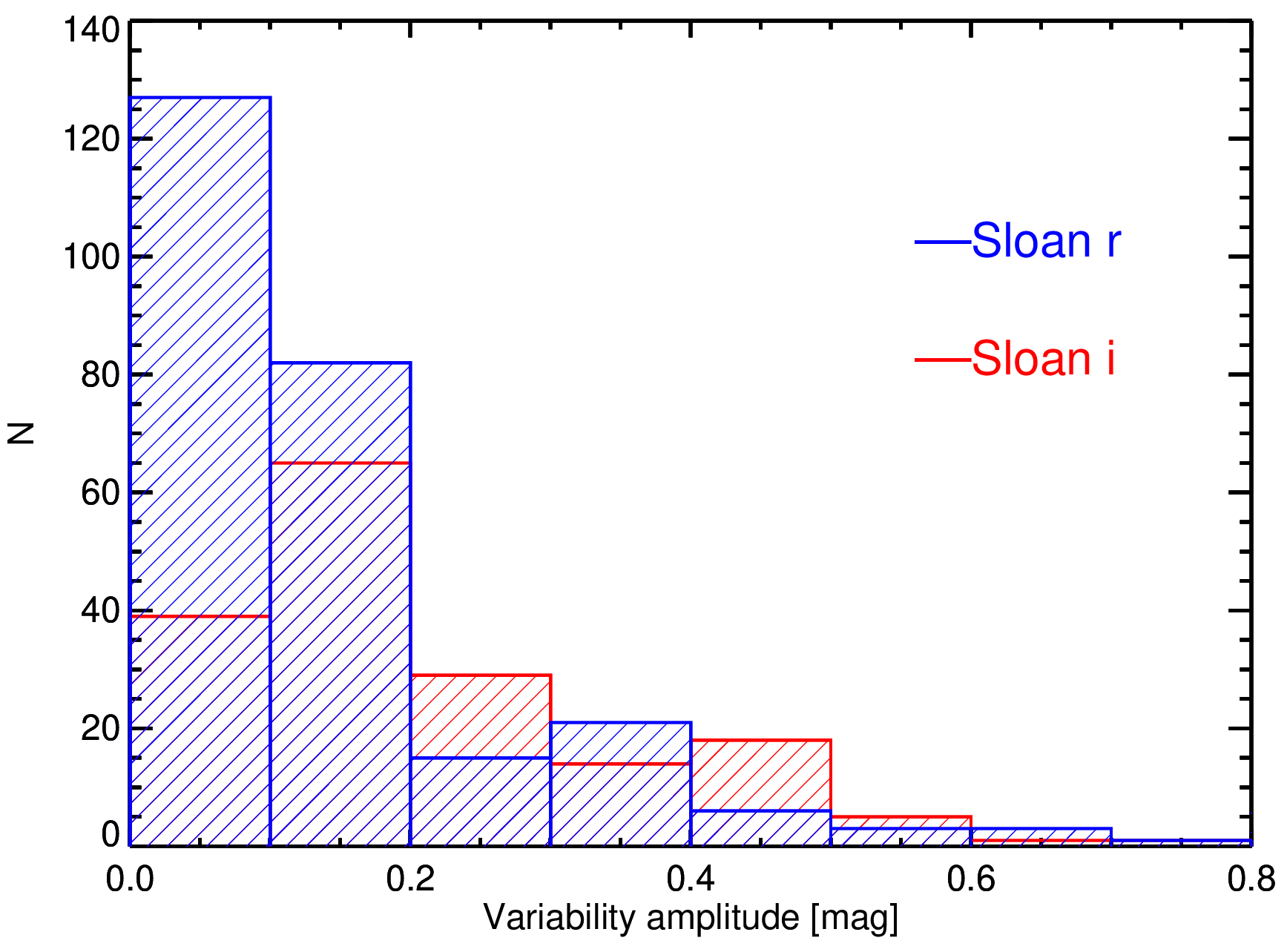}
  \caption
   {Distribution of magnitudes ({\it left}) and amplitudes ({\it right}) for 263 variable high-mass stars in $r$ (blue) and
   $i$ (red).}
   \label{fig1}
\end{figure}

The GDS comprises $\sim 16$ million stars of which $\sim 85.000$ are variable. The variability selection procedure was based on three methods: i) Stetson variability $J$-index ($J$-method, \citealt{1996PASP..108..851S}; ii) amplitude method ($\delta_{\rm A}$) which corresponds to the difference between minimum and maximum brightness in the time series, and iii) standard deviation ($\sigma_{\rm LC}$; SD-method) of the light curves. A full description of the identification of variable objects can be found in \cite{2015AN....336..590H}. In brief, since the $\delta_{A}$ and SD methods are sensitive to the scatter in the light curves,  they have to be determined as a function of brightness ($m$). We therefore used a 5\,$\sigma$ threshold applied to an $8^{th}$ degree polynomial fit $\sigma_{fit}(m)$ where $a_{fit}(m)$ was obtained from both $\delta_{A} - m$ and $\sigma_{\rm LC} - m$ diagrams, respectively. A source is considered as variable if the condition $\sigma_{\rm LC} > \sigma_{\rm fit}(m) + 5\ \sigma_{\rm SD}$ is satisfied for the SD method or if $\delta_{A} > a_{fit}(m) + 5\ \sigma_{\delta_{A}}$ is satisfied for the $\delta_{\rm A}$ method. In the case of the Stetson $J$-index, a threshold value of $0.5$ was used (\citealt{2012AJ....143..140F}). Therefore, a source is considered as variable if the condition $| J | > 0.5$ is satisfied. Due to systematic errors, the $J$-index usually finds positive detections for objects with very low amplitudes, hence, an additional condition based on the $\delta_{\rm A}$ method was also introduced. Finally, a source is considered as variable if a variability flag is raised by at least one of the previous methods.

\begin{figure}
\includegraphics[width=1.0\columnwidth]{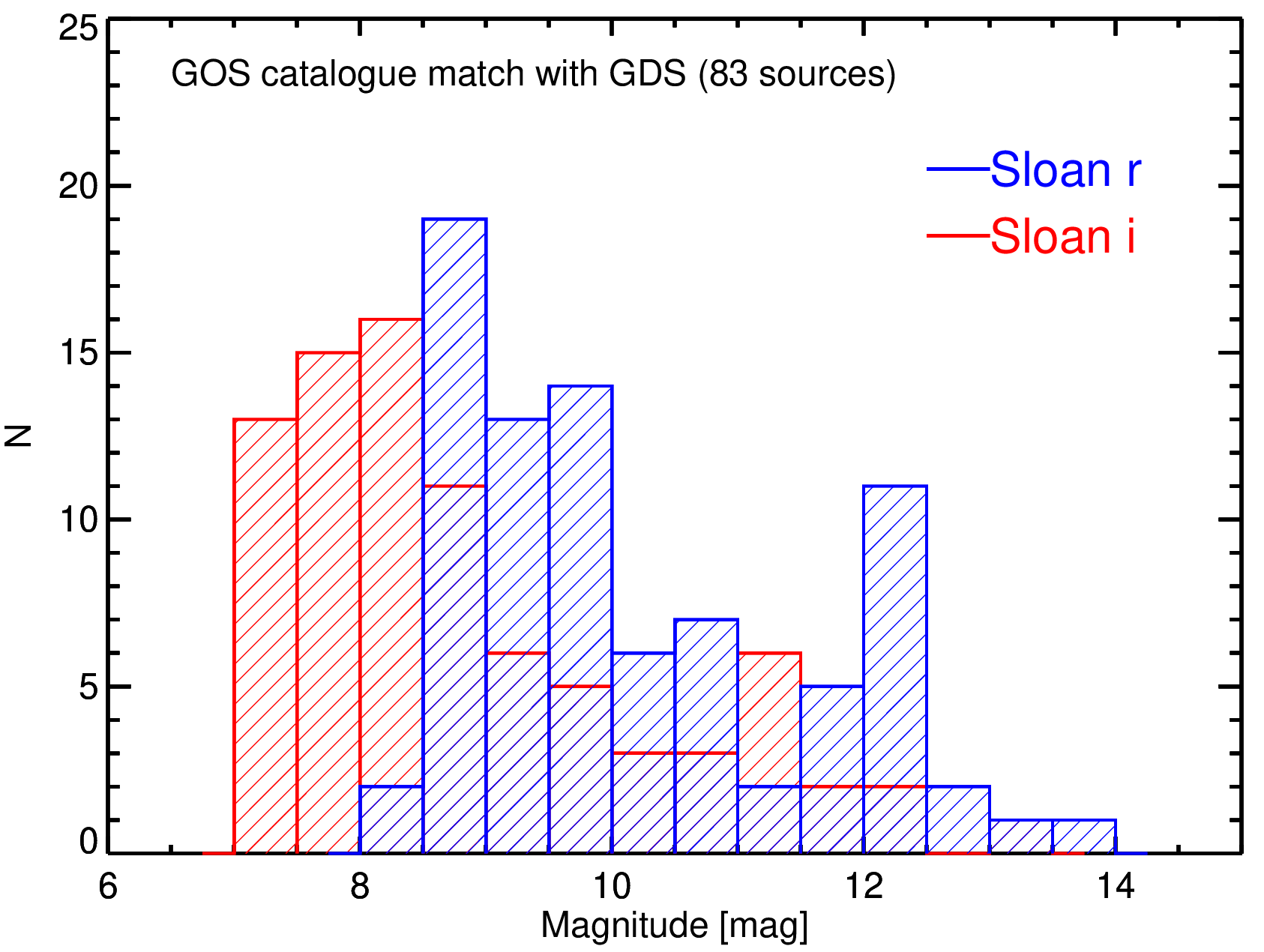}
\includegraphics[width=1.0\columnwidth]{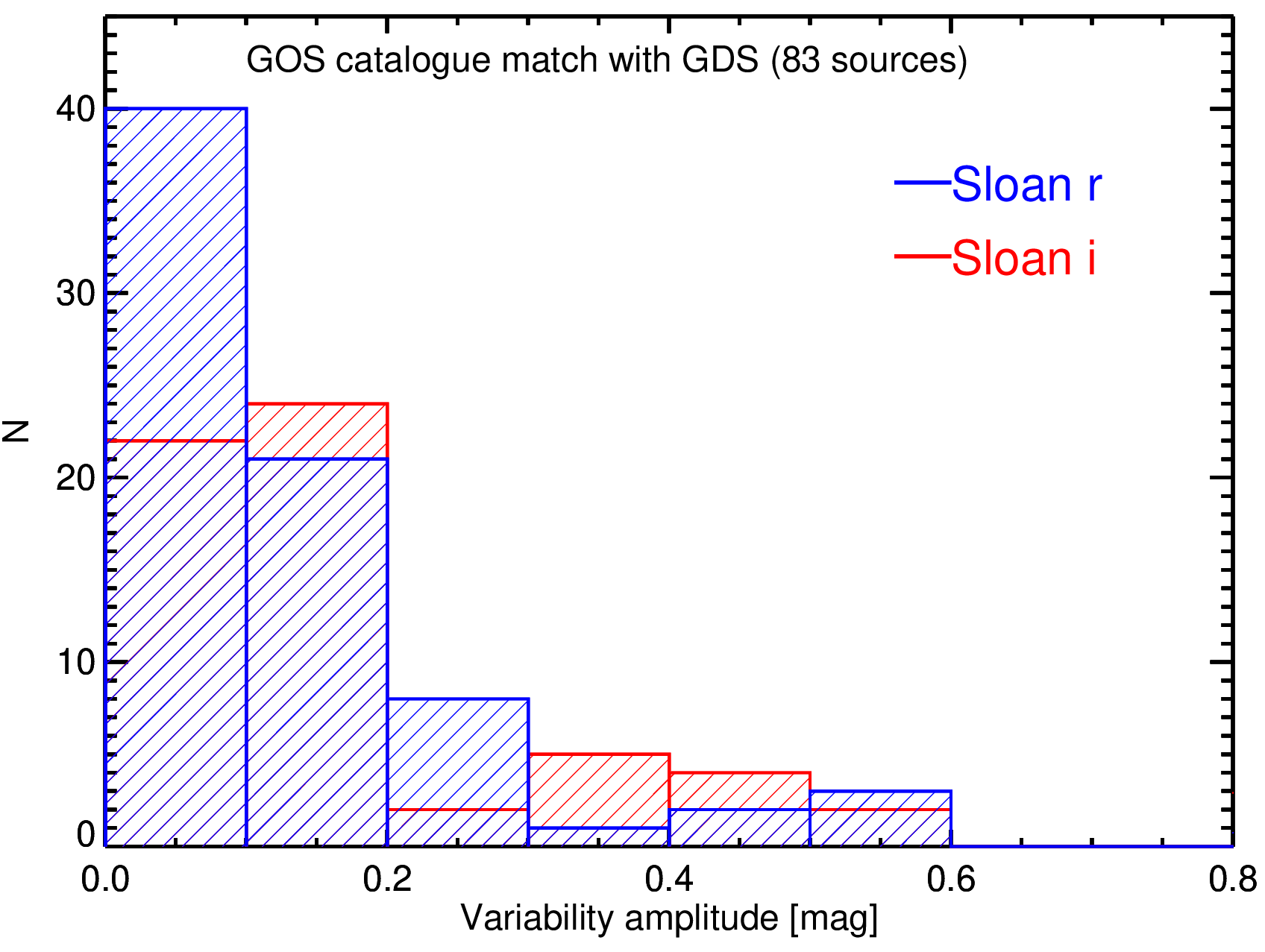}
  \caption{Same as Fig.\,1 but for the GOSC matched stars.}
   \label{fig2}
\end{figure}

By matching all detected sources with the SIMBAD catalogue we found 473 O-type stars of which 263 are variable. The matching was performed using an optimal 3" search radius, which is a good compromise between the varying positional accuracies in the SIMBAD database and a minimisation of multiple matches. Figure~\ref{fig1} shows the distribution of magnitudes and variability amplitudes for 263 O-type stars.

We performed a second match between the variable O-type stars that were previously identified in SIMBAD database (263) and the Galactic O-Star Catalogue (GOSC\footnote{Here we use the latest 4.1 version of the catalogue (June 2018), which list 594 O-type stars. See https://gosc.cab.inta-csic.es/.}, \citealt{2016ApJS..224....4M}). We find a match with 83 sources, including two EBs from Paper\,I (Pismis\,24-1 and HD\,319702), with magnitudes ranging between $\sim 7 - 12$\,mag and  $\sim 9 - 14$\,mag in the $r$ and $i$ filters, respectively. Figure~\ref{fig2} shows the distribution of magnitudes and variability amplitude for 83 identified GOSC stars. From the identified GOSC stars, we find that 25 sources show periodic variations among them 13 systems with well defined eclipses. Among all identified high-mass stars we found 41 EB systems (14\%). After the light curves have been folded with the corresponding period, a system is considered as EB when it shows well-defined primary and secondary minima. Six EBs were known before and we confirm the previous results; the remaining 35 systems are new discoveries. The three sources from Paper\,I were omitted from this statistics. Table~\ref{tab:objectlist} lists the EBs and their positions. We present the observational time span of the EB systems in the Appendix (Figure~\ref{timeline}).

\subsection{Data}

The GDS observations were performed with the 15\,cm Robotic Bochum Twin Telescope (RoBoTT) of the Universit\"atssternwarte Bochum, located near Cerro Armazones in Chile. The high-mass stars from the present work comprehend observations obtained from 268 fields along the galactic plane between 2009 and 2018 (\citealt{2015AN....336..590H}). The photometric monitoring was conducted simultaneously in the Sloan~$r$ (6230\,\AA) and $i$ (7616\,\AA) filters. For several stars, we performed additional single-epoch Johnson $UBV$ observations.

The data reduction was standardized, including bias, dark current, flatfield, astrometry and astrometric distortion corrections performed with IRAF\footnote{IRAF is distributed by the National Optical Astronomy Observatory, which is operated by the Association of Universities for Research in Astronomy (AURA) under cooperative agreement with the National Science Foundation.} in combination with SCAMP (\citealt{2006ASPC..351..112B}) and SWARP (\citealt{2002ASPC..281..228B}) routines.

The photometry was done with the DAOPHOT routine from IRAF, using an aperture radius of 4"; this choice maximized the signal-to-noise ratio and delivered the lowest absolute scatter for the fluxes. The brightness of the variable program stars was calculated relative to nearby non-variable reference stars located in the same field. The absolute photometric calibration was obtained from fluxes of about 20 standard stars (\citealt{2009AJ....137.4186L}) observed during the same nights as the science targets (see Paper\,I for more information).

The number of epochs during the monitoring campaign as well as the photometric $UBV$ results are summarized in the Appendix in Table~\ref{tab:photometryg}.

\section{Data analysis}

\subsection{Spectral types}

\begin{figure}
  \centering
  \includegraphics[angle=0,width=\columnwidth]{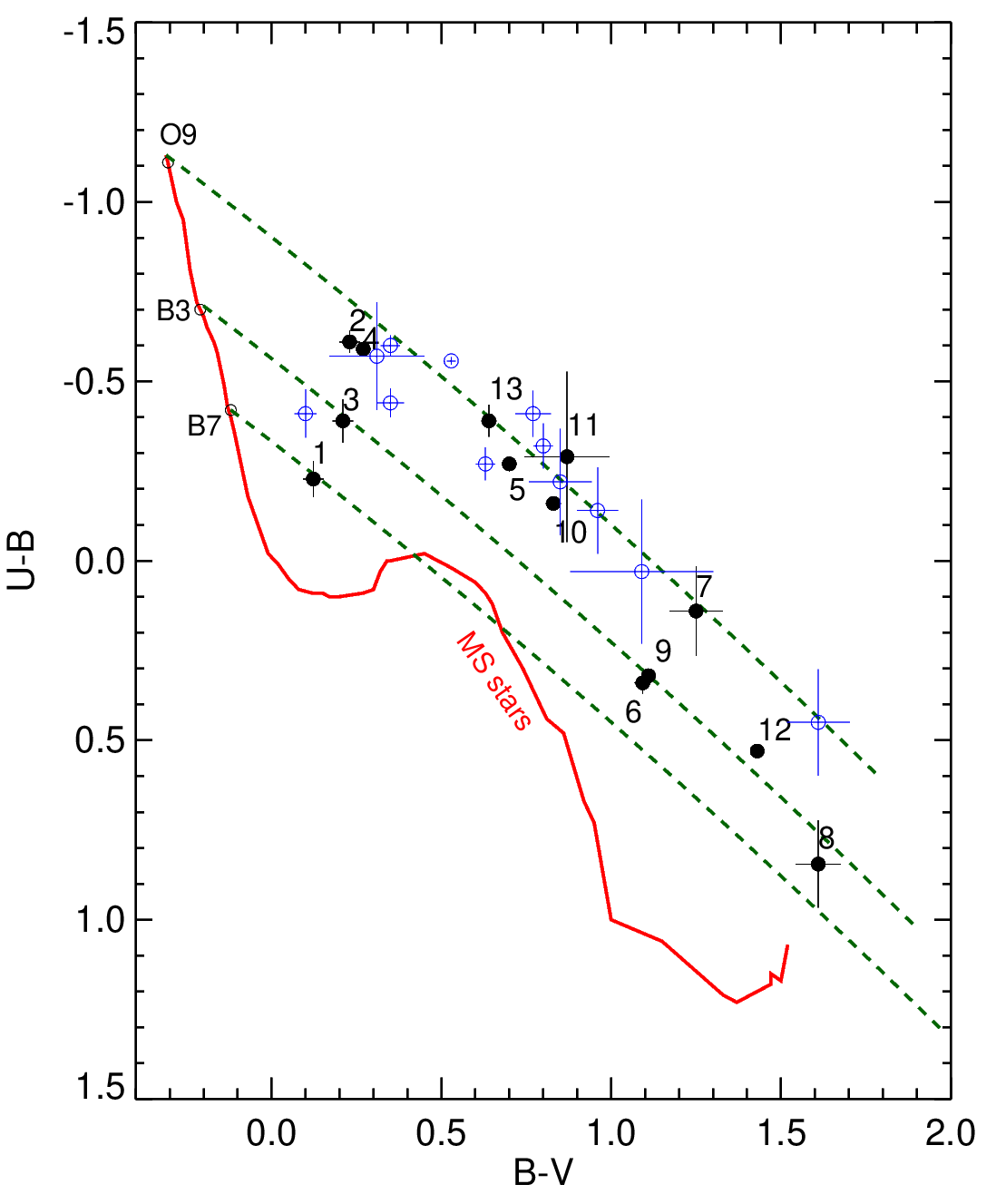}
  \caption{$UBV$ color-color diagram for stars in our sample. The black dots (No.1-13; Tab.~\ref{tab:objectlist}) correspond to new high-mass EBs with successfully recovered orbital parameters. The blue open circles (No.14-28; Tab.~\ref{tab:objectlist}) correspond to new high-mass variables where we could only determine the orbital period.
  The red curve is the locus of unreddened main-sequence stars from O9 to M4 (\citealt{1970A&A.....4..234F}).}
  \label{UBV diagram}
\end{figure}

Tab.~\ref{tab:objectlist} lists the existing spectral designations which are sometimes coarse or doubtful as described in the
original literature; nevertheless, the sample includes only high-mass stars. Therefore, we also used the single-epoch $UBV$ data to obtain a photometric classification in parallel. For the identified sub-sample of GOSC stars, we used the spectral types as listed in the latest release of the catalogue. Fig.~\ref{UBV diagram} corroborates that all objects are OB-type stars by their $UBV$ colors. When dereddening the stars via the reddening path $E_{U-B}/E_{B-V}$ = $0.72$ + $0.05E_{B-V}$ (\citealt{1956ApJ...124..367H}), we obtain spectral types from O9 to B7. The photometric solution for star No.\,1 is not unique; given, however, the spectroscopic classification of O9-B0 (\citealt{1975A&AS...21..193N}) we prefer the earliest solution.

Generally one must note that in the case of a binary the observed colors are a combination of the flux contributions from two
stars. Depending on the mass ratio $q = M_2/M_1$ of the components the combined colors will either agree with the colors of the primary ($q = 1$) or they will be redder ($q < 1$). In the latter case, the combined photometric spectral type will be later than the spectroscopic type of the primary. This problem will be discussed more thoroughly with the individual sources.

\subsection{Light curve Modelling}

The light curves were analysed assuming a standard Roche geometry based on the Wilson-Devinney (WD) code (\citealt{1971ApJ...166..605W}, \citealt{1979ApJ...234.1054W}, \citealt{1990ApJ...356..613W}). The modelling procedure is explained in detail in Paper\,I; here we describe only its main characteristics. First, the orbital periods were determined using the Lafler-Kinman algorithm (\citealt{1965ApJS...11..216L}) which was later generalized and introduced as Phase Dispersion Minimization (PDM) by \cite{1978ApJ...224..953S}. The PDM is performed on a period range from 0.1 to $\sim70$ days. The choice of the lower limit is a compromise between the minimum value of $\sim0.14$ days obtained by \cite{2014ApJ...790..157D} for ultra-short period binaries from the Catalina Survey, and the short-period limit of $\sim0.22$ days for contact binary systems reported by \cite{1992AJ....103..960R}. The upper limit has a range of $13 < P < 119$ days and is dynamically calculated so that the analyzed maximum period should be $T/2$, with $T$ the length of the time series for the particular light curve. We must note that any algorithm used to find the periods of EBs is sensitive to the total number of observations and time sampling of the light curves. An efficiency test performed on the PDM method is presented in Appendix~A. Then, the effective temperature of the primary component was calculated for an adopted spectral type taking into account the spectral information (Tab.~\ref{tab:objectlist}) and our $UBV$ data (Tab.~\ref{tab:photometryg}) assuming all stars to be on the main sequence. For the conversion between spectral type and effective temperature we used the values from \cite{1991Ap&SS.183...91T}.

The symmetrical separation of the primary and secondary minimum eclipses for all stars in the present sample suggests that these systems have fairly circular orbits. We therefore assumed zero eccentricity and synchronous rotation for all model fits. In the case of circular orbits and synchronous rotation, the Roche equipotential ($\Omega$) along with the mass ratio ($q$) describe the surface structure or Roche lobe of both components in the EB system. The geometric classification of an EB, whether the system is detached, semi-detached, contact or overcontact, depends on $\Omega$ and the commonly used fill-out factor $f$. The factor $f$ is expressed in terms of the inner ($\Omega_{in}$) and outer ($\Omega_{out}$) critical Roche equipotentials. For a detached system, the factor $f$ is defined by $f = (\Omega_{in}/\Omega) - 1$ for $\Omega_{in} < \Omega$, and will lie between $-1 < f \leq 0$. For a contact binary, $f$ takes the value $f = (\Omega_{in} - \Omega/\Omega_{in} - \Omega_{out})$ for $\Omega_{in} \geq \Omega$, and will lie between $0 \leq f \leq 1$ (\citealt{1979ApJ...234.1054W}, \citealt{1990ApJ...356..613W}, but see also \citealt{2015ComAC...2....4L}). For simplicity, and to be able to work with positive values, we defined the Roche Lobe Coefficient (RLC) $F = f + 1$ which corresponds to the values listed in Table~\ref{tab:system parameters}.

\begin{figure}
  \centering
  \includegraphics[angle=0,width=\columnwidth]{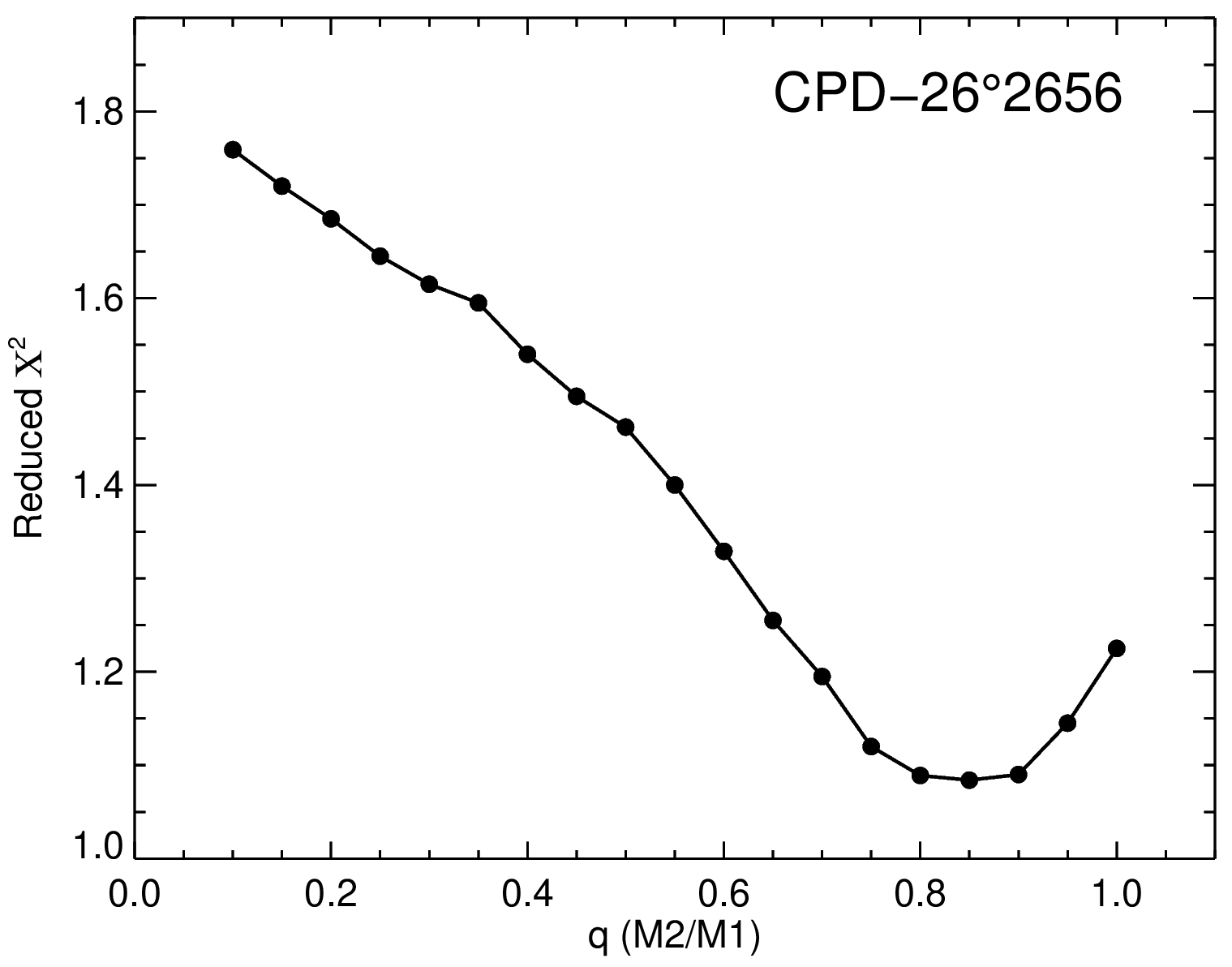}
  \caption{The behaviour of $\chi^2$ as a function of the trial mass ratios $q = M_2/M_1$ for the binary system CPD\,-26\degr\,2656.}
  \label{q diagram}
\end{figure}

\begin{table}
\begin{center}
\caption{EB systems without modelling.}
\label{tab:periods}
\begin{tabular}{clc}
\hline\hline
\noalign{\smallskip}
             \multicolumn{1}{c}{No.}
           & \multicolumn{1}{c}{Name}
           & \multicolumn{1}{c}{$P$ [d]}\\
             \hline
             \noalign{\smallskip}

14 & SS\,117				     &  $2.00930 \pm 0.00210$ \\
15 & CD\,$-$\,33\degr\,4174   &  $4.47699 \pm 0.00103$ \\
16 & CD\,$-$\,28\degr\,5257	 &  $1.50698 \pm 0.00323$ \\
17 & LS\,1221				 &  $1.14928 \pm 0.00042$ \\
18 & CD\,$-$\,51\degr\,10200  & 	$1.12509 \pm 0.00063$ \\
19 & GSC\,07380-00198		 &  $7.64607 \pm 0.00291$ \\
20 & ALS\,17569				 &  $5.26895 \pm 0.00184$ \\
21 & HD\,300214               &  $2.02429 \pm 0.00022$ \\
22 & TYC\,8958-4232-1         &  $5.10888 \pm 0.00051$ \\
23 & HD\,308974				 &  $4.74915 \pm 0.00021$ \\
24 & CPD\,$-$\,54\degr\,7198 &  $4.29379 \pm 0.00034$ \\
25 & HD\,152200              &  $8.89365 \pm 0.00081$ \\
26 & TYC\,6265-2079-1		 &  $3.67867 \pm 0.00073$ \\
27 & BD\,$-$\,024786         &  $2.52880 \pm 0.00124$ \\
28 & ALS\,15204              &  $2.23266 \pm 0.00030$ \\
\hline
\end{tabular}
\end{center}
Notes: The numbers given in the first column correspond to the numbers used in Table 1.
\end{table}

\begin{figure*}
  \centering
  \includegraphics[angle=0,width=15cm,clip=true]{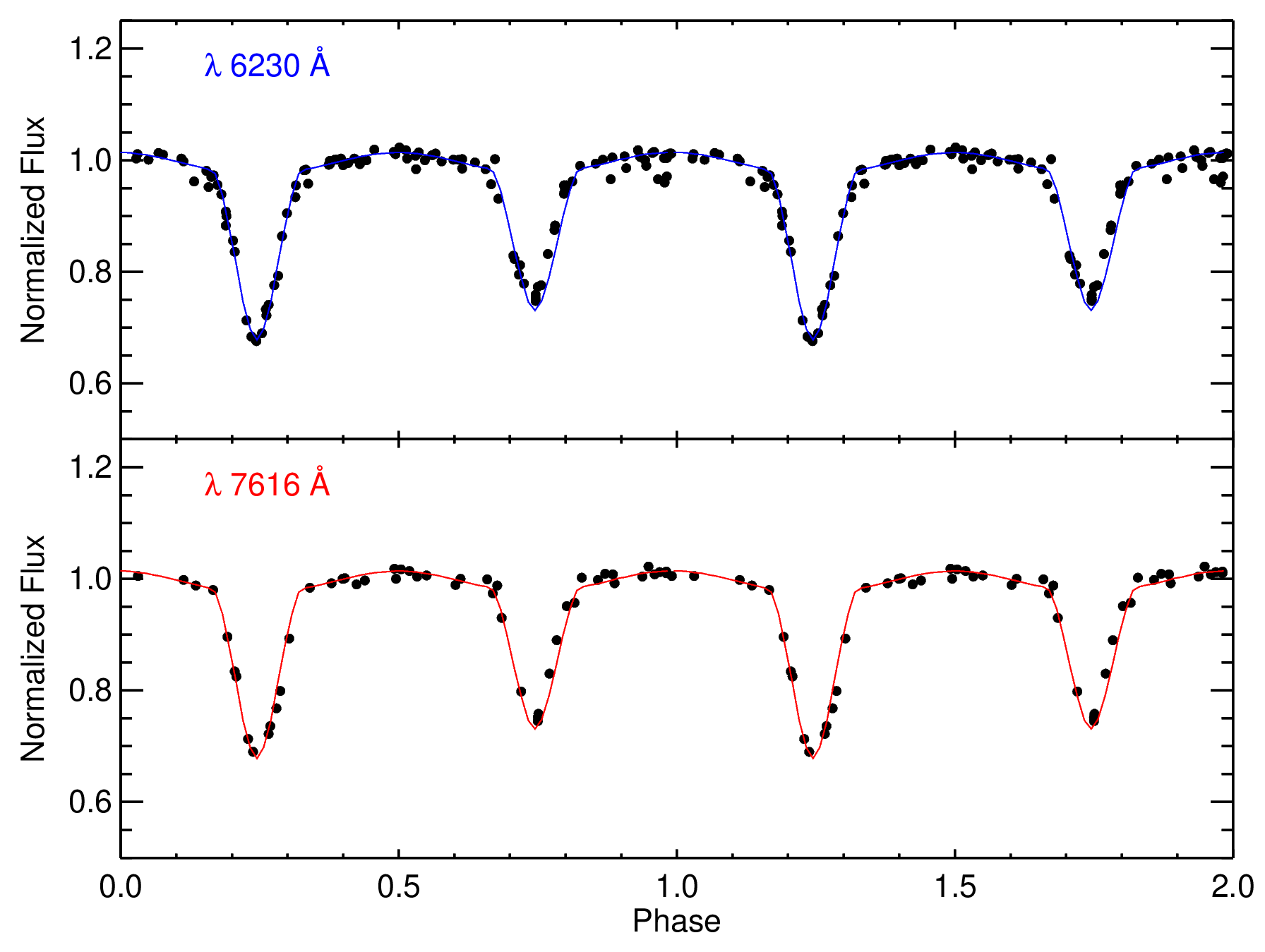}
  \caption{Observations (dots) and modelled light curves (solid curves)for the EB CPD\,$-$\,59\degr\,2603 for the Sloan $r$ (blue) and $i$ (red) filters at 6230 and 7616\,\AA\, respectively. The parameters of the model are presented in Table~\ref{tab:system parameters}.}
  \label{cpd592603_test}
\end{figure*}

Gravity-darkening coefficients $\beta_{1}= \beta_{2}=0.25$ and bolometric albedos $A_{1}=A_{2}=1.0$ were taken for early-type stars with radiative envelopes and hydrostatic equilibrium. We used a nonlinear square-root limb darkening law obtained at optical wavelengths (\citealt{1992A&A...259..227D}) with limb darkening coefficients interpolated from tables of \cite{1993AJ....106.2096V} at the given band pass. The best model was determined through repeated fits to the light curves until the minimum of $\chi^2$ was reached. During the fitting procedure the following adjustable parameters were optimized in each iteration: the effective temperature of the secondary star, the Roche lobe filling factors for the primary and secondary star, the inclination of the system and the photometric mass ratio ($q$) using the $q-$search method.

Since there are no spectroscopic mass ratios available for the systems in this study, we first searched for a reasonable starting value for $q$. Adopting mass-ratios from 0.10 to 1.0 in steps of 0.05 we obtained light curve solutions; the resulting sum of the squared deviations ($\chi^2$) for each value of $q$ was analysed. The value of $q$ corresponding to the minimum of $\chi^2$ obtained for each binary star was taken as the most plausible starting point for the mass-ratio. Figure.~\ref{q diagram} shows as an example the relation between $\chi^2$ and $q$ for the star CPD\,-26\degr\,2656. The lowest value of the residuals occurred at around $q = 0.85$. Once the starting point for the mass ratio was determined by the $q-$search method, $q$ was taken as a free parameter along with the other parameters to obtain the final solution.

%\begin{landscape}
\begin{table*}
\begin{center}
\caption{Orbital solutions and system parameters.}
\label{tab:system parameters}
\begin{tabular}{llllllll}
\hline
\hline
\noalign{\smallskip}
Results     & CPD\,-24\degr\,2836   & CPD\,-26\degr\,2656   & TYC\,6561-1765-1      & CPD\,-42\degr\,2880   & CPD\,-45\degr\,3253  & GSC\,08156-03066   & GSC\,08173-00182      \\
\noalign{\smallskip}
\hline
\noalign{\smallskip}
$P$ [d]     & $1.02 \pm 0.007$      & $2.699 \pm 0.007$     & $0.712 \pm 0.001$     & $1.90 \pm 0.009$      & $6.68 \pm 0.009$  & $2.3318 \pm 0.006$& $19.47 \pm 0.009$  \\
$T_0$       & $55971.168507$        & $55968.14481$         & $55927.26512$         & $55984.204734$        & $56236.345903$    & $56314.182400$    & $56333.182049$    \\
$M_2/M_1$   & $0.416 \pm 0.076$     & $0.856 \pm 0.102$     & $0.753 \pm 0.079$     & $0.612 \pm 0.083$     & $0.624 \pm 0.087$ & $0.996 \pm 0.096$ & $0.714 \pm 0.088$ \\
$i$         & $57\fdg3 \pm 2\fdg1$  & $79\fdg7 \pm 1\fdg6$  & $55\fdg7 \pm 1\fdg7$  & $64\fdg8 \pm 1\fdg8$  & $64\fdg1 \pm 2\fdg1$& $52\fdg4\pm1\fdg4$& $59\fdg5\pm1\fdg7$\\
RLC(1)      & $0.860 \pm 0.011$     & $0.650 \pm 0.007$     & $0.880 \pm 0.009$     & $0.730 \pm 0.008$     & $0.730 \pm 0.012$ & $0.932 \pm 0.011$ & $0.890 \pm 0.013$ \\
RLC(2)      & $0.840 \pm 0.010$     & $0.740 \pm 0.005$     & $0.642 \pm 0.005$     & $0.720 \pm 0.007$      & $0.720\pm 0.012$ & $0.928 \pm 0.009$ & $0.870 \pm 0.012$ \\
$T_1$ [K]   & $29\,230$             & $25\,570$             & $18\,445$             & $29\,230$             & $29\,230$         & $25\,570$         & $18\,445$         \\
$T_2$ [K]   & $19\,520 \pm 1510$    & $23\,250 \pm 1112$    & $13\,150 \pm 1180$    & $22\,110 \pm 1120$    & $22\,250 \pm 1302$& $24\,150 \pm 1290$& $13\,920 \pm 1260$ \\
$\Delta D_p$& $7.5\%$               & $33.5\%$              & $6.0\%$               & $8.5\%$               & $9.0\%$           & $10.0\%$          & $12.5\%$          \\
$\Delta D_s$& $5.5\%$               & $28.5\%$              & $5.0\%$               & $5.5\%$               & $6.0\%$           & $9.0\%$          & $10.5\%$          \\
\hline
\end{tabular}
\end{center}
Notes: $P$ is the orbital period (in days) obtained from the PDM analysis; The reference time $T_0$ refers to the time of the primary eclipse (HJD-2400000); $M_2/M_1$ is the mass ratio; $i$ the orbital inclination. RLC(1) and RLC(2) are the Roche Lobe Coefficients, $T_1$ and $T_2$ are the effective temperatures, and $\Delta D_p$ and $\Delta D_s$ correspond to the amplitude of the minima for the primary and secondary component, respectively.
\end{table*}
%\end{landscape}

%\begin{landscape}
\begin{table*}
\begin{center}
\contcaption{}
\label{tab:continued}
\begin{tabular}{llllllll}
\hline
\hline
\noalign{\smallskip}
Results     & TYC\,8175-685-1       & ALS\,18551            & CPD\,$-$\,39\degr\,7292   & Pismis\,24-4   &  CD\,$-$\,29\degr\,14032 & CD\,$-$\,31\degr\,5524  & V\,467\,Vel \\
\noalign{\smallskip}
\hline
\noalign{\smallskip}
$P$ [d]     & $3.4085 \pm 0.009$    & $1.360 \pm 0.006$     & $2.051 \pm 0.004$      & $4.29 \pm 0.009$      & $6.945 \pm 0.009$   & $3.2325 \pm 0.008$  & $2.753\pm0.004$ \\
$T_0$       & $56316.345486$        & $56749.08153$         & $56557.08014$          & $55827.095764$        & $56183.05922$       & $56318.310961$      & $56402.10429$   \\
$M_2/M_1$   & $0.987 \pm 0.094$     & $1.000 \pm 0.098$     & $0.885 \pm 0.061$      & $0.982 \pm 0.091$     & $0.989 \pm 0.091$   & $0.9532 \pm 0.089$  & $0.732 \pm 0.064$ \\
$i$         & $56\fdg7\pm1\fdg9$    & $59\fdg \pm 1\fdg8$  & $52\fdg3 \pm 1\fdg7$   & $61\fdg9\pm1\fdg6$    & $60\fdg8\pm1\fdg3$  & $53\fdg2\pm1\fdg8$  & $65\fdg7\pm1\fdg9$ \\
RLC(1)      & $0.993 \pm 0.014$     & $0.921 \pm 0.008$     & $0.999 \pm 0.010$      & $0.760 \pm 0.009$     & $0.762 \pm 0.010$   & $0.968 \pm 0.010$   & $0.790 \pm 0.009$  \\
RLC(2)      & $0.990 \pm 0.012$     & $0.930 \pm 0.009$     & $0.984 \pm 0.009$      & $0.960 \pm 0.008$     & $0.738 \pm 0.008$   & $0.951 \pm 0.010$   & $0.840 \pm 0.009$ \\
$T_1$ [K]   & $18\,445$             & $41\,860$             & $25\,570$              & $29\,230$             & $29\,230$           & $25\,570$           & $37\,870$			\\
$T_2$ [K]   & $18\,445 \pm 1350$    & $41\,460 \pm 1234$    & $18\,790 \pm 1310$     & $29\,230 \pm 926$     & $27\,850 \pm 1190$  & $18\,200 \pm 1230$  & $31\,640 \pm 1740$	\\
$\Delta D_p$& $14.5\%$              & $12.0\%$              & $15.4\%$               & $14.5\%$              & $6.1\%$             & $10.3\%$            & $17.3\%$	\\
$\Delta D_s$& $13.5\%$              & $12.0\%$              & $10.3\%$               & $14.5\%$              & $5.9\%$             & $8.2\%$             & $11.2\%$	\\
\hline
\end{tabular}
\end{center}
%Notes: The reference time $T_0$ refers to the time of the primary eclipse (HJD-2400000); RLC is the Roche Lobe Coefficient.
\end{table*}
%\end{landscape}

%\begin{landscape}
\begin{table*}
\begin{center}
\contcaption{}
\label{tab:continued}
\begin{tabular}{llllllll}
\hline
\hline
\noalign{\smallskip}
Results     & HD\,92607		        & CPD\,$-$\,59\degr\,2603             & CPD\,$-$\,59\degr\,2628			     & CPD\,$-$\,59\degr\,2635   & V662\,Car       & EM\,Car             & HD\,115071 \\
\noalign{\smallskip}
\hline
\noalign{\smallskip}
$P$ [d]     & $1.2959\pm0.0025$   & $2.1529\pm0.0095$     & $1.4694\pm0.01541$     & $2.2998\pm0.0051$     & $1.4135\pm0.01304$  & $3.4143\pm0.01023$  & $2.7314\pm0.00053$ \\
$T_0$       & $56324.256644$        & $56328.37082$         & $56358.01023$          & $56376.216898$        & $56344.277986$      & $56324.286944$      & $56411.094363$   \\
$M_2/M_1$   & $0.964 \pm 0.090$     & $0.896 \pm 0.101$     & $0.865 \pm 0.064$      & $0.876 \pm 0.071$     & $0.776 \pm 0.059$   & $0.816 \pm 0.068$   & $0.979 \pm 0.092$ \\
$i$         & $55\fdg8\pm1\fdg5$    & $79\fdg2 \pm 1\fdg7$  & $73\fdg4 \pm 1\fdg7$   & $75\fdg2\pm1\fdg6$    & $74\fdg3\pm1\fdg7$  & $80\fdg1\pm1\fdg8$  & $56\fdg8\pm1\fdg6$ \\
RLC(1)      & $0.972 \pm 0.012$     & $0.658 \pm 0.007$     & $0.960 \pm 0.011$      & $0.776 \pm 0.010$     & $0.963 \pm 0.013$   & $0.772 \pm 0.010$   & $0.880 \pm 0.009$  \\
RLC(2)      & $0.952 \pm 0.012$     & $0.745 \pm 0.008$     & $0.950 \pm 0.010$      & $0.687 \pm 0.008$     & $0.954 \pm 0.010$   & $0.689 \pm 0.009$   & $0.640 \pm 0.008$ \\
$T_1$ [K]   & $32\,882$             & $35\,874$             & $31\,884$              & $34\,877$             & $40\,862$           & $35\,874$           & $30\,789$			\\
$T_2$ [K]   & $31\,655 \pm 1236$    & $32\,220 \pm 1360$    & $30\,120 \pm 1310$     & $31\,090 \pm 1186$    & $31\,989 \pm 1338$  & $31\,130 \pm 1368$  & $30\,080 \pm 1980$	\\
$\Delta D_p$& $14.5\%$              & $29.4\%$              & $38.2\%$               & $28.1\%$              & $40.4\%$            & $37.4\%$            & $17.3\%$	\\
$\Delta D_s$& $14.3\%$              & $26.2\%$              & $34.3\%$               & $23.3\%$              & $30.2\%$            & $29.3\%$            & $11.2\%$	\\
\hline
\end{tabular}
\end{center}
%Notes: The reference time $T_0$ refers to the time of the primary eclipse (HJD-2400000); RLC is the Roche Lobe Coefficient.
\end{table*}
%\end{landscape}

%\begin{landscape}
\begin{table*}
\begin{center}
\contcaption{}
\label{tab:continued}
\begin{tabular}{llllll}
\hline
\hline
\noalign{\smallskip}
Results     & HD\,152219		    & CD\,$-$\,41\degr\,11042             & TYC\,7370-460-1	     & HDE\,323110		     & Pismis\,24-13   \\
\noalign{\smallskip}
\hline
\noalign{\smallskip}
$P$ [d]     & $4.24028\pm0.00222$   & $9.8972\pm0.0052$    & $2.52643\pm0.017980$   & $5.2052\pm0.0156$     & $20.1418\pm0.00064$ \\
$T_0$       & $56552.987616$        & $56555.025961$        & $57163.310116$         & $57174.301238$        & $55779.07093$       \\
$M_2/M_1$   & $0.534 \pm 0.086$     & $0.973 \pm 0.089$     & $0.776 \pm 0.059$      & $0.726 \pm 0.049$     & $0.996 \pm 0.121$   \\
$i$         & $65\fdg4\pm1\fdg7$    & $69\fdg7 \pm 1\fdg7$  & $80\fdg3 \pm 1\fdg9$   & $73\fdg8\pm1\fdg8$    & $51\fdg9\pm2\fdg9$  \\
RLC(1)      & $0.842 \pm 0.013$     & $0.790 \pm 0.009$     & $0.820 \pm 0.012$      & $0.956 \pm 0.015$     & $0.914 \pm 0.140$   \\
RLC(2)      & $0.793 \pm 0.011$     & $0.800 \pm 0.010$     & $0.790 \pm 0.009$      & $0.922 \pm 0.013$     & $0.910 \pm 0.138$   \\
$T_1$ [K]   & $30\,789$             & $32\,383$             & $38\,867$              & $31\,368$             & $38\,867$           \\
$T_2$ [K]   & $25\,560 \pm 1023$    & $32\,000 \pm 1017$    & $32\,120 \pm 1118$     & $27\,169 \pm 1146$    & $37\,890 \pm 3595$  \\
$\Delta D_p$& $16.2\%$              & $19.6\%$              & $42.3\%$               & $36.2\%$              & $10.8\%$             \\
$\Delta D_s$& $12.1\%$              & $19.3\%$              & $35.2\%$               & $31.4\%$              & $10.6\%$             \\
\hline
\end{tabular}
\end{center}
%Notes: The reference time $T_0$ refers to the time of the primary eclipse (HJD-2400000); RLC is the Roche Lobe Coefficient.
\end{table*}
%\end{landscape}

\begin{figure}
  \centering
  \includegraphics[angle=0,width=\columnwidth]{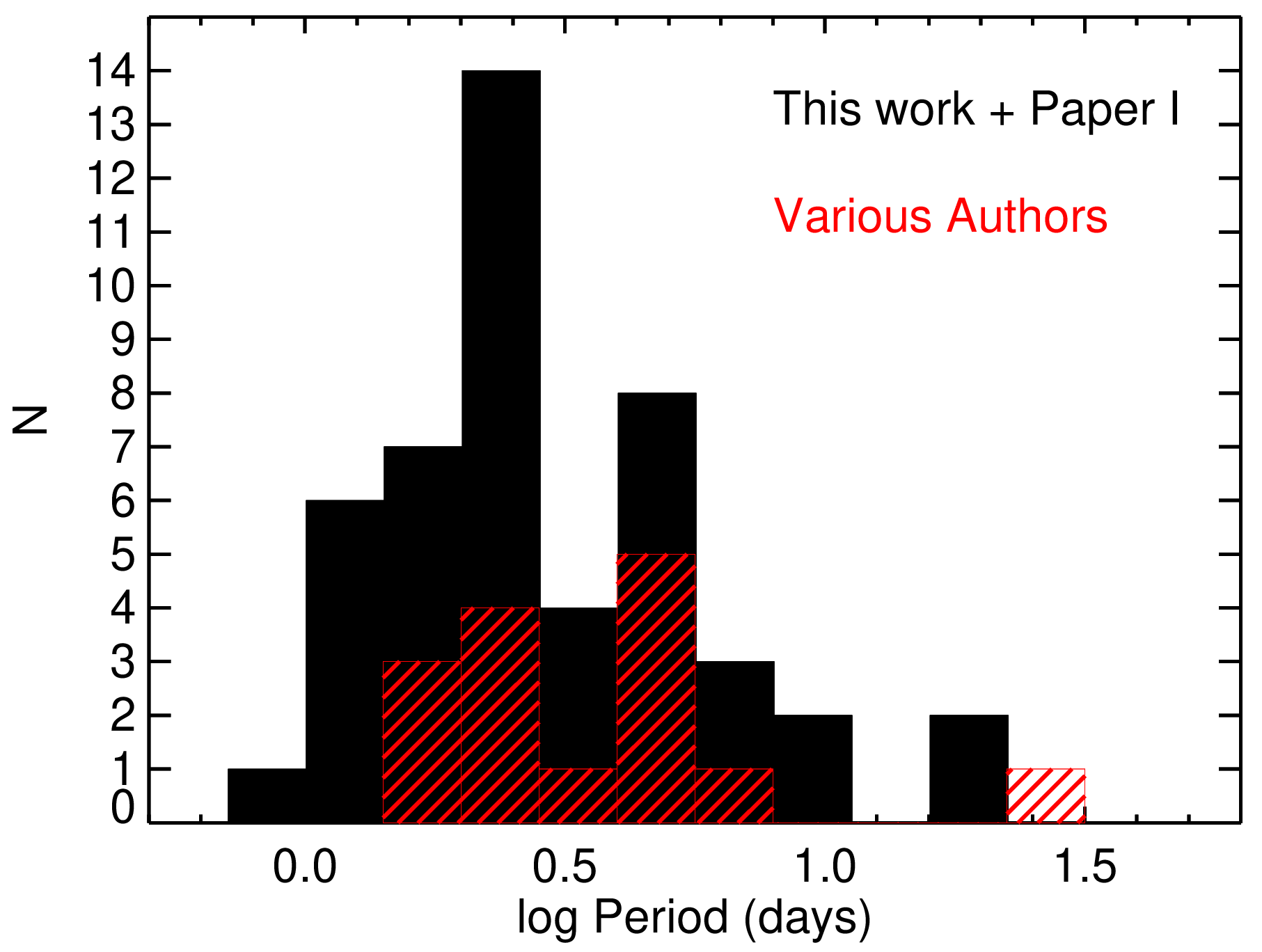}
  \caption{Distributions of orbital periods. }
  \label{histogram_period}
\end{figure}

\section{Results and discussion}
\label{sec:discussion}

For 26 EBs the quality of the light curves allows a proper modelling. Apart from these 26 high-quality light curves, there are another 15 new systems where time sampling and amount of observations allow the determination of the orbital periods but not a proper modeling. Table~\ref{tab:periods} lists these systems and our results; the corresponding light curves are shown in Figure~\ref{phase_lc} in the Appendix~E1. Despite the large scatter observed in the folded light curves, the EB nature of the systems is well defined. Other light curves for GOSC stars with doubtful EB nature are shown in the Appendix~F in Figure~\ref{phase_lc_GOSC}. Their characteristics and photometric measurements are listed in Tables \ref{tab:goscmatch} and \ref{tab:photometrygosc} respectively. Those systems are not considered further in our analysis.

As an example to illustrate the results of our modelling, we select here the EB system CPD\,$-$\,59\degr\,2603. The remaining multiple high-mass systems are discussed in a similar way in  Appendix~B where we describe their properties obtained from the light curve analysis.

CPD\,$-$\,59\degr\,2603 or V572\,Car (ID:0211 GOSC) is a member of the open cluster Trumpler\,16 in the Carina Complex (\citealt{2001MNRAS.326.1149R}). Our PDM analysis yields a period of $2.15285 \pm 0.00955642$ days, consistent with the spectroscopic period $2.15294 \pm 0.00214$ days obtained by \cite{2001MNRAS.326.1149R}. A somewhat sparser monitoring of this object was carried out by \cite{1993ARep...37..152A} between 1980-1991. By folding the spectroscopic period with the photometric data obtained by \cite{1993ARep...37..152A}, \cite{2001MNRAS.326.1149R} derived a lower limit for the inclination ($i\geq77^{\circ}$) together with the masses for the EB system. This is the first time that a high quality sampled light curve is obtained for this source. We fixed the temperature of the primary star to $T_1 = 35874$\,K corresponding to a spectral type O7.5\,V (\citealt{2001MNRAS.326.1149R}). The primary component fills up its Roche lobe at $\sim 66\%$ while the secondary at $\sim 75\%$ favouring a detached configuration with an orbital inclination $i = 79.2^{\circ} \pm 1.7$ (Fig.~\ref{model_apc}). The inclination found here is consistent with the lower limit $i\geq77 ^{\circ}$ reported by \cite{2001MNRAS.326.1149R}. We find an effective temperature of $T_2 = 32220$\,K for the secondary star corresponding to a spectral type O9.5\,V in agreement with the spectroscopic results. Figure.~\ref{cpd592603_test} shows the observed light curves in the Sloan $r$ and $i$ filters folded with the orbital period along with the best model.

The results obtained for all the EBs are summarized in Tab.~\ref{tab:system parameters}; all light curves are displayed in Appendix~D along with their best fit model.

\subsection{Comparison with previous works}

We compare the observed orbital parameters obtained in this work with previous results for high-mass EBs. Those systems are located in the associations Cassiopeia~OB6 (\citealt{2006ApJ...639.1069H}) and Cygnus OB2 (\citealt{2014ApJS..213...34K}, \citealt{2015ApJ...811...85K}), open clusters NGC\,6231, in the Scorpious~OB1 association (\citealt{2003A&A...405.1063S}, \citealt{2006MNRAS.371...67S}, \citealt{2017A&A...600A..33M}), and Trumpler\,16 in the Carina~OB1 association (\citealt{2001A&A...369..561F}, \citealt{2006MNRAS.367.1450N}). In total there are 15 known periods from previous investigations. Our study yields 47 periods -- including the three stars from Paper\,I.  The orbital period distribution covers the range $0.71 \leq P \leq 20.14$ days (Fig.~\ref{histogram_period}). Only two objects have orbital periods larger than 10 days (GSC\,08173-0018, $P = 19.47$; Pismis\,24-13, $P = 20.14$) while seven objects exhibit orbital periods shorter than 2 days. The shortest periods are found for TYC\,6561-1765-1 and CPD\,$-$\,24\degr\,2836 with 0.71 and 1.02 days, respectively. Combining our results with the existing values, half of the periods concentrate around 2.4 and 4.8 days. As can be seen in Fig.~\ref{histogram_period}, this bimodal distribution is already apparent in the individual data sets.

Using the rapid binary evolution code from \cite{2002MNRAS.329..897H}, \cite{2005A&A...442.1003S} derived the theoretical expected distributions of maximum (primary) eclipse depths ($\Delta m$) versus periods of EBs for different luminosities (their Fig.~8 with $\Delta m > 0.1$ mag). In the case of B-type systems, the expected periods range from 0.175 to 65 days. However, for O-type systems no periods shorter than 1.44 days are expected ($\log P \sim 0.16$ in their Fig.~8 and table A.1). As pointed out by \cite{2005A&A...442.1003S}, the mean eclipse probabilities are only valid as long as the systems have similar properties i.e. similar radii, luminosities and orbit sizes. Moreover, their calculations have been carried out by neglecting tidal deformations, reflection effects and assuming zero limb-darkening coefficients. This would explain the fact that in our sample we find three O-type EBs (HD\,92607; O9\,IV+O9.5\,V $P=1.29$ days, ALS\,18551; O4.5\,V+O4.5\,V $P=1.36$ days, and V662\,Car; O5\,V+O9.5\,V $P=1.41$ days) which are near-contact systems with periods shorter than the theoretical limit. CPD\,-59\degr\,2628 (O9.5\,V+O9.5\,V) has a slightly larger period ($P=1.46$ days) than the limit, but has also a near-contact configuration.

All systems from the present survey exhibit circular orbits ($e = 0$). This behaviour is a consequence of the short orbital periods because tidal forces will circularize the orbit before Roche lobe overflow (\citealt{2002MNRAS.329..897H}). A crude estimate of the orbit semi-major axis ($a$) can be obtained from the orbital period by adopting the masses for the effective temperatures and spectral type of the components. We used the masses reported in \cite{2005A&A...436.1049M} for O-type stars and \cite{2006MNRAS.371..252L} and \cite{2010AN....331..349H} for B-type stars. We find a range of $7.52 \leq a \leq 70.96\ R_\odot$ for B-type systems and $16.13 \leq a \leq 122.57\ R_\odot$ for O-type systems.

\begin{figure}
  \centering
  \includegraphics[angle=0,width=\columnwidth]{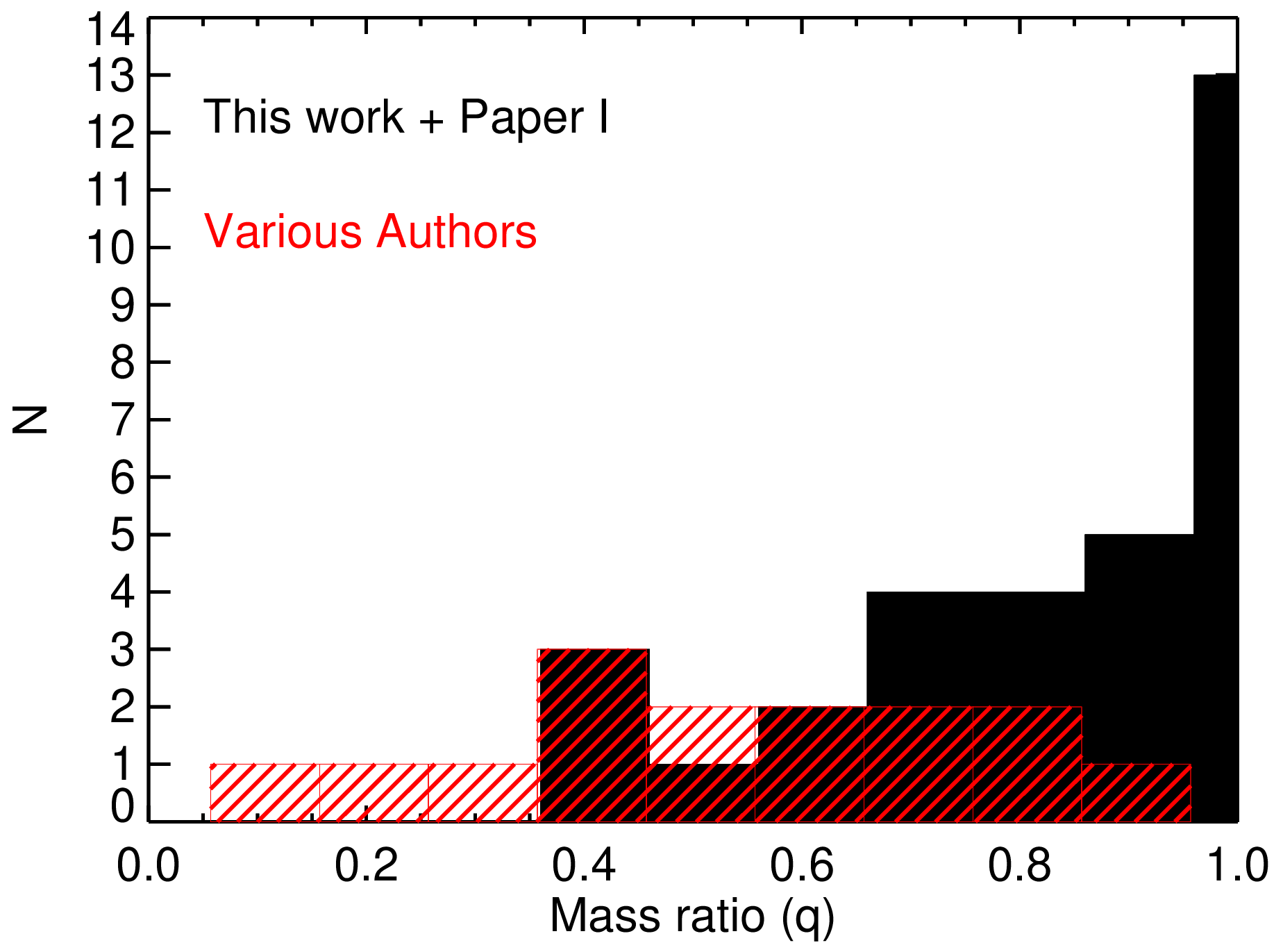}
  \caption{Distributions of mass ratio $q$.}
  \label{histogram_q}
\end{figure}

\begin{figure}
  \centering
  \includegraphics[angle=0,width=\columnwidth]{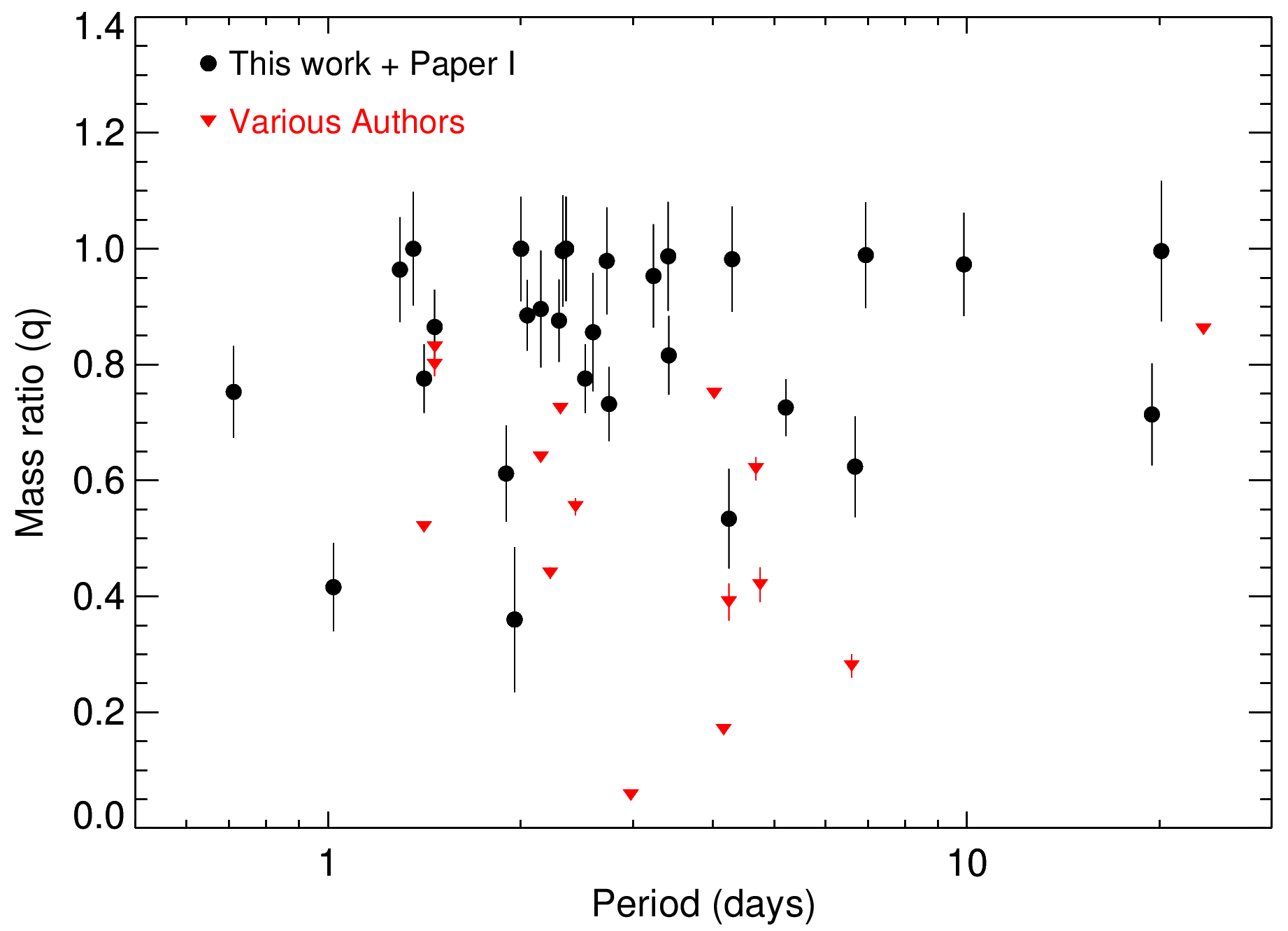}
  \caption{Mass ratio vs. orbital period.}
  \label{q_period}
\end{figure}

The distribution of mass ratios $q = M_{2}/M_{1}$ is shown in Fig.~\ref{histogram_q}. While the known mass ratios display a fairly uniform distribution between 0.05 and 0.95, the present distribution starts only at $q \sim 0.4$; 62\% of the systems have mass ratios larger than 0.8. The average value is $q = 0.8$ suggesting a large fraction of similar-mass binaries. It is likely that such systems were created during the star formation process rather than by random tidal capture. For instance, the small separation between the components ($a < 1.0$ AU) supports a Kozai cycle star formation scenario (\citealt{2007ApJ...669.1298F}), although a third component could not yet be detected in our sample. One could also speculate that the small separation between the components is an indication for an advanced stage of interaction that could lead to a merger event. This is supported by the large fraction of semi-detached and near-contact systems observed in our sample. On the other hand, semi-detached systems with short periods could be the product of a post-interaction process. Nevertheless we note that the results from the present investigation are biased by the detection method, i.e. transits, which favor compact systems with large diameters of the components. A proper discrimination between different star formation scenarios is only possible after follow-up radial velocity measurements and accurate mass determinations. Fig.~\ref{q_period} corroborates previous results that there is no correlation between mass ratio and orbital period.

\section{Conclusions}

The current results improve the existing statistics for high-mass EBs significantly. The Roche lobe analysis of 26 systems indicates that 12 systems have a detached geometry while 14 systems have a semi-detached geometry; seven of them are near-contact systems. The high fraction of near-contact systems found in this work suggests that mass transfer will likely dominate their evolution and new contact binary systems will emerge. This agrees with results by \citet{2012Sci...337..444S} who estimate that more than 70\% of high-mass stars will exchange mass with a companion and 30\% of them will turn into a binary merger.

Moreover, binary interaction can lead to a significant increase of the stellar rotation rates, and for which, fast rotators will often appear as single stars (\citealt{2011IAUS..272..531D}). In that case, if the binary interaction is responsible for the high rotational velocities observed among O-type stars, their multiplicity fraction will be strongly affected. We are engaged in a spectroscopic follow-up study of this sample, in order to investigate the relationship between fast rotation and binary interaction as well as to determine the physical parameters and to track the evolutionary state of the individual systems.

\section*{Acknowledgements}

We acknowledge support from the IdP II 2015 0002 64 and DIR/WK/2018/09 grants of the Polish Ministry of Science and Higher Education. This research made use of the NASA/IPAC Extragalactic Database (NED) which is operated by the Jet Propulsion Laboratory, California Institute of Technology, under contract with the National Aeronautics and Space Administration. This research has made use of the SIMBAD database, operated at CDS, Strasbourg, France. This research has made use of "Aladin sky atlas" developed at CDS, Strasbourg Observatory, France. We thank the anonymous referee for the constructive comments which contributed to improve this paper.

%%%%%%%%%%%%%%%%%%%%%%%%%%%%%%%%%%%%%%%%%%%%%%%%%%



\bibliographystyle{mnras}
\bibliography{mnras_highmass} % if your bibtex file is called example.bib

%\begin{thebibliography}{99}
%\end{thebibliography}

%%%%%%%%%%%%%%%%%%%%%%%%%%%%%%%%%%%%%%%%%%%%%%%%%%

%%%%%%%%%%%%%%%%% APPENDICES %%%%%%%%%%%%%%%%%%%%%

\appendix

\section{The performance of the PDM algorithm and the time span of the monitoring campaign}

\begin{figure*}
  \centering
  \includegraphics[angle=0,width=15cm,clip=true]{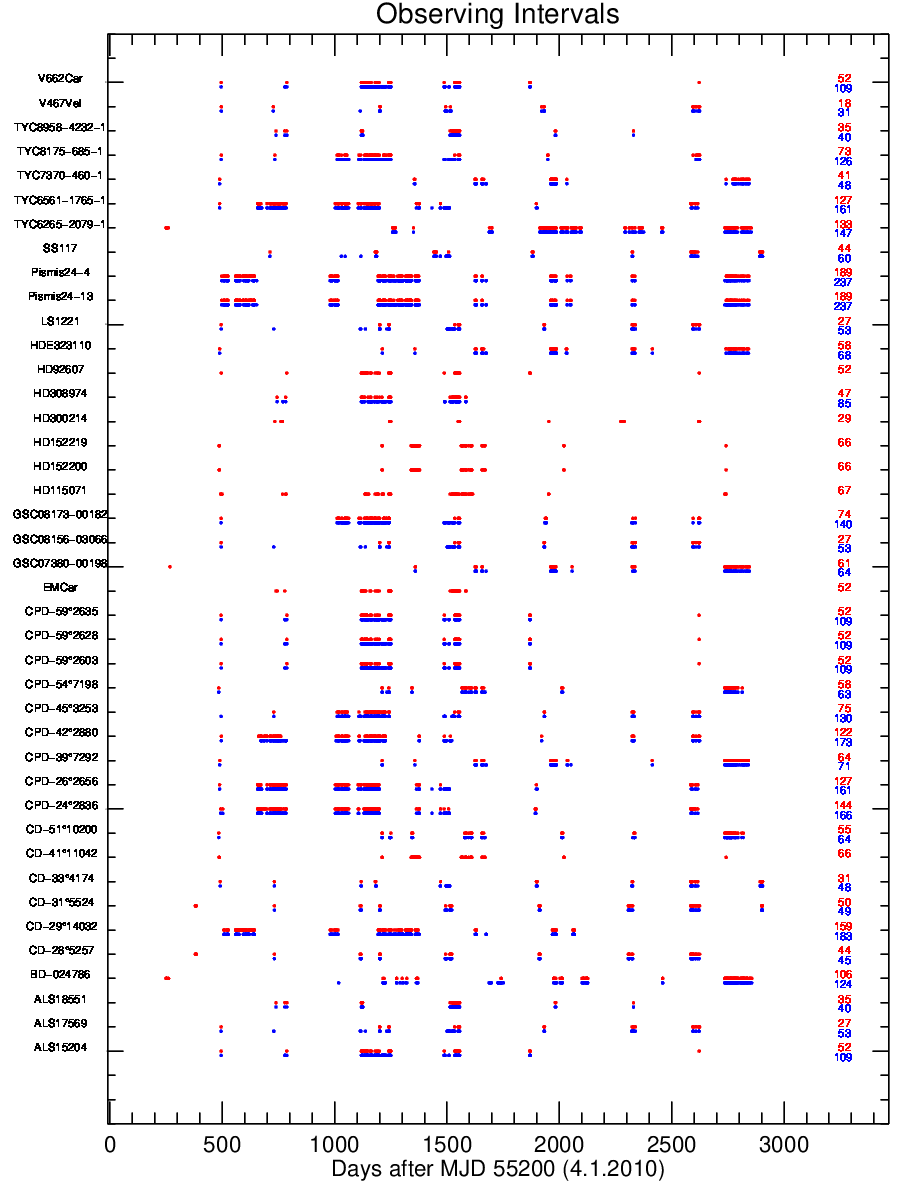}
  \caption{Timeline of the monitoring campaign for the EB systems. Observations performed in filters $r$ and $i$ are represented by blue and red dots, respectively. The numbers at the right correspond to the total amount of observations for each filter.}
  \label{timeline}
\end{figure*}

\begin{figure}
  \centering
  \includegraphics[angle=0,width=\columnwidth]{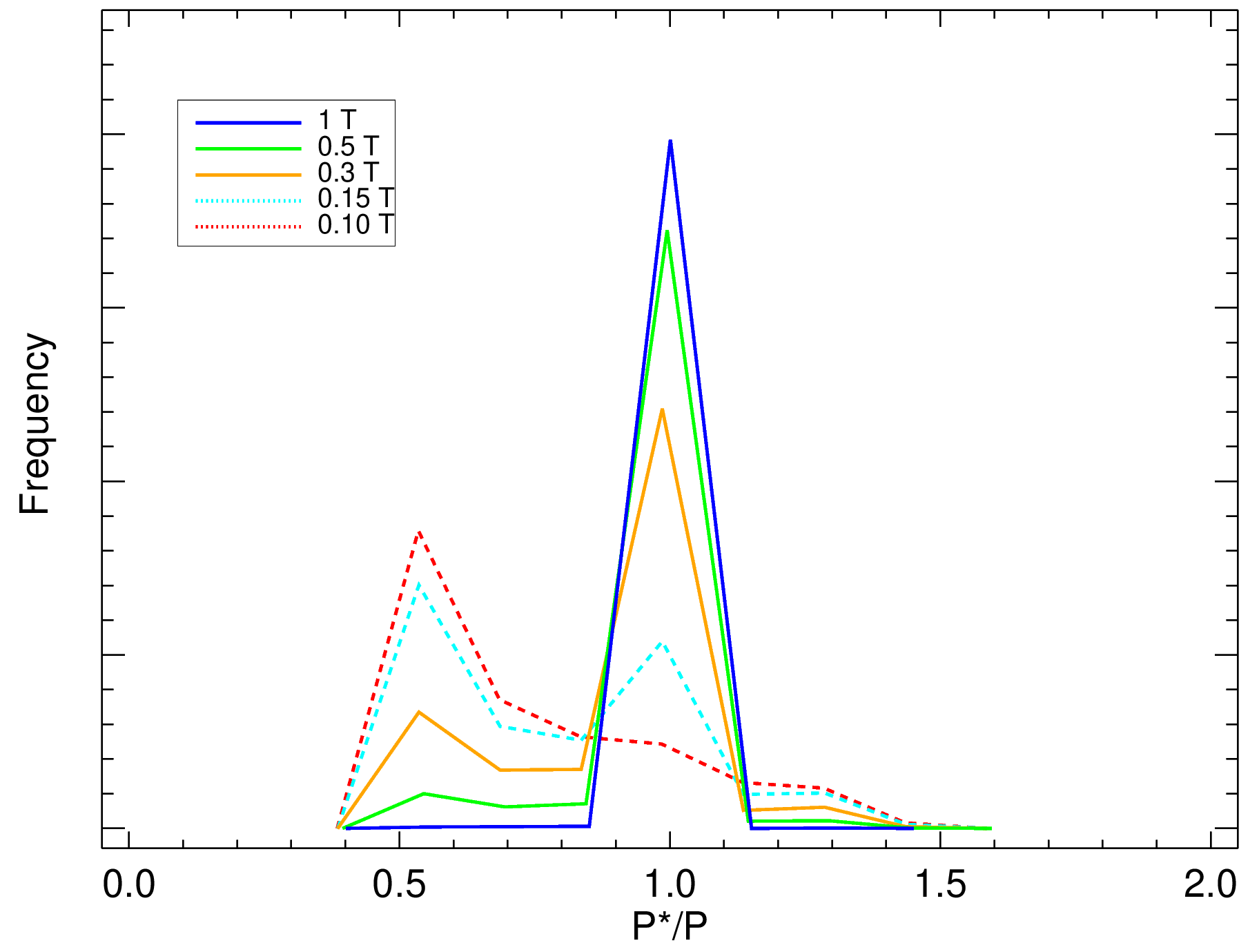}
  \caption{Recovered distributions of the period $P*$ for different time sampling and number of observations $T$.}
  \label{sampling_test}
\end{figure}

In order to test the efficiency of the PDM algorithm, and to identify potential biases with respect to the true period, we create 2000 randomly sampled light curves with different number of observations $T$ for the EBs CPD\,$-$\,59\degr\,2603. We use the sloan $r$ light curve with a total of 109 data points and which yields a period $P = 2.152851 \pm 0.009556$ as determined in section~\ref{sec:discussion}. For each of the 2000 light curves, we
calculate the period $P*$ using the PDM algorithm. The recovered $P*$ distributions are shown in Figure
\ref{sampling_test}. We find that the algorithm performs reliable even when using only $\sim30\%$ of the original data points
(brown solid line). The centroid of the distributions are clearly biased to shorter periods if the amount of observations
decrease to $\sim15\%$ (magenta dotted line), although still the algorithm is able to recover the true period in about $40\%$ of
the cases. The extreme situation can be seen if the amount of observations decrease by $\sim90\%$. For that case, the distribution is biased to almost half of the true period $P$. As already mentioned in Paper 1, the reliability of the PDM to work with only a moderate number of data points is a crucial advantage over other algorithms that strongly depends on the amount of observations (e.g. Lomb-Scargle periodogram by \citealt{1982ApJ...263..835S}, analysis of variance by \citealt{1999ApJ...516..315S}).

\section{Individual sources with EBs models}

In the following, we discuss the multiple high-mass systems in more detail and describe their properties as obtained from the light curve analysis.

\subsection{CPD\,$-$\,24\degr\,2836}

CPD\,$-$\,24\degr\,2836 (CD\,$-$\,24\degr\,5918, TYC\,6544-559-1) was catalogued as O9-B0 by \cite{1975A&AS...21..193N} with the remark that the spectrum was "disturbed by overlap". The blue magnitude is given as $B = 11.3$.
The Tycho-2 catalog (\citealt{2000A&A...355L..27H}) give photometric values $B = 11.25$ and $V = 11.17$, while our photometry yields $B = 11.40$ and $V = 11.28$. These differences might indicate a faint color variability.
Our $UBV$ values yield a photometric spectral type of B7\,V, significantly later than the classification from the objective prism observation. This large deviation may be due to both photometric errors and/or the indicated color variability. Additionally, a secondary of lower mass may enhance this effect. We therefore trust the spectroscopic classification and adopt a spectral type of B0\,V with an effective temperature of $T_1 = 29230$\,K.

The system shows well-defined eclipses favouring a detached configuration with an inclination $i = 57.3^{\circ}$ where both components fill up their Roche lobes at about 86\% and 83\% (Fig.~\ref{model_apa}). The temperature of the secondary component is $T_2 = 19520$\,K compatible with a spectral type B2.5\,V.

\subsection{CPD\,$-$\,26\degr\,2656}

The star CPD\,$-$\,26\degr\,2656 (CD\,$-$\,26\degr\,5070, TYC\,6561-87-1) was classified as a double-lined binary (B1:\,V:) by \cite{1966ArA.....4...65L} displaying photometric variability; the quoted photometry is $U = 10.64$, $B = 11.28$, and $V = 11.0$.
The star is also contained in the spectral survey of the Southern Milky Way \citep{1975A&AS...21..193N} and listed there as No.\,79; the blue magnitude is $B = 11.2$ and the spectral classification is OB:.
\cite{1981A&AS...45..193P} give a tentative spectral classification of B0\,V based on the strength of H$\gamma$ and H$\delta$.
Eventually, \cite{1983MNRAS.205..241R} list the star as B5\,V; the quoted photometry is $U = 10.57$, $B = 11.18$, and $V = 10.95$.
Our own photometry yields $U = 10.55$, $B = 11.18$, and $V = 10.95$ in agreement with the previous values and suggests a photometric spectral type of B1\,V.
Summarizing the spectroscopic results we prefer a B1-type star which agrees perfectly with its dereddened $UBV$ colors (Fig.~\ref{UBV diagram}). We therefore fixed the effective temperature of the primary star to $T_1 = 25570$\,K.

The eclipsing nature of CPD\,$-$\,26\degr\,2656 was already established by \cite{2003AcA....53..341P} using data from the All Sky Automated Survey (ASAS~II) catalogue of variable stars. The author determined a photometric period of $2.6990$ days and suggested a detached EB configuration; we obtain the same period. The light curves in both filters show similar variations with the primary and secondary minimum having different depths $\Delta D_{p} = \Delta D_{p,min} - \Delta D_{max}= 33\%$ and $\Delta D_{s} = \Delta D_{s,min} - \Delta D_{max} = 28\%$. This can be interpreted by a temperature difference of the components. The model fit of the light curve yields an effective temperature of $T_2 = 23250$\,K for the secondary component corresponding to a spectral type of about B2\,V.

The primary component fills up its Roche lobe at $\sim 65\%$ while the secondary at $\sim 74\%$ favouring a detached configuration with an orbital inclination $i = 79.7^{\circ}$ (Fig.~\ref{model_apa}).

\subsection{TYC\,6561-1765-1}

TYC\,6561-1765-1 was classified as O9.5\,V: by \cite{1981A&AS...45..193P}, their No.\,1469; however, this designation was regarded as doubtful by the authors. The Tycho-2 catalog (\citealt{2000A&A...355L..27H}) give photometric values $B = 12.30$ and $V = 11.97$. Our own photometry yields $B = 12.52$ and $V = 12.30$ suggesting some variability. Including our $U$-value of 12.13 we obtain a photometric classification of B3\,V. The photometric agreement with the Tycho-2 catalog leads us to fix the effective temperature of the primary star to $T_1 = 18445$\,K.

The light curves in both filters show similar variations with amplitudes $\Delta D_{p} = 6\%$ and $\Delta D_{s} = 5\%$ for the primary and secondary minimum respectively. The primary component fills up its Roche lobe at $\sim 88\%$ while the secondary at $\sim 71\%$ favouring a detached configuration with an orbital inclination $i = 55.7^{\circ}$ (Fig.~\ref{model_apa}). The model fit of the light curve yields an effective temperature of $T_2 = 13150$\,K for the secondary component corresponding to a spectral type of about B7\,V.

\subsection{CD\,$-$\,31\degr\,5524}

The star CD\,$-$\,31\degr\,5524 was originally classified as O9-9.5 by \cite{1975A&AS...21..193N}, their No.\,163. \cite{1992AJ....104..590O} provided $UBV$ photometry with $V = 11.37$ and colors $B-V = 0.26$ and $U-B = -0.52$. Subsequent observations performed
by \cite{1990AJ....100..737R} yield consistent photometric results with $V = 11.40$ and colors $B-V = 0.27$ and $U-B = -0.59$; the
colors are compatible with a reddened B1\,V. star (Fig.~\ref{UBV diagram}). We therefore fixed the effective temperature of the
primary star to $T_1 = 25570$\,K. The primary and secondary components fill up their Roche lobes to about 97\% and 95\%, respectively,
favouring a near-contact configuration with an orbital inclination of $i = 53.2^{\circ}$. The 3D view of the Roche geometry of the
system ($\phi = 0.54$) shows that most likely the future mass transfer will create a new contact binary system (Fig.~
\ref{model_apa}).

\subsection{CPD\,$-$\,42\degr\,2880}

Objective prism spectra of CPD\,$-$\,42\degr\,2880 (TYC\,7687-444-1, LS\,1142) yield the classification O9.5-B2 \citep{1975A&AS...21..193N}, their No.\,300, with a blue magnitude of $B = 11.7$.
The Tycho-2 catalog (\citealt{2000A&A...355L..27H}) give photometric values $B = 11.70$ and $V = 11.17$ while \cite{1977A&AS...27..343D} give $U = 11.53$, $B = 11.80$, and $V = 11.10$.
Our own photometry yields $U = 11.53$, $B = 11.80$, and $V = 11.10$ in perfect accord with the previous values and suggests a photometric spectral type of B0\,V.
This agreement with the classification from the objective prism spectrum leads us to fix the effective temperature of the primary star to $T_1 = 29230$\,K.

\begin{figure*}
  %\centering
  \includegraphics[width=0.67\columnwidth]{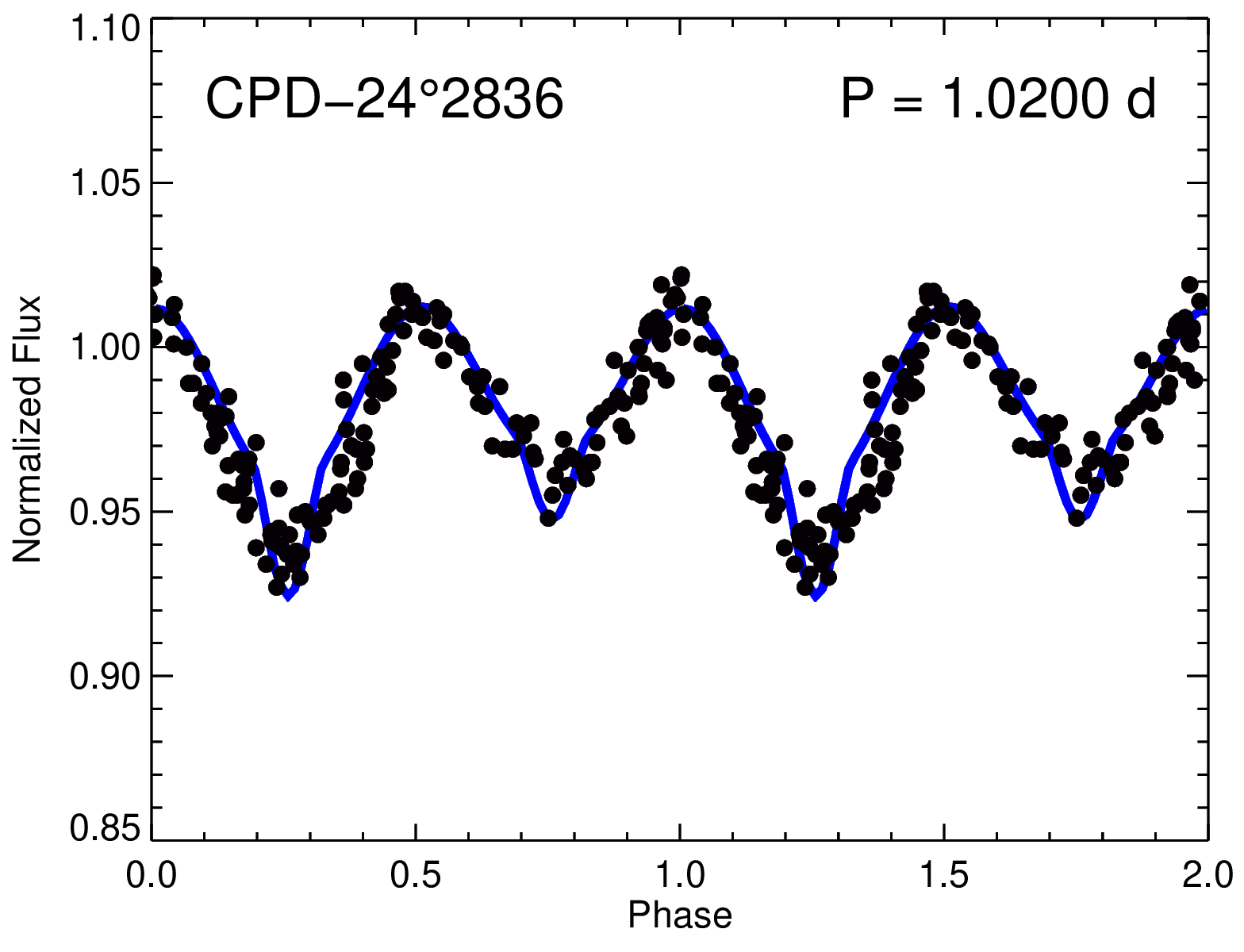}
  \includegraphics[width=0.67\columnwidth]{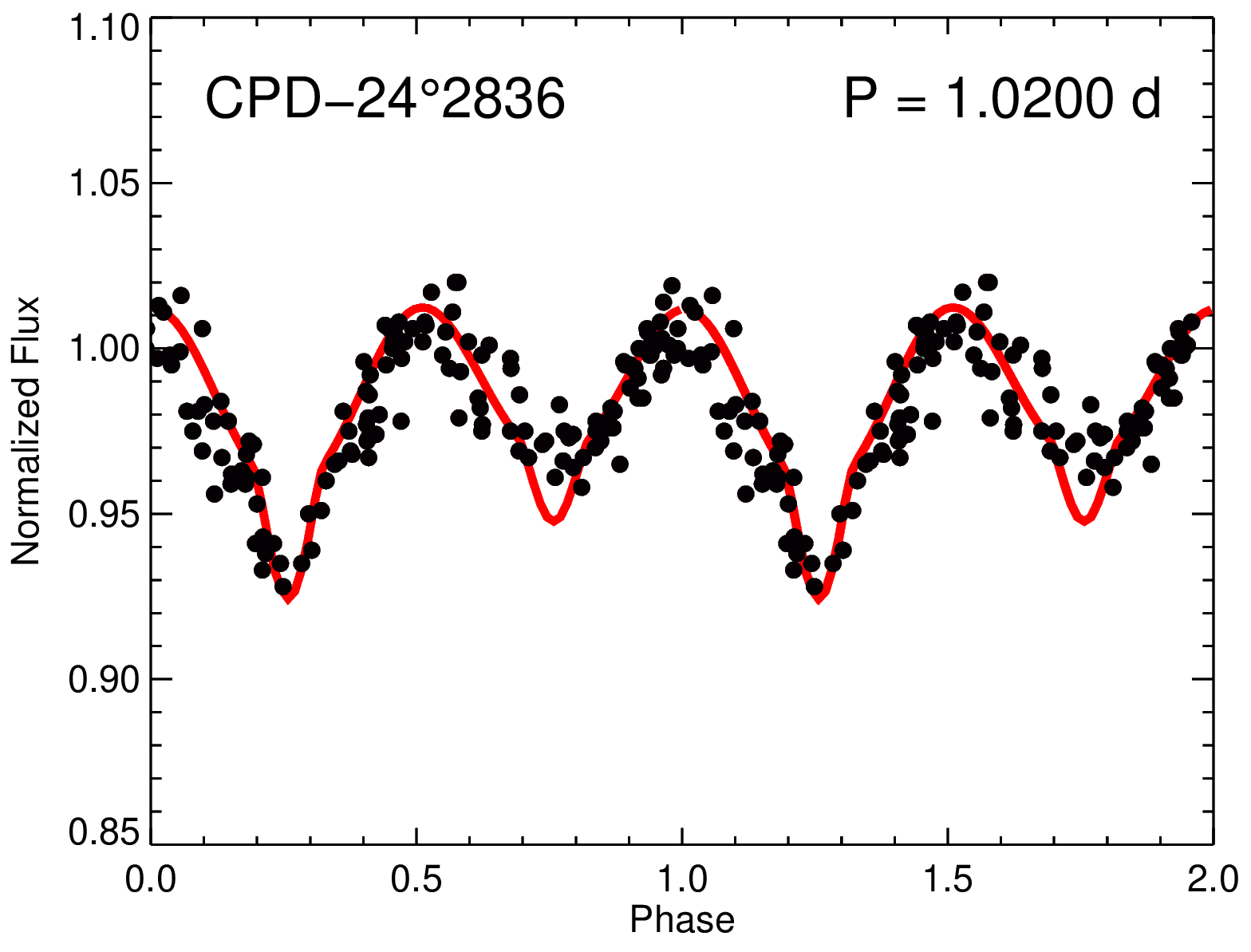}
  \includegraphics[width=0.67\columnwidth]{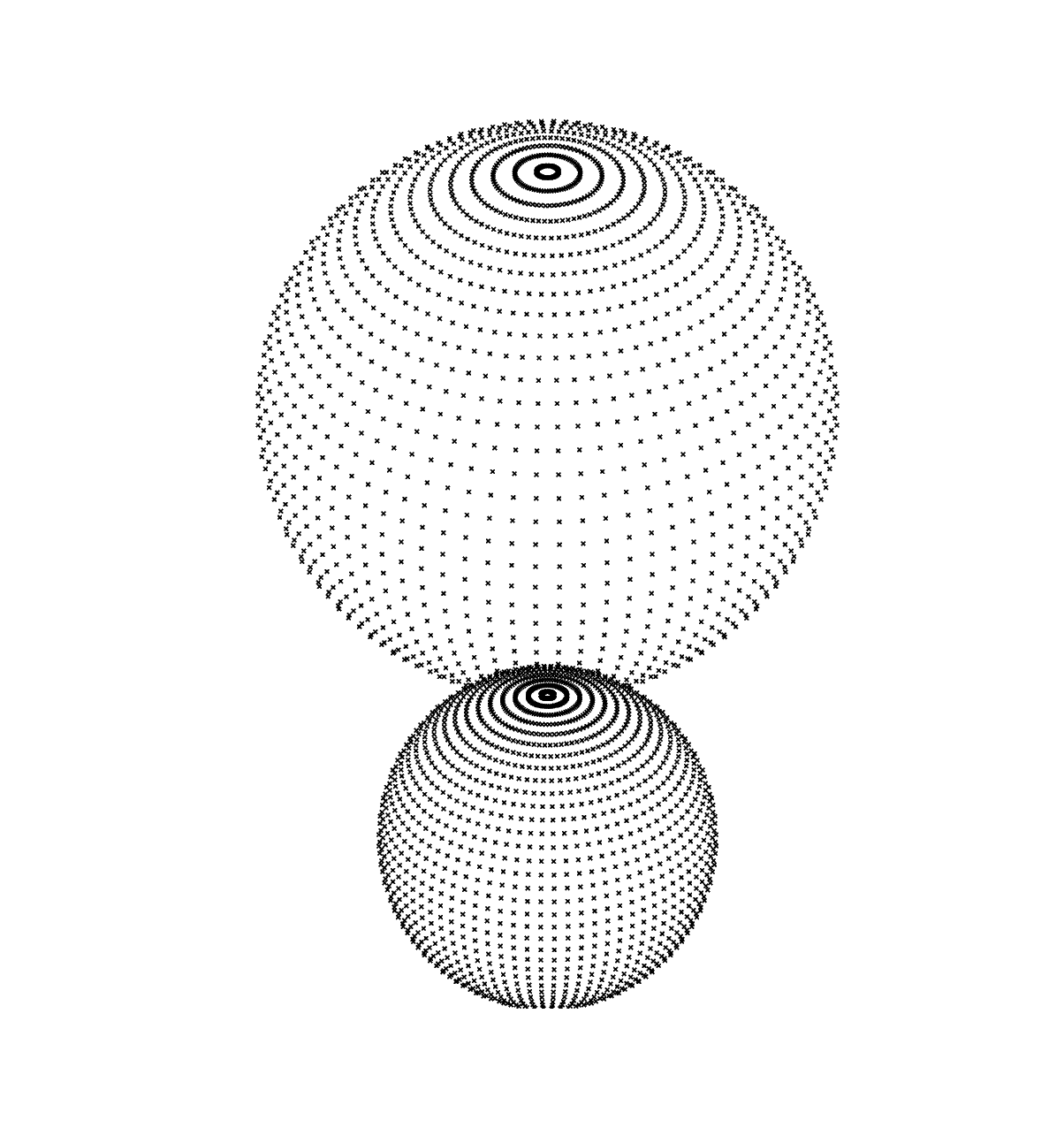}
  \includegraphics[width=0.67\columnwidth]{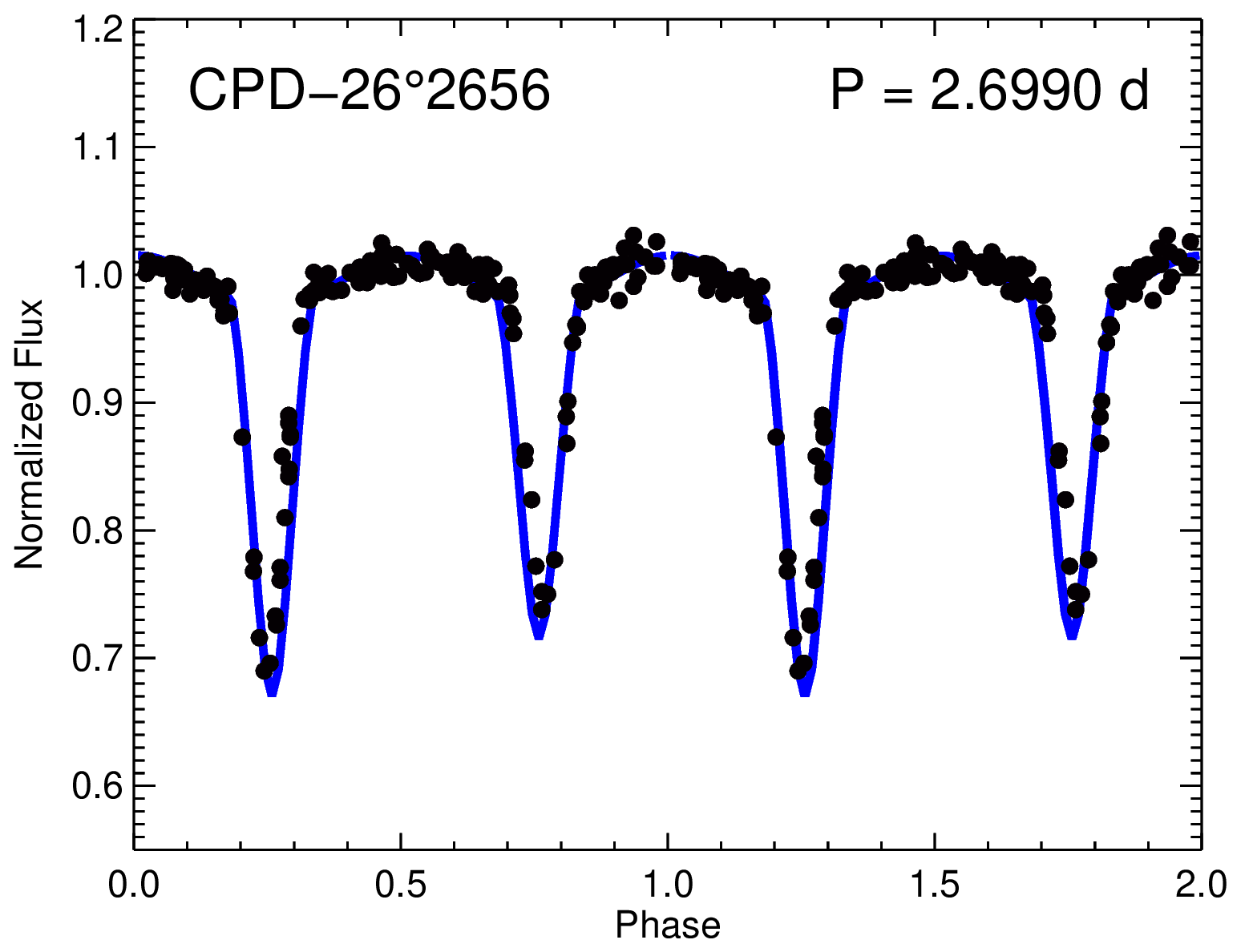}
  \includegraphics[width=0.67\columnwidth]{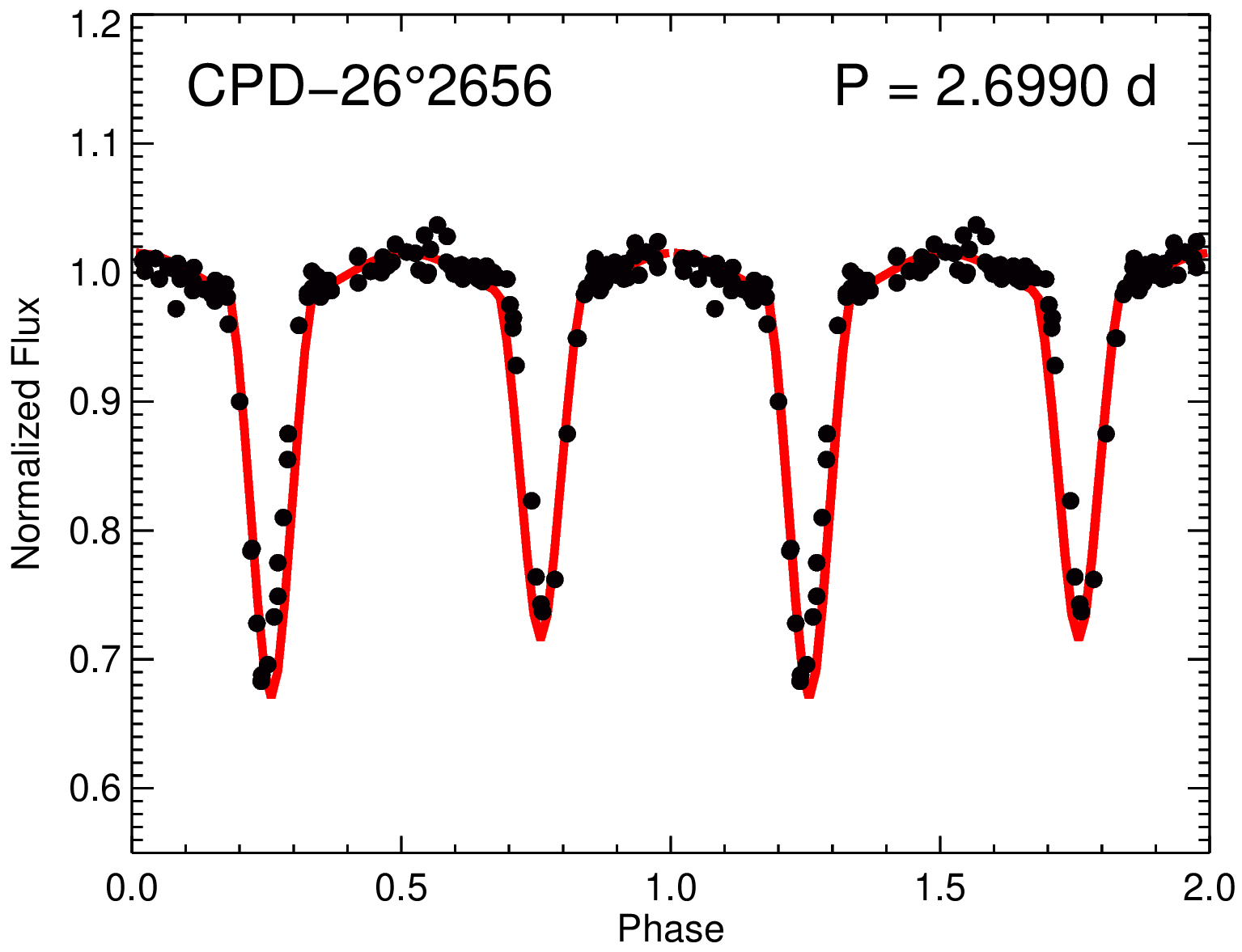}
  \includegraphics[width=0.67\columnwidth]{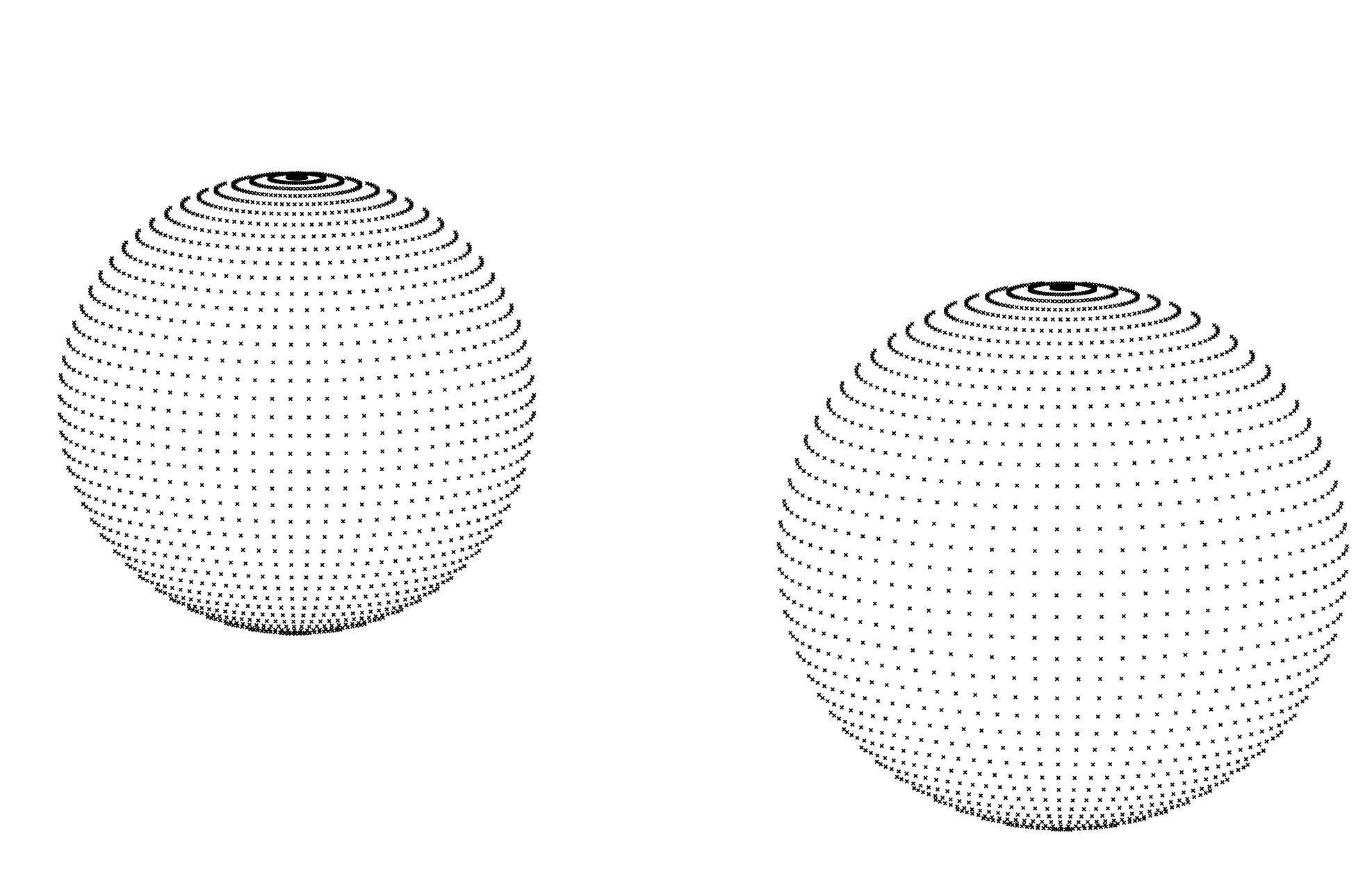}
  \includegraphics[width=0.67\columnwidth]{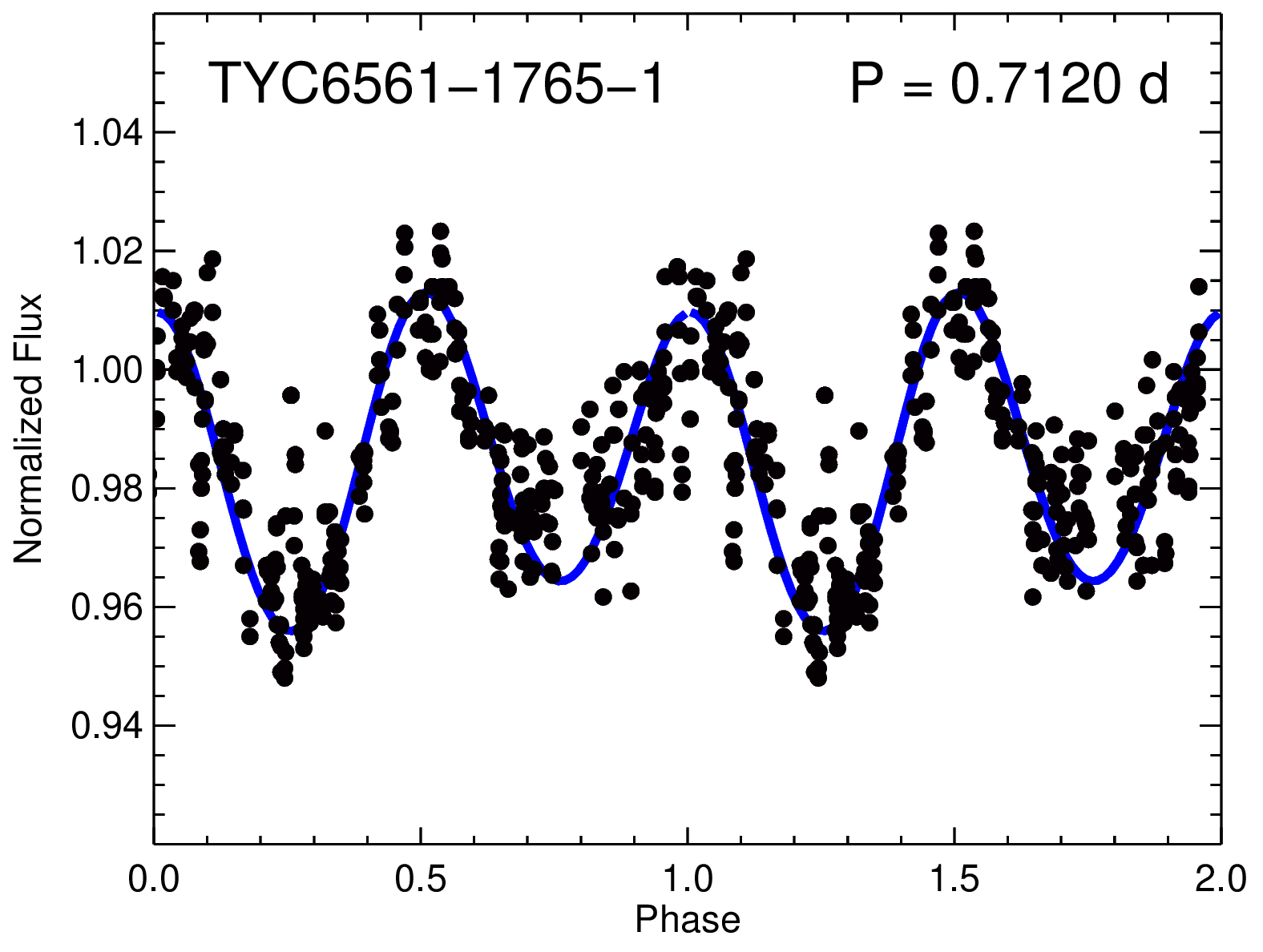}
  \includegraphics[width=0.67\columnwidth]{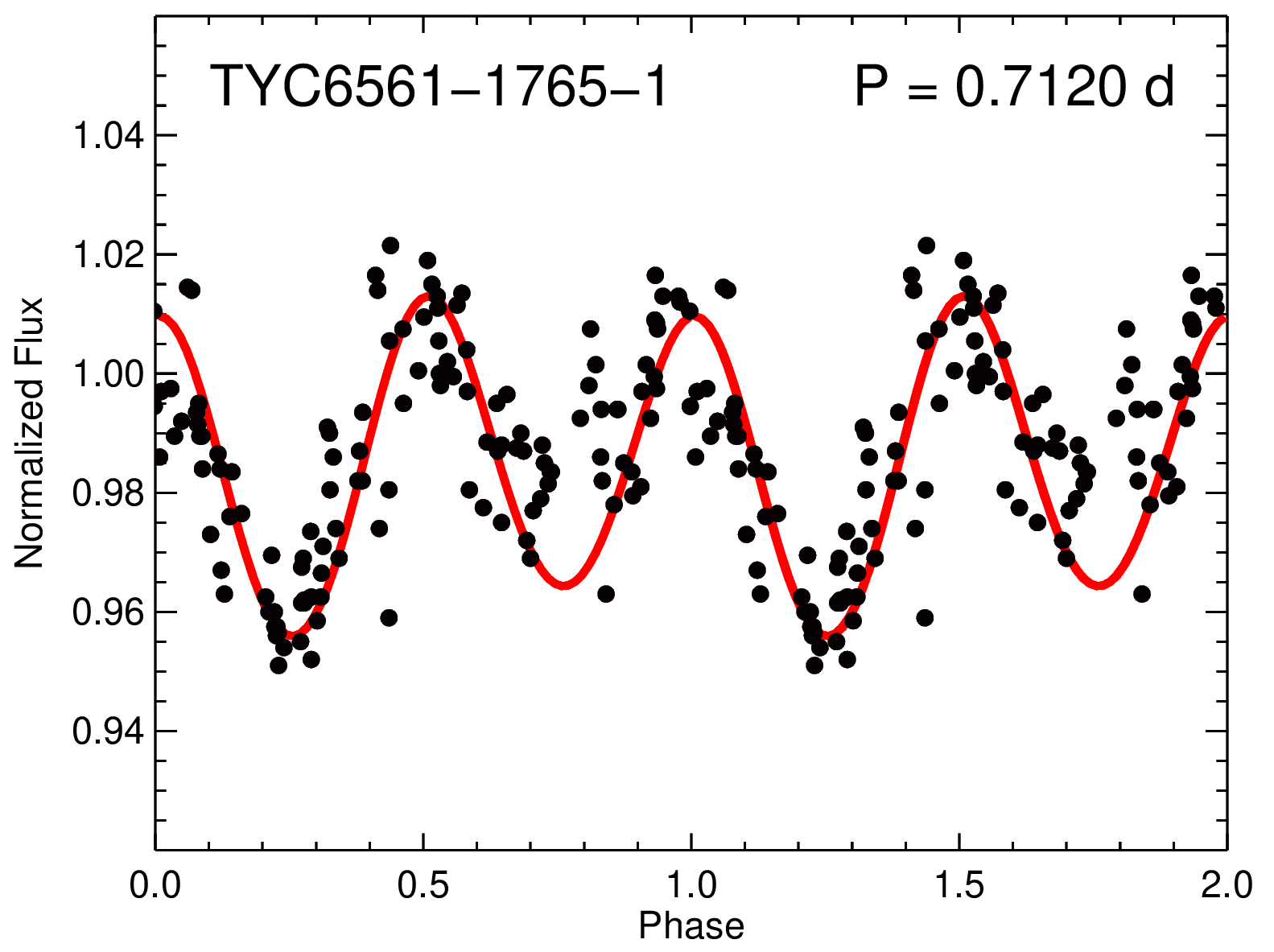}
  \includegraphics[width=0.67\columnwidth]{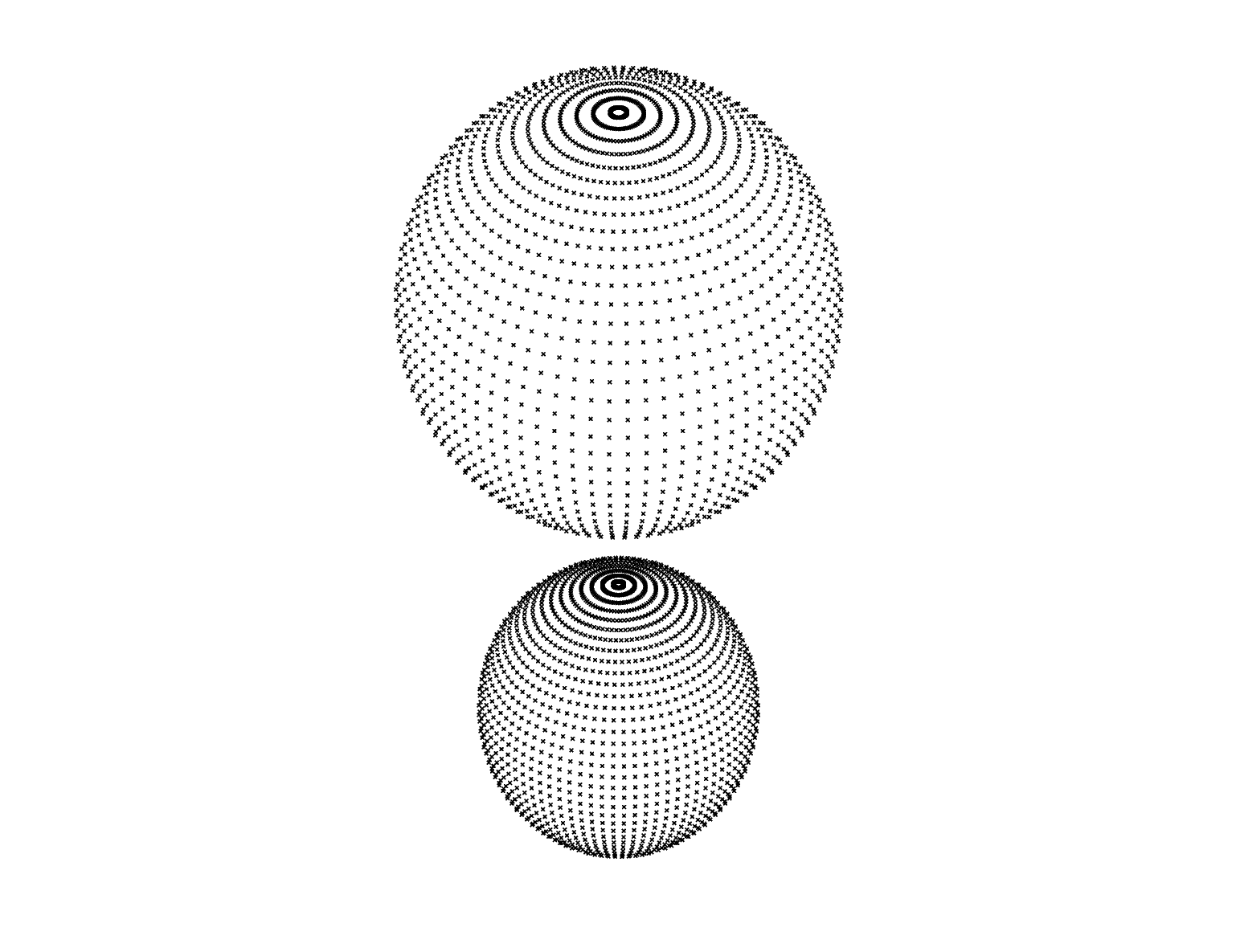}
  \includegraphics[width=0.67\columnwidth]{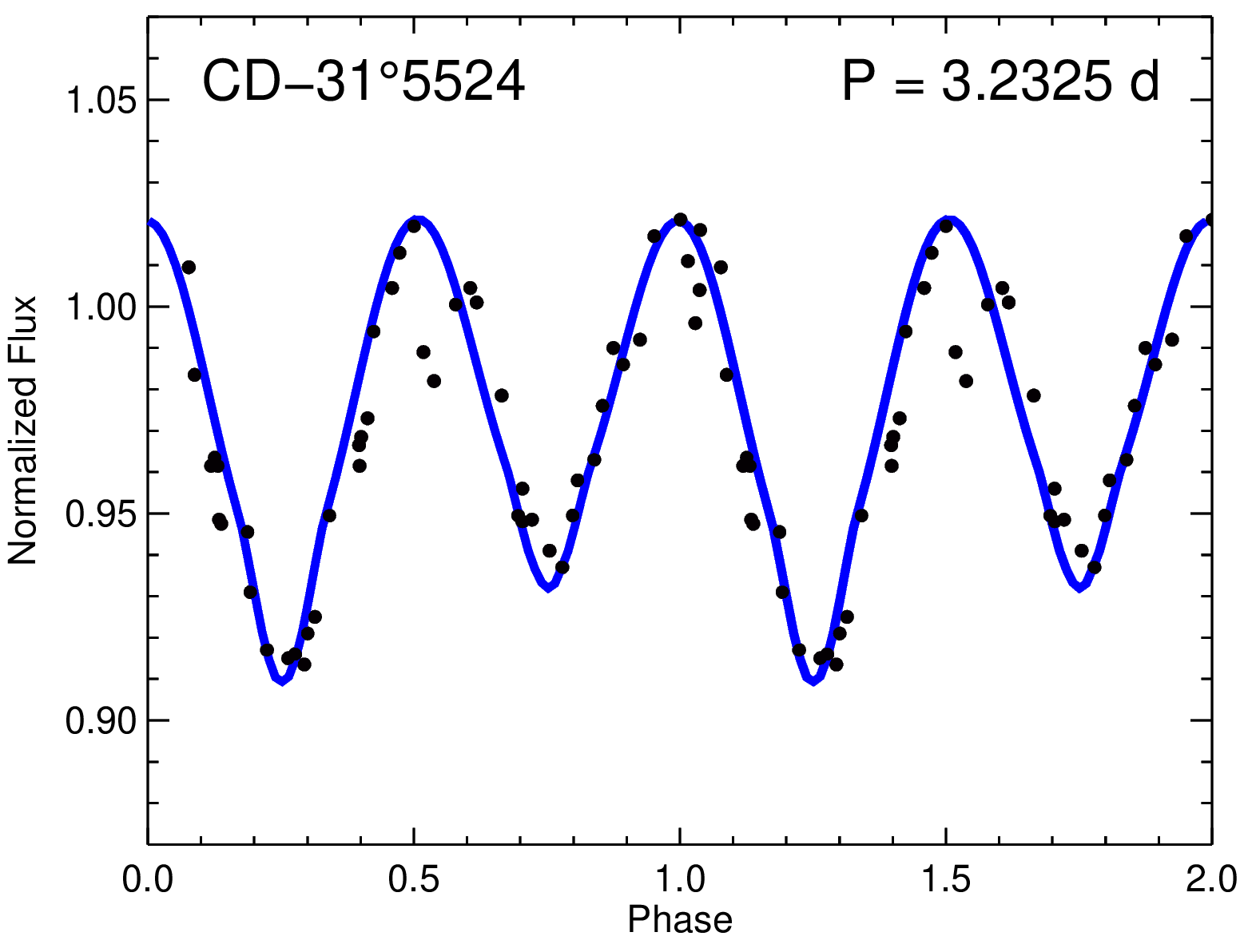}
  \includegraphics[width=0.67\columnwidth]{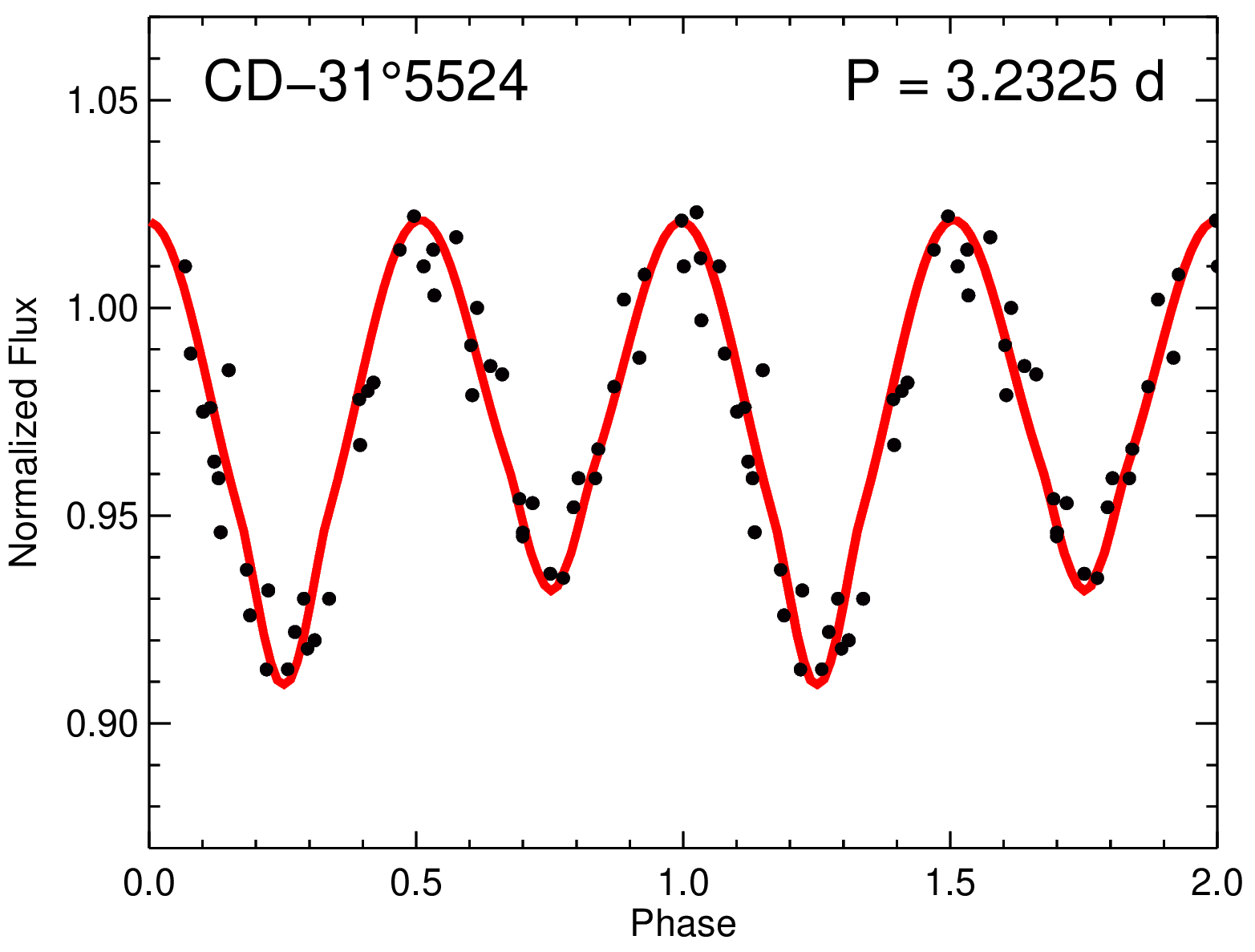}
  \includegraphics[width=0.67\columnwidth]{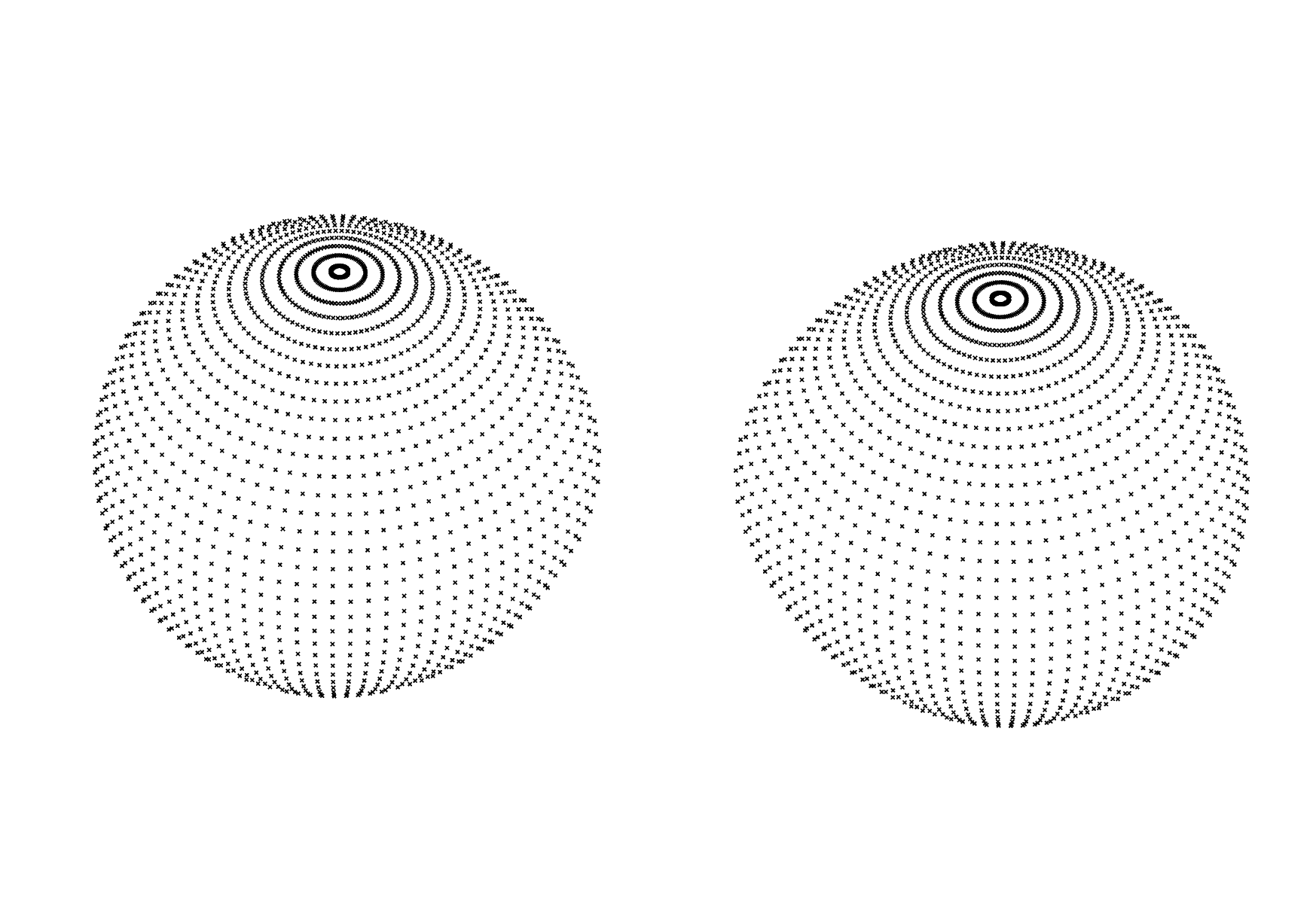}
  \includegraphics[width=0.67\columnwidth]{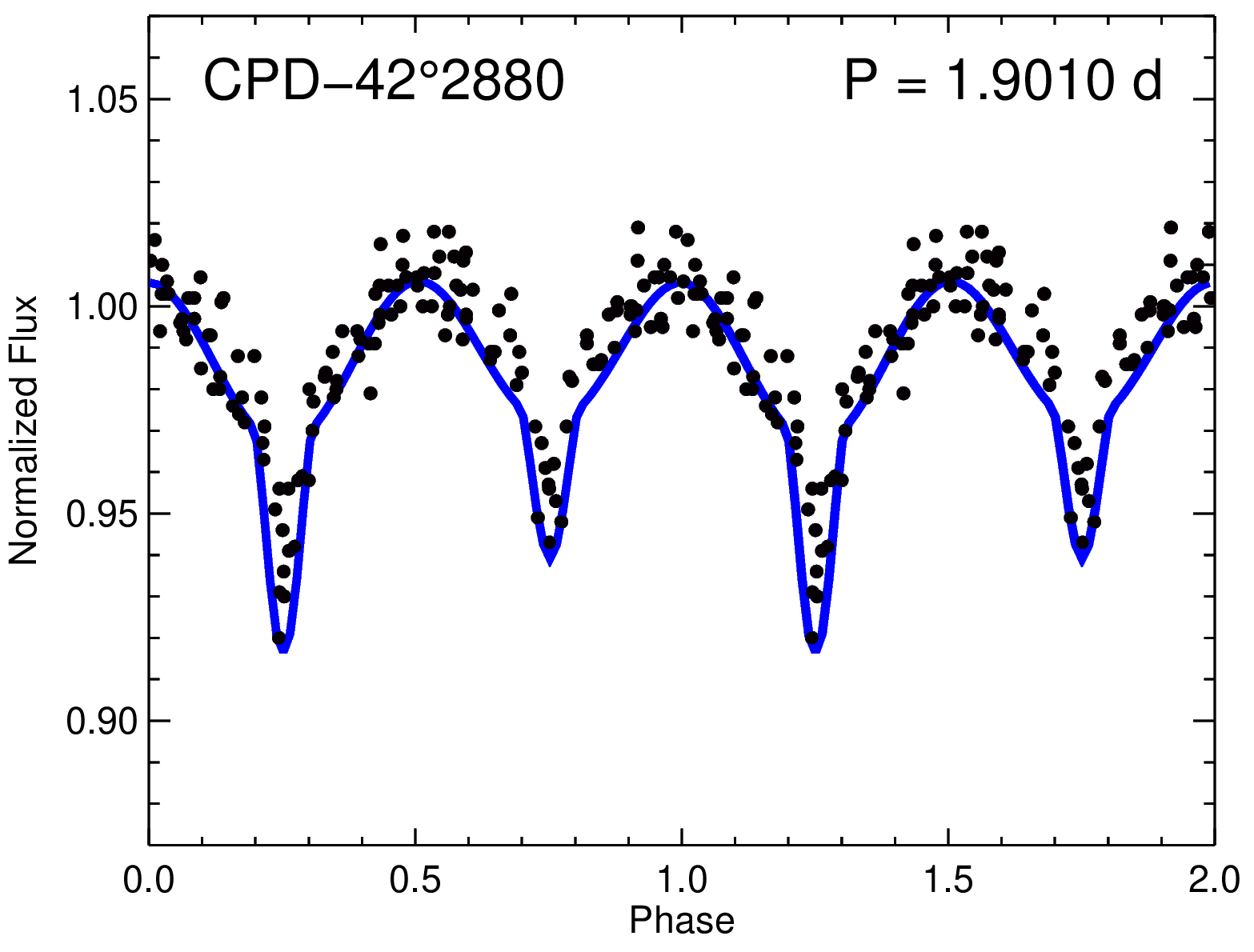}
  \includegraphics[width=0.67\columnwidth]{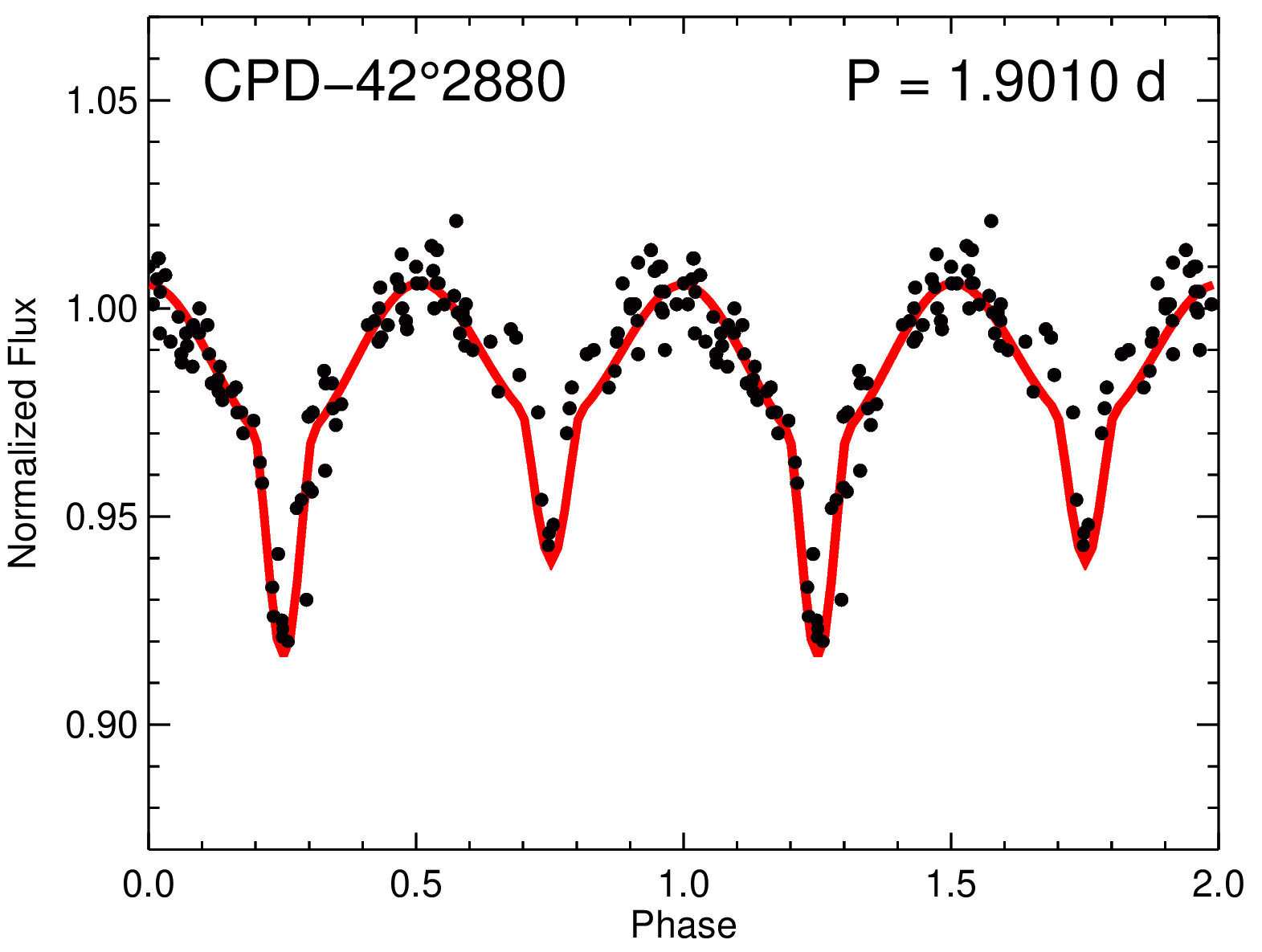}
  \includegraphics[width=0.67\columnwidth]{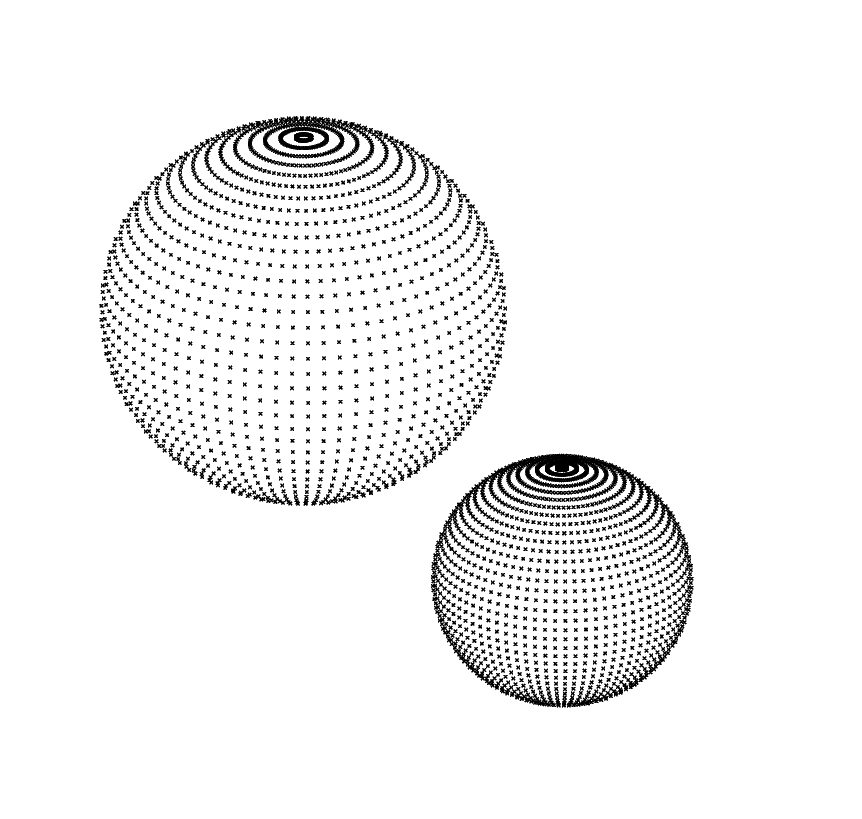}
  \caption{Light curves in $r$ (blue) and $i$ (red), and 3D view of the binary system.}
  \label{model_apa}
\end{figure*}

The system shows well-defined eclipses favouring a detached configuration, where both components fill up their Roche lobes at about 73\% and 72\%, respectively. The temperature of the secondary component is $T_2 = 22110$\,K equivalent to a B2\,V. As displayed in Fig.~\ref{model_apa}, the light curves show nearly identical amplitude variations with primary and secondary minimum eclipses at different depths ($\Delta D_{p} = 8\%$, $\Delta D_{s} = 6\%$). This behaviour is characteristic for an eclipsing binary with different effective temperatures; the primary minimum is due to the eclipse of the more luminous star by the less luminous companion. The primary and secondary minima are separated symmetrically by 0.6 in phase suggesting a circular orbit.

\subsection{CPD\,$-$\,45\degr\,3253}

CPD\,$-$\,45\degr\,3253 (CD\,$-$\,45\degr\,4706, TYC\,8152-206-1, LS\,1207) was classified as B0: by \cite{1975A&AS...21..193N}, their No.\,373; the blue magnitude is listed as $B = 12.0$.
This object is part of a survey searching for faint OB stars in the Vela region (\citealt{1977AJ.....82..474M}). These authors derive a spectral designation of OB$^0$, their star No.\,36, and obtain photometric values of $B = 12.01$ and $V = 10.77$.
\cite{1988JRASC..82..276S} derive a spectral type of O9:, also based on objective-prism spectroscopy; furthermore they list photometric values of $B = 11.99$ and $V = 10.79$.
The Tycho-2 catalog (\citealt{2000A&A...355L..27H}) give photometric values $B = 11.54$ and $V = 10.62$.
Our own photometry yields $B = 11.97$ and $V = 10.88$ suggesting some variability. Including our $U$-value of 12.31 we obtain a photometric classification of B3\,V.
As an average we fix the effective temperature of the primary star to $T_1 = 29230$\,K, corresponding to a B0\,V star.

The primary and secondary minimum eclipses are separated symmetrically by 0.11 in phase at depths of $\Delta D_{p} = 9\%$ and $\Delta D_{s} = 6\%$. The best-fitting model favours a detached configuration with an inclination $i = 64.1^{\circ}$ and requires that both components fill up their Roche lobes at about 73\% and 72\%, respectively (Fig.~\ref{model_apb}). Our calculations yield an effective temperature of $T_2 = 22250$\,K for the secondary star suggesting that this new system is most likely composed of B0\,V + B2\,V.

\subsection{GSC\,08156-03066}

GSC\,08156-03066 is part of the survey of faint OB stars performed by \cite{1977AJ.....82..474M}, their star No.\,4; the authors derived a spectral designation of OB$^0$ and obtained photometric values of $B =13.16$ and $V = 11.79$. \cite{1977AJ.....82..474M} provided $UBV$ photometry with $U = 13.26$, $B = 13.01$, and $V = 11.78$ which is compatible with the previous values and suggests a photometric spectral type of B1\,V. Our own photometry yields $U = 13.32$, $B = 13.18$, and $V = 11.93$ indicating a spectral type of O9.5, slightly earlier than the classification by \cite{1977AJ.....82..474M}. While our $B-V$ color is consistent with the result obtained by \cite{1977AJ.....82..474M}, the $U-B$ color differs by 0.11\,mag. We therefore prefer a spectral type B1\,V with an effective temperature $T_1 = 25570$\,K.

The system shows well-defined eclipses with a period of 2.3318 days. The amplitude variations for the primary and secondary minimum are $\Delta D_{p} = 10\%$ and $\Delta D_{s} = 9\%$ suggesting similar temperatures for both components. The temperature of the secondary component is $T_2 = 24150$\,K suggesting a pair of two B1\,V stars. As displayed in Fig.~\ref{model_apb}, the 3D view of the Roche geometry at phase $\phi = 0.50$ shows that the system has a semi-detached geometry where both stars are close to filling their Roche lobes.

\subsection{GSC\,08173-00182}

GSC\,08173-00182 is also located in the Vela region. \cite{1977AJ.....82..474M} give a classification of OB$^+$ but claim it as doubtful (their No.\,101). The photometric entries for this star are $B = 14.24$ and $V = 12.55$.
Our own photometry yields $B = 14.40$ and $V = 12.79$ suggesting some variability. Including our $U$-value of 15.24 we obtain a photometric classification of B5\,V.
Taking into account that main sequence stars from the OB$^+$ group tend to be of earlier spectral type we adopt a B3\,V classification for the primary. This leaves us with an effective primary temperature of $T_1 = 18445$\,K.

The primary and secondary minimum eclipses are separated symmetrically by 0.15 in phase. The model suggests a detached system (Fig.~\ref{model_apb}).
The temperature of the secondary component is $T_2 = 13920$\,K suggesting that this new system is most likely composed of B3\,V + B6\,V.

\subsection{TYC\,8175-685-1}

The star TYC\,8175-685-1 (ALS\,17580) was classified as OB$^-$ by \cite{1977AJ.....82..474M}, their No.\,143; however, this designation was regarded as doubtful by the authors. The photometric values are $B = 13.35$ and $V = 12.24$.
Muzzio (1979) provided $UBV$ photometry with $U = 13.55$, $B = 13.23$, and $V = 12.10$ which is compatible with the previous values. In the following we adopt a photometric spectral type of B3\,V and fix the effective temperature to $T_1 = 18445$\,K for the primary.

Using data from the ASAS~II archive \cite{2003AcA....53..341P} determined a photometric period of $3.4085$ days. Furthermore, he suggested that TYC\,8175-685-1 forms an eclipsing contact (or almost contact) binary configuration. We obtain the identical photometric period as reported by \cite{2003AcA....53..341P}. The amplitude variations for the primary and secondary minimum eclipses are $\Delta D_{p} = 14\%$ and $\Delta D_{s} = 13\%$, respectively, suggesting again similar temperatures for both components. This yields a B3\,V + B3\,V system.

As displayed in Fig.~\ref{model_apb}, the 3D view of the Roche geometry of the system ($\phi = 0.54$) shows that the system has a semi-detached (near-contact) geometry where both stars are close to filling their Roche lobes. Most likely the future mass transfer will create a new contact binary system.

\subsection{ALS\,18551}

ALS\,18551 is a member of the open cluster Collinder\,228 and was originally classified as an O5 star by \cite{1976A&AS...23..231W} using photometric data. We notice that the ALS\,18551 identification in SIMBAD (\citealt{2000A&AS..143....9W}) is based on a relatively large distance to the given source ($\sim$10.8"). The Website displays an image centred on the corresponding source coordinates based on Aladin Lite (\citealt{2014ASPC..485..277B}) which shows no recognizable object at the given position. Manual inspection of our coordinates shows a matching object in Aladin Lite. The SIMBAD coordinate is referenced by \cite{1976A&AS...23..231W}. Therein, the object matches (epoch B1950) and the given plate image shows the correct source.

\begin{figure*}
  %\centering
  \includegraphics[width=0.67\columnwidth]{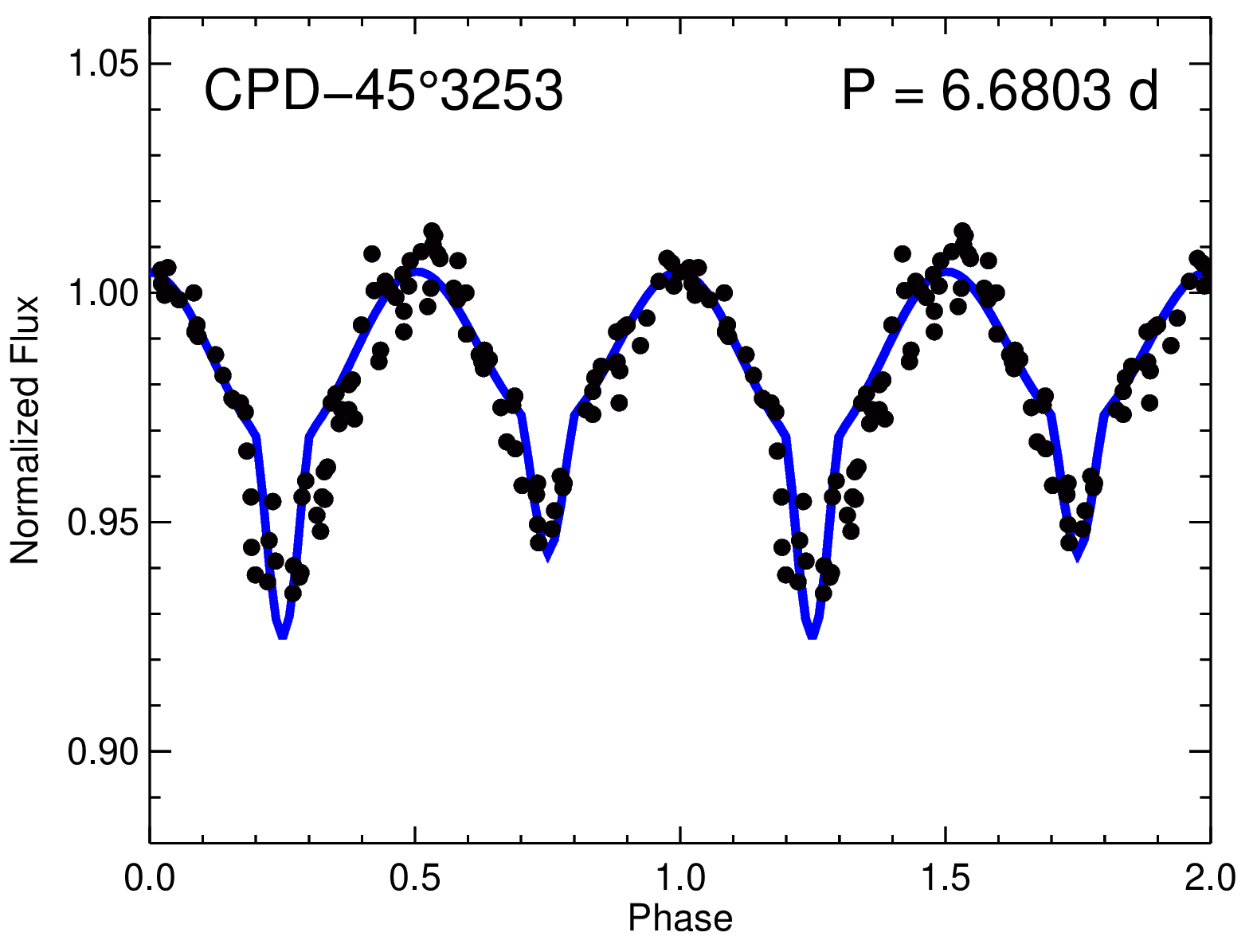}
  \includegraphics[width=0.67\columnwidth]{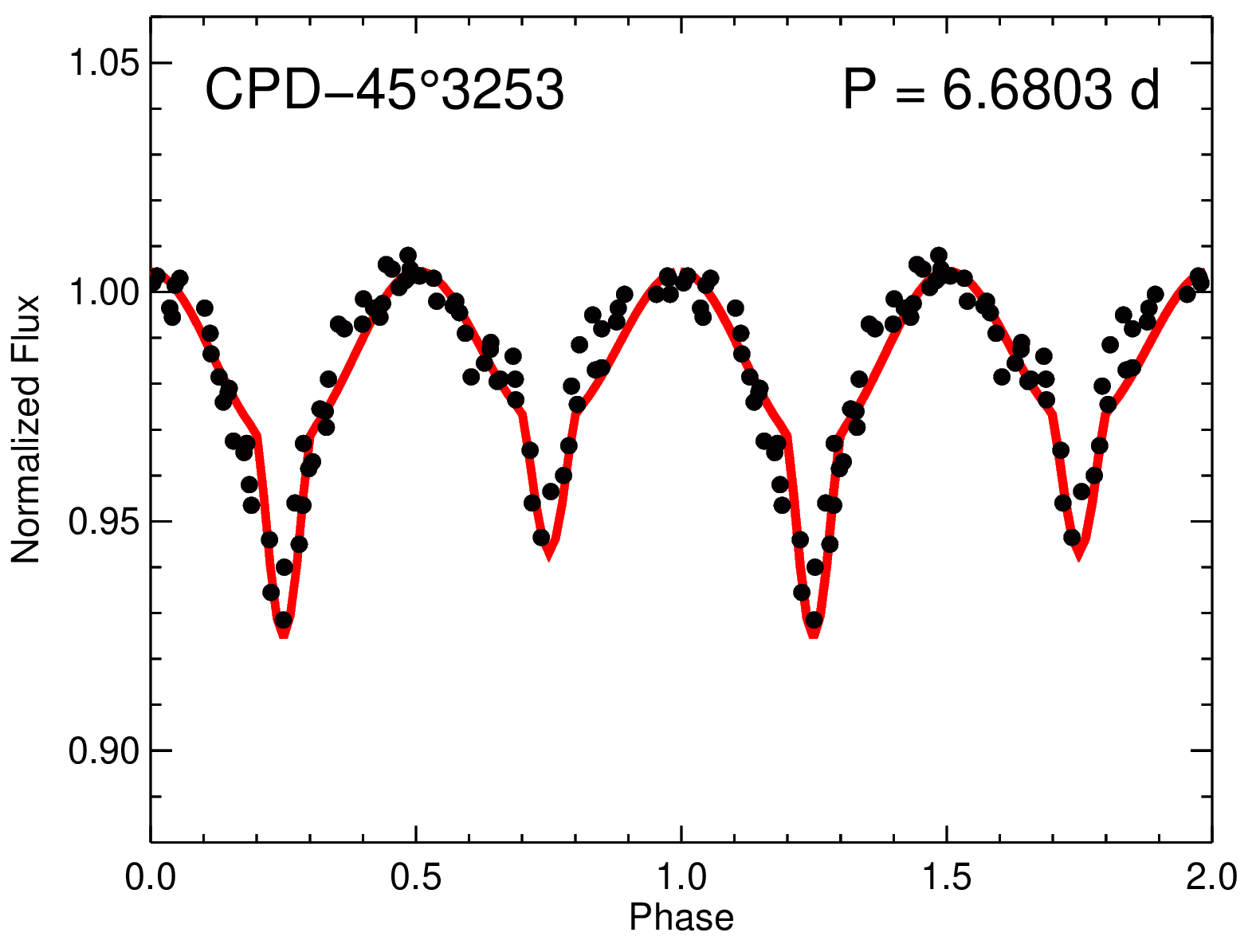}
  \includegraphics[width=0.67\columnwidth]{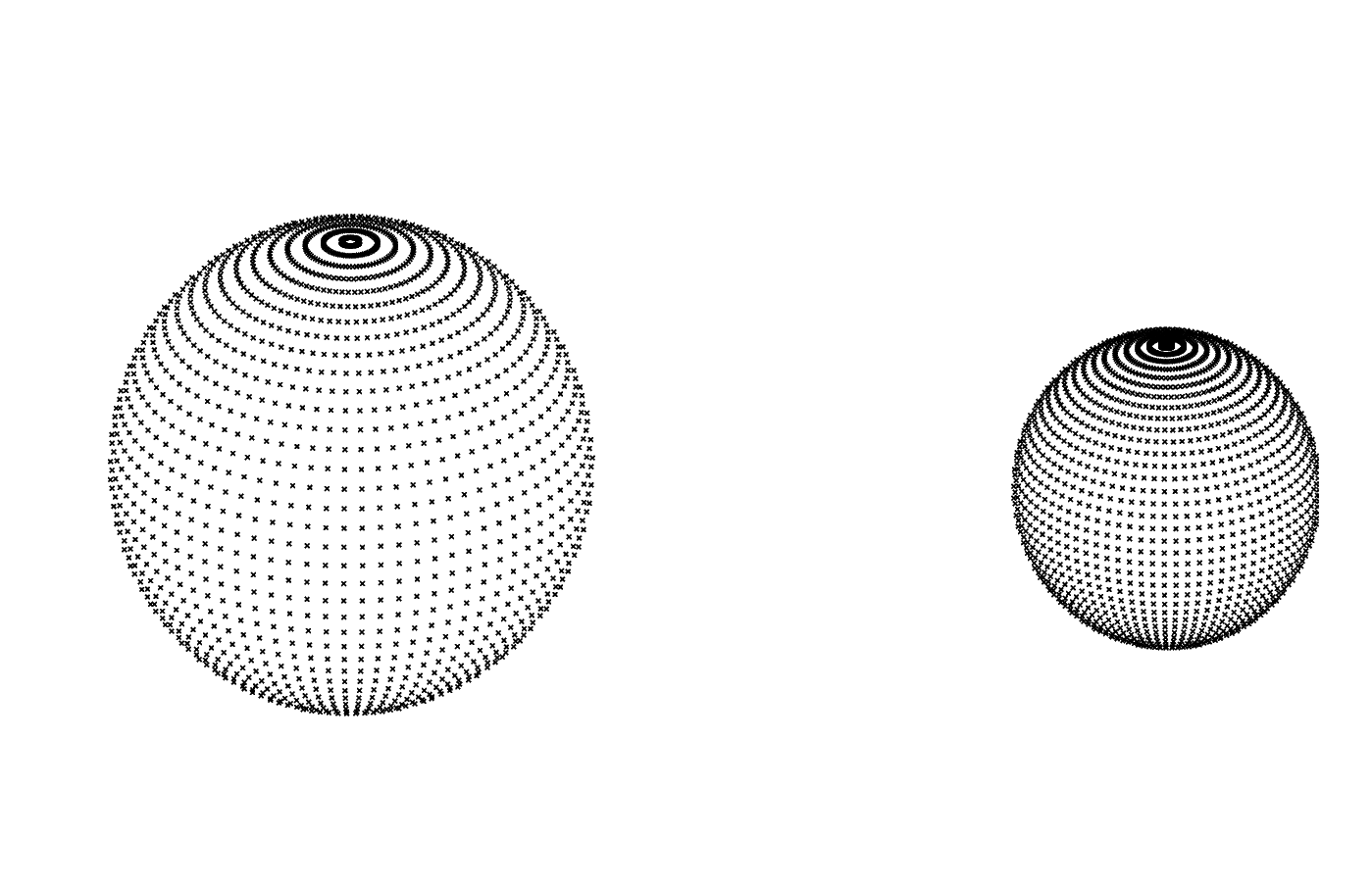}
  \includegraphics[width=0.67\columnwidth]{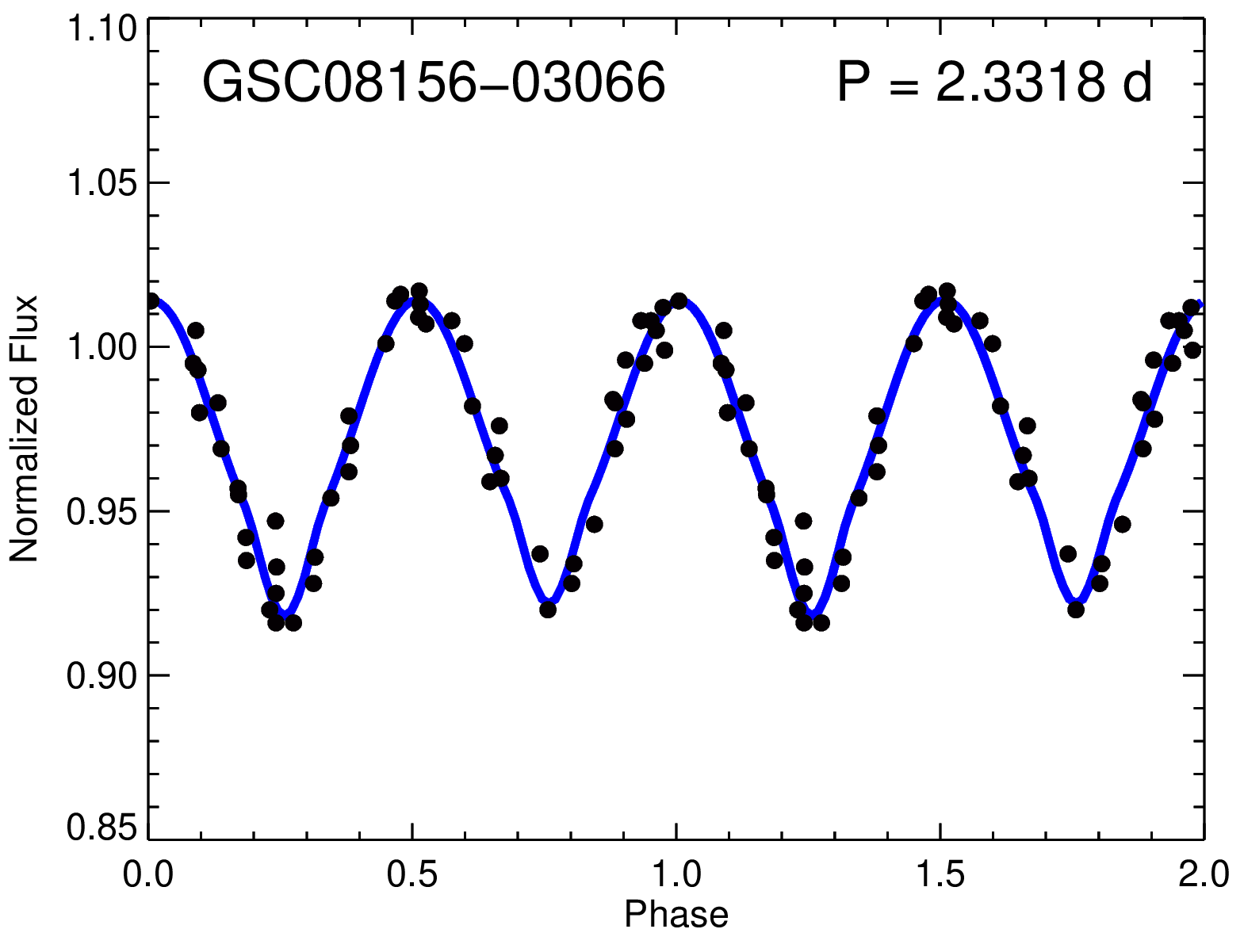}
  \includegraphics[width=0.67\columnwidth]{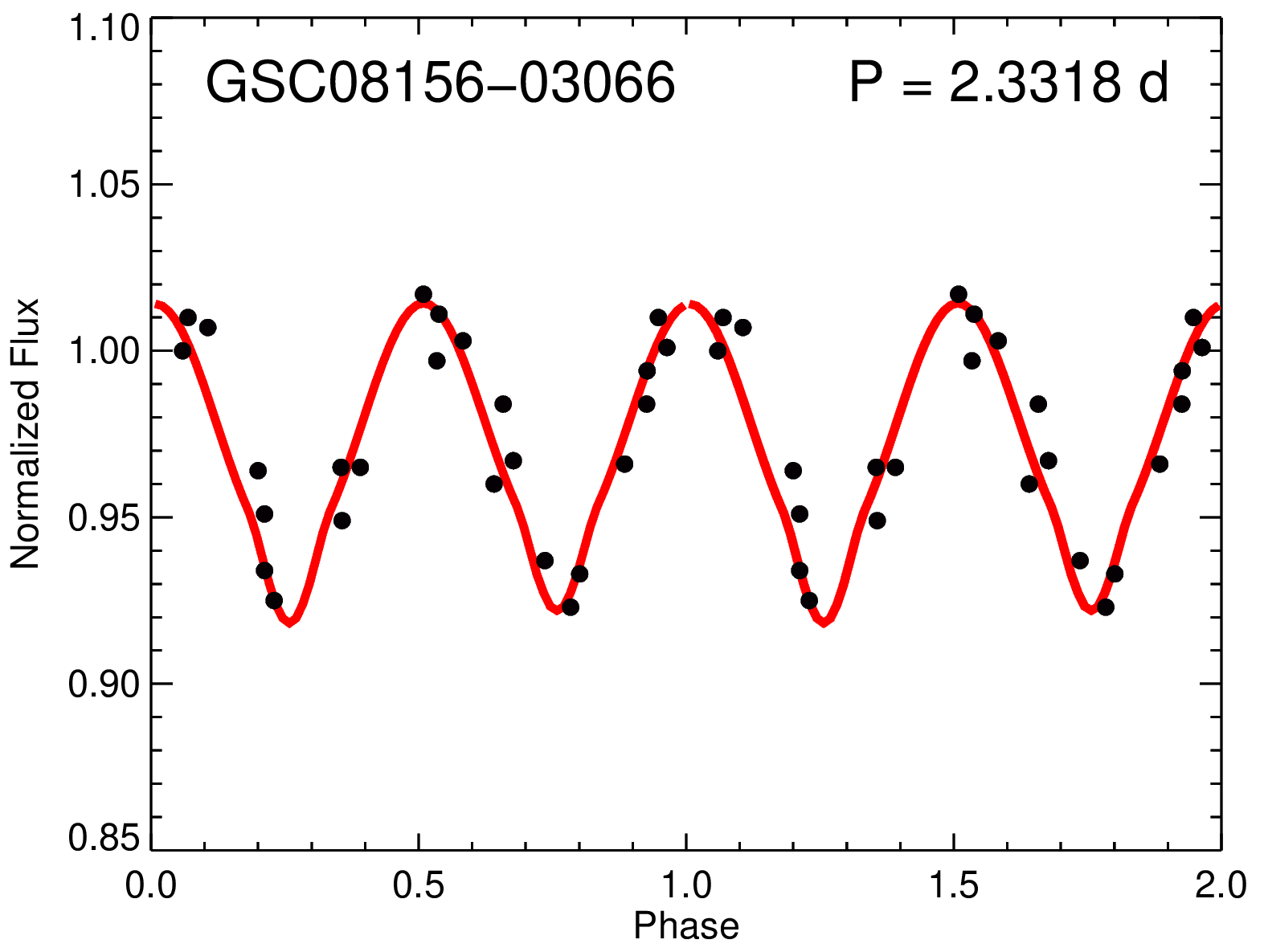}
  \includegraphics[width=0.67\columnwidth]{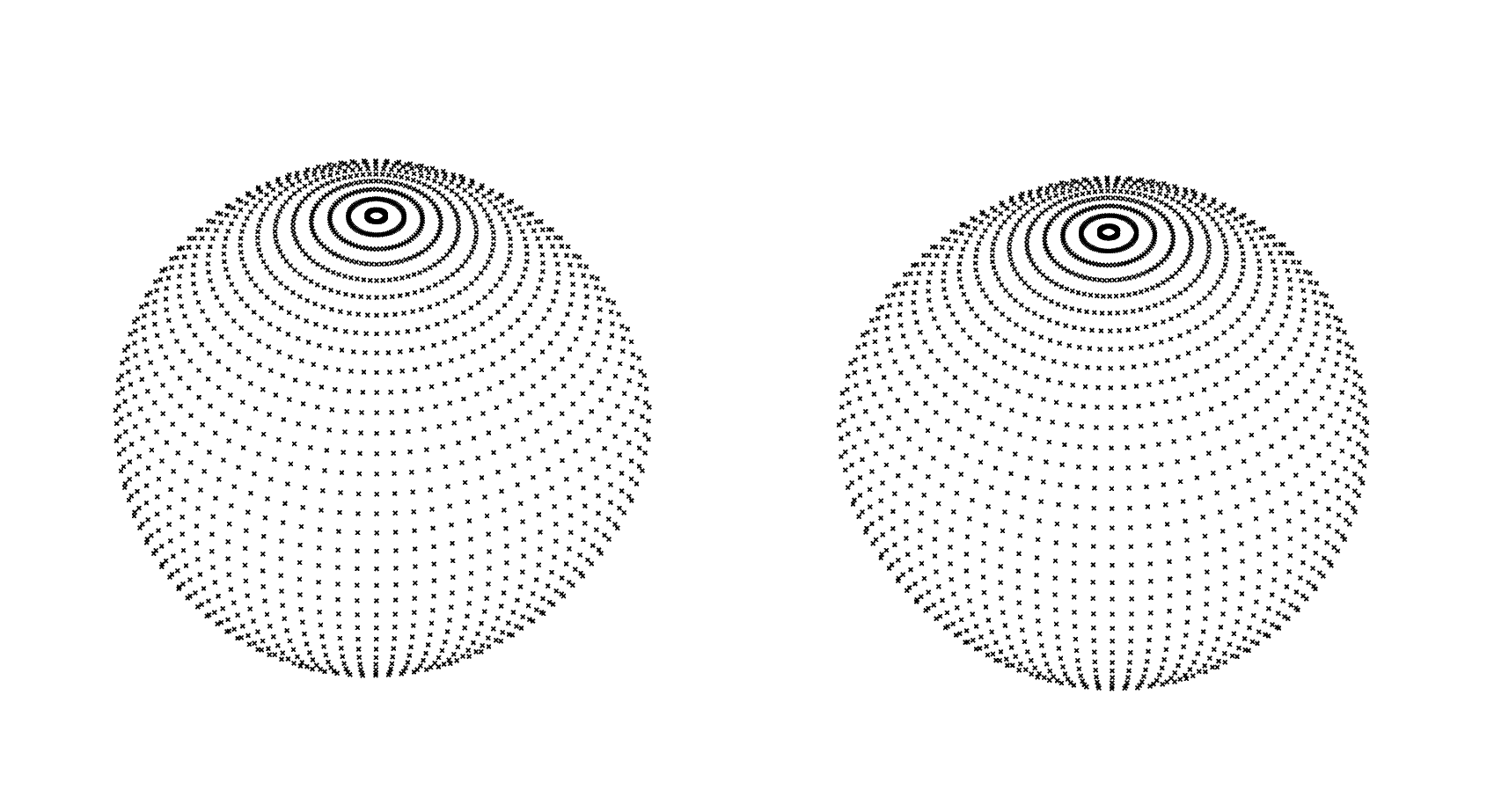}
  \includegraphics[width=0.67\columnwidth]{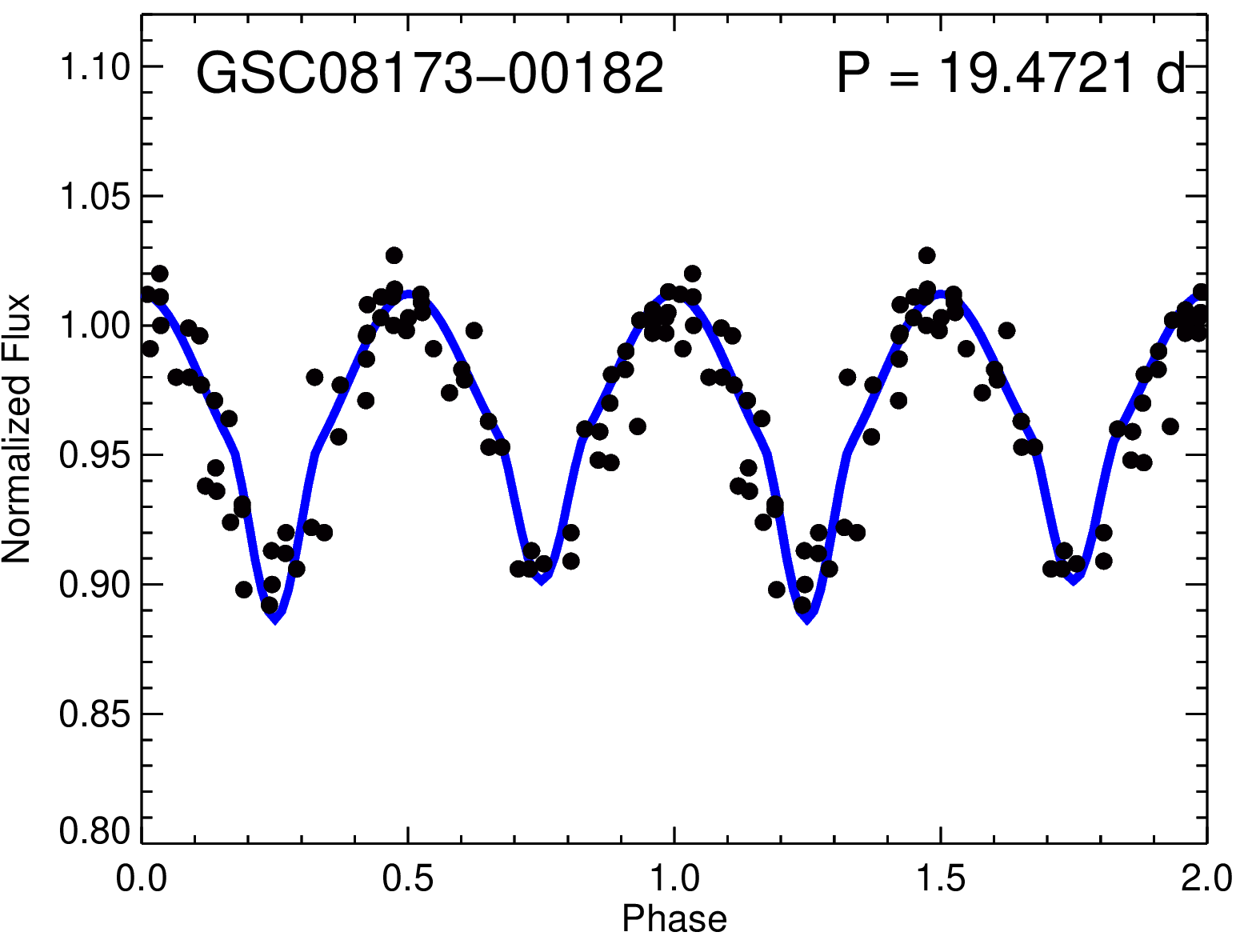}
  \includegraphics[width=0.67\columnwidth]{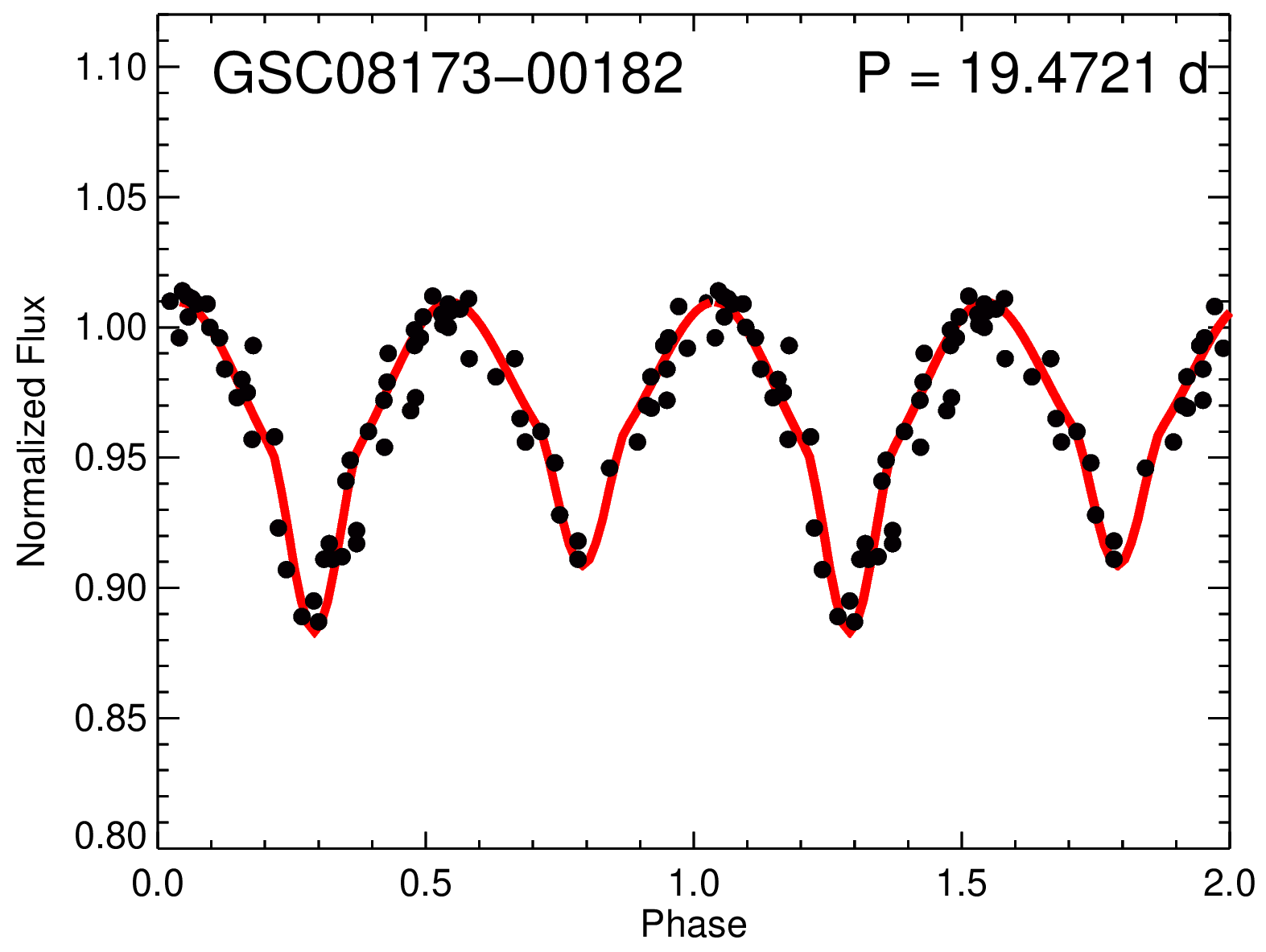}
  \includegraphics[width=0.67\columnwidth]{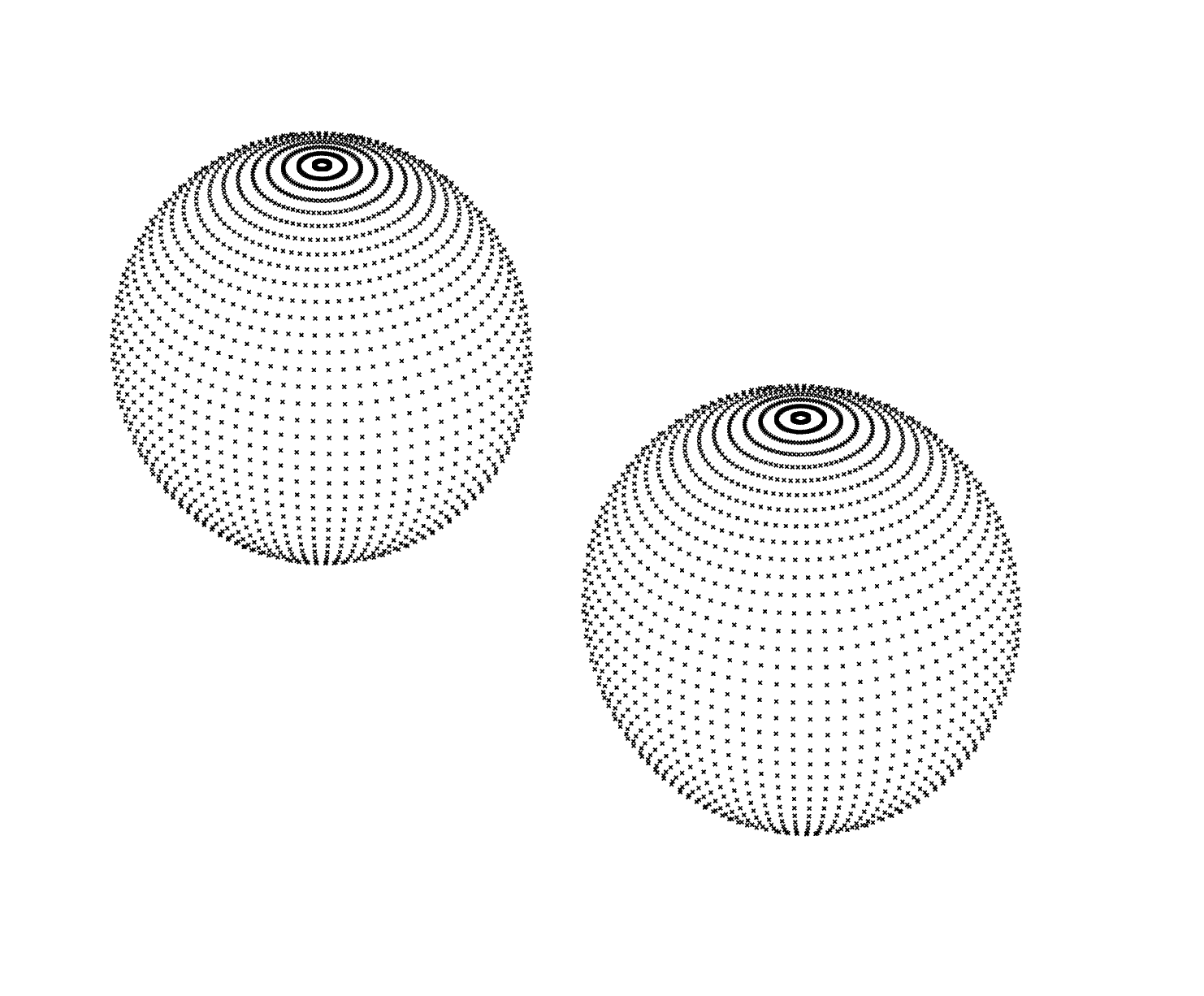}
  \includegraphics[width=0.67\columnwidth]{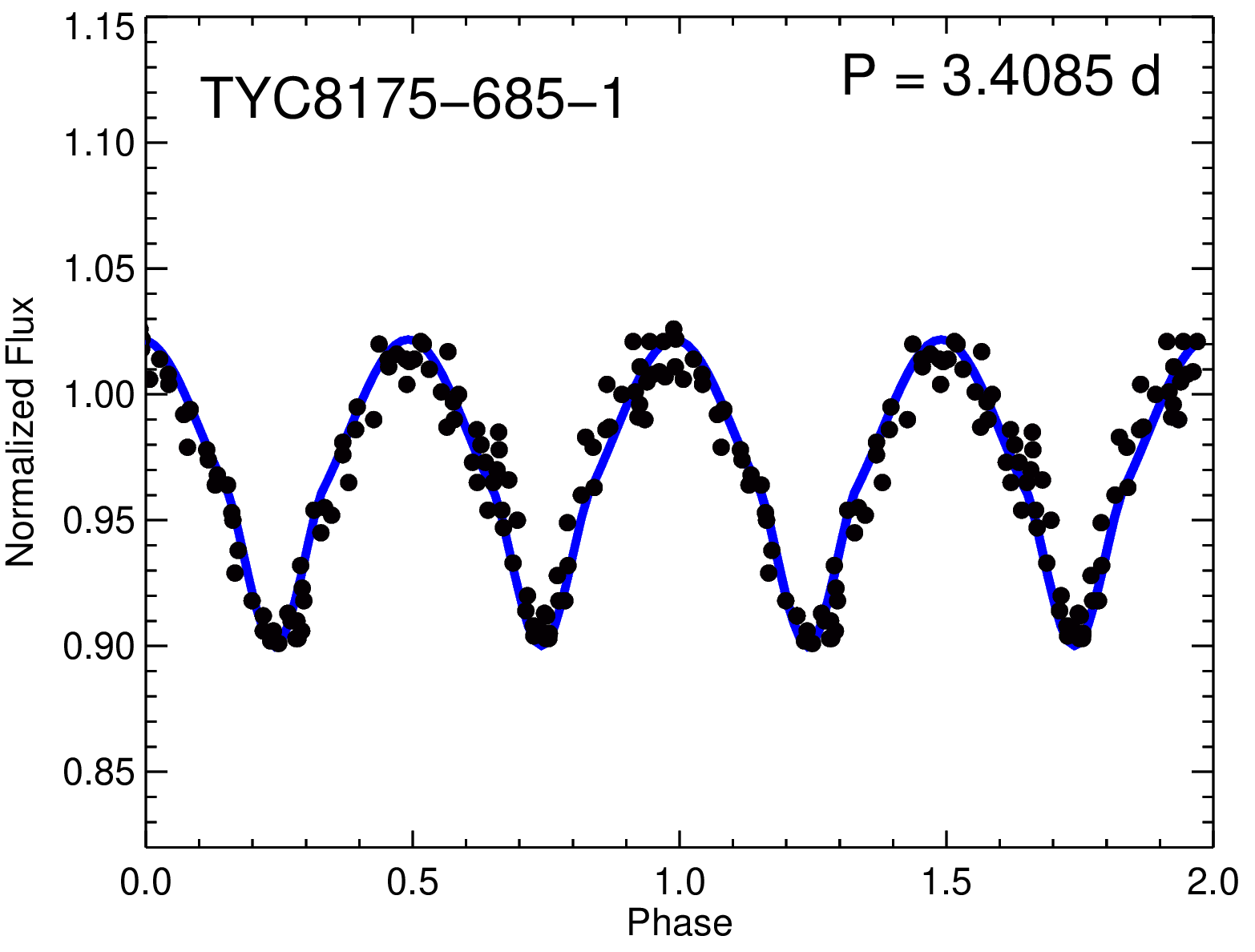}
  \includegraphics[width=0.67\columnwidth]{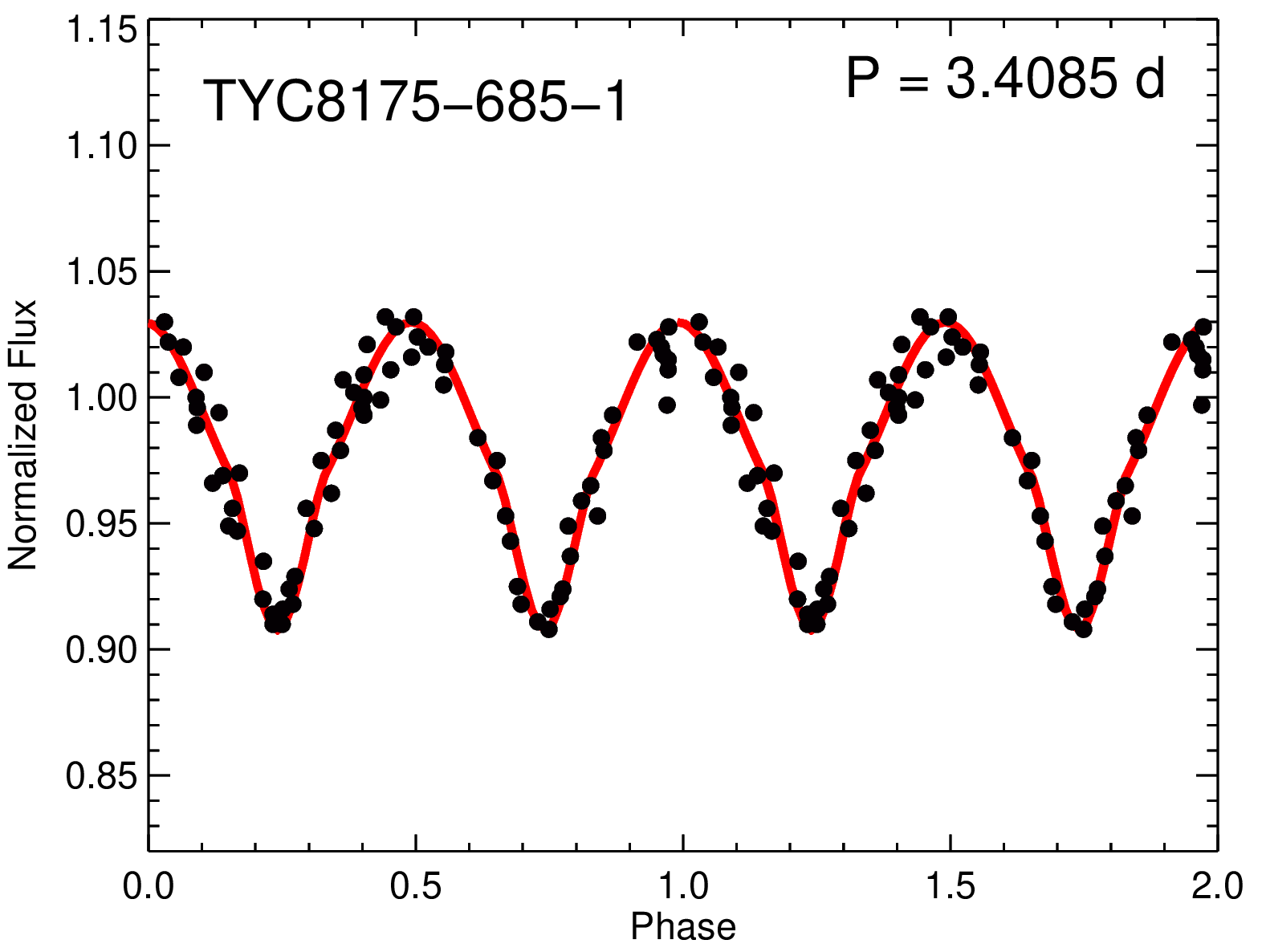}
  \includegraphics[width=0.67\columnwidth]{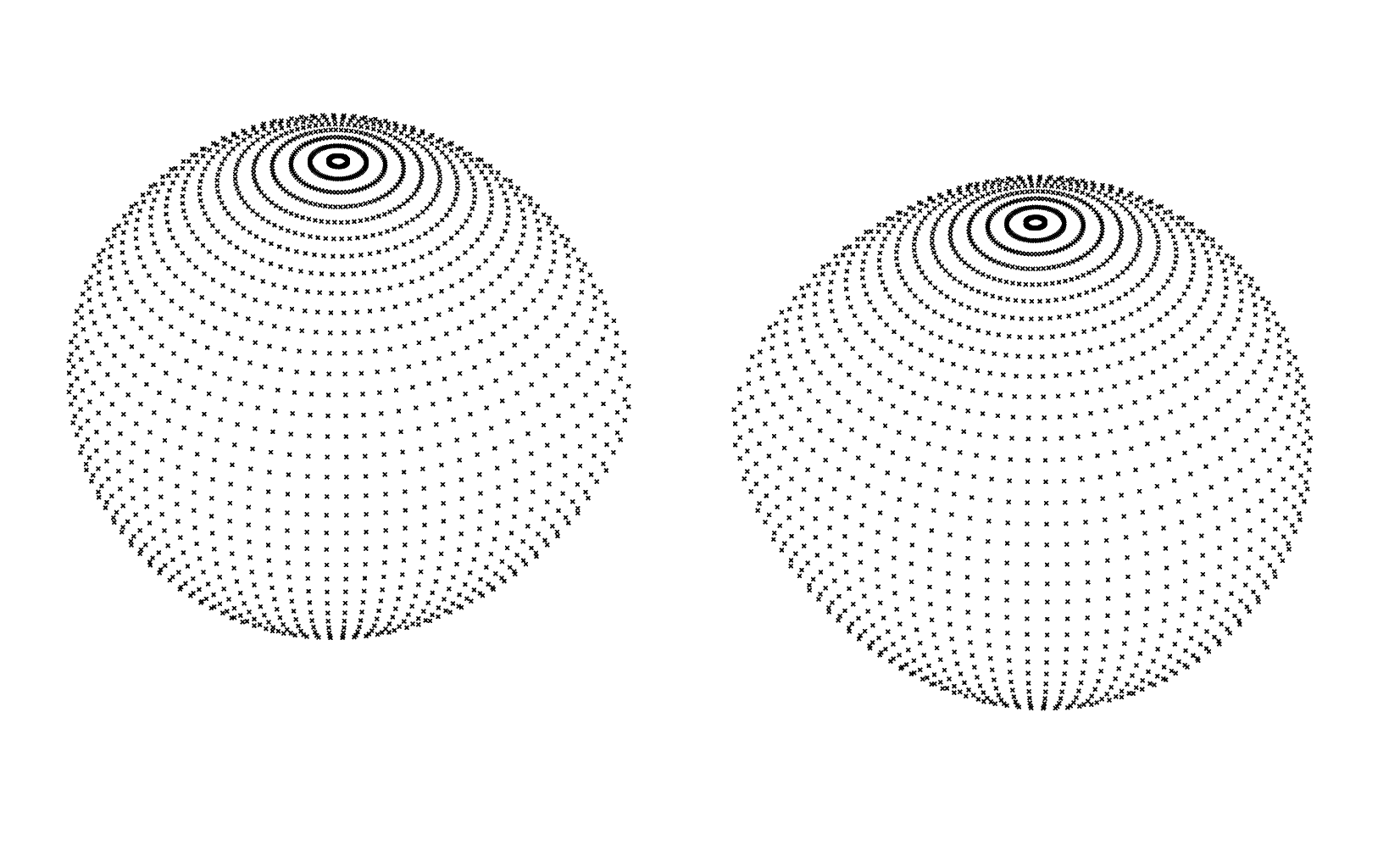}
  \includegraphics[width=0.67\columnwidth]{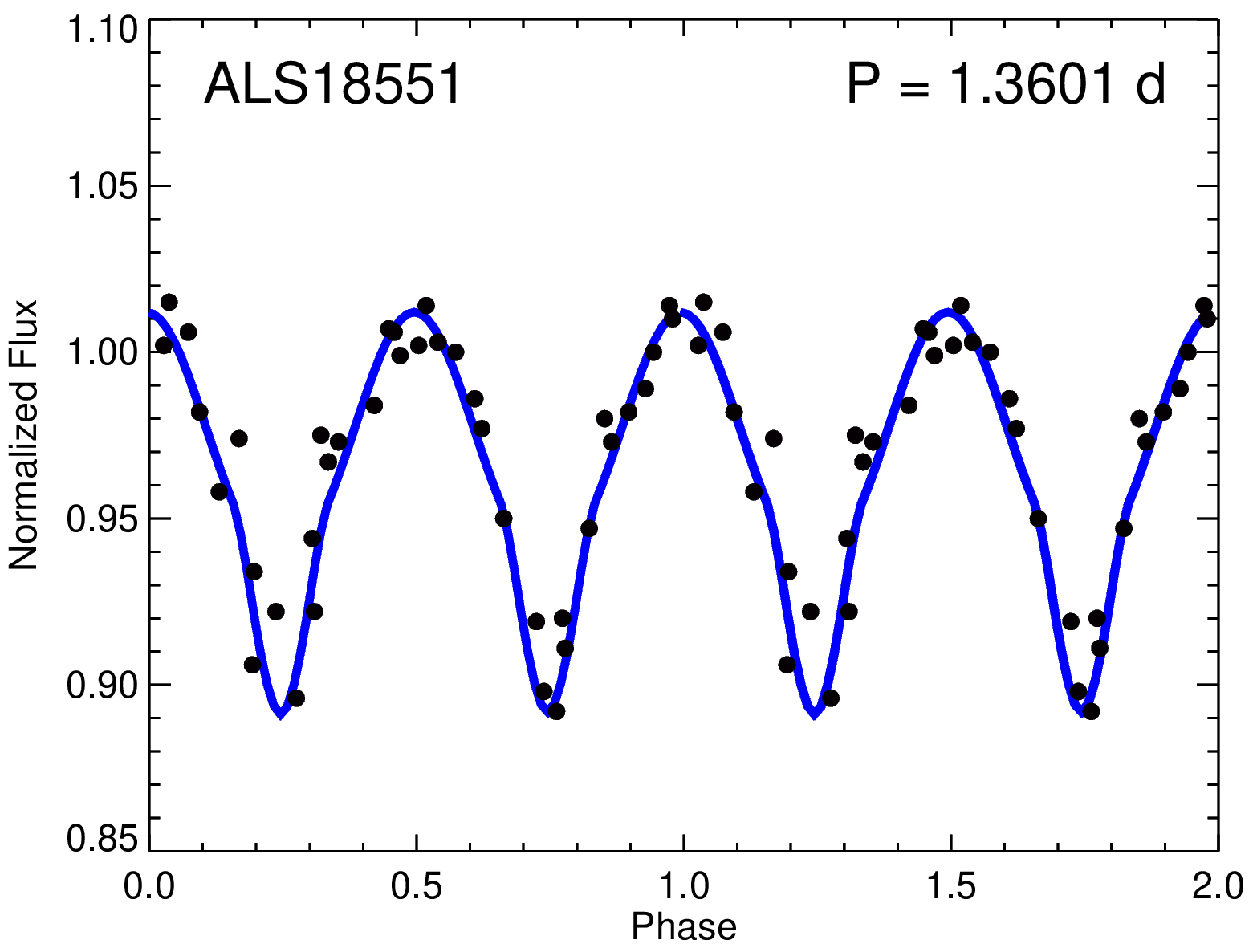}
  \includegraphics[width=0.67\columnwidth]{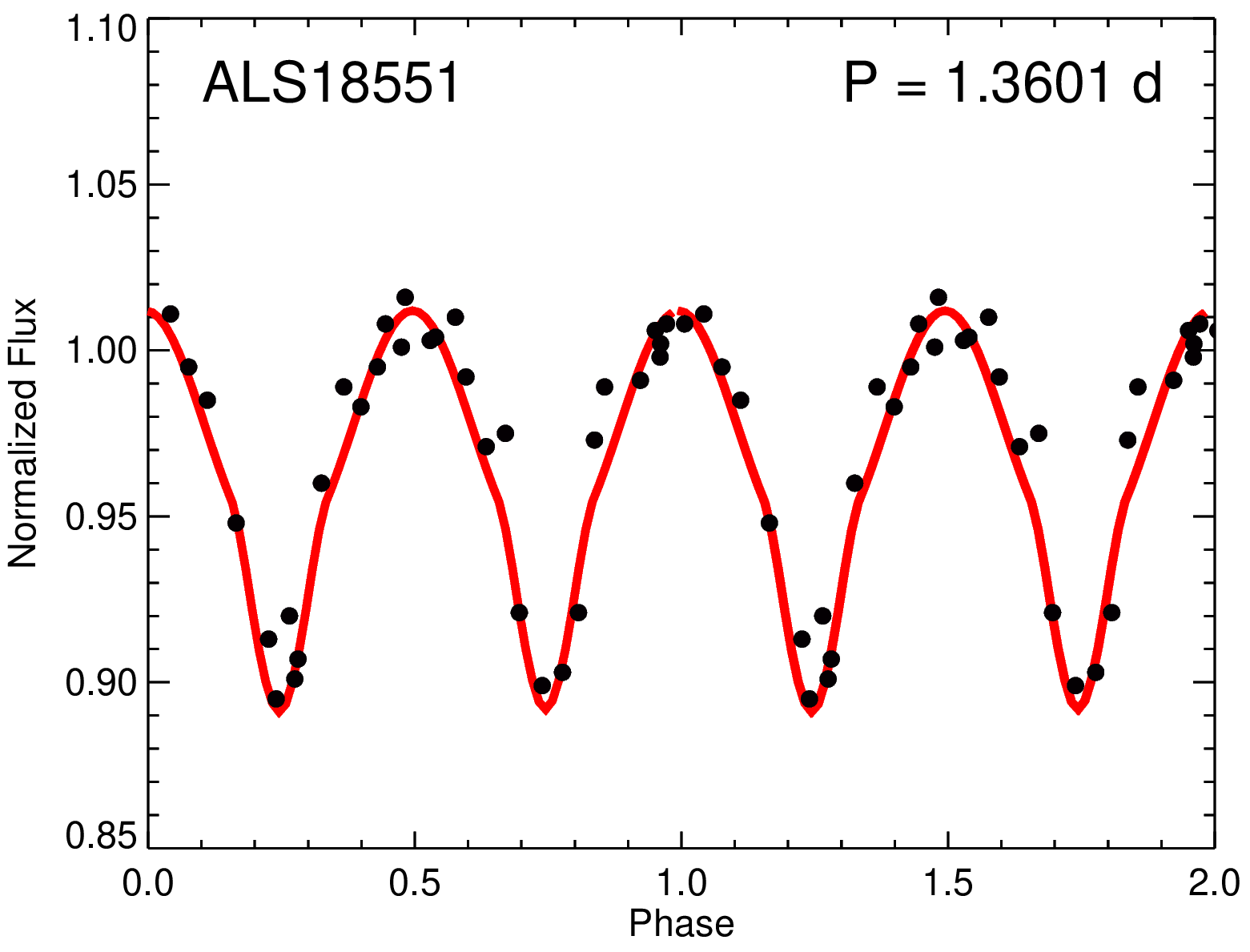}
  \includegraphics[width=0.67\columnwidth]{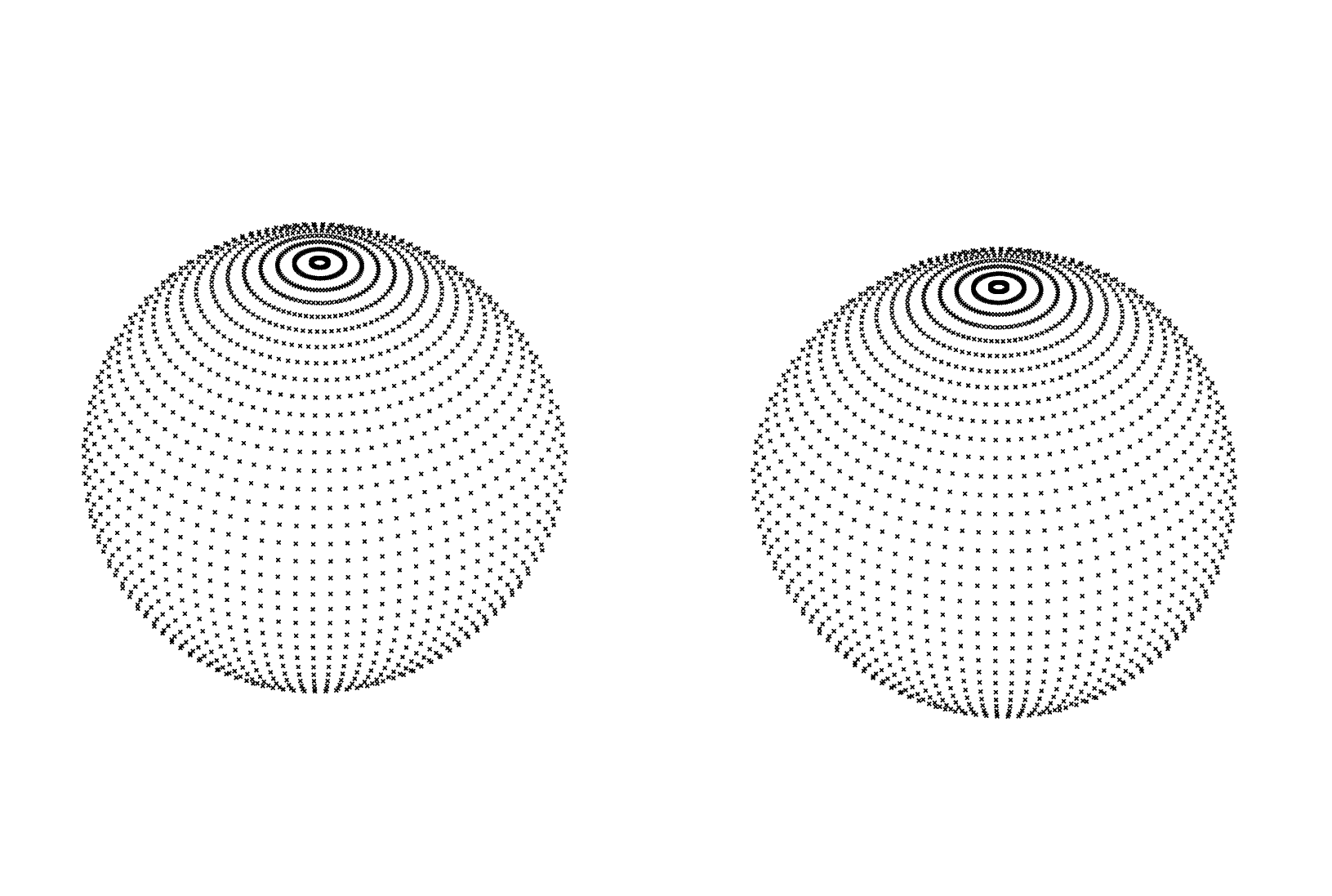}
  \caption{Same as Figure A1.}
  \label{model_apb}
\end{figure*}

High-resolution spectroscopy (\citealt{2016ApJS..224....4M}) has detected a O4.5\,V SB2 spectroscopic binary; however, no period has been published to date. From our light curves we obtain a photometric period of 1.360 days. We fixed $T_1 = 41860$\,K according to the spectral type reported by \citet{2016ApJS..224....4M} and used the observational $T_{\rm eff}$ calibration by \cite{2005A&A...436.1049M} (his Table\,4). The system shows well-defined eclipses favouring a near-contact configuration, where both components fill up their Roche lobes at about 93\% and 92\%, respectively (Fig.~\ref{model_apb}). The model fit of the light curve yields an effective temperature of the secondary star $T_2 = 41460$\,K which is consistent with the spectroscopic results obtained by \citet{2016ApJS..224....4M}. Here we confirm photometrically that the system is composed of two O4.5\,V stars.

\subsection{CPD\,$-$\,39\degr\,7292}

The star CPD\,$-$\,39\degr\,7292 was first observed by \cite{1982A&AS...50..261D} as part of a monitoring of luminous stars in the southern Milky Way. The reported photometric values are $U = 11.33$, $B = 11.53 $, and $V = 10.81$; the colors are compatible with a reddened B1\,V (Fig.~\ref{UBV diagram}). Drilling (1991) give a classification of OB$^+$ and provided $UBV$ photometry with $U = 11.51$, $B = 11.74 $, and $V = 10.96$; the colors are compatible with a reddened B0\,V star (Fig.~\ref{UBV diagram}). Subsequent spectroscopic observations by \cite{1993ApJS...89..293V} yielded a spectral type B1\,III\,n. Our own photometry yields $U = 11.59$, $B = 11.88 $, and $V = 11.01$ corresponding to a spectral type O5, significantly earlier than the spectroscopic and past photometric observations. Moreover, the three different photometric data sets indicate a color variability and therefore we prefer to adopt a spectral type B0V with an effective temperature of $T_1 = 25570$\,K. The assumed spectral type is consistent with both previous photometric results (\citealt{1982A&AS...50..261D}; \citealt{1991ApJS...76.1033D}) and the spectroscopic results by \cite{1993ApJS...89..293V}.

The amplitude variations for the primary and secondary minimum eclipses are $\Delta D_{p} = 15\%$ and $\Delta D_{s} = 10\%$, which can be interpreted through the difference between the temperatures of the components. The primary and secondary components fills up their Roche lobe about 98\% and 99\% respectively, favouring a near-contact configuration with an orbital inclination of $i = 52.5^{\circ}$. The 3D view of the Roche geometry of the system ($\phi = 0.50$) shows that most likely future mass transfer will create a new contact binary system (Fig.~\ref{model_apc}).

\subsection{Pismis\,24-4}

The star Pismis\,24-4 belongs to the young open cluster Pismis\,24 which resides within NGC\,6357 in the Sagittarius spiral arm at a distance of 1.7\,kpc (\citealt{2012A&A...539A.119F}). This well-studied region is known as a rich reservoir of young stars, containing several OB stars (\citealt{2001AJ....121.1050M}) and several shell-like H\,{\sc ii} regions (\citealt{2012A&A...538A.142R}).

Pismis\,24-4 was classified as a O9-B0? by \cite{2007ApJS..168..100W}; this tentative spectral type was inferred from a color-magnitude diagram by dereddening 2MASS colors to the 1\,Myr pre-main sequence (pre-MS) isochrone. \cite{1973A&AS...10..135M} presented the original $UBV$ photometry with $U = 15.89$, $B =15.36$, and $V = 13.93$; these colors are compatible with a reddened B3\,V star. In the following we adopt a spectral type B0\,V and fix the effective temperature of the primary to $T_1 = 29230$\,K.

The light curves show that the primary and secondary minimum have the same amplitude variations ($\sim 14\%$), characteristic for an eclipsing binary with equal effective temperatures. We therefore suggest that this new system is most likely composed of two B0\,V stars. The primary component fills up its Roche lobe about 76\% while the secondary about 96\% favouring a semi-detached configuration with an orbital inclination of 61.9$^{\circ}$. As displayed in Fig.~\ref{model_apc}, the 3D view of the Roche geometry of the system, $\phi = 0.88$ shows a transitional evolutionary state of the system for which one of the stars is close to fill its Roche lobe.

\subsection{CD\,$-$\,29\degr\,14032}

CD\,$-$\,29\degr\,14032 was classified as B1\,III by \cite{1993ApJS...89..293V}. The Tycho-2 catalog (\citealt{2000A&A...355L..27H}) give photometric values $B = 11.72$ and $V = 11.96$. Our own photometry yields $B = 11.69$ and $V = 11.05$ which suggest variability. Including our $U-$value of 11.30 we obtain a photometric classification of O9.5\,V. As an average we fix the effective temperature of the primary star to $T_1 = 29230$\,K, corresponding to a B0\,V star.

The system shows well-defined eclipses favouring a detached configuration (Fig.~\ref{model_apc}), where both components fill up their Roche lobes at about 76\% and 73\%, respectively. The best fit to the light curve yields a temperature of $T_2 = 26850$\,K for the secondary component corresponding to a spectral type B1\,V.

\subsection{CPD\,$-$\,59\degr\,2603}

CPD\,$-$\,59\degr\,2603 or V572\,Car (ID:0211 GOSC) is a member of the open cluster Trumpler\,16 in the Carina Complex (\citealt{2001MNRAS.326.1149R}). We find a period of $2.152851 \pm 0.009556$ days consistent with the spectroscopic period $2.15294 \pm 0.00214$ days obtained by \cite{2001MNRAS.326.1149R}. A somewhat sparser monitoring of this object was carried out by \cite{1993ARep...37..152A} between 1980-1991. By folding the spectroscopic period with the photometric data obtained by \cite{1993ARep...37..152A}, \cite{2001MNRAS.326.1149R} derived a lower limit for the inclination ($i\geq77^{\circ}$) together with the masses for the EB system. This is the first time that a high quality sampled light curve is obtained for this source. We fixed the temperature of the primary star to $T_1 = 35874$\,K corresponding to a spectral type O7.5\,V (\citealt{2001MNRAS.326.1149R}). The primary component fills up its Roche lobe at $\sim 66\%$ while the secondary at $\sim 75\%$ favouring a detached configuration with an orbital inclination $i = 79.2^{\circ} \pm 1.7$ (Fig.~\ref{model_apc}). The inclination found here is consistent with the lower limit $i\geq77 ^{\circ}$ reported by \cite{2001MNRAS.326.1149R}. We find an effective temperature of $T_2 = 32220$\,K for the secondary star corresponding to a spectral type O9.5\,V in agreement with the spectroscopic results.

\subsection{CPD\,$-$\,59\degr\,2628}

CPD\,$-$\,59\degr\,2628 has been catalogued by GOSC (ID:0218) with a spectral type of O9.5\,V. Therefore, we fixed the temperature of the primary star to $T_1 = 31884$\,K. The system shows well-defined eclipses with a period of $1.46\pm0.015$ days as derived with the PDM algorithm. The amplitude variations for the primary and secondary minimum are $\Delta D_{p} = 38\%$ and $\Delta D_{s} = 34\%$ suggesting slightly different temperatures for both components. The primary component fills up its Roche lobe at $\sim 96\%$, the secondary at $\sim 95\%$ favouring a near-contact configuration with an orbital inclination $i = 73.4^\circ$. We find an effective temperature of $T_2 = 30120$\,K for the secondary star corresponding to a spectral type O9.5\,V.

\subsection{CPD\,$-$\,59\degr\,2635}

The GOSC (ID:0220) gives a spectral type O8\,V; therefore we fixed the temperature of the primary star to $T_1 = 34877$\,K. The primary component fills up its Roche lobe at $\sim 78\%$, the secondary at $\sim 69\%$ favouring a detached configuration with an orbital inclination $i = 75.2^{\circ}$ (Fig.~\ref{model_apc}). We find an effective temperature of $T_2 = 31090$\,K for the secondary star corresponding to a spectral type O9.5\,V.

\subsection{V662\,Car}

The GOSC (ID:0226) gives a spectral type O5\,V; therefore we fixed the temperature of the primary star to $T_1 = 40862$\,K. The primary component fills up its Roche lobe at $\sim 96\%$, the secondary at $\sim 95\%$ favouring a near-contact configuration with an orbital inclination $i = 74.3^{\circ}$ (Fig.~\ref{model_apc}). We find an effective temperature of $T_2 = 31989$\,K for the secondary star corresponding to a spectral type O9.5\,V.

\subsection{TYC\,7370-460-1}

The GOSC (ID:0375) gives a spectral type O6\,V; therefore we fixed the temperature of the primary star to $T_1 = 38867$
\,K. The primary component fills up its Roche lobe at $\sim 82\%$, the secondary at $\sim 79\%$ favouring a detached
configuration with an orbital inclination $i = 80.3^{\circ}$ (Fig.~\ref{model_apc}). We find an effective temperature of $T_2 =
32120$\,K for the secondary star corresponding to a spectral type O9\,V.

\subsection{Pismis\,24-13}

The GOSC (ID:0398) gives a spectral type O6\,V; therefore we fixed the temperature of the primary star to $T_1 = 38867$\,K. The primary component fills up its Roche lobe at $\sim 91\%$, the secondary at $\sim 91\%$ favouring a semi-detached configuration (near contact) with an orbital inclination $i = 51.9^{\circ}$ (Fig.~\ref{model_apc}). We find an effective temperature of $T_2 = 37890$\,K for the secondary star corresponding to a spectral type O6\,V.

\subsection{HDE\,323110}

\begin{figure*}
  %\centering
  \includegraphics[width=0.67\columnwidth]{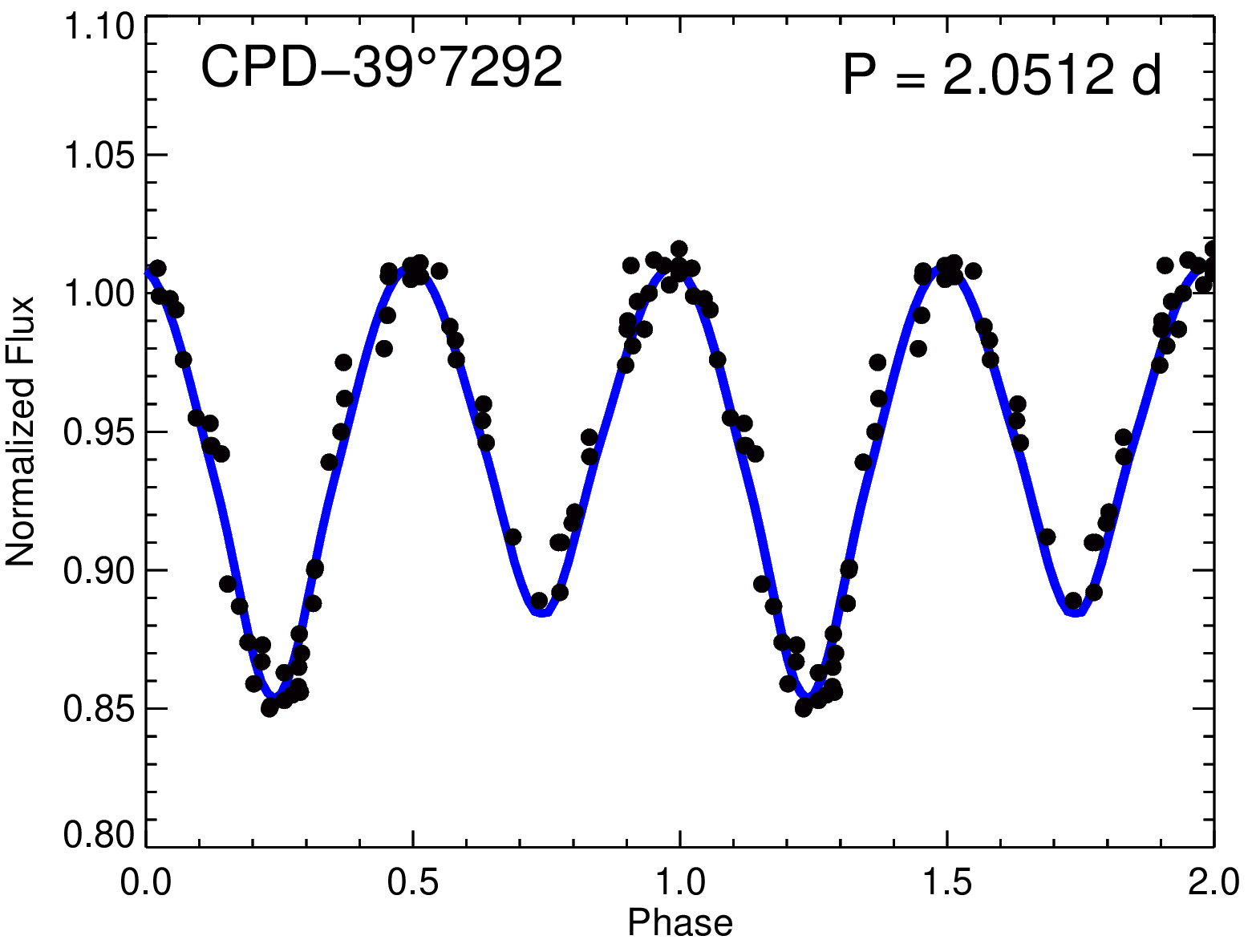}
  \includegraphics[width=0.67\columnwidth]{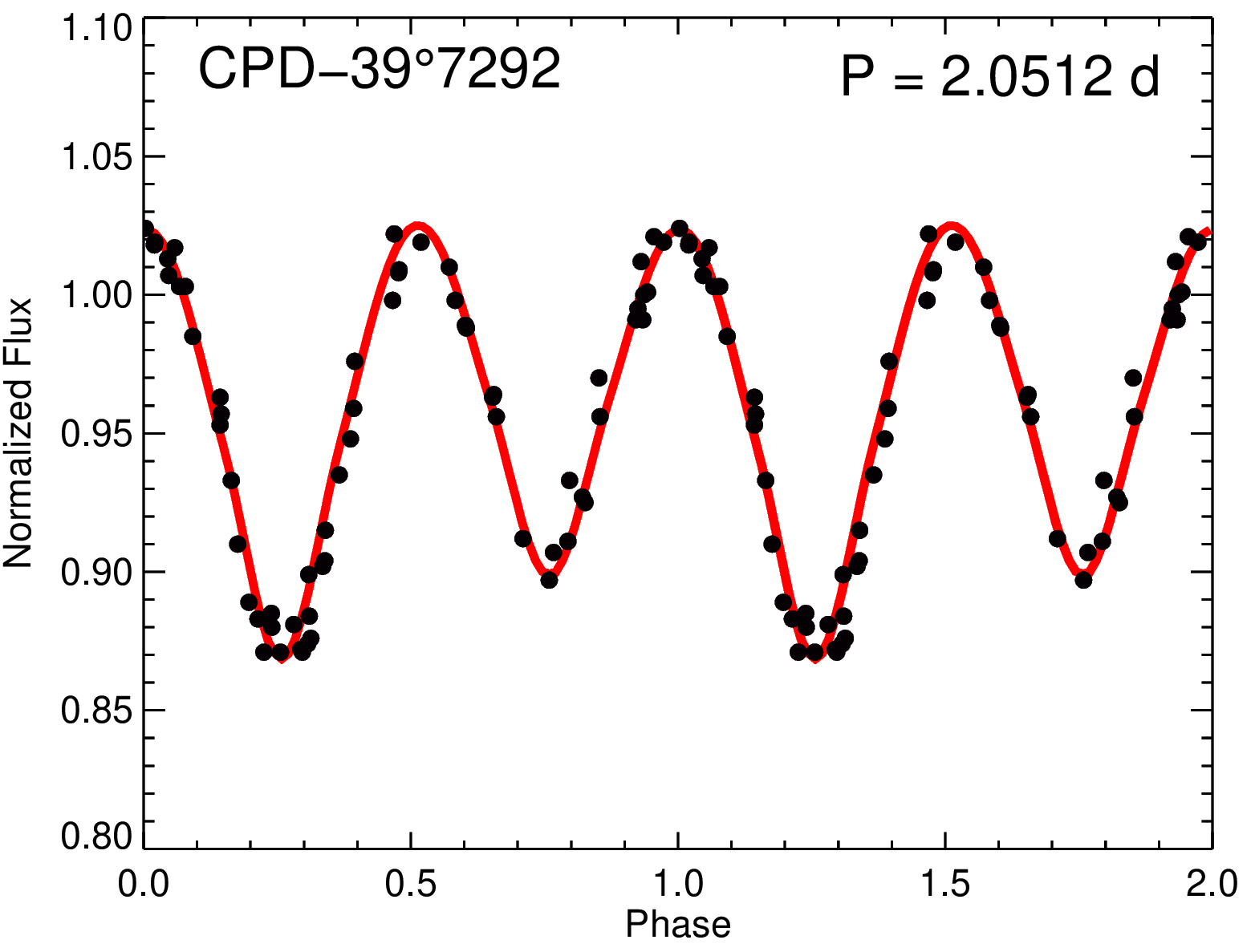}
  \includegraphics[width=0.67\columnwidth]{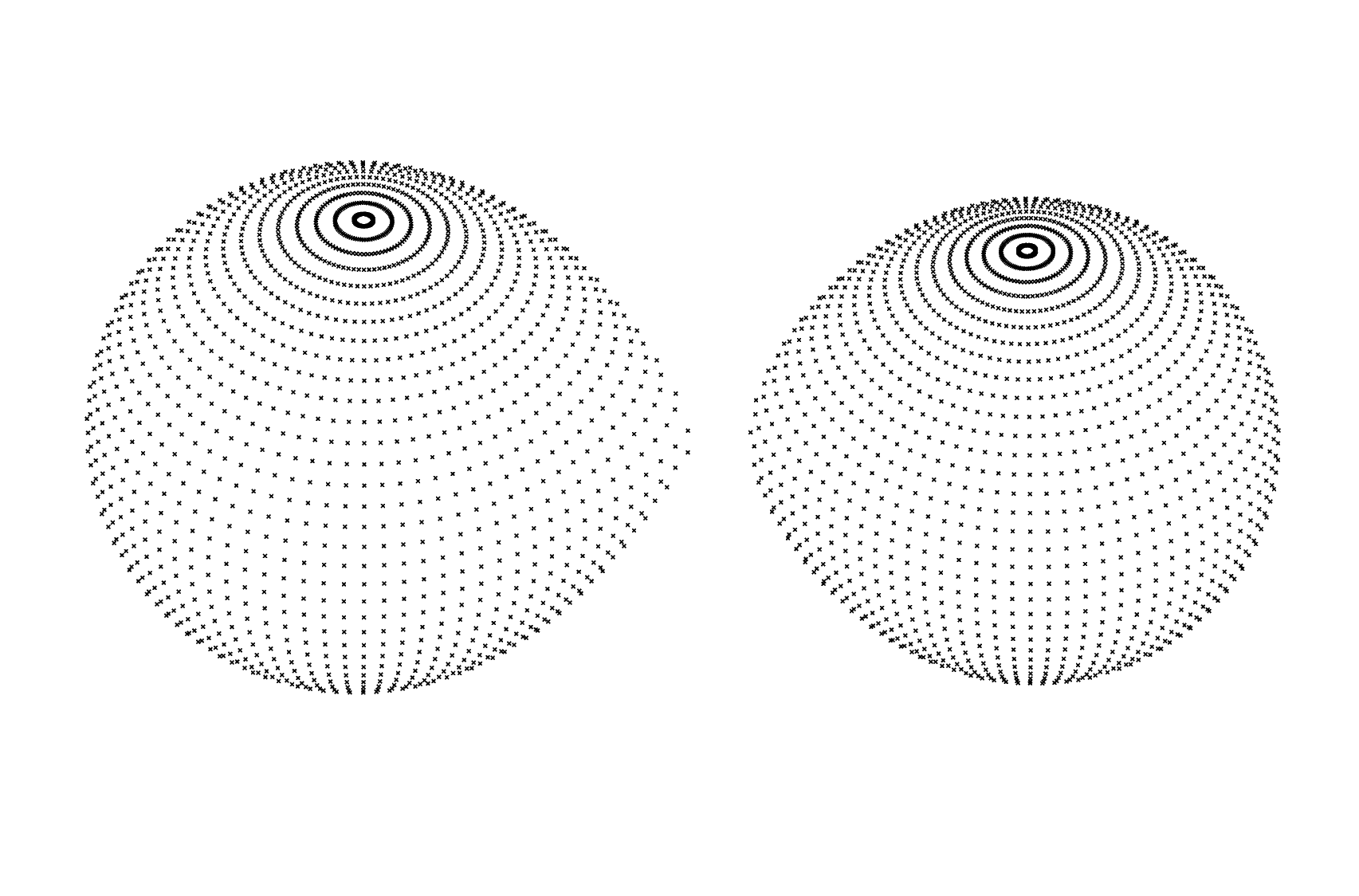}
  \includegraphics[width=0.67\columnwidth]{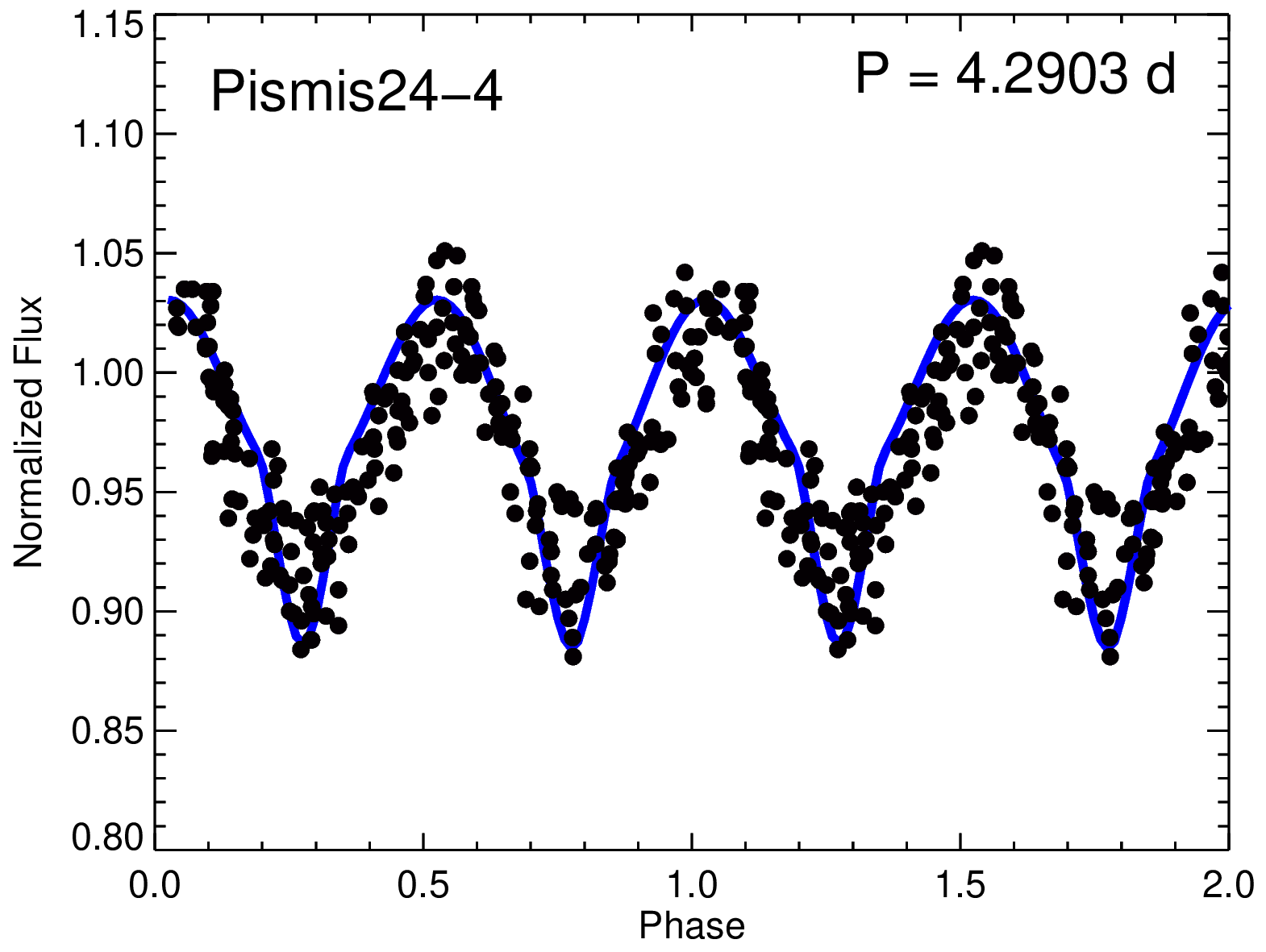}
  \includegraphics[width=0.67\columnwidth]{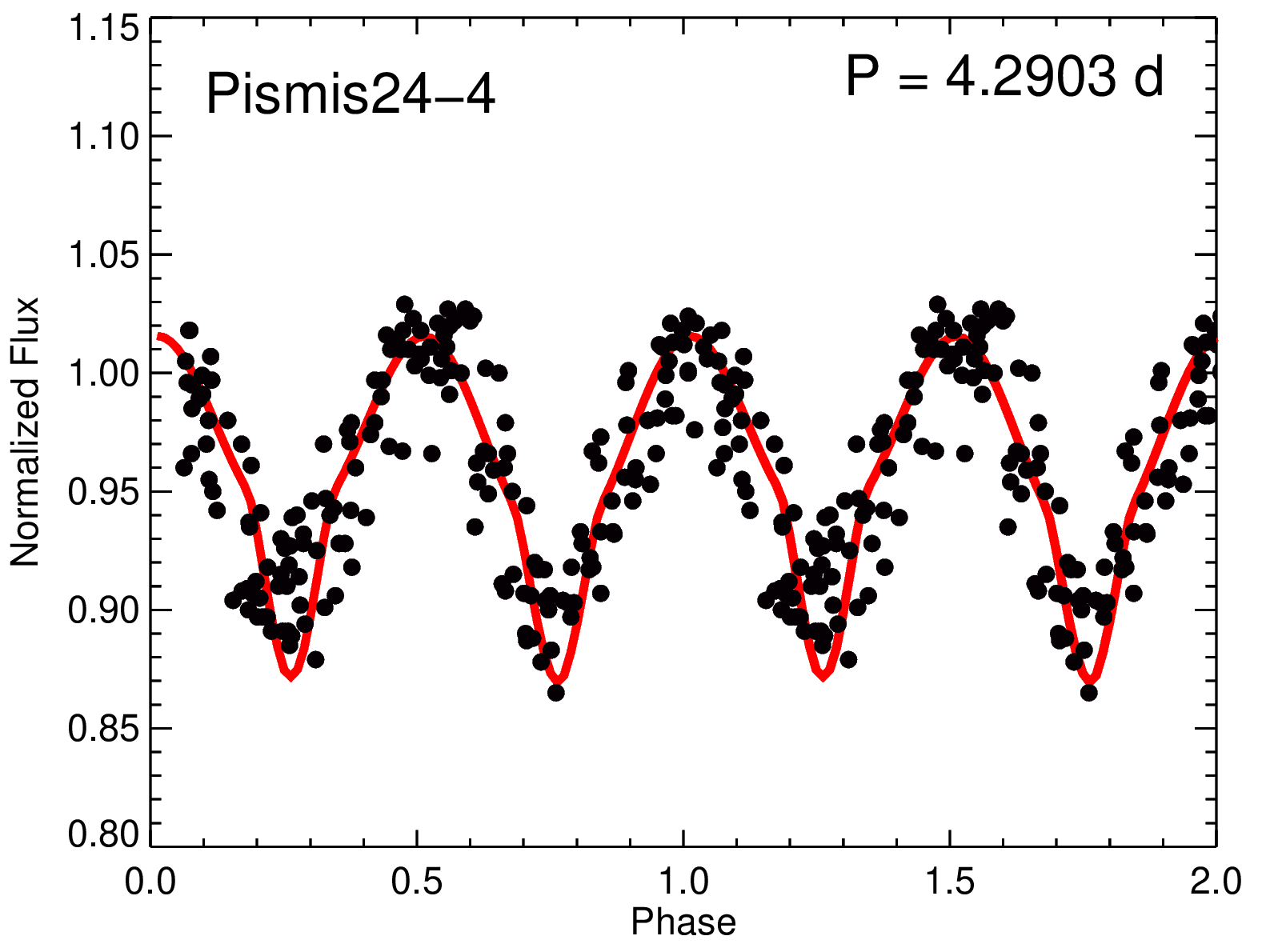}
  \includegraphics[width=0.67\columnwidth]{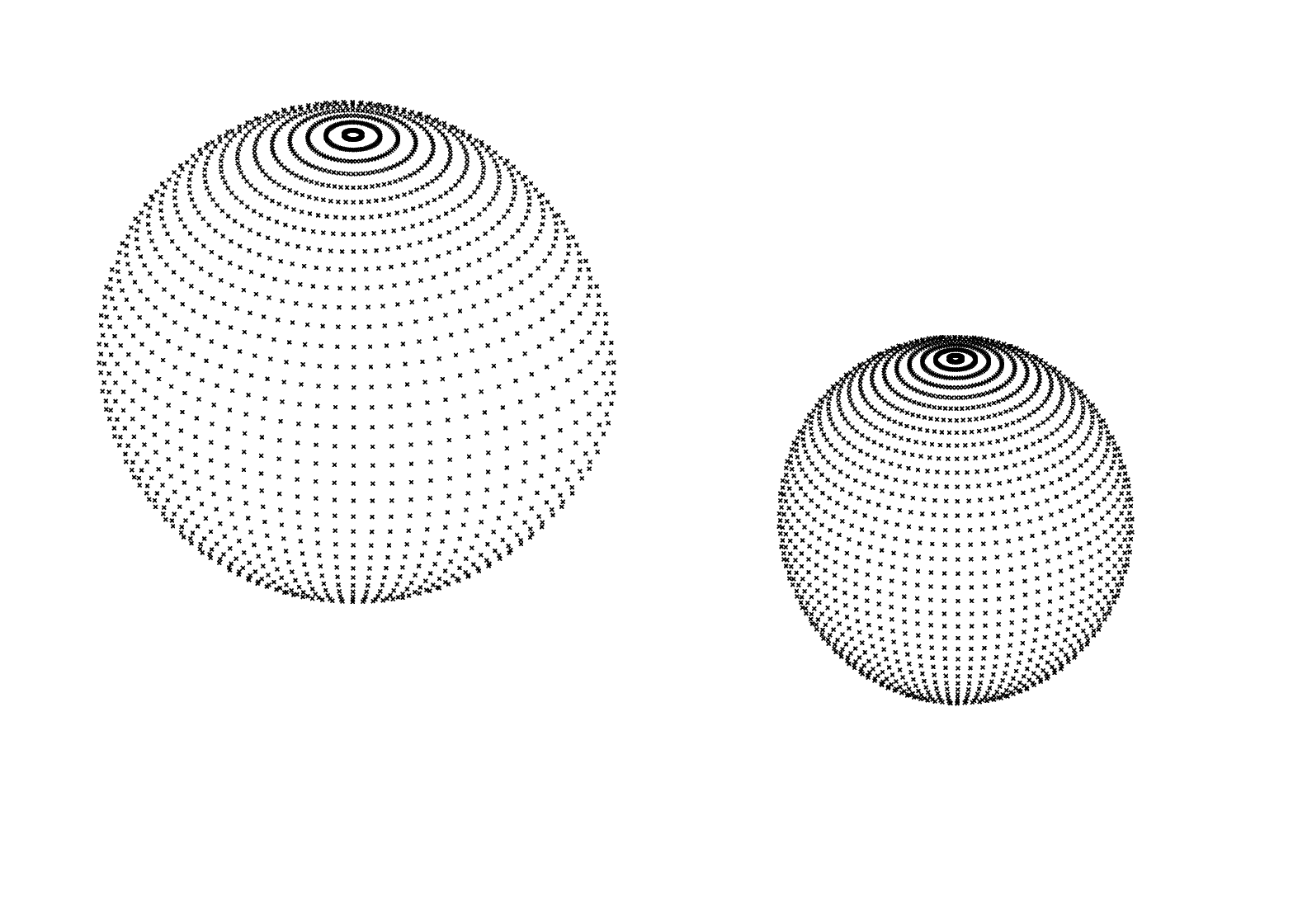}
  \includegraphics[width=0.67\columnwidth]{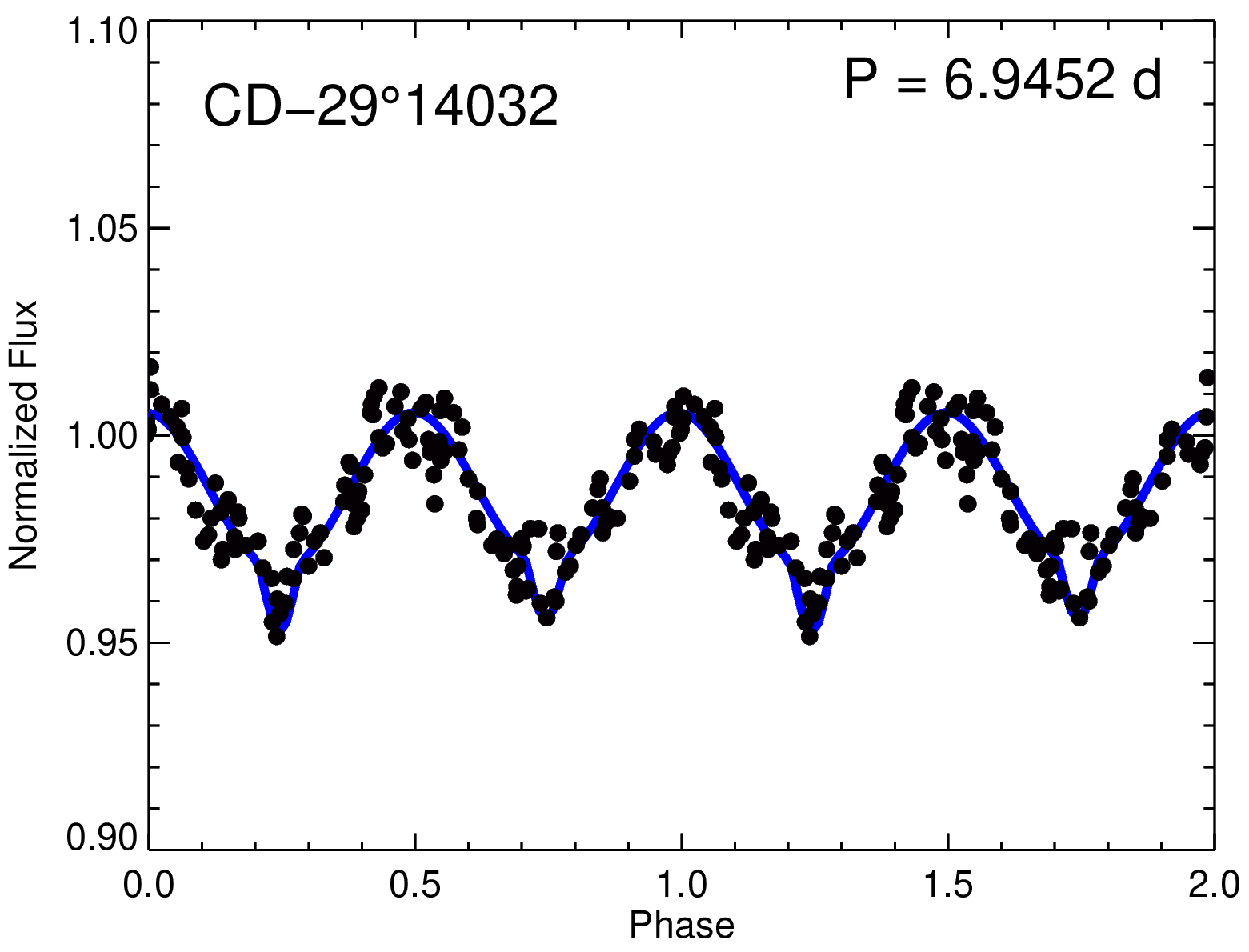}
  \includegraphics[width=0.67\columnwidth]{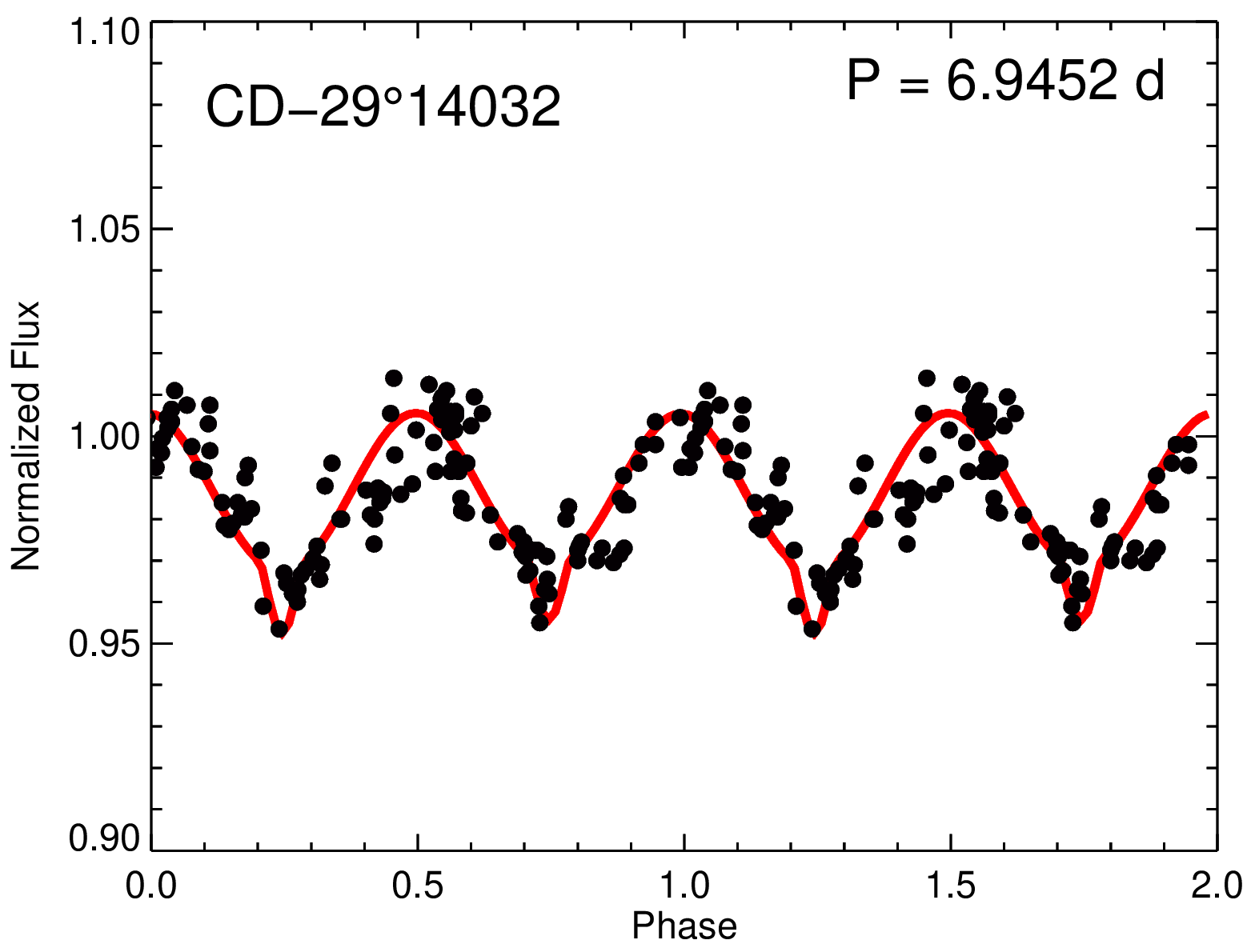}
  \includegraphics[width=0.67\columnwidth]{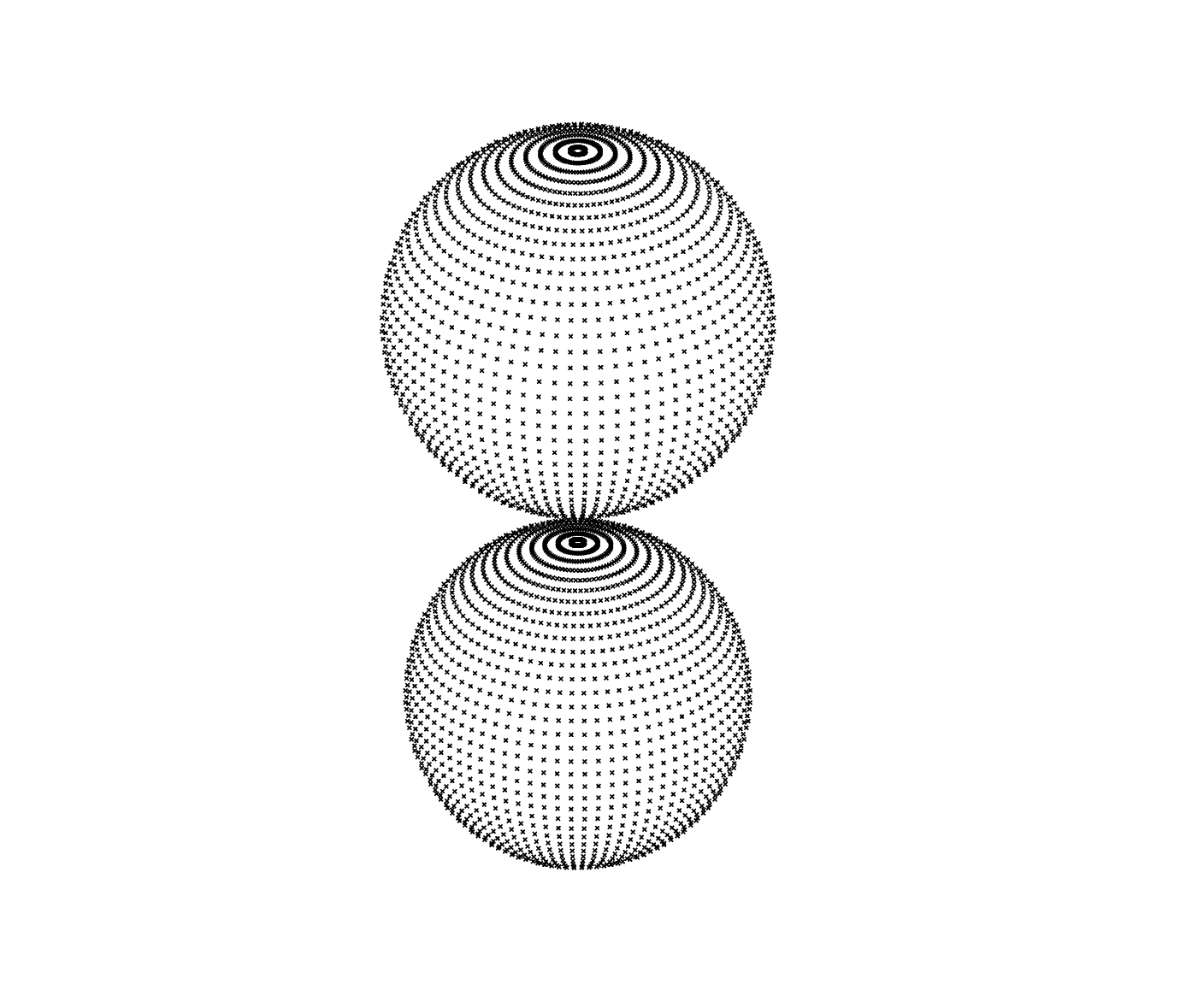}
  \includegraphics[width=0.67\columnwidth]{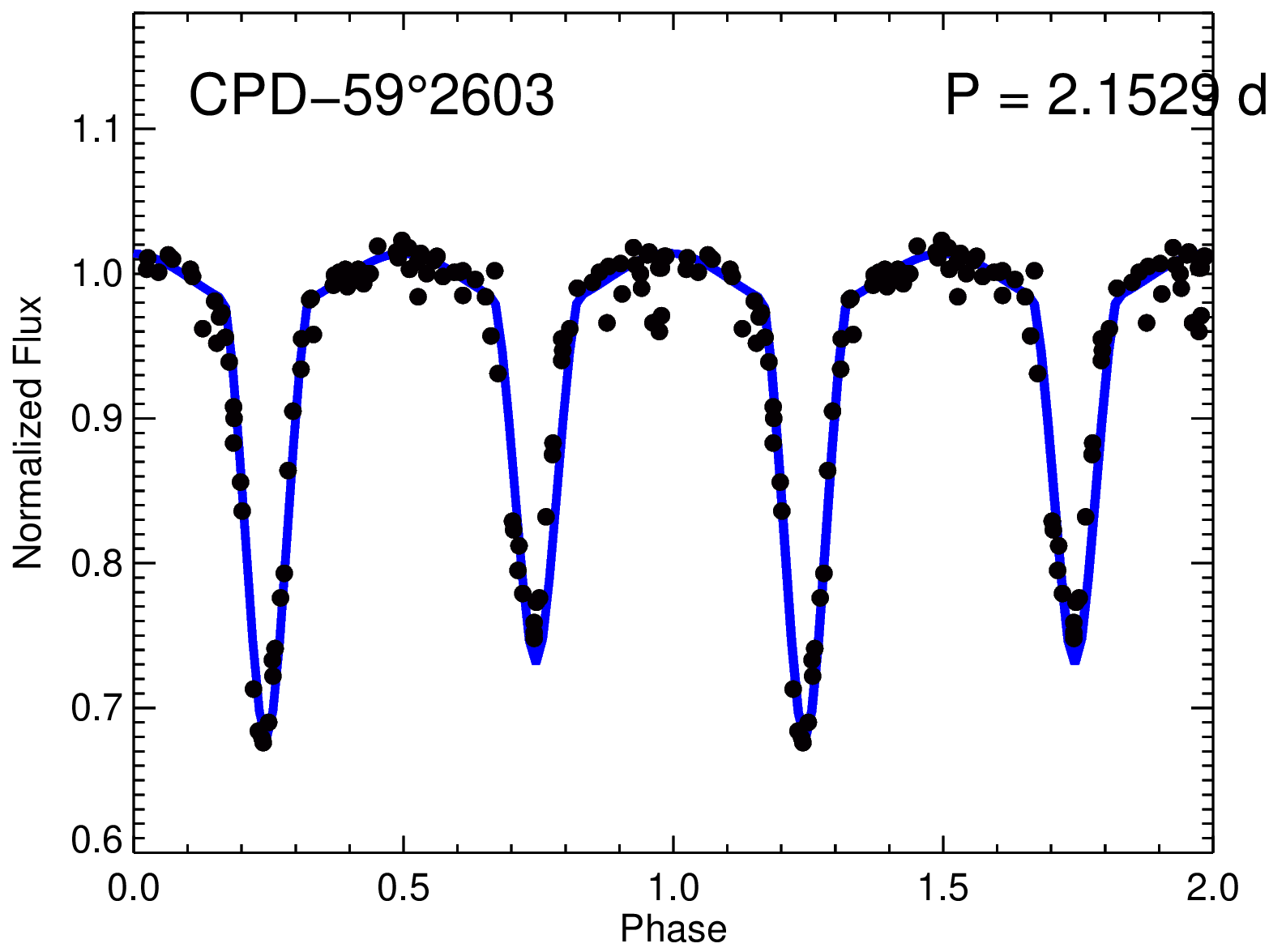}
  \includegraphics[width=0.67\columnwidth]{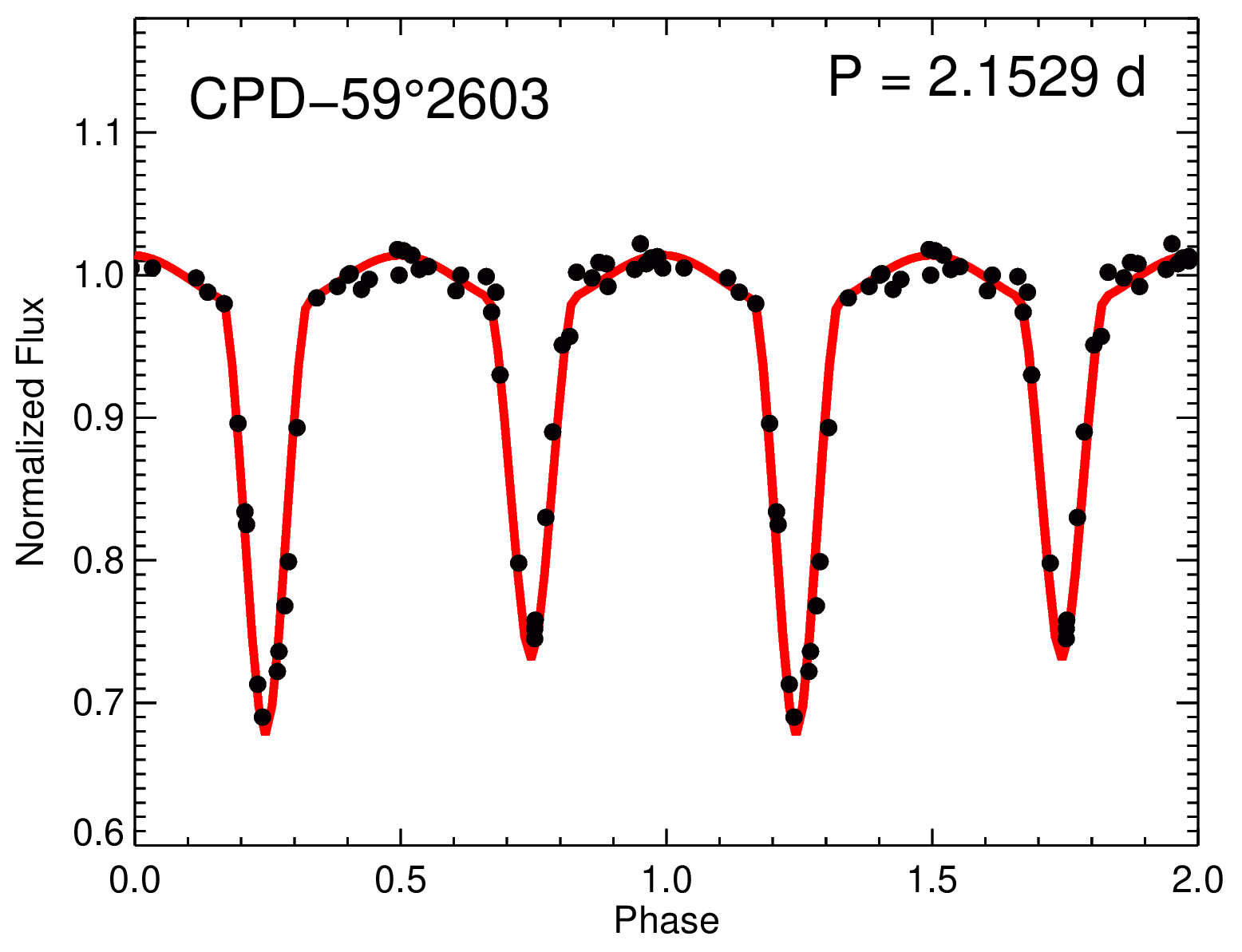}
  \includegraphics[width=0.67\columnwidth]{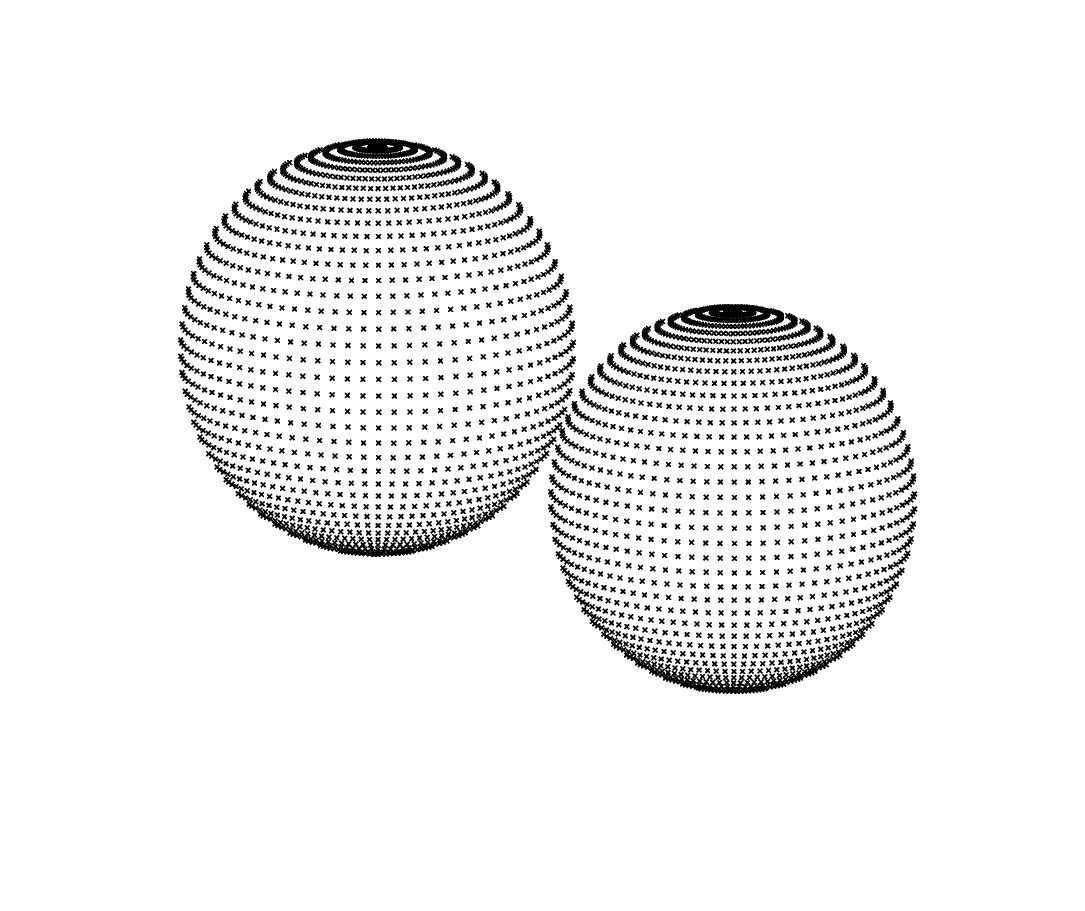}
  \includegraphics[width=0.67\columnwidth]{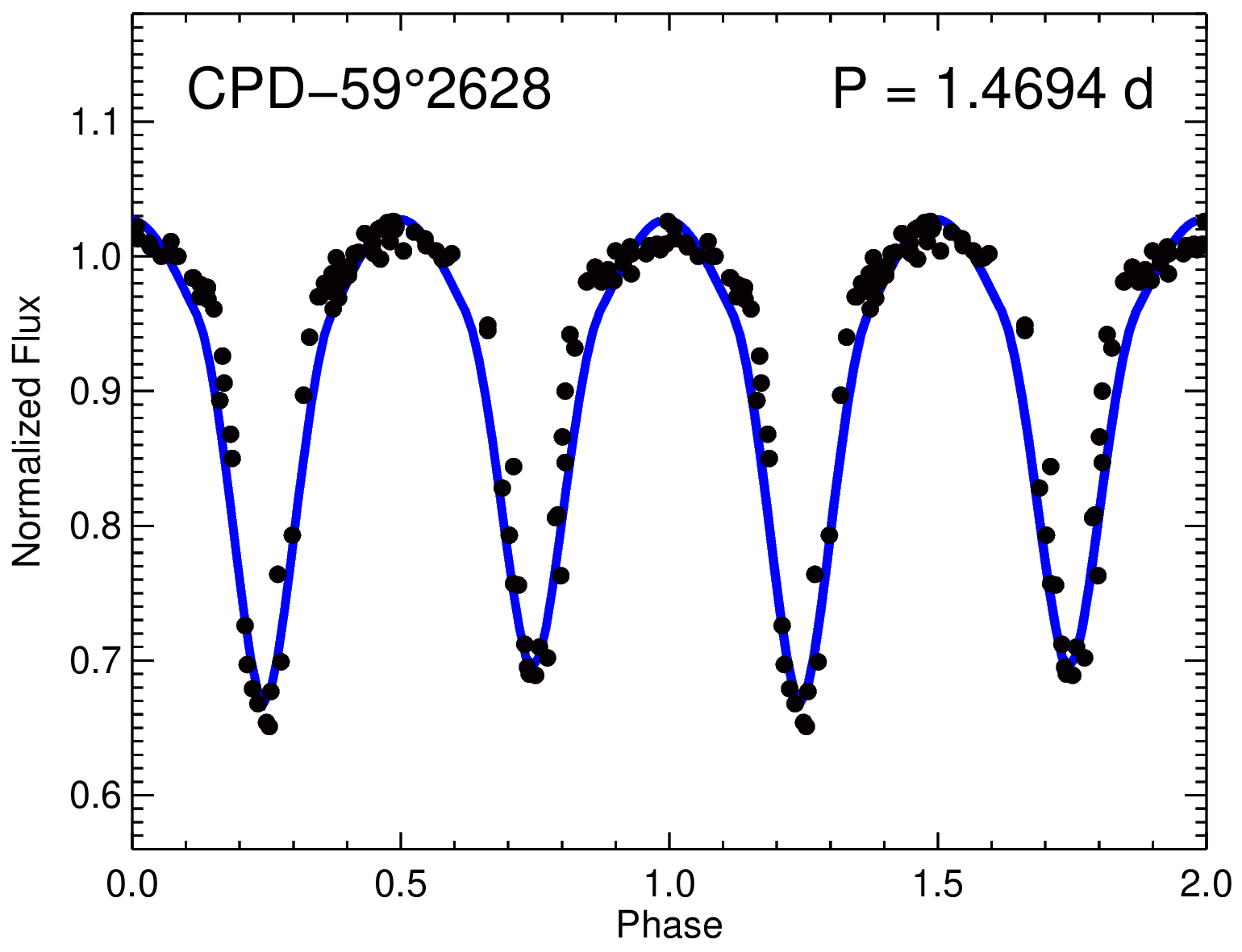}
  \includegraphics[width=0.67\columnwidth]{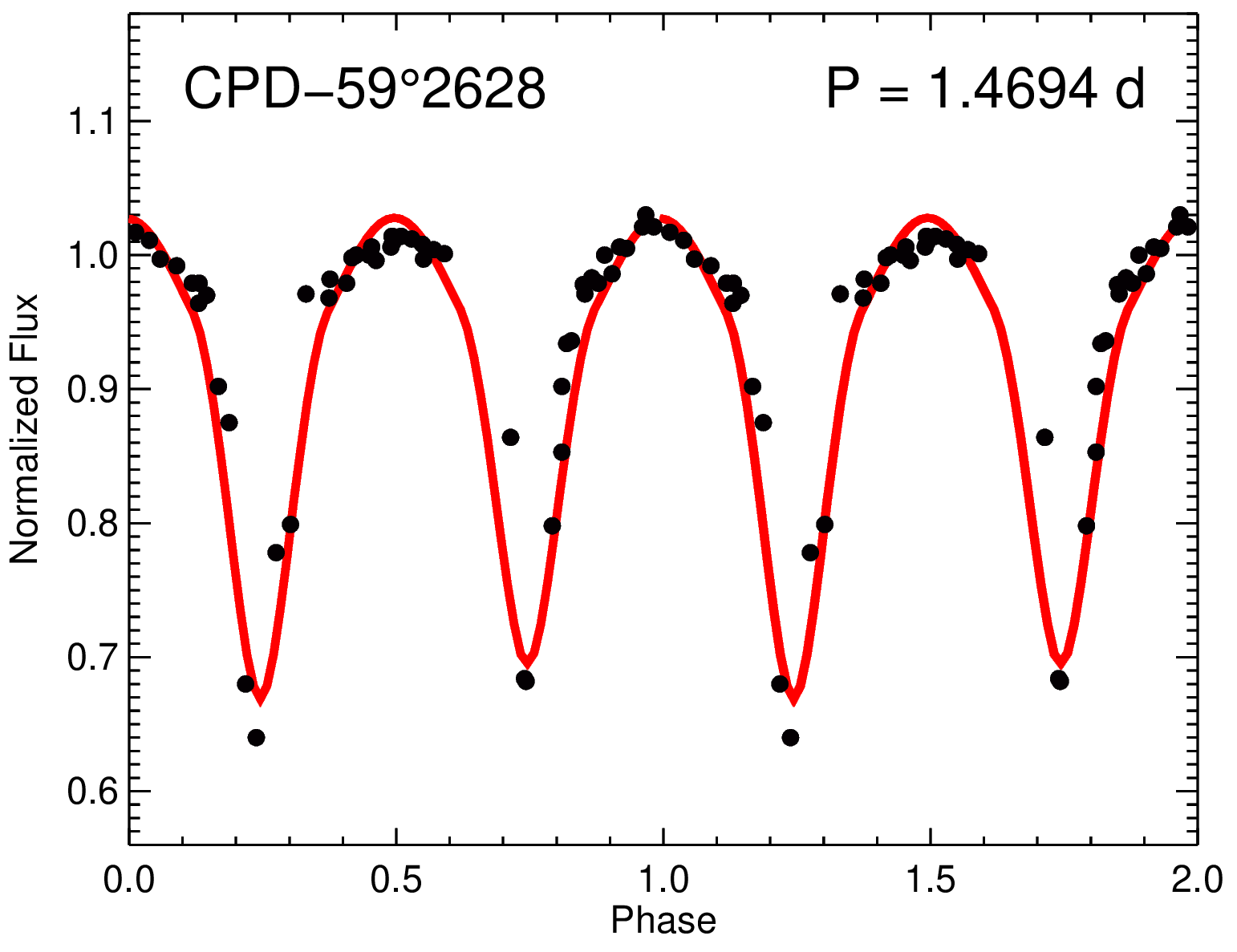}
  \includegraphics[width=0.67\columnwidth]{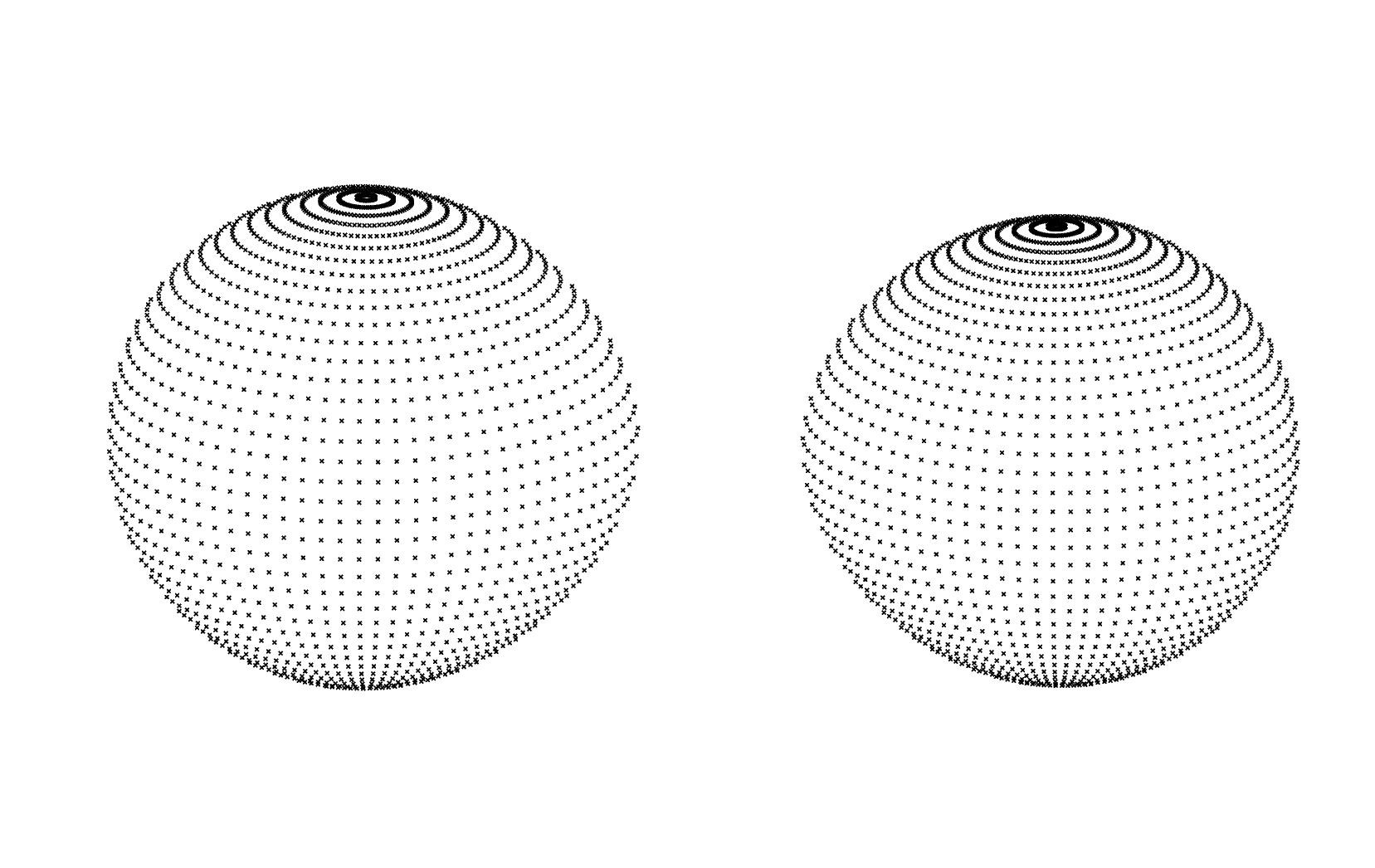}
  \caption{Same as Figure A1.}
  \label{model_apc}
\end{figure*}

We find a period of $5.20520 \pm 0.0156351$ days. A period of 5.206898 days has been reported by \cite{2003AcA....53..341P} with ASAS catalogue. The GOSC (ID:0391) gives a spectral type ON91Ia; therefore we fixed the temperature of the primary star to $T_1 = 31368$\,K. The primary component fills up its Roche lobe at $\sim 96\%$, the secondary at $\sim 92\%$ favouring a near-contact configuration with an orbital inclination $i = 73.8^{\circ}$ (Fig.~\ref{model_apc}). We find an effective temperature of $T_2 = 27169$\,K for the secondary star corresponding to a spectral type B1\,V.

\subsection{V\,467\,Vel}

The GOSC (ID:0142) gives a spectral type O6.5\,V; therefore we fixed the temperature of the primary star to $T_1 = 37870$\,K. The amplitude variations for the primary and secondary minimum eclipses are $\Delta D_{p} = 17\%$ and $\Delta D_{s} = 11\%$, which can be interpreted through the difference between the temperatures of the components. The best-fitting model favours a detached configuration with an inclination $i = 65.7^{\circ}$ (Fig.~\ref{model_apd}). We find an effective temperature of $T_2 = 31640$\,K for the secondary star suggesting that this system is most likely composed of O6.5\,V + O9.5\,V.

\subsection{HD\,92607}

The GOSC (ID:0170) gives a spectral type O9\,IV; therefore we fixed the temperature of the primary star to $T_1 = 32882$\,K. The amplitude variations for the primary and secondary minimum eclipses are $\Delta D_{p} = 14\%$ and $\Delta D_{s} = 14\%$, suggesting similar temperatures for both components. The primary component fills up its Roche lobe about 97\%, the secondary about 95\% favouring a near-contact configuration with an orbital inclination of 55.8$^{\circ}$ (Fig.~\ref{model_apd}). We find an effective temperature of $T_2 = 31655$\,K for the secondary star corresponding to a spectral type O9.5V.

\subsection{EM\,Car}

The GOSC (ID:0260) gives a spectral type O7.5\,V; therefore we fixed the temperature of the primary star to $T_1 = 35874$\,K. We find a period of $3.41430 \pm 0.0102392$. The same period of 3.4143 has been reported by \cite{2003AcA....53..341P} with ASAS catalogue. The primary component fills up its Roche lobe at $\sim 77\%$, the secondary at $\sim 69\%$ favouring a detached configuration with an orbital inclination $i = 80.1^{\circ}$ (Fig.~\ref{model_apd}). We find an effective temperature of $T_2 = 31130$\,K for the secondary star corresponding to a spectral type O9.5\,V.

\subsection{HD\,115071}

The GOSC (ID:0290) gives a spectral type O9.5\,III; therefore we fixed the temperature of the primary star to $T_1 =
30789$\,K. The primary component fills up its Roche lobe at $\sim 88\%$, the secondary at $\sim 64\%$ favouring a semi-
detached configuration with an orbital inclination $i = 56.8^{\circ}$ (Fig.~\ref{model_apd}). We find an effective temperature of
$T_2 = 30080$\,K for the secondary star corresponding to a spectral type O9.5\,III.

\subsection{HD\,152219}

For HD\,152219 we find a period of $4.24028 \pm 0.00222341$ days which is identical to the spectroscopic period reported by \cite{2006MNRAS.371...67S}. These authors mention that eclipses are observed but their light curve does not have sufficient data to perform a more quantitative analysis. Our light curves shown very well defined eclipses. We fixed the temperature of the primary star to $T_1 = 30789$\,K corresponding to a spectral type O9.5\,III. The primary component fills up its Roche lobe at $\sim 84\%$, the secondary at $\sim 79\%$ favouring a semi-detached configuration with an orbital inclination $i = 65.4^{\circ}$ (Fig.~\ref{model_apd}). We find an effective
temperature of $T_2 = 25560$\,K for the secondary star corresponding to a spectral type B1\,V.

\subsection{CD\,$-$\,41\degr\,11042}

\begin{figure*}
  %\centering
  \includegraphics[width=0.67\columnwidth]{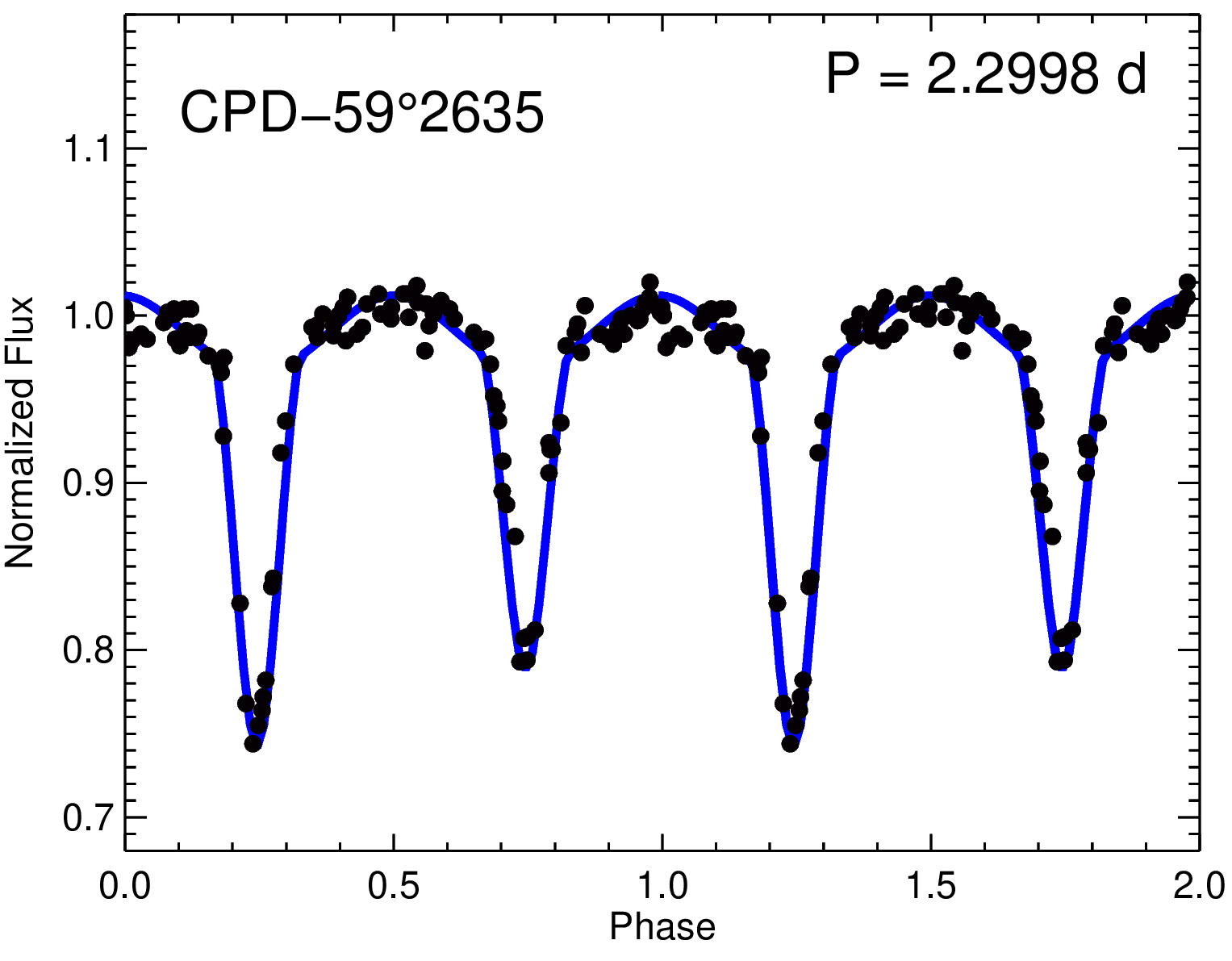}
  \includegraphics[width=0.67\columnwidth]{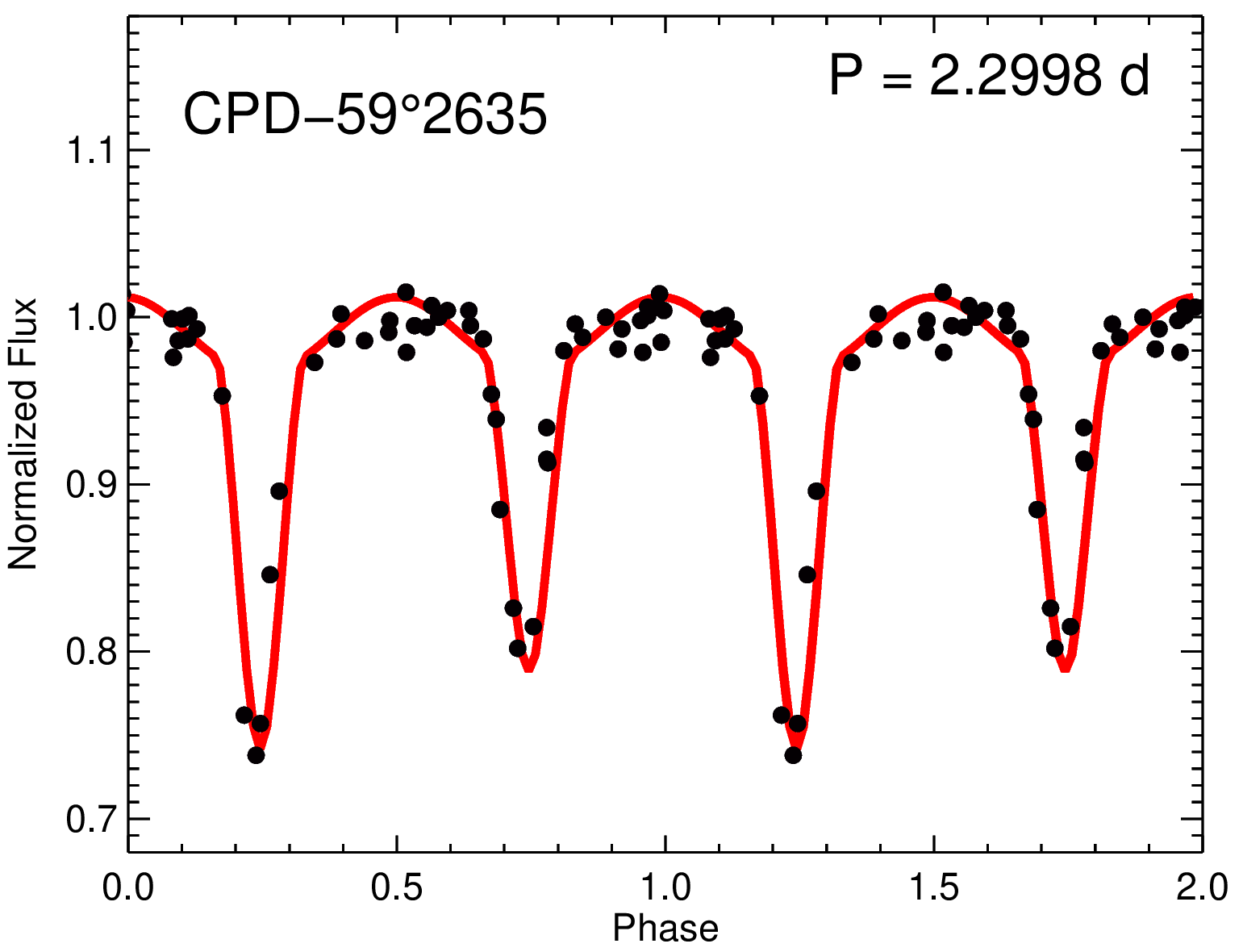}
  \includegraphics[width=0.67\columnwidth]{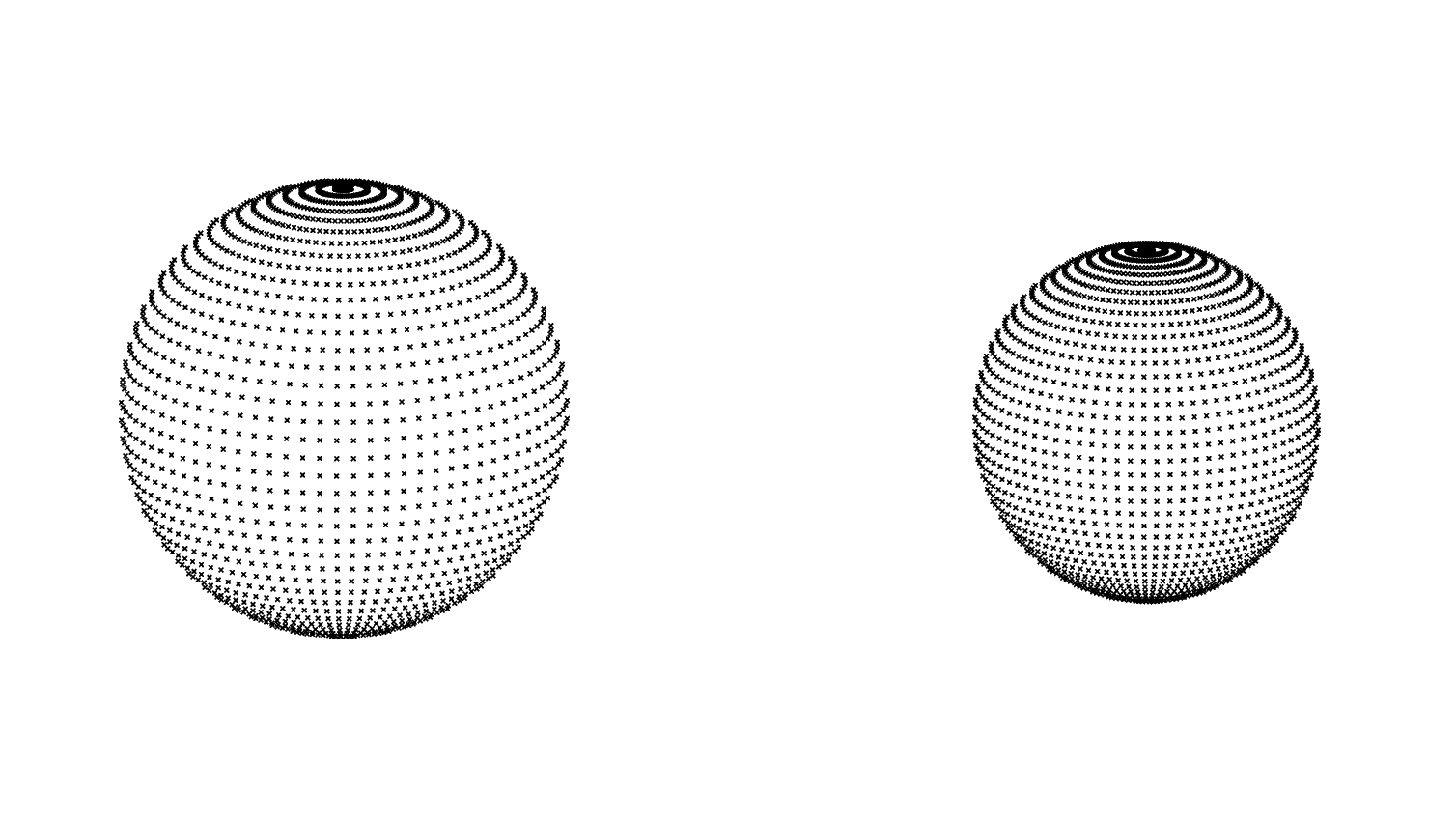}
  \includegraphics[width=0.67\columnwidth]{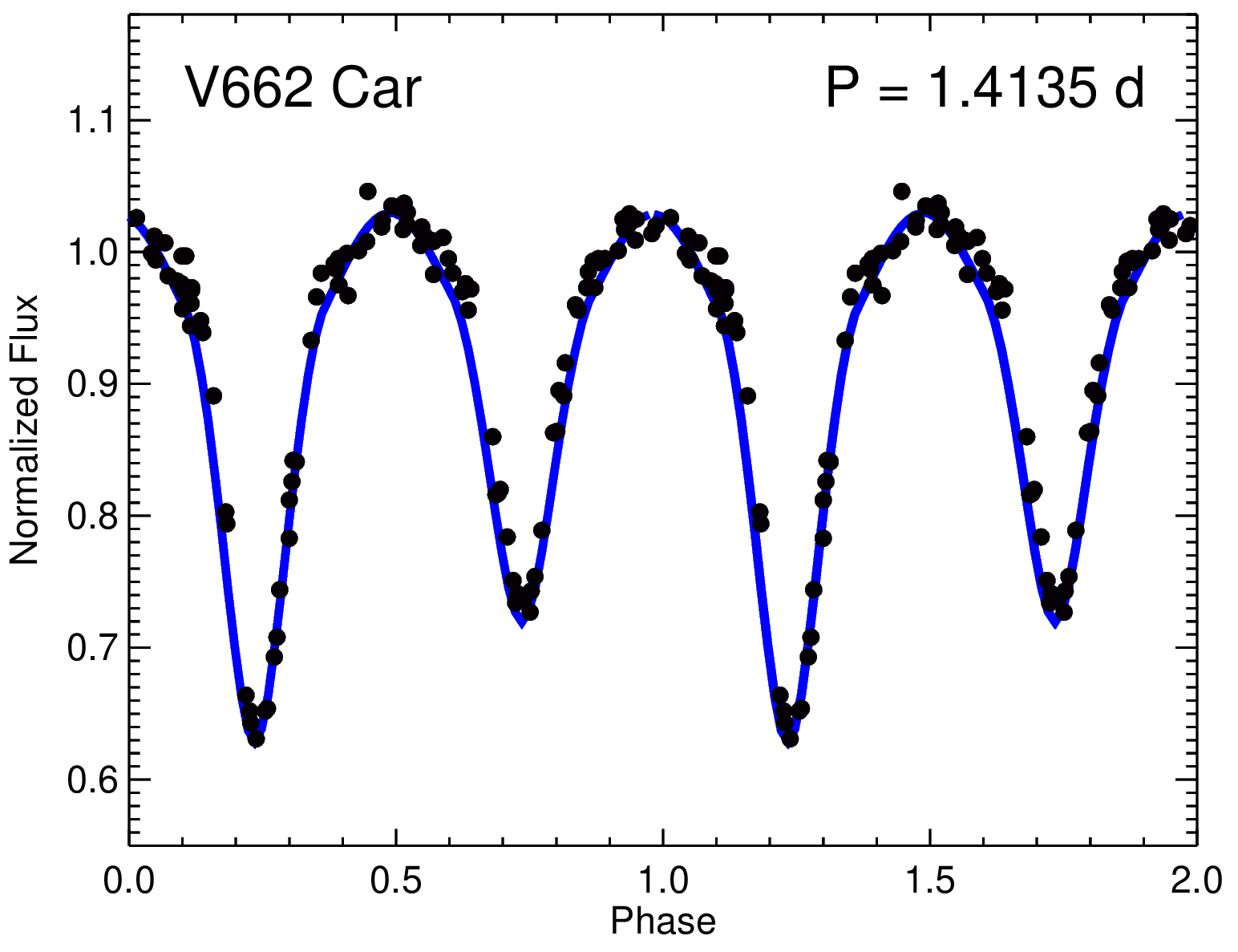}
  \includegraphics[width=0.67\columnwidth]{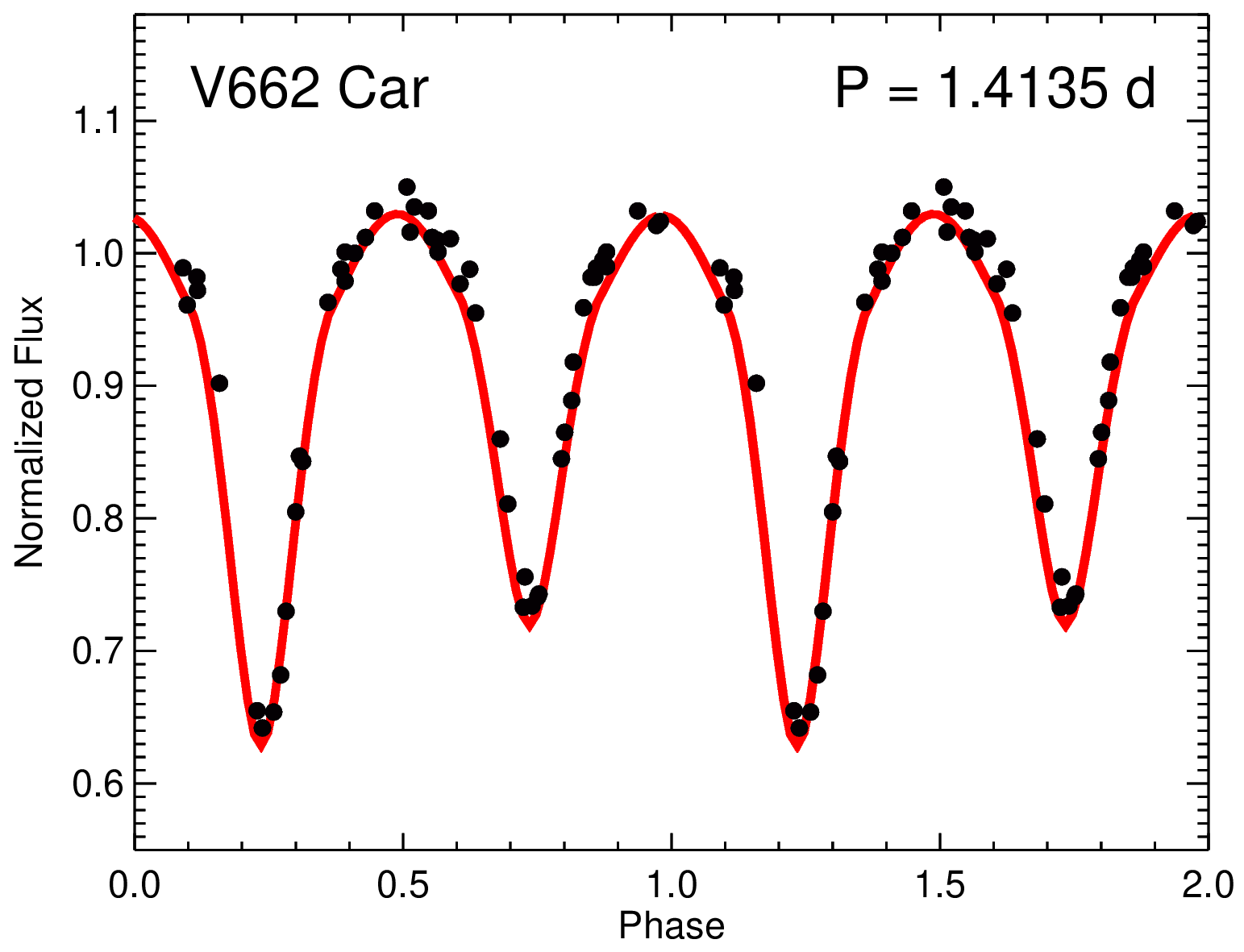}
  \includegraphics[width=0.67\columnwidth]{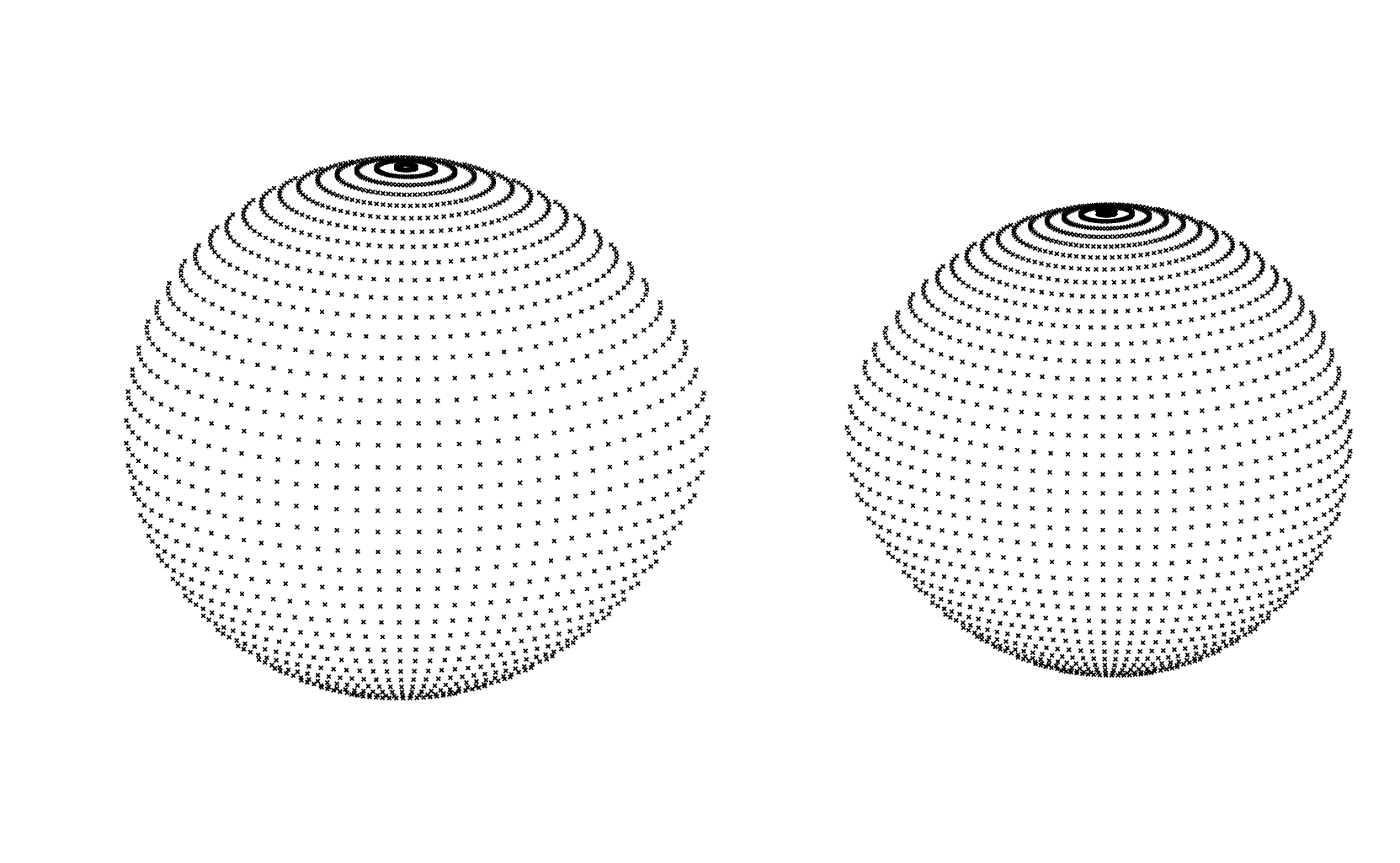}
  \includegraphics[width=0.67\columnwidth]{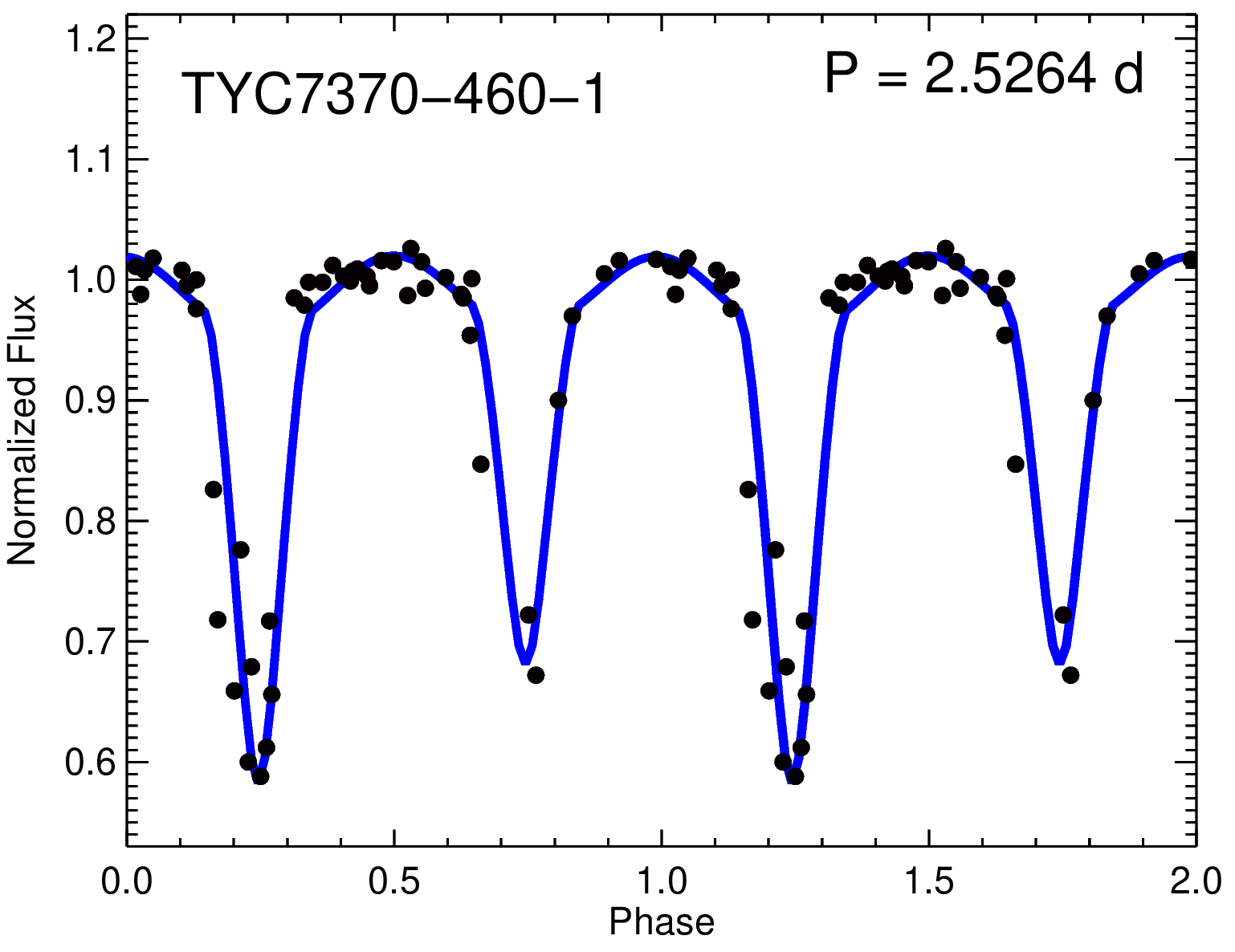}
  \includegraphics[width=0.67\columnwidth]{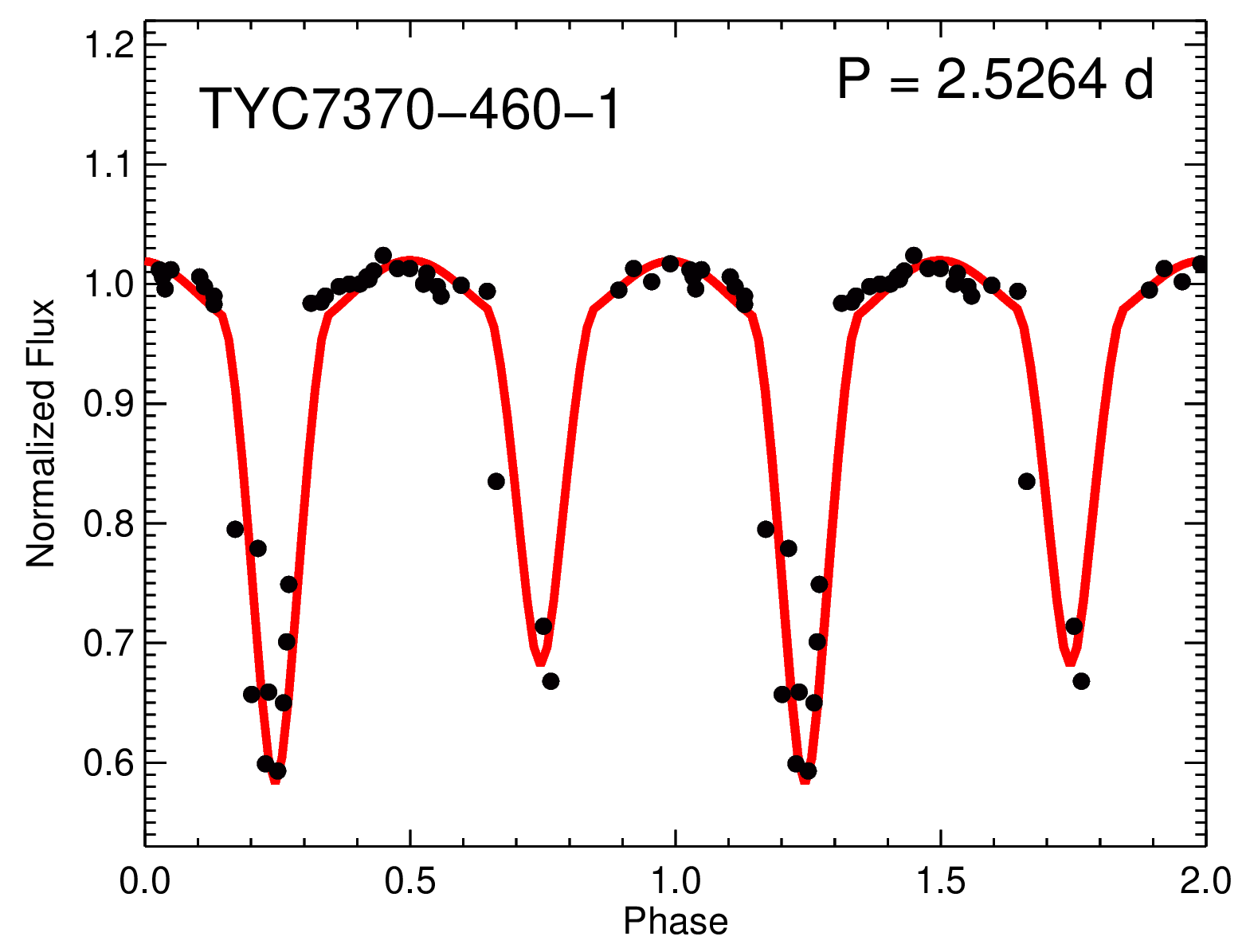}
  \includegraphics[width=0.67\columnwidth]{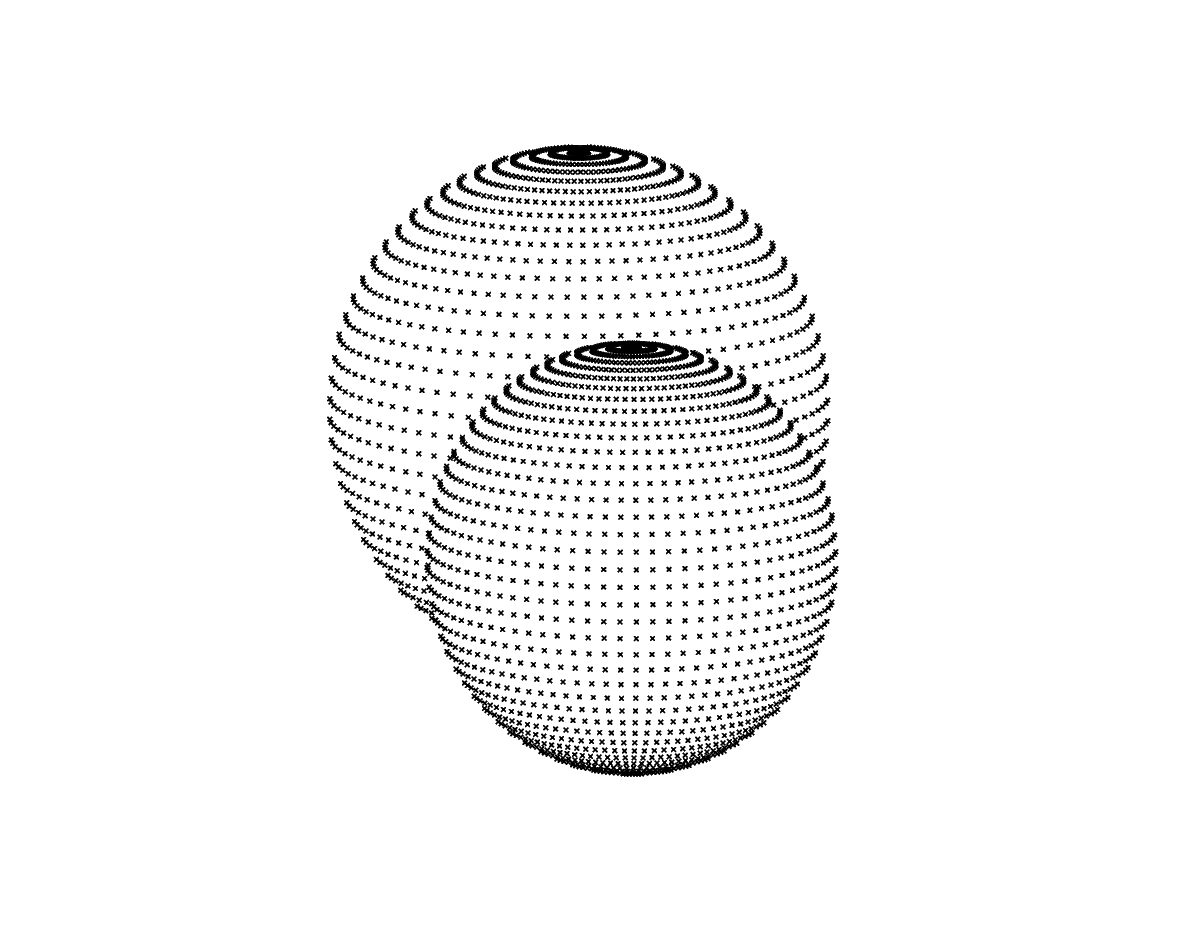}
  \includegraphics[width=0.67\columnwidth]{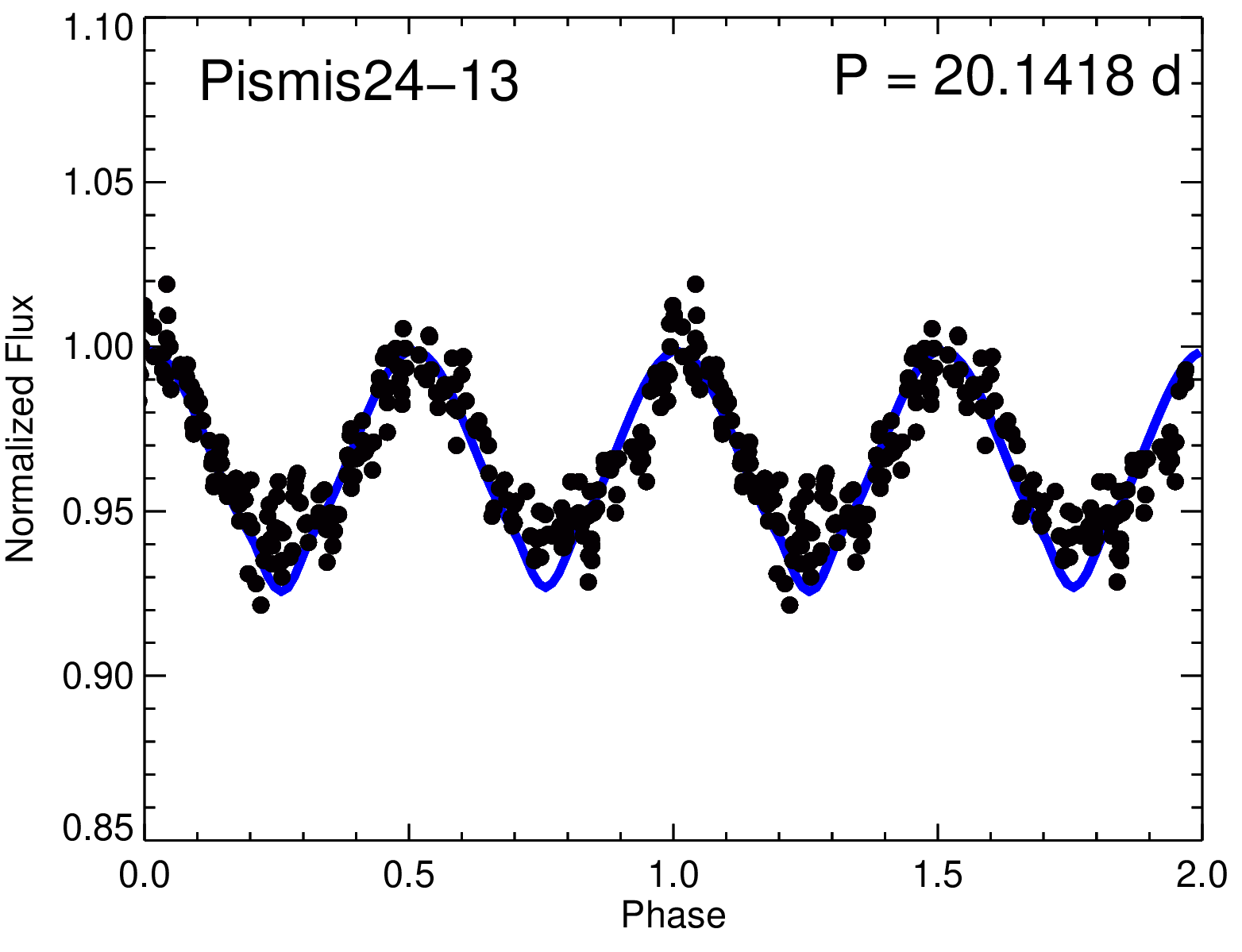}
  \includegraphics[width=0.67\columnwidth]{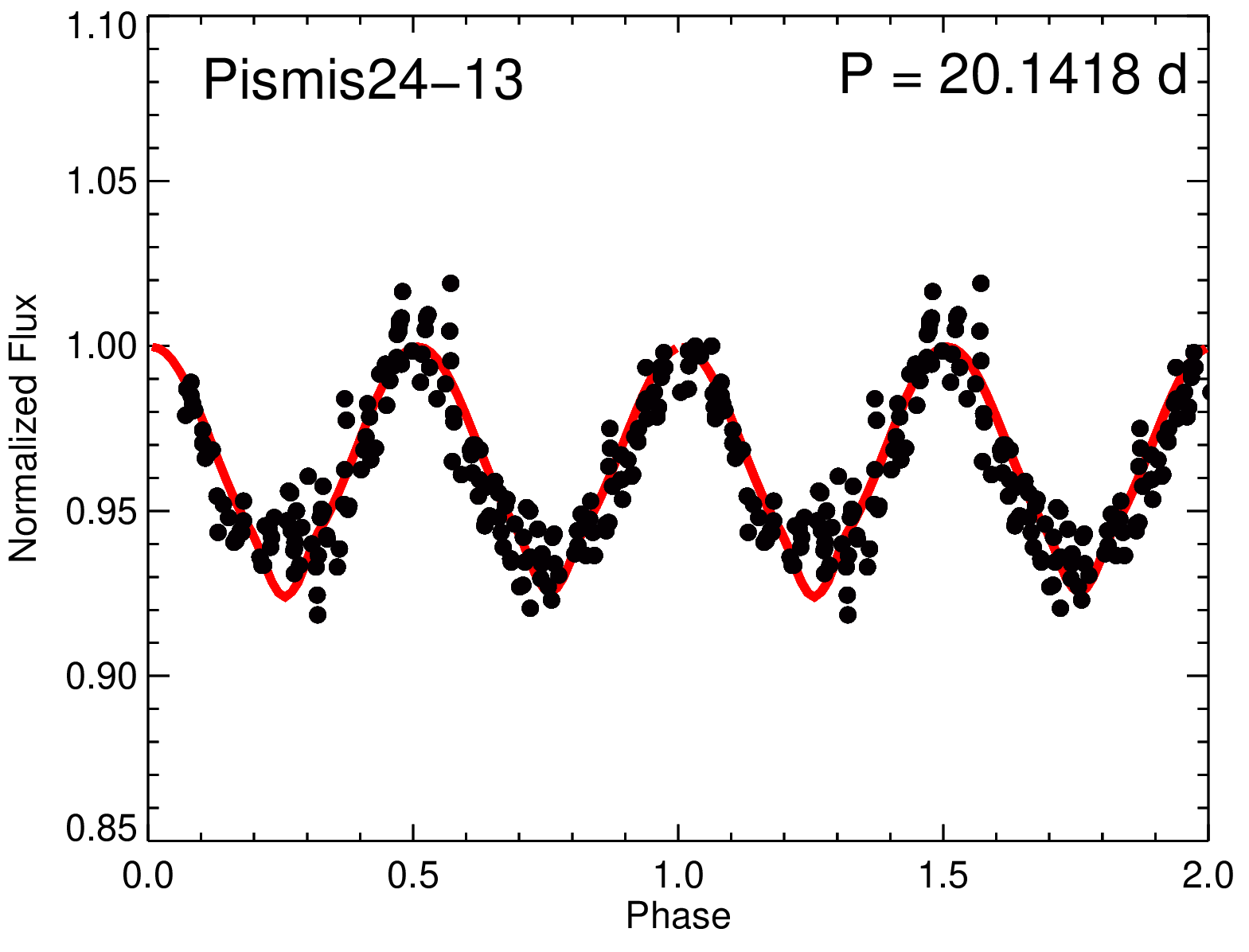}
  \includegraphics[width=0.67\columnwidth]{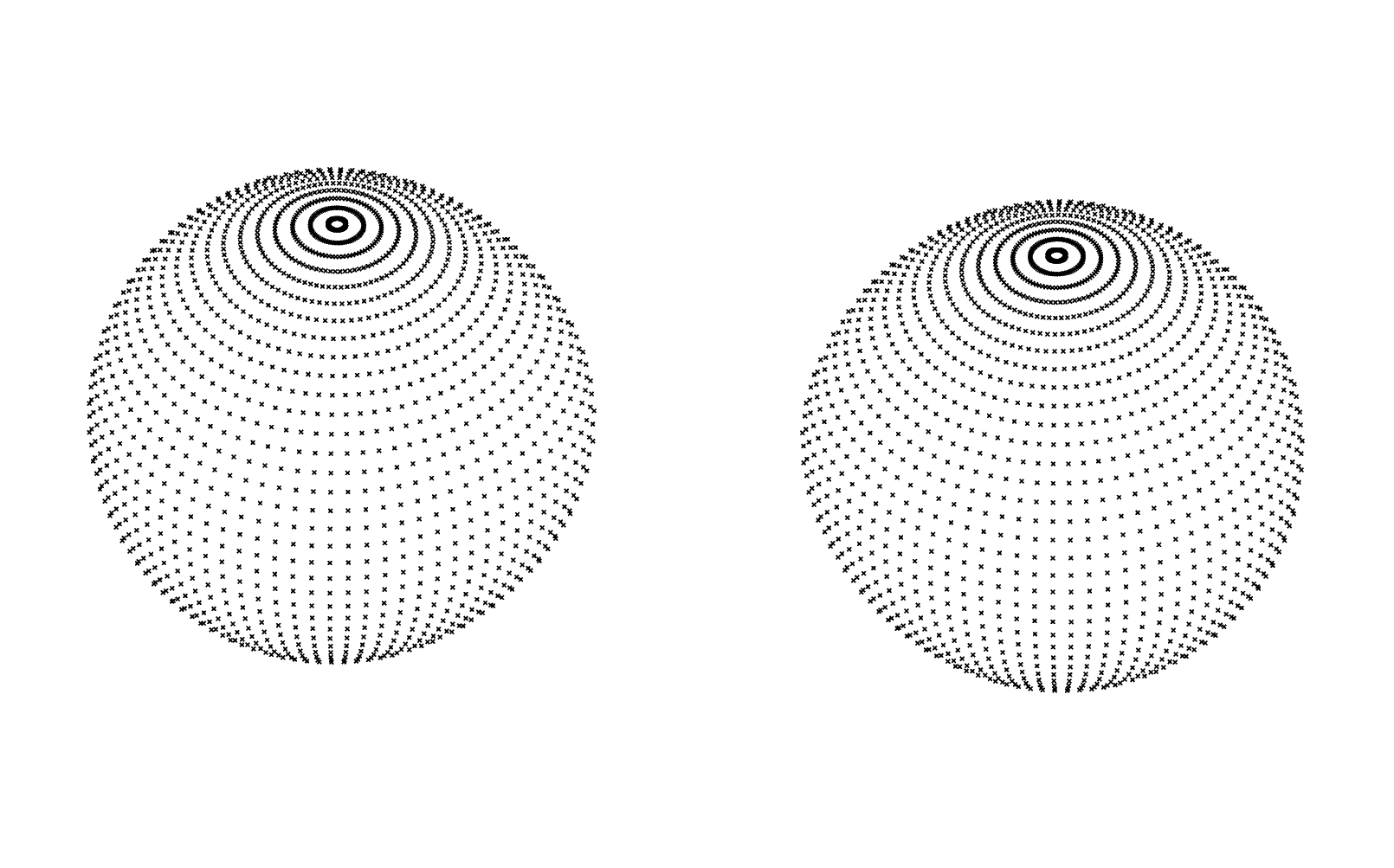}
  \includegraphics[width=0.67\columnwidth]{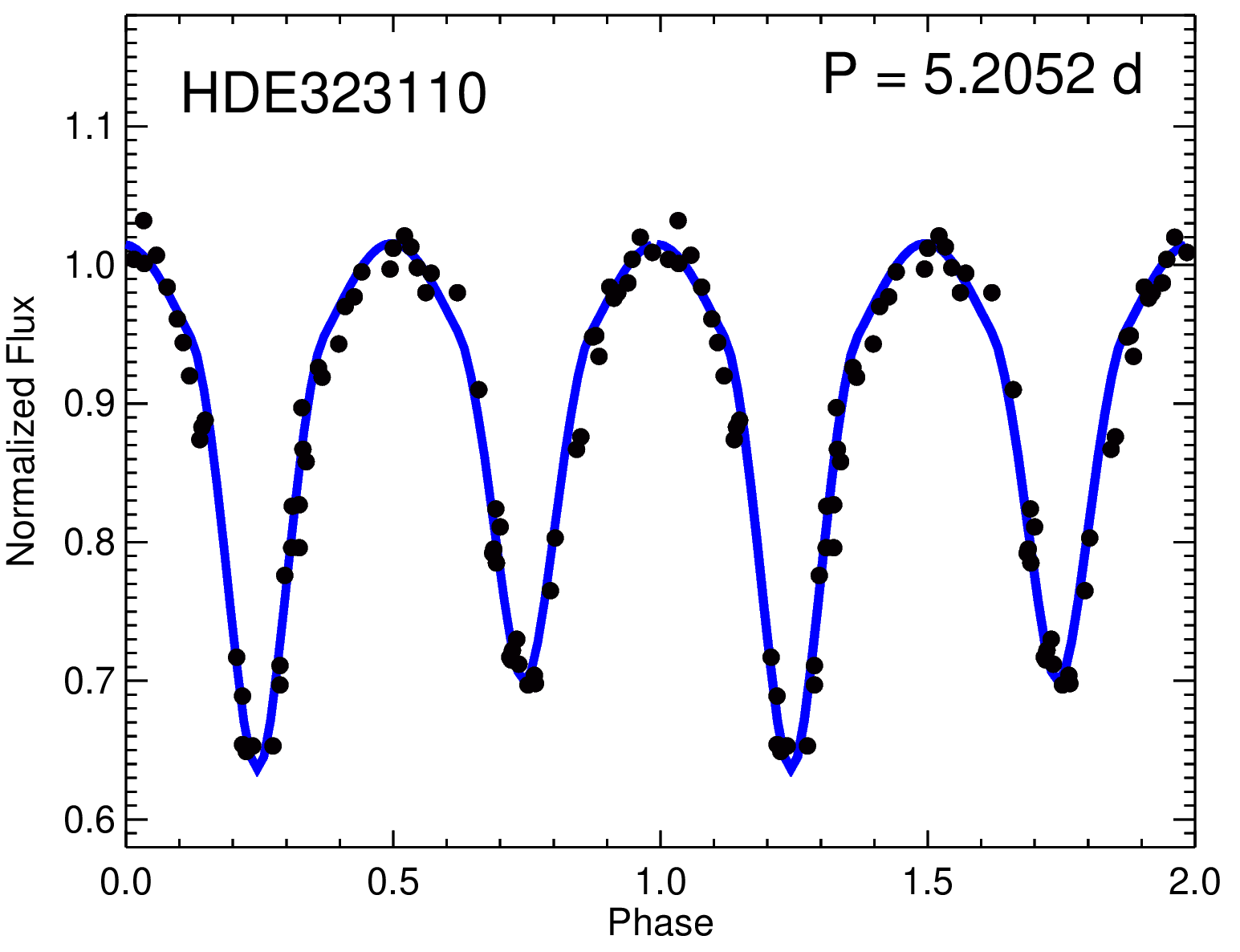}
  \includegraphics[width=0.67\columnwidth]{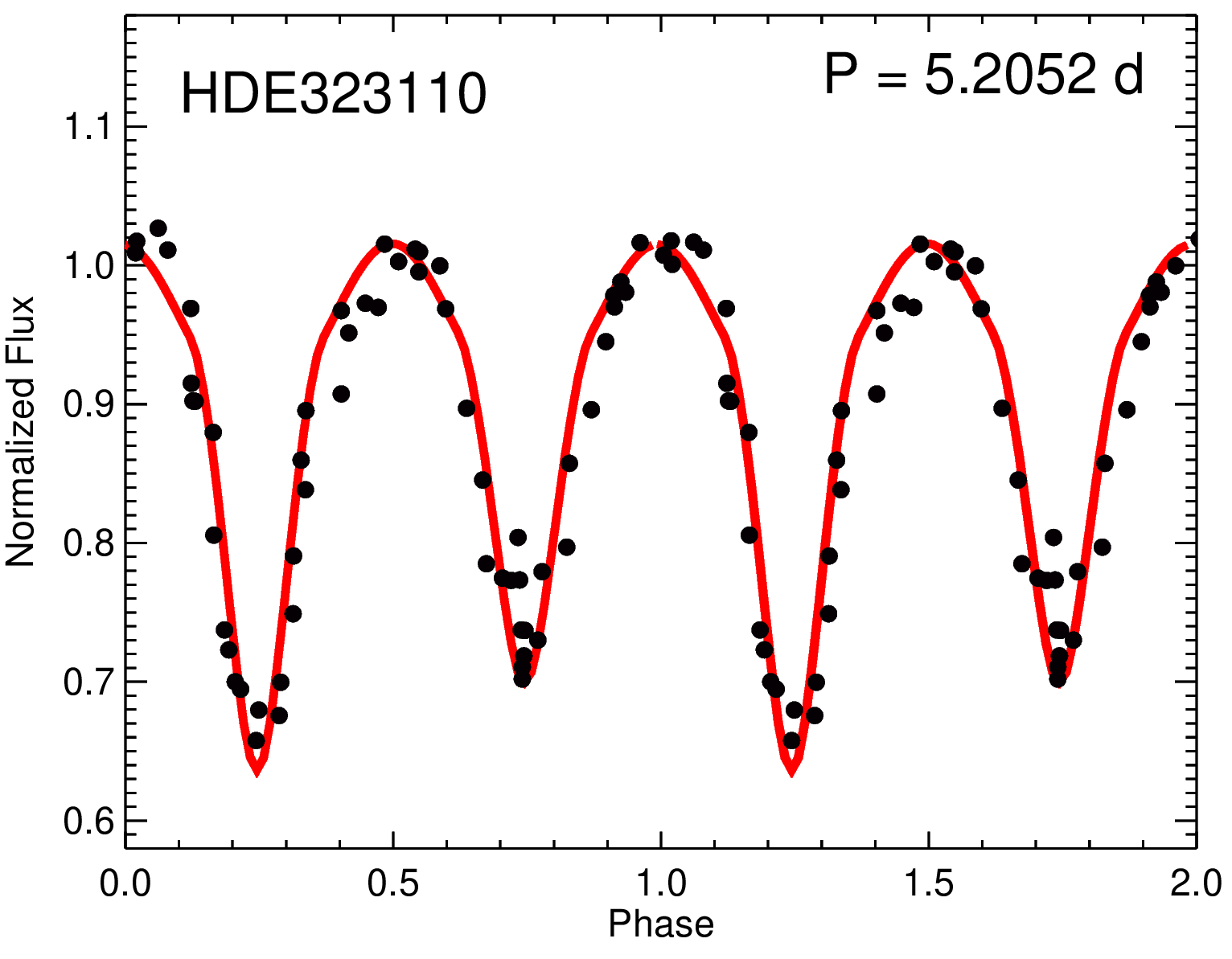}
  \includegraphics[width=0.67\columnwidth]{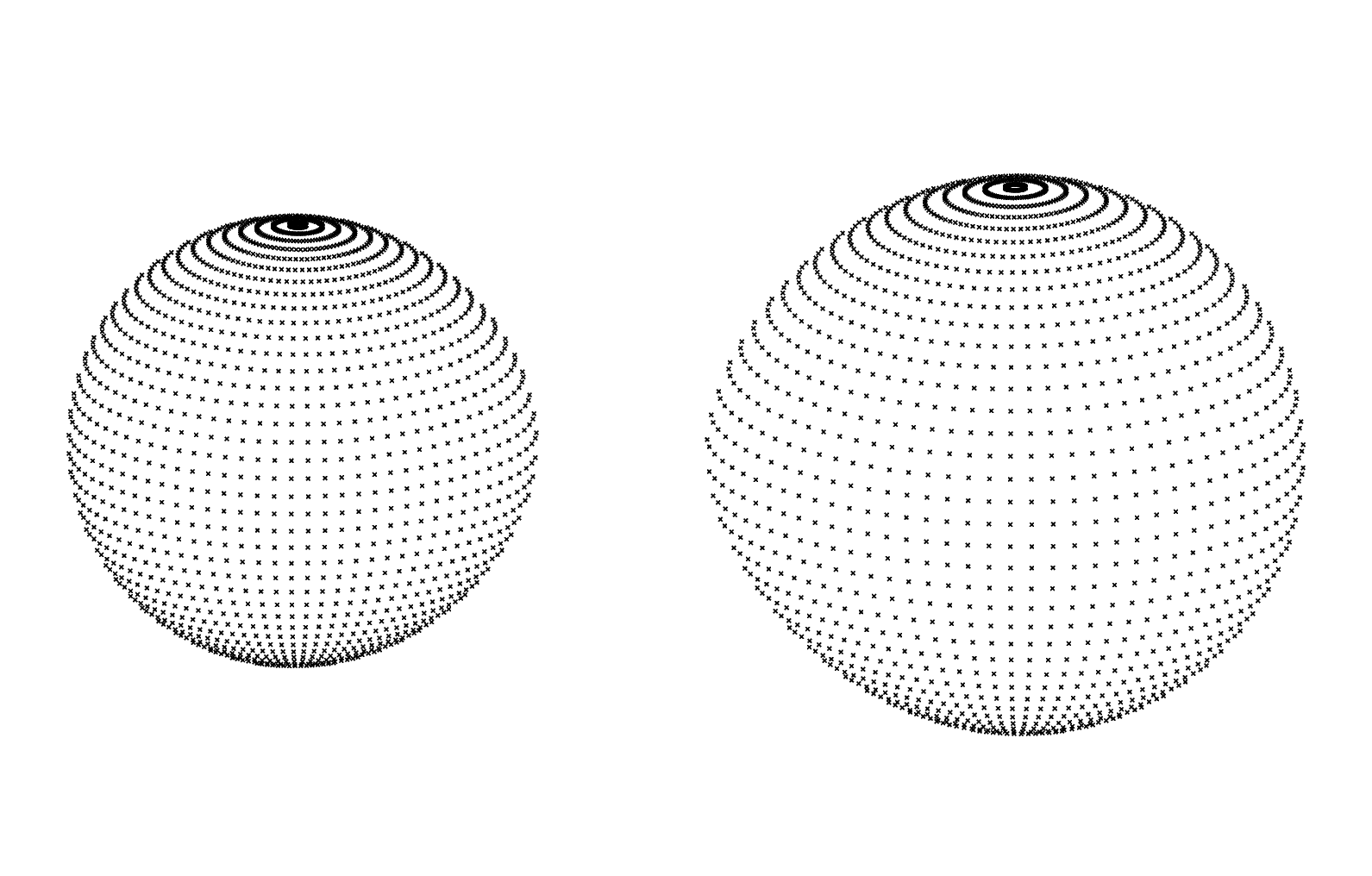}
  \caption{Same as Figure A1.}
  \label{model_apd}
\end{figure*}

For CD\,$-$\,41\degr\,11042 (V1034\,Sco) the GOSC (ID:0351) gives a spectral type O9.2\,IV; therefore we fixed the temperature of the primary star to $T_1 = 32383$\,K. The primary component fills up its Roche lobe at $\sim 79\%$, the secondary at $\sim 80\%$ favouring a semi-detached configuration with an orbital inclination $i = 69.7^{\circ}$ (Fig.~\ref{model_apd}). We obtain an effective temperature of $T_2 = 32000$\,K for the secondary star corresponding to a spectral type O9\,IV.

\section{Individual notes for other sources from GOSC}

\begin{table*}
\begin{center}
\caption{Characteristics of GOSC stars with doubtful eclipsing binary nature.}
\label{tab:goscmatch}
\begin{tabular}{clllcl}
\hline\hline
\noalign{\smallskip}
             \multicolumn{1}{c}{No.}
           & \multicolumn{1}{c}{Name}
           & \multicolumn{1}{c}{RA}
           & \multicolumn{1}{c}{Dec}
           & \multicolumn{1}{c}{spectral designations}
           & \multicolumn{1}{c}{references}\\
             \multicolumn{1}{c}{}
           & \multicolumn{1}{c}{}
           & \multicolumn{2}{c}{J2000}
           & \multicolumn{1}{c}{}
           & \multicolumn{1}{c}{}\\
             \hline
             \noalign{\smallskip}
1  & HD\,52533\,A 				& 07 01 27.05  & -03 07 03.28  & O8.5IV      	  & \cite{2007ApJ...655..473M} \\
2  & HD\,64315\,AB  			& 07 52 20.28  & -26 25 46.69  & O5.5V, O7V  	  & -- \\
%3  & V\,467\,Vel    			& 08 43 49.81  & -46 07 08.78  & O6.5V       	  & --       \\
%4  & HD\,92607  				& 10 40 12.43  & -59 48 10.10  & O9IV        	  & --     \\
3  & HDE\,303312 				& 10 43 30.84  & -59 29 23.80  & O9.7IV           & --       \\
4  & HD\,93161\,A         		& 10 44 08.84  & -59 34 34.49  & O7.5V, O9V       & \cite{2005MNRAS.359..688N} \\
5  & HD\,93205        			& 10 44 33.74  & -59 44 15.46  & O3.5V, O8V       & \cite{2000ApJ...529..463A} \\
%8  & CPD\,$-$\,59\degr\,2603    & 10 44 47.31  & -59 43 53.23  & O7.5V, B0V(n)    & \cite{2001MNRAS.326.1149R} \\
%9  & CPD\,$-$\,59\degr\,2628   	& 10 45 08.23  & -59 40 49.48  & O9.5V, B0.5V(n)  & --       \\
%10 & CPD\,$-$\,59\degr\,2635 	& 10 45 12.72  & -59 44 46.17  & O8V, O9.5V  	  & --   \\
%11 & V662\,Car             		& 10 45 36.32  & -59 48 23.37  & O5V, B0:V        & --       \\
6 & ALS\,18553				 	& 10 58 37.77  & -61 08 00.35  & O6II		      & --       \\
7 & THA\,35$-$\,II$-$\,153  	& 10 59 00.81  & -61 08 50.24  & O3.5I            & --     \\
%14 & EM\,Car	 				& 11 12 04.50  & -61 05 42.94  & O7.5V, O7.5V     & \cite{2003AcA....53..341P}    \\
8 & HD\,101131\,AB				& 11 37 48.44  & -63 19 23.51  & O5.5V, O8:V      & \cite{2002ApJ...574..957G} \\
9 & HD\,101190                 & 11 38 09.91  & -63 11 48.61  & O6IV             & -- \\
10 & HD\,101223 				& 11 38 22.77  & -63 12 02.80  & O8V              & --		\\
%18 & HD\,115071		 			& 13 16 04.80  & -62 35 01.47  & O9.5III, B0Ib    & --       \\
11 & HD\,120678 				& 13 52 56.41  & -62 43 14.24  & O9.5V            & --       \\
%20 & HD\,152219					& 16 53 55.61  & -41 52 51.47  & O9.5III          & \cite{2006MNRAS.371...67S} \\
12 & HD\,152218              	& 16 53 59.99  & -41 42 52.83  & O9IV, B0:V:      & \cite{2008NewA...13..202S}     \\
%22 & CD\,$-$\,41\degr\,11042	& 16 54 19.85  & -41 50 09.36  & O9.2IV, B1:V     & --       \\
%23 & TYC\,7370-460-1   			& 17 18 15.40  & -34 00 05.94  & O6V, O8V         & --       \\
%24 & HDE\,323110 				& 17 21 15.79  & -37 59 09.58  & ON9Ia            & \cite{2003AcA....53..341P} \\
%25 & Pismis\,24-13              & 17 24 45.79  & -34 09 39.94  & O6V              & --       \\
\hline
\end{tabular}
\end{center}
Notes:
The two spectral types in column (5) indicate the existence of a secondary component in the unresolved spectrum from the GOSC. The references in column (6) correspond to previous studies carried out on the specific sources and which are not related to the information from the GOSC.
\end{table*}

\subsection{HD\,52533\,A}

HD\,52533\,A or ALS\,9251 (ID:0115 GOSC) shows an $i$ variability amplitude $\delta_{Ai} = 0.3598$\,mag. We find a period of $21.9652 \pm 0.00267$ days but the eclipses are not well constrained (Fig.~\ref{phase_lc_GOSC}). \cite{2007ApJ...655..473M} reported a period of $22.186 \pm 0.0002$ days but light curves have not been published.

\subsection{HD\,64315\,AB}

HD\,64315\,AB or CPD\,$-$\,26\degr\,2698 (ID:0129 GOSC) shows variability amplitudes $\delta_{Ar} = 0.1098$ and $\delta_{Ai} = 0.1091$ mag in $r$ and $i$, respectively. We find a period of $23.4548 \pm 0.000876$ days. The absence of a secondary eclipse suggest another type of variability.

\subsection{HDE\,303312}

HDE\,303312 or TYC\,5322-2139-1 (ID:0175 GOSC). We find a period of $6.27386 \pm 0.000653807$ days. A primary eclipse can be seen in the folded light curve (Fig.~\ref{phase_lc_GOSC}) with an amplitude variation $\Delta_{p} = 15$\%. Probably the secondary component is too faint to be detected.

\subsection{HD\,93161\,A}

HD\,93161\,A or CPD\,$-$\,58\degr\,2631A (ID:0193 GOSC) shows an amplitude variation $\delta_{Ai} = 0.139$ mag. We find a period of $8.56620 \pm 0.0006$ days but eclipses are not well constrained. Our photometric period is identical to the spectroscopic period $8.566 \pm 0.004$ reported by \cite{2005MNRAS.359..688N}.

\subsection{HD\,93205}

HD\,93205 or V* V560 Car (ID:0203 GOSC) shows a variability amplitude $\delta_{Ai} = 0.032$ mag. A period of $6.08$ days has been reported by \cite{2000ApJ...529..463A}. The authors measured a variability amplitude $\sim 0.02$\,mag using a narrow-band continuum filter centered at $5140 \pm 90$\,\AA\, at the Lowell 0.6\,m telescope. We did not find any periodicity during the time span of our monitoring campaign.

\subsection{ALS\,18553}

ALS\,18553 (ID:0244 GOSC) shows variability amplitudes $\delta_{Ar} = 0.090$ and $\delta_{Ai} = 0.087$ mag. We find a period of $6.11892 \pm 0.000290$ days. A primary eclipse can be seen in the folded light curve with an amplitude variation $\Delta_{p} = 7$\% (Fig.~\ref{phase_lc_GOSC}). The secondary component is not detected.

\subsection{THA35-II-153}

THA35-II-153 (ID:0248 GOSC) shows variability amplitudes $\delta_{Ar} = 0.219$ and $\delta_{Ai} = 0.201$ mag. We find a period of $8.00419 \pm 0.00114904$ days. A primary minimum is detected with an amplitude variation $\Delta_{p} = 15$\%. A secondary minimum with amplitude $\Delta_{s} = 5$\% is observed in $r$ with albeit larger scatter than in $i$. The small amount of observations during the primary minimum and the large scatter observed in the secondary minimum does not allow us to classify this system as EB.

\subsection{HD\,101131\,AB}

The star HD\,101131\,AB or V* V1051 Cen (ID:0270 GOSC) is a member of the young open cluster IC\,2944 (\citealt{1987AJ.....93..868W}). The system is classified as a doubled-lined spectroscopic binary with an elliptical orbit, and a period of $9.64659 \pm 0.0051301$ (\citealt{2002ApJ...574..957G}). We obtain a period of $9.64659 \pm 0.0051301$ days identical to the spectroscopic period reported by \cite{2002ApJ...574..957G}. The primary minimum is well defined while the secondary has a smaller amplitude. Our light curves lack sufficient observations to perform further analysis.

\subsection{HD\,101190}

HD\,101190 or CPD\,$-$\,62\degr\,2163 (ID:0272 GOSC) shows a variability amplitude $\delta_{Ai} = 0.102$ mag. We find a period of $22.2565 \pm 0.000393694$ days. The primary eclipse is well defined but the secondary has a smaller amplitude.

\subsection{HD\,101223}

HD\,101223 or CPD\,$-$\,62\degr\,2171 (ID:0275 GOSC) shows a variability amplitude $\delta_{Ar} = 0.095$ mag. We find a period of $2.66557 \pm 0.000603550$ days. The primary and secondary minimum eclipses are well defined and separated symmetrically by 0.15 in phase (Fig.~\ref{phase_lc_GOSC}).

\subsection{HD\,120678}

HD\,120678 or CPD\,$-$\,62\degr\,3703 (ID:0300 GOSC) shows a variability amplitude $\delta_{Ai} = 0.184$ mag. We find a period of $17.3780 \pm 0.000188527$ days. The primary minimum is not fully covered during the time span of our monitoring campaign and the secondary minimum has a larger scatter. We cannot conclude on the binary nature of the system.

\subsection{HD\,152218}

HD\,152218 or CPD\,$-$\,41\degr\,7713 (ID:0342) is located at the core of the Sco OB\,1 in the open cluster NGC\,6231 (\citealt{2008NewA...13..202S}). A spectroscopic period of 5.604 days has been reported by \cite{2008NewA...13..202S}. We find a period of $3.64521 \pm 0.000338727$ which differs significantly from the spectroscopic period. The EB system appears to be detached but the eclipses are not well constrained. Therefore, we do not consider the system for further analysis.

\section{High-quality light curves}

\begin{figure*}
  %\centering
  \includegraphics[width=0.67\columnwidth]{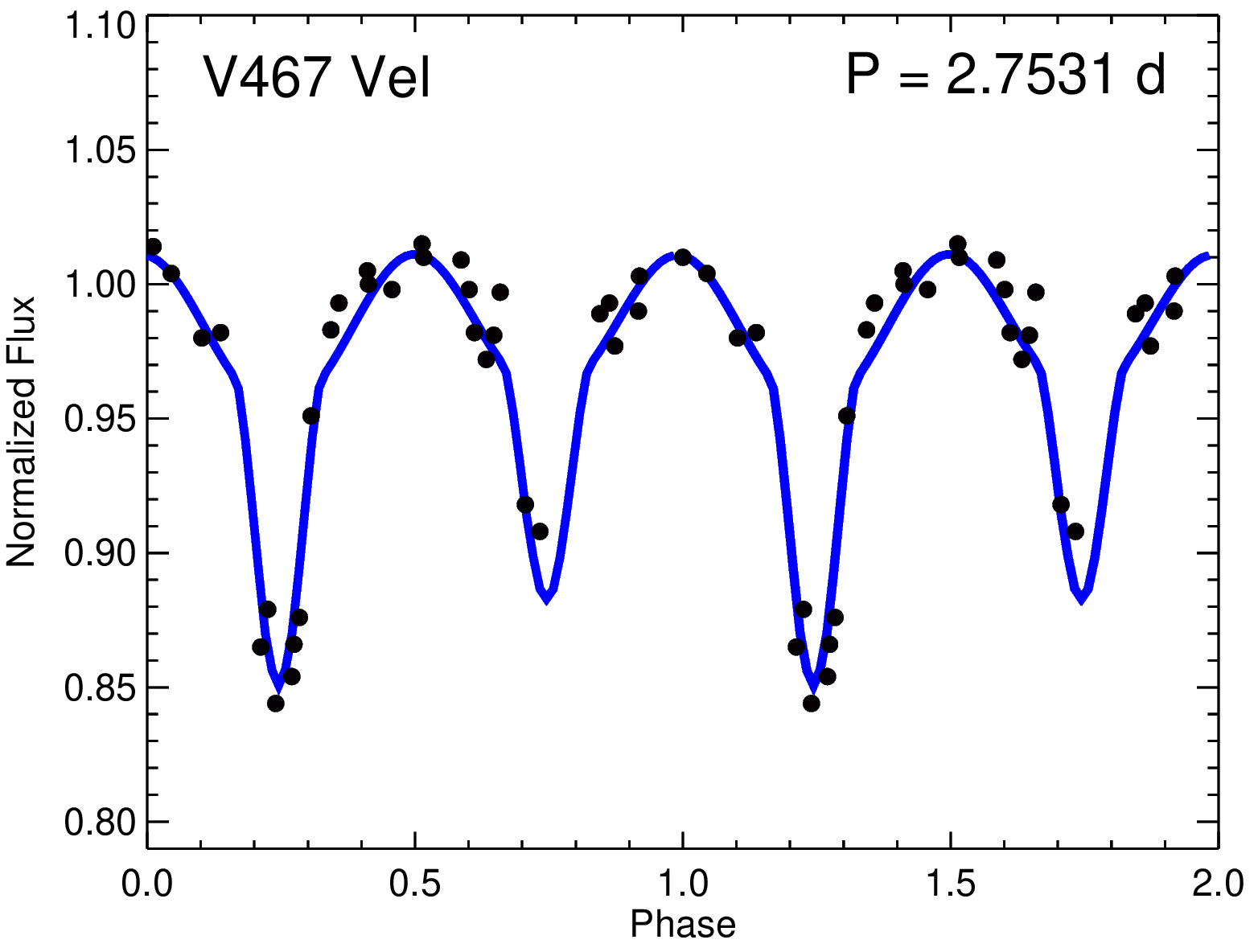}
  \includegraphics[width=0.67\columnwidth]{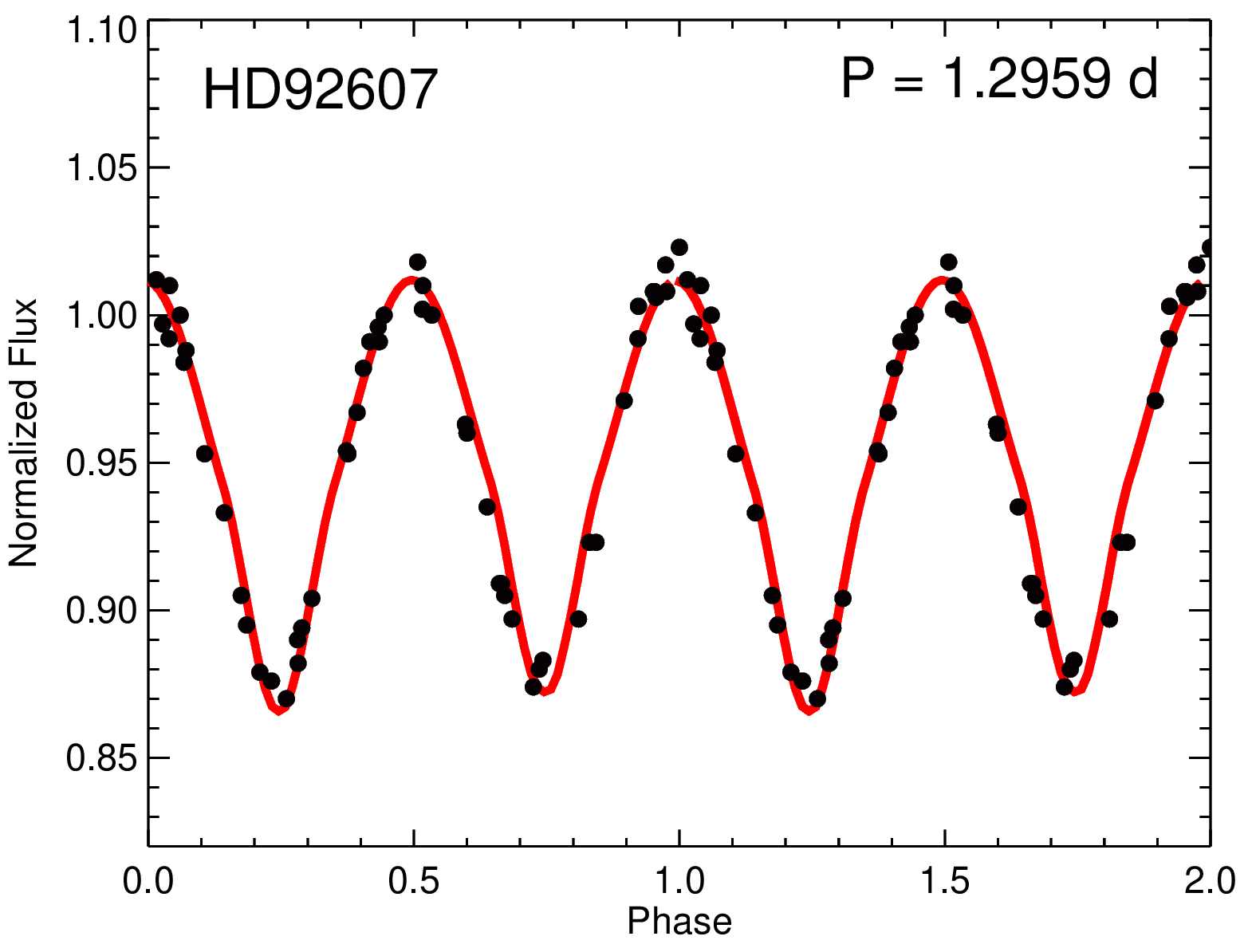}
  \includegraphics[width=0.67\columnwidth]{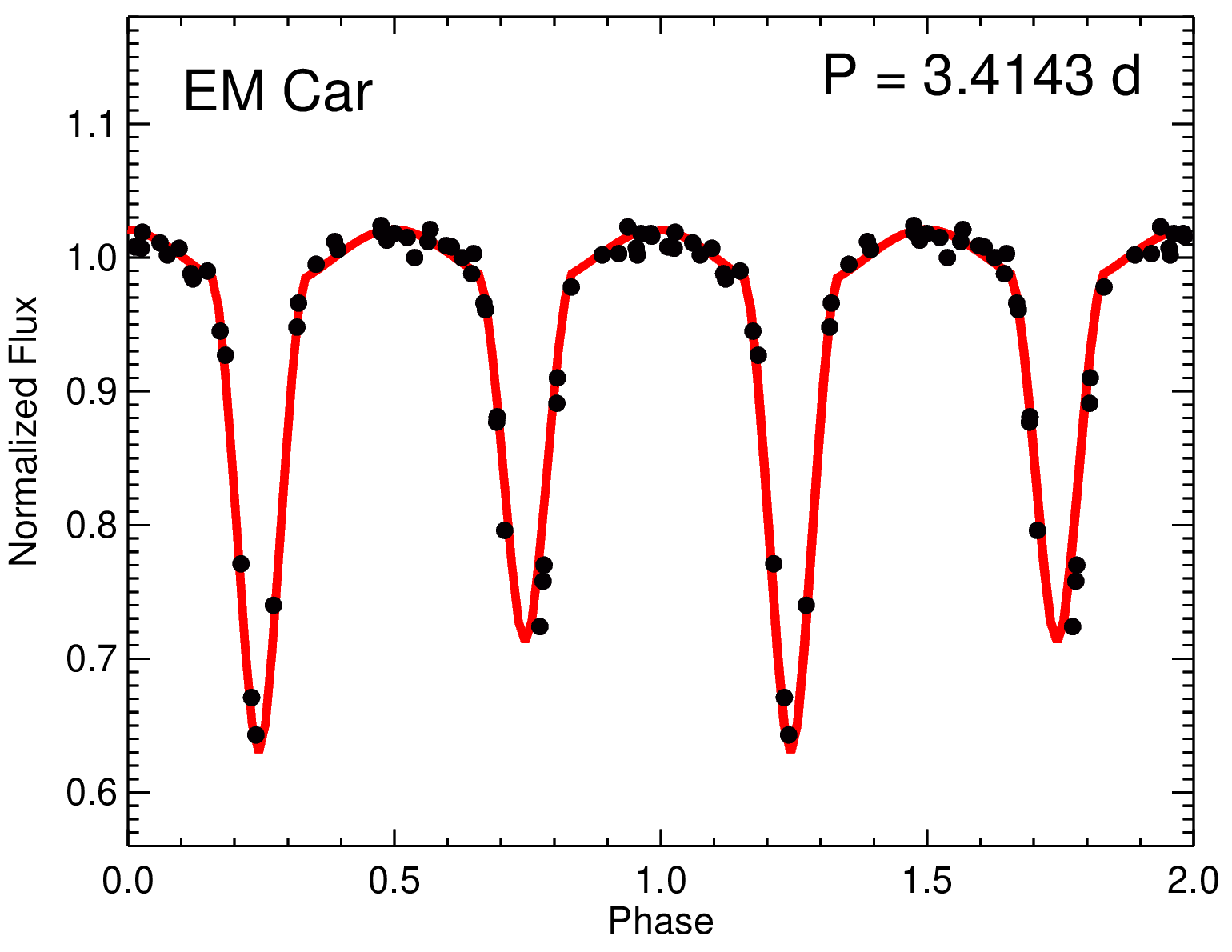}
  \includegraphics[width=0.67\columnwidth]{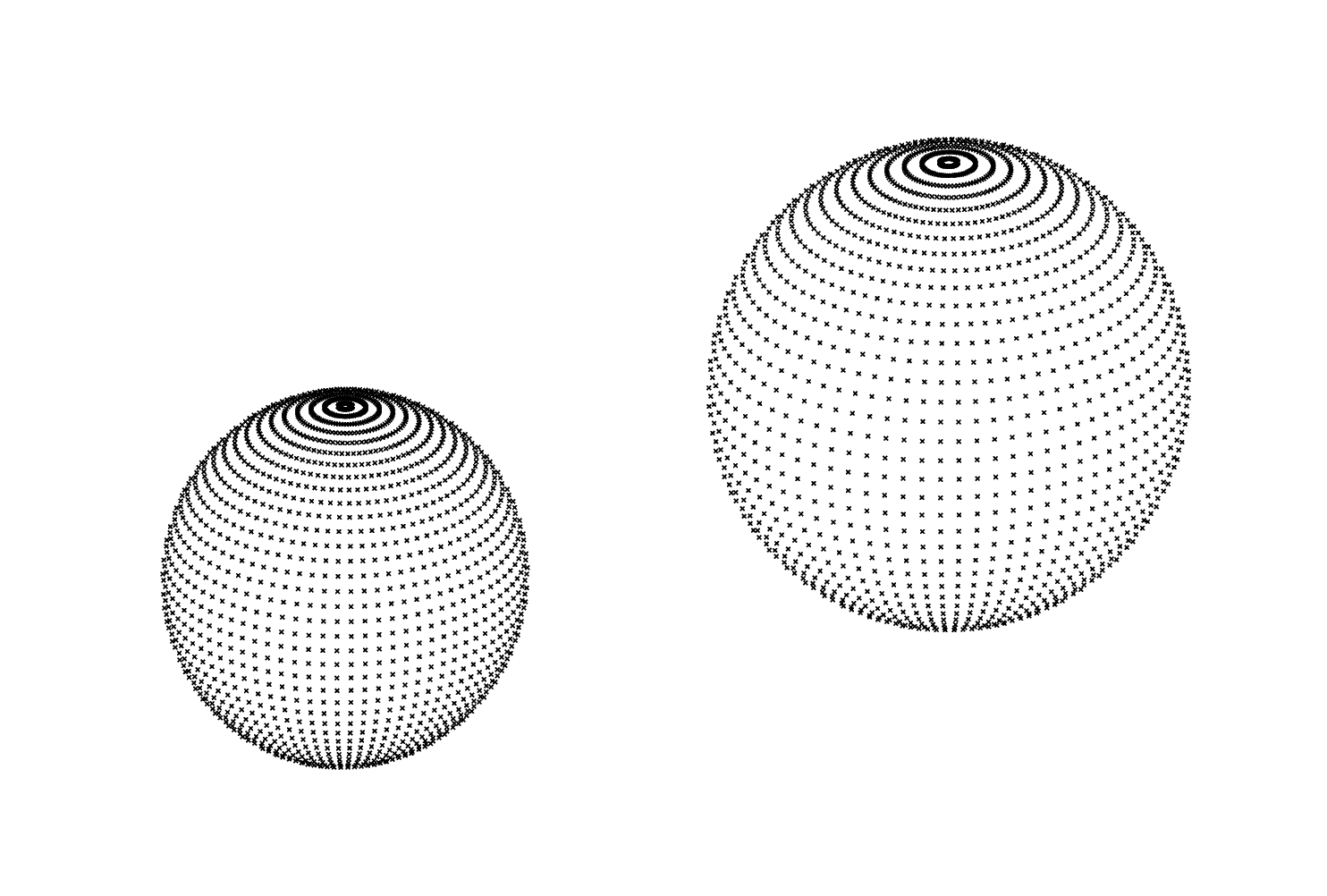}
  \includegraphics[width=0.67\columnwidth]{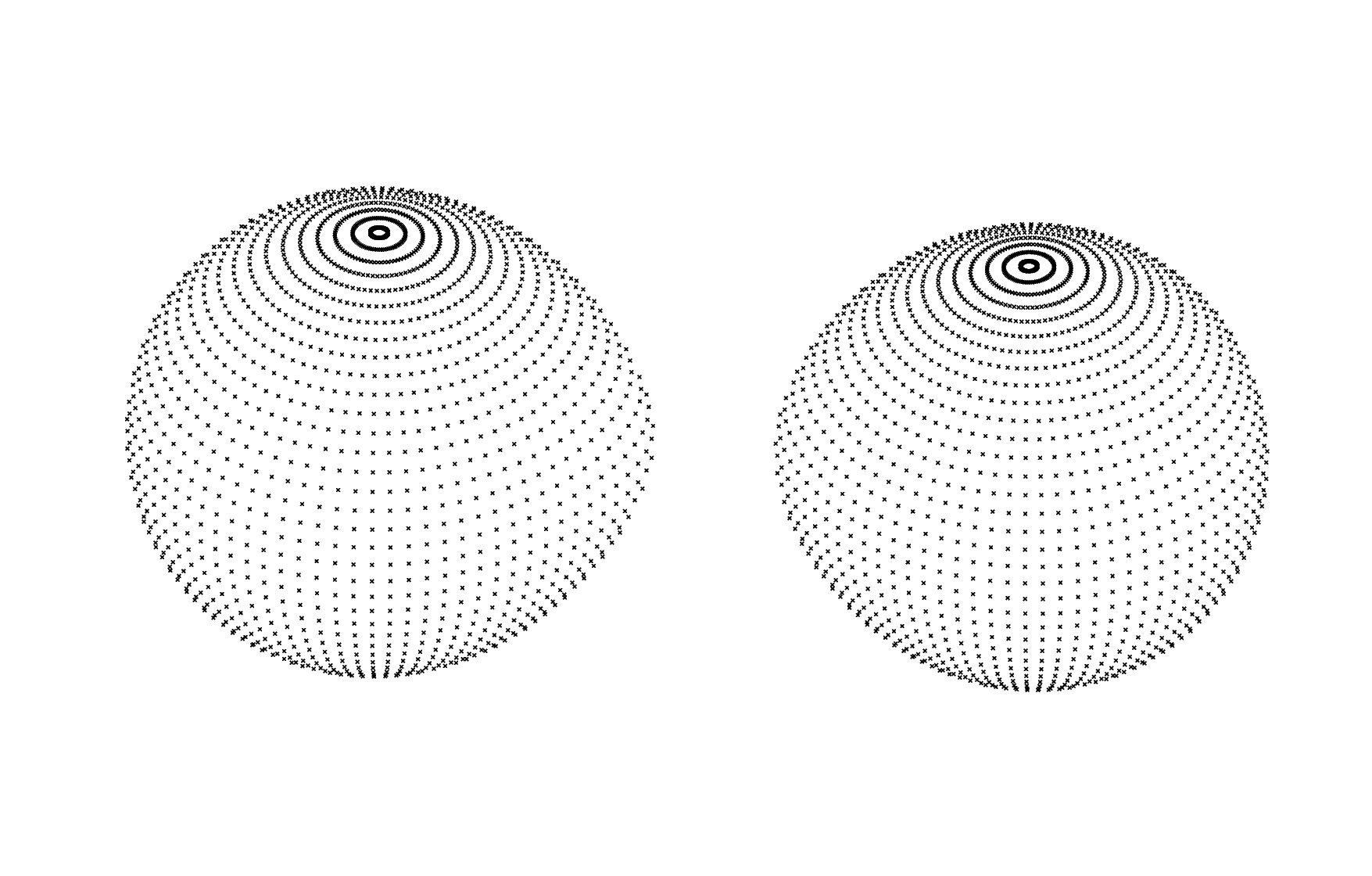}
  \includegraphics[width=0.67\columnwidth]{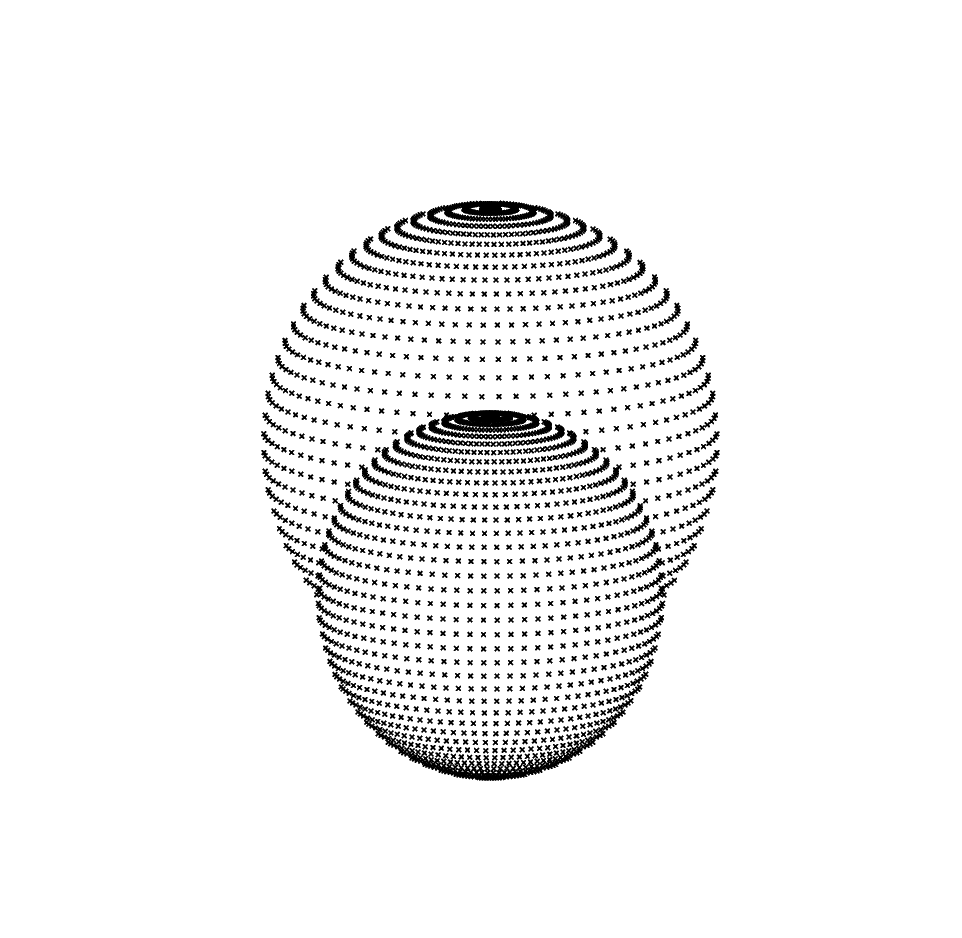}
  \includegraphics[width=0.67\columnwidth]{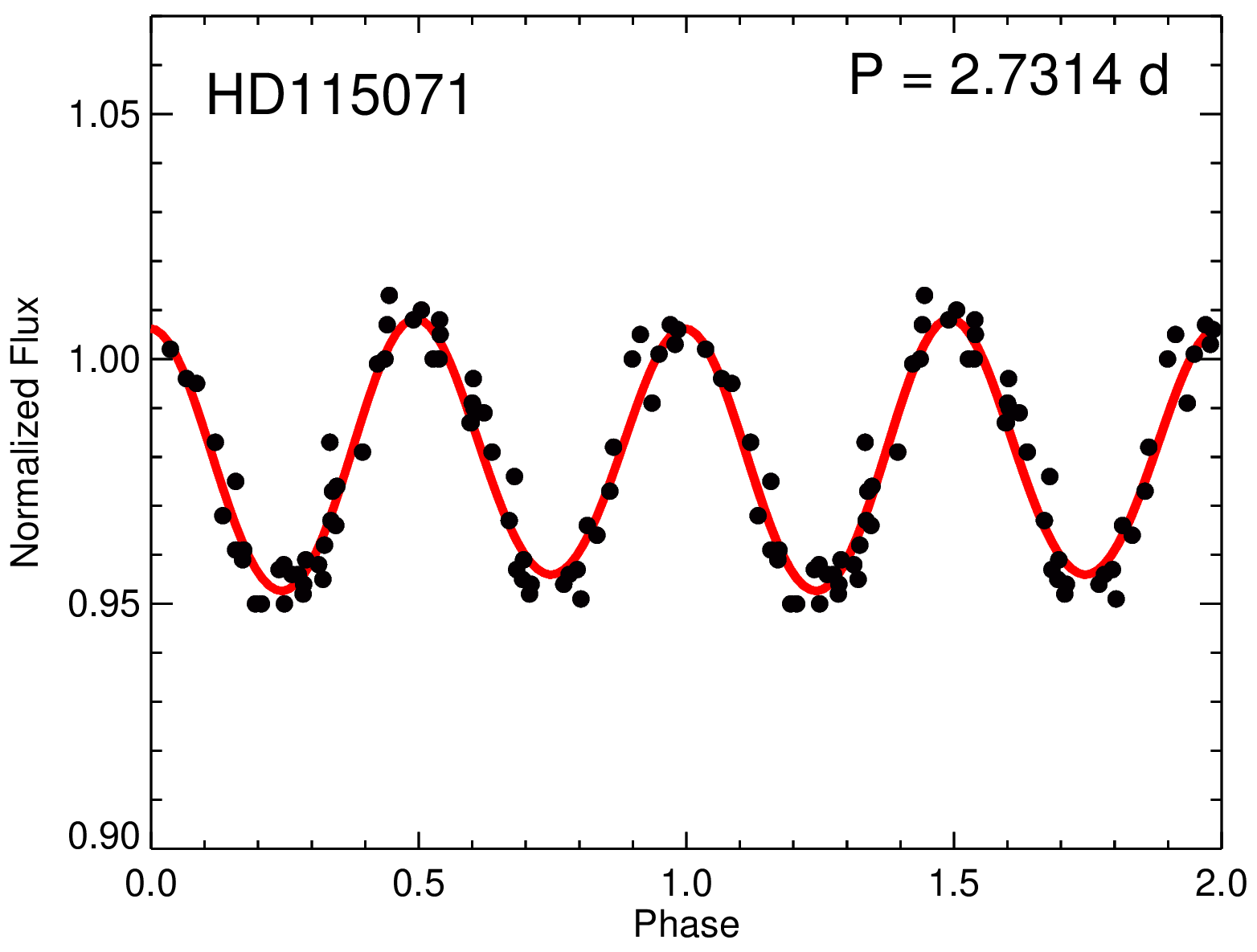}
  \includegraphics[width=0.67\columnwidth]{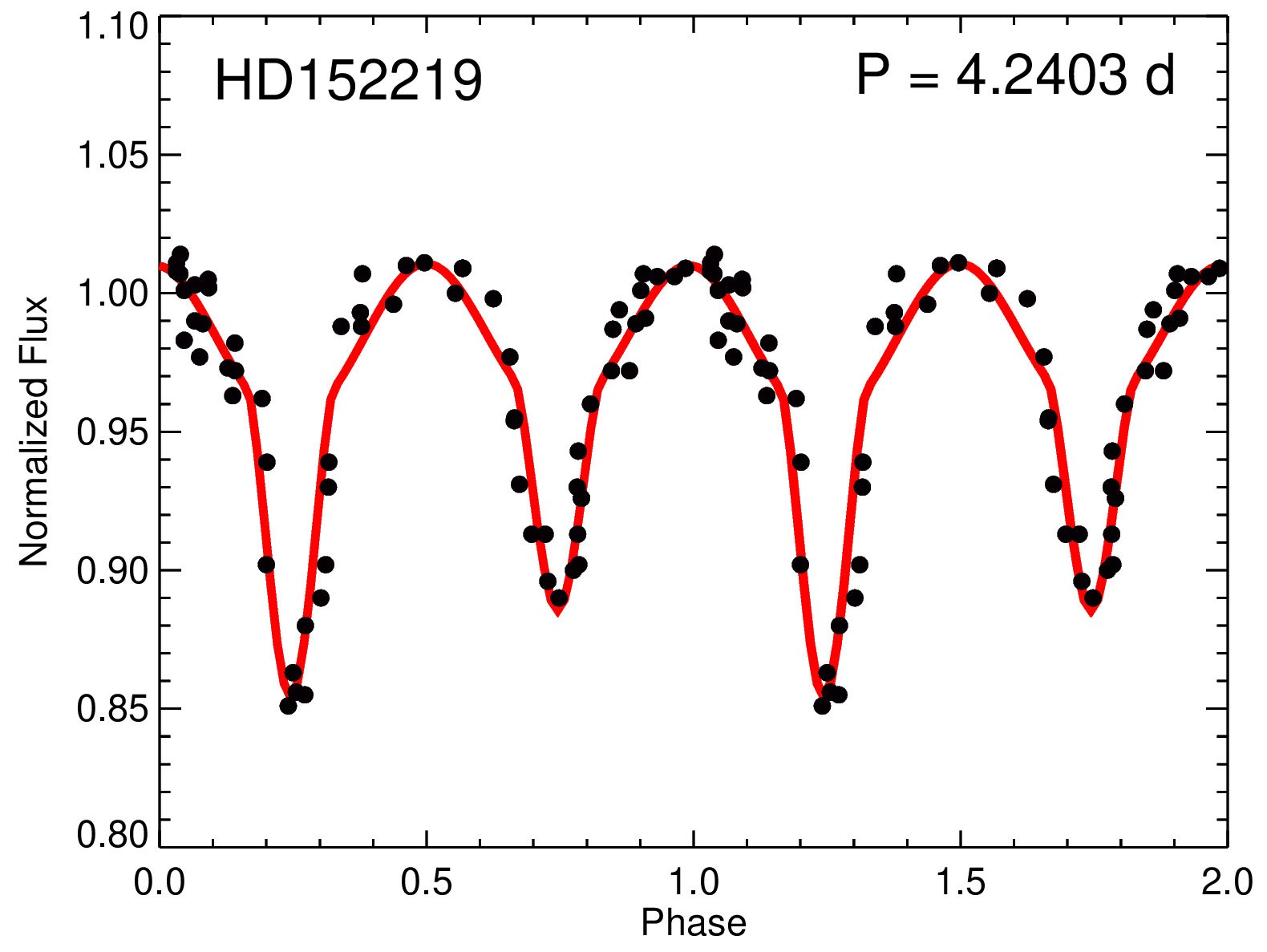}
  \includegraphics[width=0.67\columnwidth]{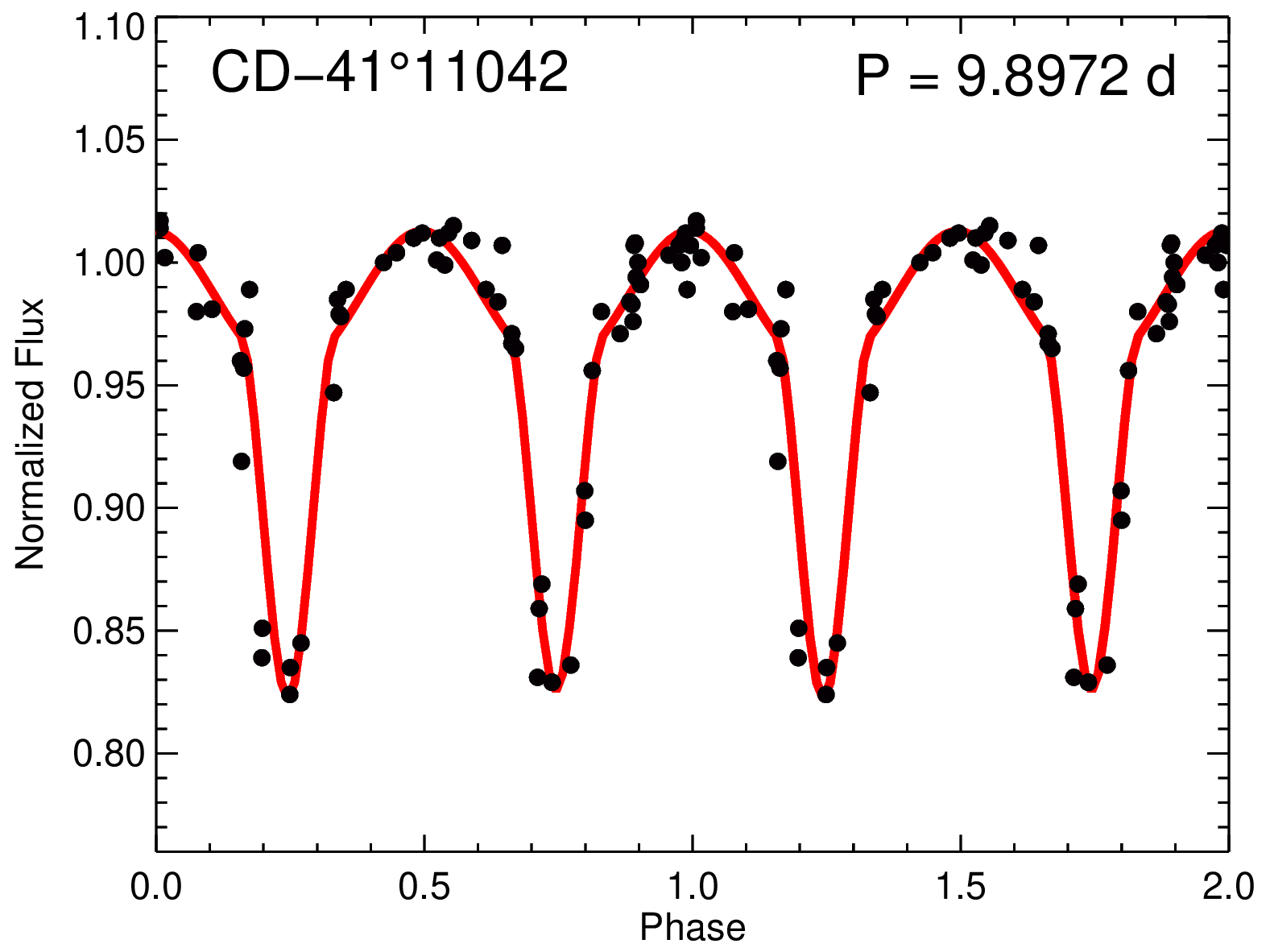}
  \includegraphics[width=0.67\columnwidth]{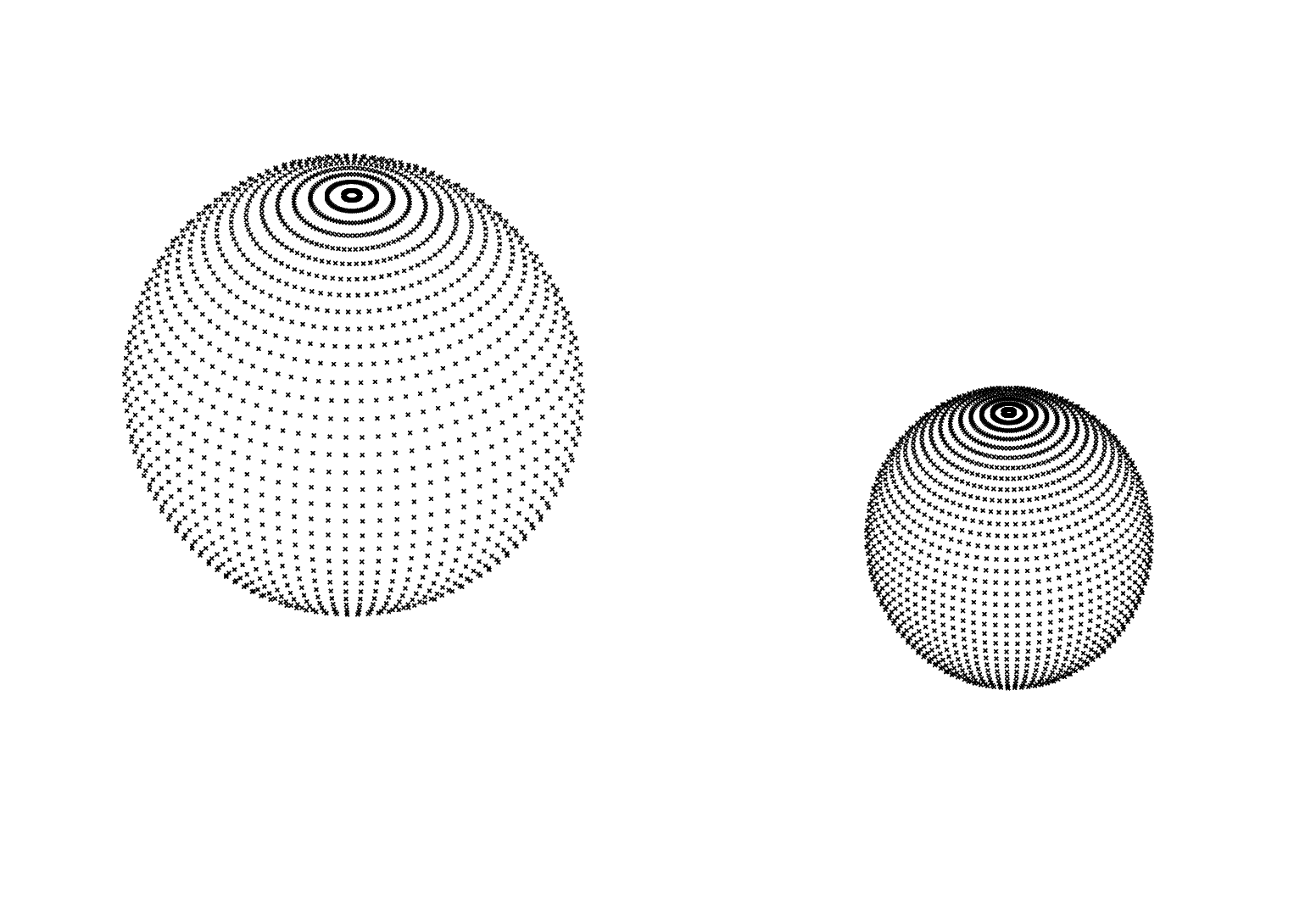}
  \includegraphics[width=0.67\columnwidth]{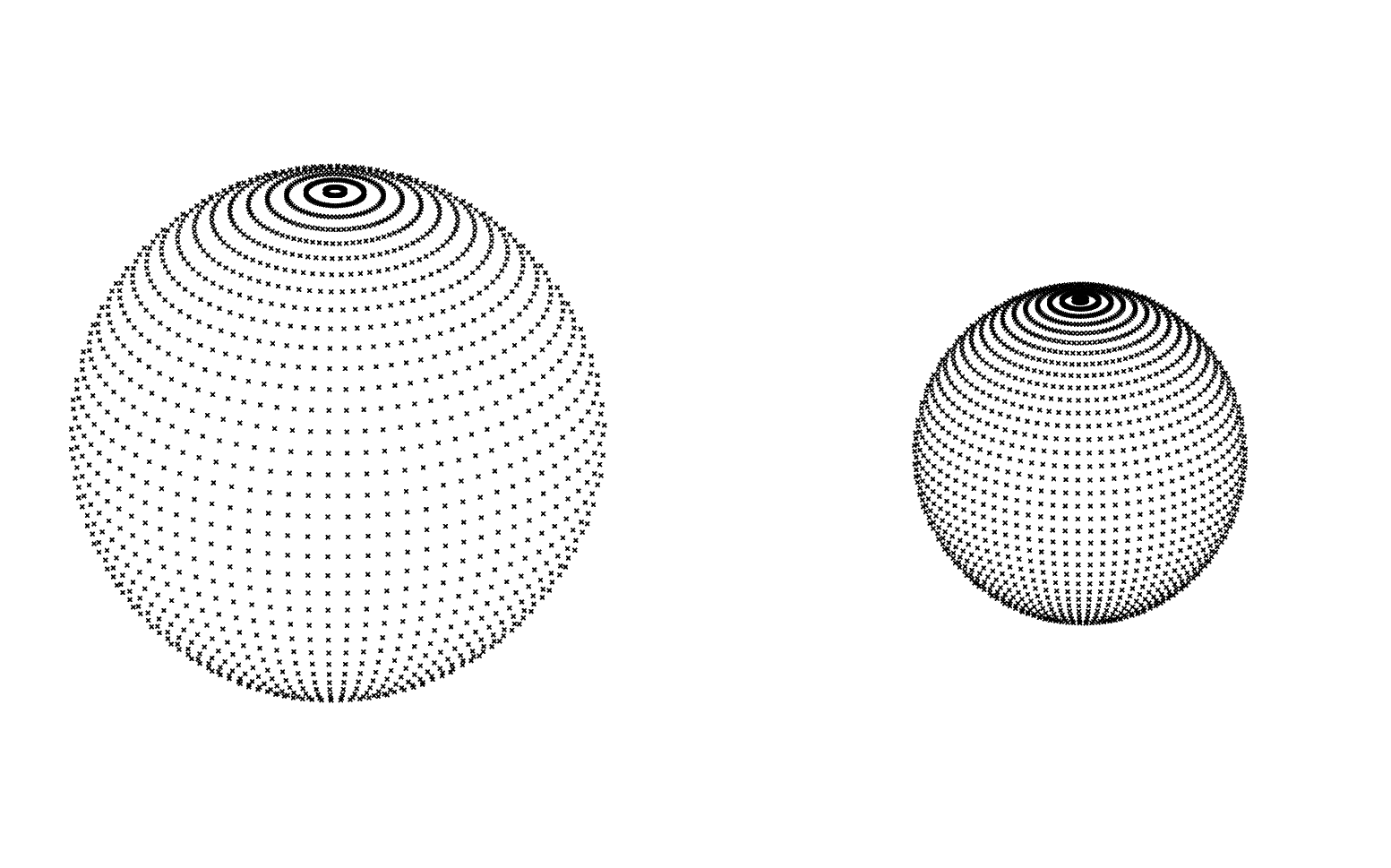}
  \includegraphics[width=0.67\columnwidth]{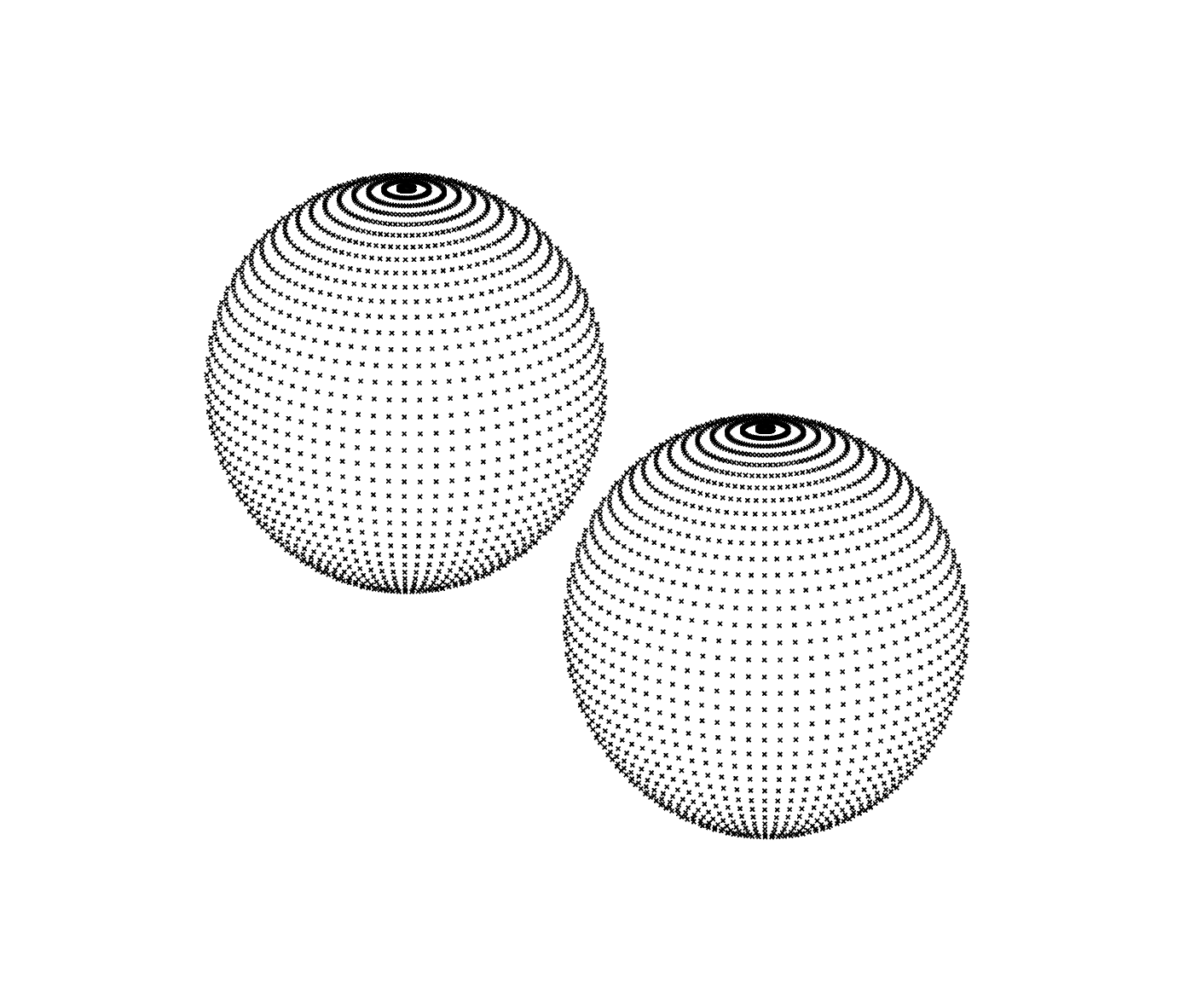}
  \caption{Same as Figure A1 but light curves and models for only one filter.}
  \label{model_ape}
\end{figure*}

\section{Light curves for periodic systems without modelling}

\begin{figure*}
  \centering
  \includegraphics[width=15cm,clip=true]{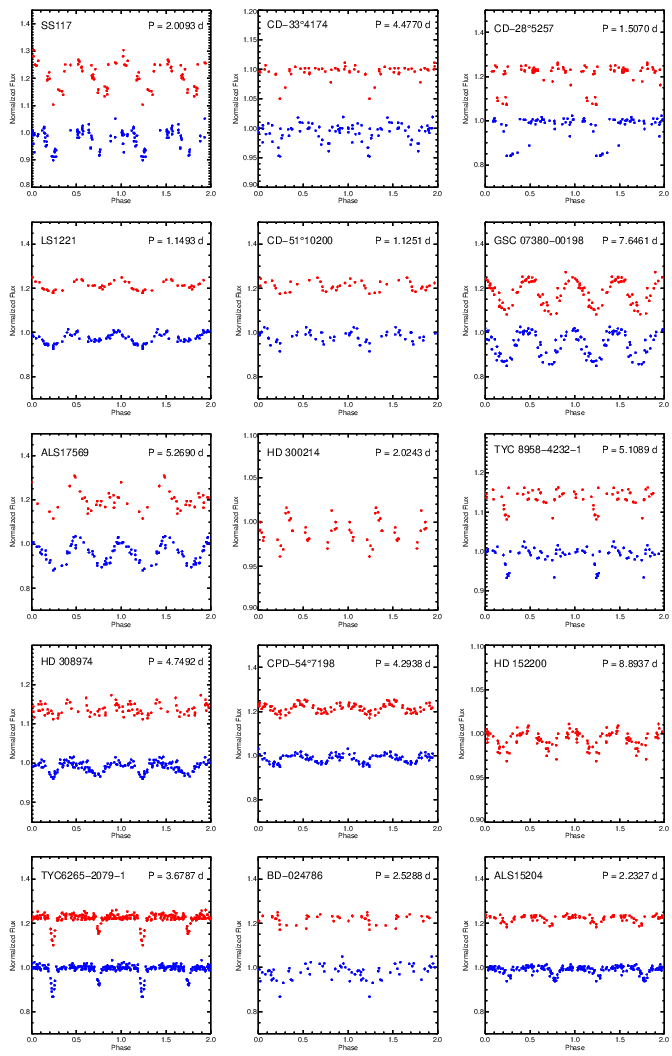}
  \caption{Light curves in $r$ (blue) and $i$ (red) for periodic systems without modelling.
  }
  \label{phase_lc}
\end{figure*}

\section{Light curves for additional GOSC stars}

\begin{figure*}
  \centering
  \includegraphics[width=15cm,clip=true]{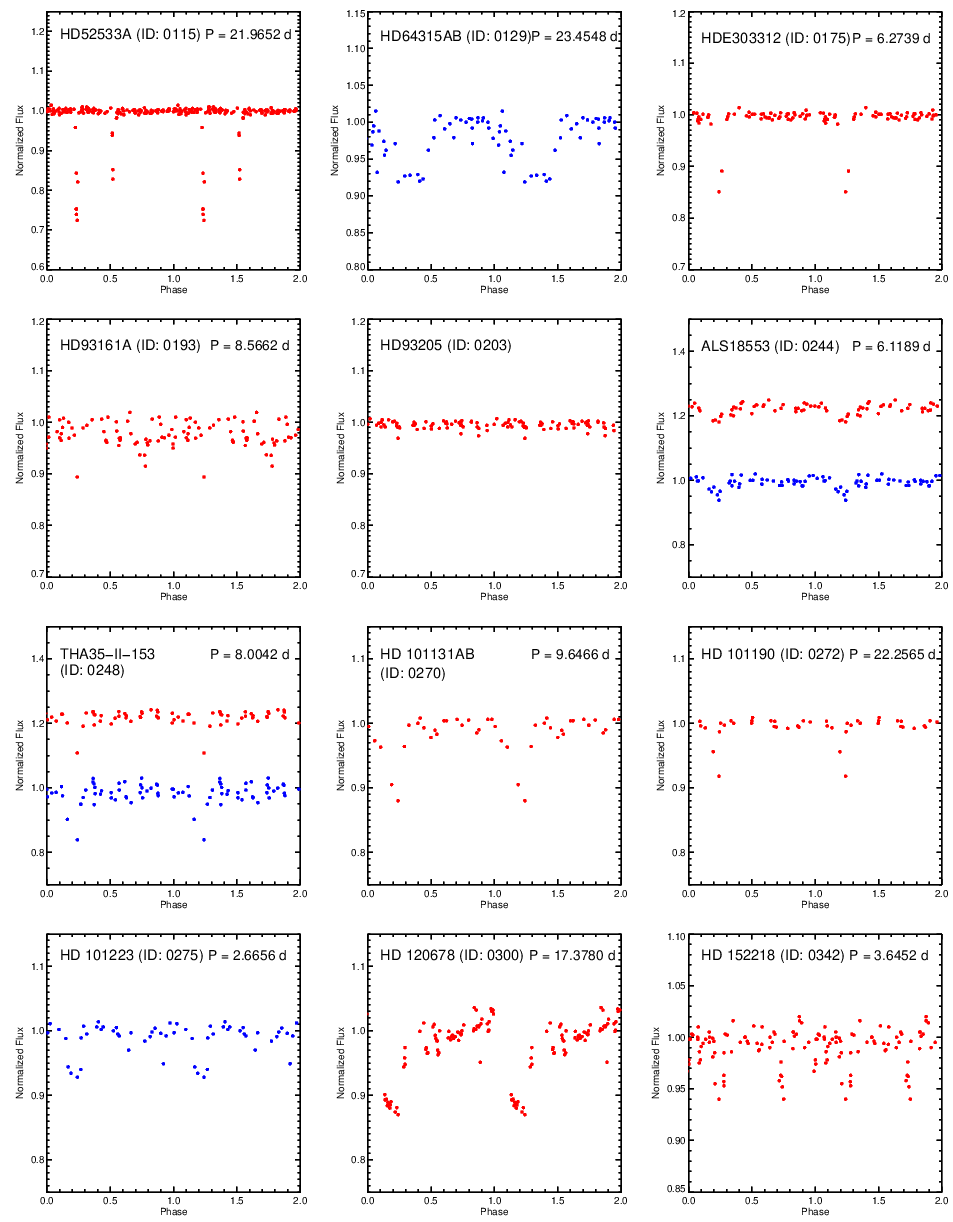}
  \caption{Light curves in $r$ (blue) and $i$ (red) for objects in the GOSC and with doubtful EB nature.
  }
  \label{phase_lc_GOSC}
\end{figure*}

\begin{landscape}
\begin{table}
\begin{center}
\caption{Photometric observations of high-mass eclipsing binaries.}
\label{tab:photometryg}
\begin{tabular}{clccccccccccc}
\hline\hline
\noalign{\smallskip}
             \multicolumn{1}{c}{No.}
           & \multicolumn{1}{c}{Name}
           & \multicolumn{2}{c}{No. of epochs}
           & \multicolumn{1}{c}{$U$}
           & \multicolumn{1}{c}{$B$}
           & \multicolumn{1}{c}{$V$}
           & \multicolumn{1}{c}{$r$}
           & \multicolumn{1}{c}{$i$}
           & \multicolumn{1}{c}{$\delta_{r}$}
           & \multicolumn{1}{c}{$\delta_{i}$}
           & \multicolumn{1}{c}{Spec. Type}
           & \multicolumn{1}{c}{$UBV$ references}\\
             \multicolumn{1}{c}{}
           & \multicolumn{1}{c}{}
           & \multicolumn{1}{c}{$r$}
           & \multicolumn{1}{c}{$i$}
           & \multicolumn{1}{c}{mag}
           & \multicolumn{1}{c}{mag}
           & \multicolumn{1}{c}{mag}
           & \multicolumn{1}{c}{mag}
           & \multicolumn{1}{c}{mag}
           & \multicolumn{1}{c}{mag}
           & \multicolumn{1}{c}{mag}
           & \multicolumn{1}{c}{(from $UBV$)}
           & \multicolumn{1}{c}{}\\
             \hline
             \noalign{\smallskip}
1 & SS\,117				     &  60 &  44 &   --            & $12.30\pm0.03$  & $12.02\pm0.06$ & $11.85\pm0.011$ & $11.79\pm0.012$ & 0.17 & 0.22 & -	& \cite{2013AJ....145...44Z}\\
2 & CPD\,$-$\,24\degr\,2836  & 166 & 144 & $11.17\pm0.016$ & $11.40\pm0.008$ & $11.28\pm0.005$ & $11.25\pm0.011$ & $11.35\pm0.009$ & 0.10 & 0.10 & B7 & this work   \\
3 & CD\,$-$\,33\degr\,4174  &  48 &  31 & $12.41\pm0.013$ & $12.40\pm0.013$ & $11.28\pm0.010$ & $10.86\pm0.009$ & $10.36\pm0.006$ & 0.07 & 0.07 & - & \cite{1992AJ....104..590O}       \\
4 & CPD\,$-$\,26\degr\,2656  & 161 & 127 & $10.55\pm0.010$ & $11.18\pm0.007$ & $10.95\pm0.005$ & $10.91\pm0.010$ & $10.94\pm0.008$ & 0.42 & 0.42 & B1 & this work   \\
5 & TYC\,6561-1765-1         & 161 & 127 & $12.13\pm0.017$ & $12.52\pm0.009$ & $12.30\pm0.008s$ & $12.19\pm0.011$ & $12.26\pm0.015$ & 0.11 & 0.12 & B3 & this work    \\
6 & CD\,$-$\,28\degr\,5257	 &  45 &  44 & $10.90\pm0.028$ & $11.59\pm0.022$ & $11.34\pm0.013$ & $11.33\pm0.011$ & $11.31\pm0.010$ & 0.21 & 0.22 & - & Havlen (1972) \\
7 & CD\,$-$\,31\degr\,5524   &  49 &  50 & $11.08\pm0.020$ & $11.67\pm0.013$ & $11.40\pm0.010$ & $11.35\pm0.010$ & $11.32\pm0.011$ & 0.12 & 0.15 & B0 & \cite{2005AJ....130.1652R}  \\
8 & V\,467\,Vel $^*$               & 31  & 18  & $10.60\pm0.016$ & $11.26\pm0.011$ & $10.88\pm0.014$ & $10.83\pm0.008$ & $10.71\pm0.007$ & 0.21 & 0.20 & O6   & \cite{1991ApJS...76.1033D}  \\
9 & CPD\,$-$\,42\degr\,2880  & 173 & 122 & $11.53\pm0.017$ & $11.80\pm0.012$ & $11.10\pm0.011$ & $10.83\pm0.010$ & $10.62\pm0.008$ & 0.12 & 0.10 & B0 & Denoyelle (1977)  \\
10 & CPD\,$-$\,45\degr\,3253  & 130 &  75 & $12.31\pm0.028$ & $11.97\pm0.006$ & $10.88\pm0.005$ & $10.25\pm0.010$ & $9.68\pm0.007$  & 0.11 & 0.10 & B3 & this work   \\
11 & GSC\,08156-03066         &  53 &  27 & $13.32\pm0.031$ & $13.18\pm0.012$ & $11.93\pm0.008$ & $11.34\pm0.009$ & $10.69\pm0.007$ & 0.12 & 0.11 & O9.5 & this work  \\
12 & ALS\,17569				 &  53 &  27 & $14.96\pm0.005$ & $14.86\pm0.010$ & $14.08\pm0.019$ & $13.83\pm0.022$ & $13.51\pm0.027$ & 0.17 & 0.21 & - & \cite{1979AJ.....84..639M}       \\
13 & LS\,1221				 &  53 &  27 & $12.75\pm0.005$ & $12.80\pm0.010$ & $11.85\pm0.019$ & $11.48\pm0.009$ & $11.04\pm0.008$ & 0.10 & 0.08 & - & \cite{1979AJ.....84..639M}        \\
14 & GSC\,08173-00182         & 140 &  74 & $15.24\pm0.034$ & $14.40\pm0.021$ & $12.79\pm0.008$ & $11.99\pm0.009$ & $11.15\pm0.009$ & 0.15 & 0.14 & B5   & this work   \\
15 & TYC\,8175-685-1          & 126 &  73 & $13.55\pm0.017$ & $13.23\pm0.015$ & $12.10\pm0.011$ & $11.72\pm0.010$ & $11.21\pm0.010$ & 0.15 & 0.16 & B3   & \cite{1977AJ.....82..474M}\\
16 & HD\,300214              &  -- &  29 &  $9.98\pm0.017$ &  $9.75\pm0.012$ &  $8.66\pm0.011$ & -- & $7.77\pm0.006$ & -- & 0.06 & -    & Denoyelle (1977)     \\
17 & HD\,92607 $^*$   			  & --  & 52  & $7.36\pm0.013$  & $8.23\pm0.009$  & $8.23\pm0.012$ & -- & $8.36\pm0.007$ & -- & 0.17 & O9   & Forte (1976)          \\
18 & ALS\,15204              & 109 & 52  & $10.93$ & $11.49$ & $10.96\pm0.010$ & $10.20\pm0.009$ & $9.86\pm0.009$  & 0.08 & 0.07 & - & \cite{2012AJ....143...41H} \\
19 & CPD\,$-$\,59\degr\,2603 $^*$   & 109 & 52  & $8.17\pm0.020$  & $8.96\pm0.010$  & $8.82\pm0.010$ & $8.83\pm0.014$ & $8.89\pm0.007$ & 0.42 & 0.40 & O9   & \cite{1993AJ....105..980M}       \\
20 & CPD\,$-$\,59\degr\,2628 $^*$   & 109 & 52  & 8.78  & 9.63  & 9.52 & $9.56\pm0.009$ & $9.54\pm0.008$ & 0.57 & 0.48 & O9   & \cite{1963PASP...75..492F} \\
21 & CPD\,$-$\,59\degr\,2635 $^*$  & 109 & 52  & 8.31  & 9.08  & 9.31 & $9.20\pm0.011$ & $9.17\pm0.008$ & 0.37 & 0.33 & O8   & \cite{1963PASP...75..492F}           \\
22 & V662\,Car $^*$ 				  & 109 & 52  & 12.74 & 12.70 & 12.10 & $11.70\pm0.010$ & $11.24\pm0.011$ & 0.50 & 0.48 & B6   & Antokhin et al. (2008)            \\
23 & ALS18551                &  40 &  35 & $13.60\pm0.032$ & $13.76\pm0.014$ & $12.93\pm0.009$ & $10.23\pm0.009$ & $9.85\pm0.010$  & 0.10 & 0.09 & B0 & this work \\
24 & TYC\,8958-4232-1        &  40 &  35 &   --  & $11.56\pm0.09$ & $11.04\pm0.09$ & $10.23\pm0.009$ & $9.85\pm0.010$ & 0.10 & 0.09 & - & H{\o}g et al. (2000)    \\
25 & EM\,Car $^*$				  & --  & 52  & 8.06  & 8.73  & 8.42  & -- & $8.36\pm0.009$ & -- & 0.48 & O8	& \cite{1969MNRAS.143..273F}  \\
26 & HD\,308974				 &  85 &  47 &   --  & $11.44\pm0.07$ & $11.11\pm0.08$ & $11.54\pm0.010$ & $11.73\pm0.015$ & 0.06 & 0.06 & - & H{\o}g et al. (2000)    \\
27 & HD\,115071 $^*$               & --  & 67  & $7.47\pm0.015$  & $8.20\pm0.016$  & $7.95\pm0.024$  & -- & $7.98\pm0.011$ & -- & 0.10 & O9.5   & Westerlund \& Garnier (1989)	   \\
28 & CPD\,$-$\,54\degr\,7198 &  63 &  58 & $10.56\pm0.021$ & $10.99\pm0.010$ & $10.64\pm0.010$ & $10.62\pm0.010$ & $10.63\pm0.010$ & 0.09 & 0.09 & B2   & this work            \\
29 & CD\,$-$\,51\degr\,10200 &  64 &  55 & $12.21\pm0.010$ & $12.39\pm0.007$ & $11.99\pm0.007$ & $10.76\pm0.008$ & $10.38\pm0.008$ & 0.12 & 0.08 & - & \cite{1987MNRAS.229..227F}	   \\
30 & HD\,152200              &  -- &  66 &  $7.74\pm0.066$ &  $8.49\pm0.016$ & $8.38\pm0.010$ & -- & $8.53\pm0.007$ & -- & 0.09 & - & \cite{1998AJ....115..734S}    \\
31 & HD\,152219 $^*$				  & --  & 66  & $6.95\pm0.046$  & $7.70\pm0.035$  & $7.54\pm0.013$  & -- & $7.74\pm0.009$ & -- & 0.21 & O9.5   & Baume et al. (1999)       \\
32 & CD\,$-$\,41\degr\,11042 $^*$  & --  & 66  & $7.73\pm0.062$  & $8.42\pm0.018$  & $8.23\pm0.013$  & -- & $8.32\pm0.007$ & -- & 0.37 & B1   & \cite{1998AJ....115..734S}    \\
33 & CPD\,$-$\,39\degr\,7292 &  71 & 64  & $11.59\pm0.017$ & $11.88\pm0.012$ & $11.01\pm0.005$ & $10.68\pm0.008$ & $10.32\pm0.008$ & 0.20 & 0.19 & O5 & this work  \\
34 & TYC\,7370-460-1 $^*$	      & 48  & 41  & --    & 13.24 & 11.79 & $11.17\pm0.008$ & $10.36\pm0.007$ & 0.55 & 0.54 & --  & H{\o}g et al. (2000)    \\
35 & HDE\,323110 $^*$              & 68  & 58  & $10.95\pm0.021$ & $10.83\pm0.013$ & $9.73\pm0.020$  & $9.25\pm0.010$ & $8.74\pm0.008$ & 0.46 & 0.46 & B1  & Dachs et al. (1982)        \\
36 & Pismis\,24-4             & 237 & 189 & $15.89$ & $15.36$ & $13.93$ & $13.21\pm0.015$ & $12.42\pm0.017$ & 0.22 & 0.22 & B3 & Moffat \& Vogt (1973) \\
37 & Pismis\,24-13 $^*$            & 237 & 189 & 14.60 & 14.29 & 12.84 & $12.01\pm0.011$ & $11.18\pm0.009$ & 0.13 & 0.13 & O9.5  & Neckel (1984)   \\
38 & GSC\,07380-00198		 &  64 &  61 & $14.52$ & $13.76$ & $12.12$ & $11.23\pm0.008$ & $10.35\pm0.008$ & 0.19 & 0.21 & - & \cite{1977ApJ...215..106M} \\
39 & CD\,$-$\,29\degr\,14032 & 183 & 159 & $11.30\pm0.027$ & $11.69\pm0.007$ & $11.05\pm0.007$ & $10.95\pm0.010$ & $10.76\pm0.009$ & 0.18 & 0.16 & O9 & this work  \\
40 & TYC\,6265-2079-1		 & 147 & 133 & $11.30\pm0.019$ & $11.66\pm0.006$ & $10.96\pm0.004$ & $10.68\pm0.009$ & $10.33\pm0.009$ & 0.19 & 0.18 & -& \cite{1996AJ....112.2855R}  \\
41 & BD\,$-$\,024786         & 124 & 106 & $12.16\pm0.037$ & $12.39\pm0.008$ & $11.57\pm0.006$ & $11.31\pm0.009$ & $10.89\pm0.009$ & 0.17 & 0.18 & O9   & this work  \\
\hline
\end{tabular}
\end{center}
Notes: Columns (10) and (11) correspond to the variability amplitude ({$\delta$}) for filters $r$ and $i$. The reported photometry correspond to the mean non-eclipsed magnitude. $^*$ stars from the Galactic O-Star Catalogue (\citealt{2016ApJS..224....4M}) identified as EBs in the present work.
\end{table}
\end{landscape}

\begin{landscape}
\begin{table}
\begin{center}
\caption{GOSC photometric observations of stars with doubtful eclipsing binary nature.}
\label{tab:photometrygosc}
\begin{tabular}{clccccccccccc}
\hline\hline
\noalign{\smallskip}
             \multicolumn{1}{c}{No.}
           & \multicolumn{1}{c}{Name}
           & \multicolumn{2}{c}{No. of epochs}
           & \multicolumn{1}{c}{$U$}
           & \multicolumn{1}{c}{$B$}
           & \multicolumn{1}{c}{$V$}
           & \multicolumn{1}{c}{$r$}
           & \multicolumn{1}{c}{$i$}
           & \multicolumn{1}{c}{$\delta_{r}$}
           & \multicolumn{1}{c}{$\delta_{i}$}
           & \multicolumn{1}{c}{Spec. Type}
           & \multicolumn{1}{c}{$UBV$ references}\\
             \multicolumn{1}{c}{}
           & \multicolumn{1}{c}{}
           & \multicolumn{1}{c}{$r$}
           & \multicolumn{1}{c}{$i$}
           & \multicolumn{1}{c}{mag}
           & \multicolumn{1}{c}{mag}
           & \multicolumn{1}{c}{mag}
           & \multicolumn{1}{c}{mag}
           & \multicolumn{1}{c}{mag}
           & \multicolumn{1}{c}{mag}
           & \multicolumn{1}{c}{mag}
           & \multicolumn{1}{c}{(from $UBV$)}
           & \multicolumn{1}{c}{}\\
             \hline
             \noalign{\smallskip}
1 & HD\,52533\,A              & --  & 132 & $6.65\pm0.010$  & $7.61\pm0.006$  & $7.70\pm0.005$  & -- & $8.03\pm0.006$ & -- & 0.36 & O9   & \cite{1983AJ.....88..439L}      \\
2 & HD\,64315\,AB             & 37  & 27  & 8.67  & 9.46  & 9.25  & $9.18\pm0.008$ & $9.17\pm0.005$ & 0.11 & 0.11 & O5   & Feinstein \& Vazquez (1989)  \\
%3 & V\,467\,Vel               & 31  & 18  & $10.60\pm0.016$ & $11.26\pm0.011$ & $10.88\pm0.014$ & $10.83\pm0.008$ & $10.71\pm0.007$ & 0.21 & 0.20 & O6   & \cite{1991ApJS...76.1033D}  \\
%4 & HD\,92607   			  & --  & 52  & $7.36\pm0.013$  & $8.23\pm0.009$  & $8.23\pm0.012$ & -- & $8.36\pm0.007$ & -- & 0.17 & O9   & Forte (1976)          \\
3 & HDE\,303312  			  & 109 & --  & 9.63  & 10.29 & 9.97 & $9.88\pm0.008$ & $9.82\pm0.009$ & 0.21 & 0.19 & O9   & Vazquez et al. (1996)     \\
4 & HD\,93161\,A   			  & --  & 52  & 8.01  & 8.76  & 8.56 & -- & $7.76\pm0.009$ & -- & 0.14 & O9   & Vazquez et al. (1996)            \\
5 & HD\,93205         		  & --  & 52  & $6.90\pm0.020$  & $7.84\pm0.010$  & $7.76\pm0.010$ & -- & $7.89\pm0.009$ & -- & 0.04 & O3 & \cite{1993AJ....105..980M}            \\
%8 & CPD\,$-$\,59\degr\,2603   & 109 & 52  & $8.17\pm0.020$  & $8.96\pm0.010$  & $8.82\pm0.010$ & $8.83\pm0.014$ & $8.89\pm0.007$ & 0.42 & 0.40 & O9   & \cite{1993AJ....105..980M}       \\
%9 & CPD\,$-$\,59\degr\,2628   & 109 & 52  & 8.78  & 9.63  & 9.52 & $9.56\pm0.009$ & $9.54\pm0.008$ & 0.57 & 0.48 & O9   & \cite{1963PASP...75..492F} \\
%10 & CPD\,$-$\,59\degr\,2635  & 109 & 52  & 8.31  & 9.08  & 9.31 & $9.20\pm0.011$ & $9.17\pm0.008$ & 0.37 & 0.33 & O8   & \cite{1963PASP...75..492F}           \\
%11 & V662\,Car 				  & 109 & 52  & 12.74 & 12.70 & 12.10 & $11.70\pm0.010$ & $11.24\pm0.011$ & 0.50 & 0.48 & B6   & Antokhin et al. (2008)            \\
6 & ALS\,18553               & 40  & 35  & $14.02\pm0.036$ & $14.04\pm0.024$ & $12.83\pm0.016$ & $12.27\pm0.011$ & $11.61\pm0.014$ & 0.09 & 0.07 & O5   & Wramdemark (1976)\\
7 & THA\,35$-$\,II$-$\,153   & 40  & 35  & --    & $13.57\pm0.05$ & $13.30\pm0.08$ & $12.03\pm0.010$ & $11.20\pm0.012$ & 0.22 & 0.15 & --   & \cite{2018yCat.1345....0G}  \\
%14 & EM\,Car				  & --  & 52  & 8.06  & 8.73  & 8.42  & -- & $8.36\pm0.009$ & -- & 0.48 & O8	& \cite{1969MNRAS.143..273F}  \\
8 & HD\,101131\,AB           & --  & 22  & $6.28\pm0.010$  & $7.16\pm0.009$  & $7.14\pm0.005$  & -- & $7.38\pm0.015$ & -- & 0.15 & O5   & Ardeberg \& Maurice (1977)       \\
9 & HD\,101190	              & --  & 22  & 6.46  & 7.31  & 7.27  & -- & $7.53\pm0.014$ & -- & 0.10 & O6	& \cite{1983ApJS...51..321S}        \\
10 & HD\,101223				  & 29  & --  & $8.10\pm0.010$  & $8.86\pm0.009$  & $8.70\pm0.005$  & $8.72\pm0.015$ & $8.76\pm0.009$ & 0.09 & 0.04 & O9   & Ardeberg \& Maurice (1977)        \\
%18 & HD\,115071               & --  & 67  & $7.47\pm0.015$  & $8.20\pm0.016$  & $7.95\pm0.024$  & -- & $7.98\pm0.011$ & -- & 0.10 & O9.5   & Westerlund \& Garnier (1989)	   \\
11 & HD\,120678		          & --  & 69  & $7.29\pm0.016$  & $8.20\pm0.011$  & $8.14\pm0.014$  & -- & $7.71\pm0.011$ & -- & 0.18 & O5   & \cite{1991ApJS...76.1033D} \\
%20 & HD\,152219				  & --  & 66  & $6.95\pm0.046$  & $7.70\pm0.035$  & $7.54\pm0.013$  & -- & $7.74\pm0.009$ & -- & 0.21 & O9.5   & Baume et al. (1999)       \\
12 & HD\,152218               & --  & 66  & 7.03  & 7.76  & 7.59  & -- & $7.70\pm0.010$ & -- & 0.09 & O9 & Heske \& Wendker(1984)     \\
%22 & CD\,$-$\,41\degr\,11042  & --  & 66  & $7.73\pm0.062$  & $8.42\pm0.018$  & $8.23\pm0.013$  & -- & $8.32\pm0.007$ & -- & 0.37 & B1   & \cite{1998AJ....115..734S}    \\
%23 & TYC\,7370-460-1	      & 48  & 41  & --    & 13.24 & 11.79 & $11.17\pm0.008$ & $10.36\pm0.007$ & 0.55 & 0.54 & --  & H{\o}g et al. (2000)    \\
%24 & HDE\,323110              & 68  & 58  & $10.95\pm0.021$ & $10.83\pm0.013$ & $9.73\pm0.020$  & $9.25\pm0.010$ & $8.74\pm0.008$ & 0.46 & 0.46 & B1  & Dachs et al. (1982)        \\
%25 & Pismis\,24-13            & 237 & 189 & 14.60 & 14.29 & 12.84 & $12.01\pm0.011$ & $11.18\pm0.009$ & 0.13 & 0.13 & O9.5  & Neckel (1984)   \\
\hline
\end{tabular}
\end{center}
Notes: Columns (10) and (11) correspond to the variability amplitude ({$\delta$}) for filters $r$ and $i$. The reported photometry correspond to the mean non-eclipsed magnitude.
\end{table}
\end{landscape}

%%%%%%%%%%%%%%%%%%%%%%%%%%%%%%%%%%%%%%%%%%%%%%%%%%
% Don't change these lines
\bsp	% typesetting comment
\label{lastpage}
\end{document}